\magnification \magstep1
\raggedbottom
\openup 2\jot
\voffset6truemm
\hsize=15truecm
\vsize=23truecm
\def\II{{\rm1\!\hskip-1pt I}}
\def\cstok#1{\leavevmode\thinspace\hbox{\vrule\vtop{\vbox{\hrule\kern1pt
\hbox{\vphantom{\tt/}\thinspace{\tt#1}\thinspace}}
\kern1pt\hrule}\vrule}\thinspace}
\centerline {\bf DIRAC OPERATOR AND SPECTRAL GEOMETRY}
\hskip8cm{\ hep-th/9704016}
\vskip 1cm
\centerline {\it Giampiero Esposito}
\vskip 1cm
\noindent
Istituto Nazionale di Fisica Nucleare, Sezione di Napoli,
Mostra d'Oltremare Padiglione 20, 80125 Napoli, Italy;
\vskip 0.3cm
\noindent
Universit\`a degli Studi di Napoli Federico II,
Dipartimento di Scienze Fisiche, Mostra d'Oltremare
Padiglione 19, 80125 Napoli, Italy.
\vskip 1cm
\noindent
{\bf Abstract.} This paper is devoted to mathematical and 
physical properties of the Dirac operator and spectral
geometry. Spin-structures in Lorentzian and
Riemannian manifolds, and the global theory of 
the Dirac operator, are first analyzed. Elliptic 
boundary-value problems, index problems for closed
manifolds and for manifolds with boundary, Bott
periodicity and K-theory are then presented. This makes
it clear why the Dirac operator is the most fundamental,
in the theory of elliptic operators on manifolds. The
topic of spectral geometry is developed by studying
non-local boundary conditions of the Atiyah-Patodi-Singer
type, and heat-kernel asymptotics for operators of Laplace
type on manifolds with boundary. The emphasis is put on
the functorial method, which studies the behaviour of
differential operators, boundary operators and heat-kernel
coefficients under conformal rescalings of the background
metric. In the second part, a number of relevant physical
applications are studied: non-local boundary conditions for
massless spin-1/2 fields, massless spin-3/2 potentials on
manifolds with boundary, geometric theory of massive
spin-3/2 potentials, local boundary conditions in quantum
supergravity, quark boundary conditions, one-loop quantum
cosmology, conformally covariant operators and Euclidean
quantum gravity.
\vskip 100cm
\centerline {\it CHAPTER ONE}
\vskip 1cm
\centerline {\bf DIRAC OPERATOR IN PHYSICS AND MATHEMATICS}
\vskip 1cm
\noindent
{\bf Abstract.} This introductory chapter begins with a
review of Dirac's original derivation of the relativistic
wave equation for the electron. The emphasis in then put
on spin-structures in Lorentzian and Riemannian manifolds,
and on the global theory of the Dirac operator,
to prepare the ground for the topics
studied in the following chapters.
\vskip 100cm
\centerline {\bf 1.1 The Dirac equation}
\vskip 1cm
In the development of theoretical physics, the Dirac operator
finds its natural place in the attempt to obtain a
relativistic wave equation for the electron. Although we
shall be mainly concerned with elliptic operators in the
rest of our book, we thus devote this section to Dirac's
original argument (Dirac 1928, Dirac 1958).  
Indeed, it is well known that the 
relativistic Hamiltonian of a particle of mass $m$ is
$$
H=c \sqrt{m^{2}c^{2}+p_{1}^{2}+p_{2}^{2}+p_{3}^{2}}
\; \; \; \; ,
\eqno (1.1.1)
$$
and this leads to the wave equation
$$
\left[p_{0}-\sqrt{m^{2}c^{2}+p_{1}^{2}
+p_{2}^{2}+p_{3}^{2}} \right]\psi=0
\; \; \; \; ,
\eqno (1.1.2)
$$
where the $p_{\mu}$ should be regarded as operators:
$$
p_{\mu} \equiv i {\hbar} {\partial \over \partial x^{\mu}}
\; \; \; \; .
\eqno (1.1.3)
$$
Thus, multiplication of Eq. (1.1.2) on the left by the
operator
$$
p_{0}+\sqrt{m^{2}c^{2}+p_{1}^{2}+p_{2}^{2}+p_{3}^{2}}
$$
leads to the equation
$$
\Bigr[p_{0}^{2}-m^{2}c^{2}-p_{1}^{2}-p_{2}^{2}
-p_{3}^{2} \Bigr]\psi=0
\; \; \; \; ,
\eqno (1.1.4)
$$
which, being relativistically invariant, is a more
appropriate starting point for a relativistic theory. Of 
course, equations (1.1.2) and (1.1.4) are not completely
equivalent: every solution of Eq. (1.1.2) is also, by
construction, a solution of Eq. (1.1.4), whereas the
converse does not hold. Only the solutions of Eq. (1.1.4)
corresponding to positive values of $p_{0}$ are also
solutions of Eq. (1.1.2).

However, Eq. (1.1.4), being quadratic in $p_{0}$, is not
of the form desirable in quantum mechanics, where, since
the derivation of Schr\"{o}dinger's equation
$$
i{\hbar}{\partial \psi \over \partial t}=H \psi
\; \; \; \; ,
\eqno (1.1.5)
$$
one is familiar with the need to obtain wave equations
which are linear in $p_{0}$. To combine relativistic
invariance with linearity in $p_{0}$, and to obtain 
equivalence with Eq. (1.1.4), one looks for a wave 
equation which is rational and linear in $p_{0},p_{1},
p_{2},p_{3}$:
$$
\Bigr[p_{0}-\alpha_{1}p_{1}-\alpha_{2}p_{2}
-\alpha_{3}p_{3}-\beta \Bigr]\psi=0
\; \; \; \; ,
\eqno (1.1.6)
$$
where $\alpha$ and $\beta$ are independent of $p$. Since
we are studying the (idealized) case when the electron moves
in the absence of electromagnetic field, all space-time
points are equivalent, and hence the operator in square
brackets in Eq. (1.1.6) is independent of $x$. This implies
in turn that $\alpha$ and $\beta$ are independent of $x$,
and commute with $p$ and $x$ operators. At a deeper level,
$\alpha$ and $\beta$ make it possible to obtain a relativistic
description of the spin of the electron, i.e. an angular 
momentum which does not result from the translational motion
of the electron. 

At this stage, one can multiply Eq. (1.1.6) on the left by
the operator
$$
p_{0}+\alpha_{1}p_{1}+\alpha_{2}p_{2}
+\alpha_{3}p_{3}+\beta \; \; \; \; .
$$
This leads to the equation 
$$
\eqalignno{
\; & \biggr[p_{0}^{2}-\sum_{i=1}^{3}\alpha_{i}^{2}p_{i}^{2}
-\sum_{i \not = j}(\alpha_{i} \; \alpha_{j}
+\alpha_{j} \; \alpha_{i})p_{i}p_{j} \cr
&-\sum_{i=1}^{3}(\alpha_{i} \; \beta
+\beta \; \alpha_{i})p_{i}
-\beta^{2}\biggr] \psi =0 \; \; \; \; .
&(1.1.7)\cr}
$$
The equations (1.1.4) and (1.1.7) agree if $\alpha_{i}$ 
and $\beta$ satisfy the conditions
$$
\alpha_{i}^{2}=1 \; \; \forall i =1,2,3
\; \; \; \; ,
\eqno (1.1.8)
$$
$$
\alpha_{i} \; \alpha_{j}+\alpha_{j} \; \alpha_{i}=0 \; \; 
{\rm if} \; \; i \not = j
\; \; \; \; ,
\eqno (1.1.9)
$$
$$
\beta^{2}=m^{2}c^{2}
\; \; \; \; ,
\eqno (1.1.10)
$$
$$
\alpha_{i} \; \beta + \beta \; \alpha_{i}=0 \; \; 
\forall i=1,2,3
\; \; \; \; .
\eqno (1.1.11)
$$
Thus, on setting
$$
\beta=\alpha_{0} mc
\; \; \; \; ,
\eqno (1.1.12)
$$
it is possible to re-express the properties (1.1.8)--(1.1.11)
by the single equation
$$
\alpha_{a} \; \alpha_{b}+\alpha_{b} \; \alpha_{a}
=2 \delta_{ab} \; \; 
\forall a,b=0,1,2,3
\; \; \; \; .
\eqno (1.1.13)
$$
So far, we have found that, if Eq. (1.1.13) holds, the
wave equation (1.1.6) is equivalent to Eq. (1.1.4). Thus,
one can {\it assume} that Eq. (1.1.6) is the correct
relativistic wave equation for the motion of an electron in
the absence of a field. However, Eq. (1.1.6) is not entirely
equivalent to Eq. (1.1.2), but, as Dirac first pointed out,
it allows solutions corresponding to negative as well as
positive values of $p_{0}$. The former are relevant for the
theory of positrons, and will not be discussed in our 
monograph, which focuses instead on mathematical problems
in the theory of elliptic operators.

To obtain a representation of four anticommuting $\alpha$'s,
as in Eq. (1.1.13), one has to consider $4 \times 4$ matrices.
Following Dirac, it is convenient to express the $\alpha$'s
in terms of generalized Pauli matrices (see below), 
here denoted by $\sigma_{1},\sigma_{2},\sigma_{3}$, and of 
a second set of anticommuting matrices, $\rho_{1},\rho_{2},
\rho_{3}$ say. Explicitly, one may take
$$
\alpha_{1}=\rho_{1} \; \sigma_{1}
\; \; \; \; ,
\eqno (1.1.14)
$$
$$
\alpha_{2}=\rho_{1} \; \sigma_{2}
\; \; \; \; ,
\eqno (1.1.15)
$$
$$
\alpha_{3}=\rho_{1} \; \sigma_{3}
\; \; \; \; ,
\eqno (1.1.16)
$$
$$
\alpha_{0}=\rho_{3}
\; \; \; \; ,
\eqno (1.1.17)
$$
where (Dirac 1958)
$$
\sigma_{1}=\pmatrix{0&1&0&0 \cr
1&0&0&0 \cr 
0&0&0&1 \cr
0&0&1&0 \cr}
\; \; \; \; ,
\eqno (1.1.18)
$$
$$
\sigma_{2}=\pmatrix{0&-i&0&0 \cr
i&0&0&0 \cr
0&0&0&-i \cr
0&0&i&0 \cr}
\; \; \; \; ,
\eqno (1.1.19)
$$
$$
\sigma_{3}=\pmatrix{1&0&0&0 \cr
0&-1&0&0 \cr
0&0&1&0 \cr
0&0&0&-1 \cr}
\; \; \; \; ,
\eqno (1.1.20)
$$
$$
\rho_{1}=\pmatrix{0&0&1&0 \cr
0&0&0&1 \cr
1&0&0&0 \cr
0&1&0&0 \cr}
\; \; \; \; ,
\eqno (1.1.21)
$$
$$
\rho_{2}=\pmatrix{0&0&-i&0 \cr
0&0&0&-i \cr
i&0&0&0 \cr
0&i&0&0 \cr}
\; \; \; \; ,
\eqno (1.1.22)
$$
$$
\rho_{3}=\pmatrix{1&0&0&0 \cr
0&1&0&0 \cr
0&0&-1&0 \cr
0&0&0&-1 \cr}
\; \; \; \; .
\eqno (1.1.23)
$$
Note that both the $\rho$'s and the $\sigma$'s are Hermitian,
and hence the $\alpha$'s are Hermitian as well. In this 
formalism for the electron, the wave function has four
components, and they all depend on the four $x$'s only. 
Unlike the non-relativistic formalism with spin, one has
two extra components which reflect the ability of Eq. (1.1.6)
to describe negative-energy states.

By virtue of (1.1.14)--(1.1.17), Eq. (1.1.6) may be
re-expressed as
$$
\Bigr[p_{0}-\rho_{1} {\vec \sigma} \cdot {\vec p}
-\rho_{3} mc \Bigr]\psi=0
\; \; \; \; .
\eqno (1.1.24)
$$
The generalization to the case when an external
electromagnetic field is present is not difficult. For this
purpose, it is enough to bear in mind that 
$$
p_{b}=m v_{b}+{e\over c}A_{b}
\; \; \; \; \forall b=0,1,2,3
\; \; \; \; ,
\eqno (1.1.25)
$$
where $A_{b}$ are such that $A_{b}dx^{b}$ is the connection
one-form (or ``potential") of the theory. By raising indices
with the metric of the background space-time, one obtains
the corresponding four-vector
$$
A^{b}=g^{bc} \; A_{c}
\; \; \; \; .
\eqno (1.1.26)
$$
The desired wave equation of the relativistic theory of
the electron in an external electromagnetic field 
turns out to be
$$
\left[p_{0}+{e\over c}A_{0}-\rho_{1} \Bigr({\vec \sigma},
{\vec p}+{e\over c}{\vec A} \Bigr)
-\rho_{3} mc \right] \psi=0
\; \; \; \; .
\eqno (1.1.27)
$$
With this notation, the wave function should be regarded as
a column vector with four rows, while its ``conjugate imaginary",
say ${\overline \psi}^{\dagger}$, is a row vector, 
i.e. a $1 \times 4$ matrix, and obeys the equation
$$
{\overline \psi}^{\dagger}\left[p_{0}+{e\over c}A_{0}
-\rho_{1} \Bigr({\vec \sigma},{\vec p}
+{e\over c}{\vec A} \Bigr)
-\rho_{3} mc \right]=0
\; \; \; \; ,
\eqno (1.1.28)
$$
where the momentum operators operate to the right.
\vskip 1cm
\centerline {\bf 1.2 Spinor Fields in Lorentzian Manifolds}
\vskip 1cm
A space-time $(M,g)$ consists, by definition, of a connected,
four-dimensional, Hausdorff $C^{\infty}$ manifold $M$, jointly
with a Lorentz metric $g$ on $M$, and a time orientation 
given by a globally defined timelike vector field 
$X: M \rightarrow TM$. Indeed, two Lorentzian metrics on $M$ 
are identified if they are related by the action of the
diffeomorphism group of $M$. One is thus dealing with equivalence
classes of pairs $(M,g)$ as a mathematical model of the space-time
manifold (Hawking and Ellis 1973). If a Levi-Civita connection, say
$\nabla$, is given on $M$, the corresponding Riemann curvature
tensor is defined by
$$
R(X,Y)Z \equiv \nabla_{X} \nabla_{Y}Z -\nabla_{Y}\nabla_{X}Z
-\nabla_{[X,Y]}Z
\; \; \; \; .
\eqno (1.2.1)
$$
A curved space-time is a space-time manifold whose curvature
tensor does not vanish, and represents, according to
Einstein's general relativity, a good mathematical model
of the world we live in (Hawking and Ellis 1973). We are
now aiming to give an elementary introduction to the theory 
of spinor fields in a curved space-time, as a first step
towards the theory of spinor fields in Riemannian 
manifolds (whose metric is positive-definite,
instead of being Lorentzian).

For this purpose, recall that a {\it vierbein} is a
collection of four oriented mutually orthogonal vectors
$e^{{\hat a}}$, ${\hat a}=0,1,2,3$, whose components
$e_{a}^{\; \; {\hat a}}$ are related to the metric tensor
$g_{ab}$ by
$$
g_{ab}(x)=e_{a}^{\; \; {\hat c}}(x) \;
e_{b}^{\; \; {\hat d}}(x) \; \eta_{{\hat c}{\hat d}}
\; \; \; \; ,
\eqno (1.2.2)
$$
where $\eta_{{\hat c}{\hat d}}={\rm diag}(1,-1,-1,-1)$
is the Minkowski metric. The relation (1.2.2) is clearly
preserved if the vierbein is replaced by its rotated
form (Isham 1978):
$$
e_{a}^{\; \; {\hat c}}(x) \rightarrow
e_{a}^{\; \; {\hat b}}(x) \; 
\Omega_{{\hat b}}^{\; \; {\hat c}}(x)
\; \; \; \; ,
\eqno (1.2.3)
$$
where $\Omega_{{\hat b}}^{\; \; {\hat c}}$ is an
{\it arbitrary} differentiable function on space-time,
taking values in the group $SO(3,1)$. By definition,
spinor fields in curved space-time are required to 
transform as Dirac spinors under this local gauge group.
Thus, one {\it assumes} that $SL(2,C)$-valued functions
exist, say $S$, such that
$$
\Lambda(S(x))=\Omega(x)
\; \; \; \; ,
\eqno (1.2.4)
$$
where $\Lambda$ is the two-to-one covering map of 
$SL(2,C)$ onto $SO(3,1)$. In the Dirac representation of
$SL(2,C)$, one has the matrix relation
$$
S^{-1} \; \gamma_{{\hat a}} \; S
=\gamma_{{\hat b}} \; \Omega_{{\hat a}}^{\; \; {\hat b}}
\; \; \; \; ,
\eqno (1.2.5)
$$
and the spinor fields transform as
$$
\psi(x) \rightarrow S(x) \; \psi(x)
\; \; \; \; .
\eqno (1.2.6)
$$
Covariant derivatives of spinor fields are defined according
to the rules
$$
\nabla_{a} \psi \equiv \Bigr(\partial_{a}+i B_{a} \Bigr)\psi
\; \; \; \; ,
\eqno (1.2.7)
$$
$$
\nabla_{a} {\overline \psi} \equiv 
\partial_{a}{\overline \psi}
-i {\overline \psi} B_{a}
\; \; \; \; ,
\eqno (1.2.8)
$$
where $B_{a}$, called the spin-connection, has the
transformation property
$$
B_{a} \rightarrow S B_{a} S^{-1} 
+i \Bigr(\partial_{a}S \Bigr) S^{-1}
\; \; \; \; ,
\eqno (1.2.9)
$$
and is defined by
$$
B_{a} \equiv {1\over 4}B_{a{\hat c}{\hat d}}
\; \Bigr[\gamma^{{\hat c}},\gamma^{{\hat d}}\Bigr]
\; \; \; \; .
\eqno (1.2.10)
$$
A suitable form of the components $B_{a{\hat c}{\hat d}}$
is obtained from Christoffel symbols according to the rule
$$
B_{a{\hat c}{\hat d}}=\Gamma_{\; \; ab}^{c}
\; e_{{\hat c}}^{\; \; b} \; e_{{\hat d}c}
+e_{{\hat d}b} \; e_{\; \; {\hat c},a}^{b}
\; \; \; \; .
\eqno (1.2.11)
$$
By virtue of (1.2.7)--(1.2.11), the gauge-invariant
Lagrangian for a (massive) Dirac field is
$$
{\cal L}=({\rm det}e)\left[{1\over 2}\Bigr({\overline \psi}
\; \gamma_{{\hat c}} \; \nabla_{a}\psi
-\Bigr(\nabla_{a}{\overline \psi}\Bigr)
\gamma_{{\hat c}} \; \psi \Bigr)e^{a{\hat c}}
-m {\overline \psi} \psi \right]
\; \; \; \; .
\eqno (1.2.12)
$$
Some remarks are now in order (Isham 1978).
\vskip 0.3cm
\noindent
(i) Equation (1.2.4) does not hold, in general. The possibility
to lift the arbitrary function $\Omega$ from $SO(3,1)$, which
is not simply connected, to its simply connected covering 
group $SL(2,C)$, depends on the global topological structure 
of $M$. In other words, a necessary and sufficient condition
for the existence of the $SL(2,C)$-valued function $S$ 
occurring in Eq. (1.2.4) is that (Spanier 1966)
$$
\Omega_{*} \; \pi_{1}(M)=0
\; \; \; \; ,
\eqno (1.2.13)
$$
where $\pi_{1}(M)$ is the first homotopy group of
space-time. The meaning of Eq. (1.2.13) is that the
$SO(3,1)$ vierbein rotation $\Omega$ can be covered by a
spinor gauge transformation if and only if {\it any} circle
in $M$ has an image in $SO(3,1)$ that can be contracted
to a point. Indeed, if
$$
\rho: \theta \in [0,2\pi] \rightarrow \rho(\theta)
\eqno (1.2.14)
$$
is a circle map in $M$ (i.e. $\rho(0)=\rho(2\pi)$), then
$$
\Omega: \theta \in [0,2\pi] \rightarrow
\Omega(\rho(\theta))
\eqno (1.2.15)
$$
is a circle map in $SO(3,1)$, and if $\rho$ belongs to the
equivalence class $[\rho] \in \pi_{1}(M)$ of homotopically
related maps, then 
$$
\Omega_{*} [\rho]
$$ 
denotes the homotopy
class, in $\pi_{1}(SO(3,1))$, of the map (1.2.15). Thus, if
$M$ is not simply connected, there may be functions $\Omega$ 
for which the condition (1.2.13) is not respected.
\vskip 0.3cm
\noindent
(ii) The spin-connections corresponding to non-liftable
vierbein rotations $\Omega$ are obtained by gauge rotations
similar, in form, to (1.2.9), but with $\Omega$ replacing $S$.
Thus, the gauge rotation can no longer be cancelled by a
compensating transformation of spinor fields. This leads
in turn to a number of possible Lagrangians, each of which
is $SL(2,C)$ gauge-invariant, where the derivative
terms in (1.2.12) are replaced by
$$
\nabla_{a}\psi \rightarrow
\Bigr(\partial_{a}+i \Omega B_{a} \Omega^{-1}
-\Omega \partial_{a} \Omega^{-1} \Bigr) \psi
\; \; \; \; ,
\eqno (1.2.16) 
$$
$$
\nabla_{a} {\overline \psi} \rightarrow
\Bigr(\partial_{a}-i \Omega B_{a} \Omega^{-1} 
+ \Omega \partial_{a} \Omega^{-1} \Bigr)
{\overline \psi}
\; \; \; \; ,
\eqno (1.2.17)
$$
and $e^{a{\hat c}}$ is replaced by 
$i \Omega_{\; \; {\hat d}}^{{\hat c}} \; e^{a{\hat d}}$,
with $\Omega$ ranging over the equivalence classes of
gauge functions. The number of different Lagrangians
turns out to coincide with the number of elements in the
cohomology group $H^{1}(M;Z_{2})$. This is quite relevant
for the path-integral approach to quantum gravity (cf.
section 6.1), where one might have to ``sum" over 
inequivalent vierbein gauge classes, as well as over
metric tensors.

A spinor structure in a Lorentzian four-manifold can
be defined as a principal $SL(2,C)$ bundle, say $E$,
jointly with a bundle map $f$ from $E$ onto the principal
$SO(3,1)$ bundle $\xi$ of oriented orthonormal frames
(the vierbeins of the literature on general relativity).
The map $f$ has to be compatible with both the $SL(2,C)$
and $SO(3,1)$ actions, in that
$$
f(pA)=f(p) \Lambda(A)
\; \; \; \; ,
\eqno (1.2.18)
$$
where $p$ is {\it any} point in the bundle space of $E$,
$A$ is any $SL(2,C)$ group element, and $pA$ is the 
spin-frame into which $p$ is taken under the action
of $A$. Moreover, $f(p) \Lambda(A)$ is the orthonormal
frame into which $f(p)$ is taken by the $SO(3,1)$ 
element $\Lambda(A)$. {\it Spinor fields} are then the
cross-sections of the vector bundle associated with the
Dirac representation of $SL(2,C)$ on $C^{4}$. If 
spin-structures do not exist, one can, however, resort
to the formalism of spin-$c$ structures (Avis and Isham 1980). 
\vskip 1cm
\centerline {\bf 1.3 Spin-Structures: Riemannian Case}
\vskip 1cm
In the course of defining spin-structures on Riemannian
manifolds, we want to give a precise formulation of the
following basic ideas. One is given an oriented 
Riemannian manifold, say $M$. The tangent bundle of
$M$, $TM$, has the rotation group $SO(n)$ as structural
group. If it is possible to replace $SO(n)$ by its 
two-fold covering group $Spin(n)$ (cf. appendix 1.A
for the case $n=3$), one then says that $M$ can be
given a spin-structure.

Following Milnor 1963, we start from a principal
bundle, say $\xi$, over a Riemannian manifold $B$,
with total space denoted by $E(\xi)$. The structural
group of $\xi$ is assumed to be $SO(n)$, where $n$
can be {\it any} positive integer, including
$n=\infty$. By definition, a {\it spin-structure}
on $\xi$ is a pair $(\eta,f)$ consisting of
\vskip 0.3cm
\noindent
(i) A principal bundle $\eta$ over $B$, with total
space $E(\eta)$ and structural group coinciding with
$Spin(n)$, i.e. the double covering of $SO(n)$;
\vskip 0.3cm
\noindent
(ii) A map $f: E(\eta) \rightarrow E(\xi)$ such that
$$
f \; \rho_{1}=\rho_{2} (f \times \lambda)
\; \; \; \; ,
\eqno (1.3.1)
$$
where $\lambda$ is the homomorphism from $Spin(n)$ to
$SO(n)$, $\rho_{1}$ is the right action of $Spin(n)$
and $\rho_{2}$ is the right action of $SO(n)$. In other
words, there exist two equivalent ways to reach the 
total space of the given principal bundle $\xi$ over 
$B$. First, one goes from $E(\eta) \times Spin(n)$ to
$E(\eta)$ via right translation, then using the projection
map from $E(\eta)$ to $E(\xi)$. However, one can also
use the projection map $f \times \lambda$ from
$E(\eta) \times Spin(n)$ to $E(\xi) \times SO(n)$.
At this stage, one takes advantage of right translation
to go from $E(\xi) \times SO(n)$ to the total space of $\xi$.
A naturally occurring question is how to identify (or
distinguish among) spin-structures. The answer is contained
in the definition according to which a second 
spin-structure on $\xi$, say $(\eta',f')$ should be identified
with $(\eta,f)$ if there exists an isomorphism $h$ from
$\eta '$ to $\eta$ such that $f \cdot h = f'$.

In the particular case when $n=2$, one deals with 
$Spin(2)$, the two-fold covering group of the circle 
$SO(2)$; when $n=1$, one deals with $Spin(1)$, a cyclic
group of order 2. Useful examples are given by the
tangent bundle of the two-sphere $S^{2}$. This has a
unique spin-structure $(\eta,f)$, where $E(\eta)$ is a
three-sphere. The tangent bundle of the circle $S^{1}$
has instead two distinct spin-structures.

If the reader is more familiar with (or prefers) 
cohomology classes, he may follow Hirsch and Milnor
by giving the following alternative definition:
\vskip 0.3cm
\noindent
A spin-structure on $\xi$ is a cohomology class 
$\sigma \in H^{1}\Bigr(E(\xi);Z_{2}\Bigr)$ (here, of
course, $Z_{2}$ denotes the integers modulo 2), whose
restriction to each fibre is a generator of the cyclic
group $H^{1}({\rm Fibre};Z_{2})$. 
\vskip 0.3cm
\noindent
In other words, the underlying idea is that each
cohomology class belonging to $H^{1}\Bigr(E(\xi);Z_{2}
\Bigr)$ determines a two-fold covering of $E(\xi)$. It
is precisely this two-fold covering space that should be
taken as the total space $E(\eta)$. The condition involving
the restriction to each fibre ensures that each fibre is
covered by a copy of $Spin(n)$, i.e. the unique two-fold
covering of $SO(n)$. It can be shown that the two definitions
are completely equivalent, and hence one is free to choose
among the two.

Remarkably, a necessary and sufficient condition for an
$SO(n)$ bundle to be endowed with a spin-structure is
that its second Stiefel-Whitney class, $w_{2}(\xi)$, 
should vanish. Further to this, if $w_{2}(\xi)=0$, the
number of distinct spin-structures on $\xi$ is equal to
the number of elements in $H^{1}(B;Z_{2})$. Indeed, if
$B$ is connected, this property follows from the exact
sequence
$$
0 \rightarrow H^{1}(B;Z_{2}) \rightarrow
H^{1}(E(\xi);Z_{2}) \rightarrow H^{1}(SO(n);Z_{2})
\rightarrow H^{2}(B;Z_{2})
\; \; \; \; ,
$$
which can be extracted, in turn, from the spectral sequence
of the fibration $\xi$ (Milnor 1963).

By definition, a {\it spin-manifold} consists of an
oriented Riemannian manifold $M$, jointly with a 
spin-structure on $TM$, the tangent bundle of $M$.
More explicitly, if $FM$ denotes the space of oriented 
orthonormal $n$-frames on $M$, one can think of the
spin-structure as being a cohomology class 
$\sigma \in H^{1}(FM;Z_{2})$, whose restriction to each
fibre is non-trivial (for all integer values of $n$
greater than 1). The resulting spin-manifold is then
denoted by $(M,\sigma)$. In particular, if $M$ is
simply connected, so that $\sigma$ is uniquely 
determined, one simply says that $M$ itself is a 
spin-manifold.

Dirac spinor fields on the $n$-dimensional manifold
$M^{n}$ are defined to be {\it cross-sections} of the
vector bundle $E^{\tau}$ over $M^{n}$ with typical
fibre ${\cal C}^{2[n/2]}$ and structure group 
$Spin(n)$, where $\tau$ is the fundamental spinor
representation of $Spin(n)$ (see section 1.4).

When one makes explicit calculations, it is
necessary to express a spinor field and its covariant
derivative with respect to a (local) orthonormal frame
$\left \{ {\bf e}_{a} \right \}_{a=1,...,n}$ on
$M^{n}$ in components. Indeed, a metric connection 
$\omega$ induces a spin-connection on $M^{n}$, i.e.
a connection on the spin-structure $Spin(M^{n})$, 
and a covariant derivative
for the cross-sections. Given a spinor, say $\psi$, the
components of the spinor $\nabla_{{\bf e}_{a}}\psi$ in a set
of linearly independent local cross-sections of $E^{\tau}$
(such a set is called a local spinor frame, and is denoted
by $\left \{ \theta_{A} \right \}$) are given by
$$
\Bigr(\nabla_{{\bf e}_{a}}\psi \Bigr)^{A}
={\bf e}_{a}(\psi^{A})
-{1\over 2}\omega_{abc}
\Bigr(\Sigma^{bc}\Bigr)_{\; \; B}^{A} \; \psi^{B}
\; \; \; \; ,
\eqno (1.3.2)
$$
where $\omega_{abc}$ are the connection coefficients in the
frame ${\bf e}_{a}$. As is well known, such coefficients
are defined by the condition
$$
\nabla_{{\bf e}_{a}}{\bf e}_{b}
=\omega_{abc}{\bf e}_{c}
\; \; \; \; .
\eqno (1.3.3)
$$
The matrices $\Sigma^{ab}$ are the generators of the
representation $\tau$ of $Spin(n)$. They can be
expressed by the equation (cf. (1.2.10))
$$
\Sigma^{ab}={1\over 4} \Bigr[\Gamma^{a},\Gamma^{b}\Bigr]
\; \; \; \; ,
\eqno (1.3.4)
$$
where $\Gamma^{a}$ are Dirac matrices satisfying
(cf. (1.1.13))
$$
\Gamma^{a} \; \Gamma^{b}+\Gamma^{b} \; \Gamma^{a}
=2 \delta^{ab} \II
\; \; \; \; .
\eqno (1.3.5)
$$
With this notation, $\II$ is the unit matrix, the indices
$a,b$ range from 1 through $n$, and the Dirac matrices
should be thought of as carrying yet further indices $A,B$:
$$
\Gamma^{a}=\Bigr(\Gamma^{a}\Bigr)_{\; \; B}^{A}
\; \; \; \; ,
$$
where $A$ and $B$ range from 1 through $2^{[n/2]}$.
The action of the Dirac operator on a spinor is then defined
by the equation
$$
\nabla \psi=\Gamma^{a} \nabla_{{\bf e}_{a}} \; \psi
\; \; \; \; .
\eqno (1.3.6)
$$
At this stage, it is appropriate to build a global theory
of the Dirac operator in the elliptic case. This is performed in
the following section (cf. section 2.4).
\vskip 1cm
\centerline {\bf 1.4 Global Theory of the Dirac Operator}
\vskip 1cm
In Riemannian four-geometries, the 
{\it total} Dirac operator may be defined
as a first-order elliptic operator mapping smooth sections of a
complex vector bundle to smooth sections of the same bundle.
Its action on the sections (i.e. the spinor fields) is given by
composition of Clifford multiplication 
with covariant differentiation.
To understand these statements, we first summarize the properties of
connections on complex vector bundles, and we then use the 
basic properties of spin-structures which enable one to understand how
to construct the vector bundle relevant for the theory of the Dirac
operator.

A complex vector bundle (Chern 1979) is a bundle whose fibres are
isomorphic to complex vector spaces. Denoting by $E$ the total
space, by $M$ the base space, one has the projection map
$\pi : E \rightarrow M$ and the sections $s: M \rightarrow E$
such that the composition of $\pi$ with $s$ yields the identity
on the base space: $\pi \cdot s = {\rm id}_{M}$. The sections $s$ 
represent the physical fields in our applications. Moreover, denoting
by $T$ and $T^{*}$ the tangent and cotangent bundles of $M$ respectively,
a connection $\nabla$ is a map from the space of smooth
sections of $E$, say $\Gamma(E)$ (also denoted by $C^{\infty}(M,E)$), 
to the space of smooth sections of the 
tensor-product bundle $\Gamma(T^{*} \otimes E)$:
$$
\nabla: \Gamma(E) \rightarrow \Gamma(T^{*} \otimes E)
\; \; \; \; ,
$$
such that the following properties hold:
$$
\nabla(s_{1}+s_{2})=\nabla s_{1}+\nabla s_{2}
\; \; \; \; ,
\eqno (1.4.1)
$$
$$
\nabla(fs)=df \otimes s + f \nabla s
\; \; \; \; ,
\eqno (1.4.2)
$$
where $s_{1},s_{2},s \in \Gamma(E)$ and $f$ is any $C^{\infty}$
function. The action of the connection $\nabla$ is expressed 
in terms of the connection matrix $\theta$ as
$$
\nabla s = \theta \otimes s
\; \; \; \; .
\eqno (1.4.3)
$$
If one takes a section $s'$ related to $s$ by
$$
s'=h \; s
\; \; \; \; ,
\eqno (1.4.4)
$$
in the light of (1.4.2)--(1.4.4) one finds by comparison that
$$
\theta' h=d \; h + h \; \theta
\; \; \; \; .
\eqno (1.4.5)
$$
Moreover, the transformation law of the curvature matrix
$$
\Omega \equiv d \theta -\theta \wedge \theta
\; \; \; \; ,
\eqno (1.4.6)
$$
is found to be
$$
\Omega' = h \; \Omega \; h^{-1}
\; \; \; \; .
\eqno (1.4.7)
$$

We can now describe in more detail the 
spin-structures and the corresponding
complex vector bundle acted upon by the total Dirac operator.
Let $X$ be a compact oriented differentiable $n$-dimensional
manifold (without boundary) on which a Riemannian metric
is introduced. Let $Q$ be the principal tangential
$SO(n)$-bundle of $X$. As we know from section 1.3,
a spin-structure of $X$ is a principal
$Spin(n)$-bundle $P$ over $X$ 
together with a covering map
${\widetilde \pi}: P \rightarrow Q$ such that the following
commutative structure exists. 
Given the Cartesian product 
$P \times Spin(n)$, one first reaches $P$ by the right 
action of $Spin(n)$ on $P$, and one finally arrives at
$Q$ by the projection map $\widetilde \pi$. This is equivalent
to first reaching the Cartesian product 
$Q \times SO(n)$ by the map ${\widetilde \pi} \times \rho$, and
finally arriving at $Q$ by the right action of $SO(n)$ on $Q$.
Of course, by $\rho$ we denote the double covering 
$Spin(n) \rightarrow SO(n)$.
In other words, $P$ and $Q$ as above are principal fibre bundles
over $X$, and one has a commutative diagram with 
$P \times Spin(n)$ and $P$ on the top,
and $Q \times SO(n)$ and $Q$ on the bottom. The projection map
from $P \times Spin(n)$ to $Q \times SO(n)$ is
${\widetilde \pi} \times \rho$, and the projection map from
$P$ to $Q$ is $\widetilde \pi$. Horizontal arrows should be drawn
to denote the right action of $Spin(n)$ on $P$
on the top, and of $SO(n)$ on $Q$ on the bottom. 
This coincides with the picture resulting from Eq. (1.3.1),
apart from the different notation used therein.

The group $Spin(n)$ has a complex representation space 
$\Sigma$ of dimension $2^{n}$ called the spin-representation. 
If $G \in Spin(n), x \in {\Re}^{n}, u \in \Sigma$, one has
therefore
$$
G(xu)=GxG^{-1} \cdot G(u)=\rho(G) x \cdot G(u)
\; \; \; \; ,
\eqno (1.4.8)
$$
where $\rho : Spin(n) \rightarrow SO(n)$ is the covering 
map as we said before. If $X$ is even-dimensional, i.e. $n=2l$,
the representation is the direct sum of two irreducible 
representations $\Sigma^{\pm}$ of dimension $2^{n-1}$. 
If $X$ is a $Spin(2l)$ manifold with principal bundle $P$,
one can form the associated complex vector bundles
$$
E^{+} \equiv P \times \Sigma^{+}
\; \; \; \; ,
\eqno (1.4.9)
$$
$$
E^{-} \equiv P \times \Sigma^{-}
\; \; \; \; ,
\eqno (1.4.10)
$$
$$
E \equiv E^{+} \oplus E^{-}
\; \; \; \; .
\eqno (1.4.11)
$$
Sections of these vector bundles are spinor fields on $X$.

The {\it total} Dirac operator is a first-order elliptic
differential operator $D: \Gamma(E) \rightarrow \Gamma(E)$
defined as follows. Recall first that the Riemannian metric
defines a natural $SO(2l)$ connection, and this may be used
to give a connection for $P$. One may therefore consider the
connection $\nabla$ at the beginning of this section, i.e. a
linear map from $\Gamma(E)$ to $\Gamma(T^{*} \otimes E)$.
On the other hand, the tangent and cotangent bundles of $X$ are
isomorphic, and one has the map from $\Gamma (T \otimes E)
\rightarrow \Gamma(E)$ induced by {\it Clifford multiplication}
(see section 2.1) 
The total Dirac operator $D$ is defined to be 
the {\it composition} of these two maps. Thus, in terms of an
orthonormal base $e_{i}$ of $T$, one has {\it locally}
$$
Ds = \sum_{i} e_{i} (\nabla_{i}s)
\; \; \; \; ,
\eqno (1.4.12)
$$
where $\nabla_{i}s$ is the covariant derivative of 
$s \in \Gamma(E)$ in the direction $e_{i}$, and $e_{i}( \; )$
denotes Clifford multiplication. Moreover, the total
Dirac operator $D$ induces two operators
$$
D^{+}: \Gamma(E^{+}) \rightarrow \Gamma(E^{-})
\; \; \; \; ,
\eqno (1.4.13)
$$
$$
D^{-}: \Gamma(E^{-}) \rightarrow \Gamma(E^{+})
\; \; \; \; ,
\eqno (1.4.14)
$$
each of which is elliptic. It should be emphasized that ellipticity
of the total and partial Dirac operators only holds in Riemannian
manifolds, whereas it does not apply to the Lorentzian manifolds
of general relativity and of the original Dirac's theory of
spin-${1\over 2}$ particles (cf. sections 1.1 and 1.2).
\vskip 10cm
\centerline {\bf 1.A Appendix}
\vskip 1cm
This appendix describes in detail the simplest example
of two-to-one homomorphism which is relevant for the theory of
spin and spin-structures. Our presentation
follows closely the one in Wigner 
1959, which relies in turn on a method 
suggested by H. Weyl. We begin with some elementary 
results in the theory of matrices, which turn out 
to be very useful for our purposes.

(i) A matrix which transforms every real vector into a real
vector is itself real, i.e. all its elements are real. If
this matrix is applied to the $k$th unit vector (which has
$k$th component $=1$, all others vanishing), the result is
the vector which forms the $k$th row of the matrix. Thus,
this row must be real. But this argument can be applied to
all $k$, and hence all the rows of the matrix must be real.

(ii) It is also well known that a matrix {\cal O} is complex
orthogonal if it preserves the scalar product of two
arbitrary vectors, i.e. if
$$
({\vec a},{\vec b})=({\cal O}{\vec a},{\cal O}{\vec b})
\; \; \; \; .
\eqno (1.A.1)
$$
An equivalent condition can be stated in terms of one
arbitrary vector: a matrix $\cal O$ is complex orthogonal
if the length of every single arbitrary vector, say
$\vec v$, is left unchanged under transformation by 
$\cal O$. Consider now two arbitrary vectors $\vec a$
and $\vec b$, and write ${\vec v}={\vec a}+{\vec b}$. Then
our condition for the complex orthogonality of $\cal O$ is
$$
({\vec v},{\vec v})=({\cal O}{\vec v},{\cal O}{\vec v})
\; \; \; \; .
\eqno (1.A.2)
$$
By virtue of the symmetry of the scalar product:
$({\vec a},{\vec b})=({\vec b},{\vec a})$, this yields
$$
\eqalignno{
\; & ({\vec a}+{\vec b},{\vec a}+{\vec b})
=({\vec a},{\vec a})+({\vec b},{\vec b})
+2({\vec a},{\vec b}) \cr
&=({\cal O}{\vec a},{\cal O}{\vec a})
+({\cal O}{\vec b},{\cal O}{\vec b})
+2({\cal O}{\vec a},{\cal O}{\vec b})
\; \; \; \; .
&(1.A.3)\cr}
$$
However, complex orthogonality also implies that
$({\vec a},{\vec a})=({\cal O}{\vec a},{\cal O}{\vec a})$ and
$({\vec b},{\vec b})=({\cal O}{\vec b},{\cal O}{\vec b})$. 
It then follows from (1.A.3) that
$$
({\vec a},{\vec b})=({\cal O}{\vec a},{\cal O}{\vec b})
\; \; \; \; ,
\eqno (1.A.4)
$$
which implies that $\cal O$ is complex orthogonal. It can
be shown, in a similar way, that $\cal U$ is unitary if only
$({\vec v},{\vec v})=({\cal U}{\vec v},{\cal U}{\vec v})$
holds for every vector.

By definition, a matrix which leaves each real vector real,
and leaves the length of every vector unchanged, is a
{\it rotation}. 
Indeed, when all lengths are equal in the
original and transformed figures, the angles also must be
equal; hence the transformation is merely a rotation.

(iii) We now want to determine the general form of a 
two-dimensional unitary matrix
$$
{\bf u}= \pmatrix{a&b \cr c&d \cr}
\eqno (1.A.5)
$$
of determinant $+1$ by considering the elements of the product
$$
{\bf u} {\bf u}^{\dagger}=\II
\; \; \; \; .
\eqno (1.A.6)
$$ 

Recall that the $\dagger$ operation means taking
the complex conjugate and then the transposed of the 
original matrix. Thus, the condition (1.A.6) implies that
$$
a^{*}c+b^{*}d=0
\; \; \; \; ,
\eqno (1.A.7)
$$
which leads to $c=-b^{*}d/a^{*}$. The insertion of this
result into the condition of unit determinant:
$$
ad-bc=1
\; \; \; \; ,
\eqno (1.A.8)
$$
yields $\Bigr(aa^{*}+bb^{*}\Bigr)d/a^{*}=1$. Moreover, 
since $aa^{*}+bb^{*}=1$ from (1.A.6), it follows that
$d=a^{*}$ and $c=-b^{*}$. The general two-dimensional unitary
matrix with unit determinant is hence
$$
{\bf u}=\pmatrix{a&b \cr -b^{*}& a^{*}\cr}
\; \; \; \; ,
\eqno (1.A.9)
$$
where, of course, we still have to require that
$aa^{*}+bb^{*}=1$. Note that, if one writes
$a=y_{0}+iy_{3}$ and $b=y_{1}+iy_{2}$, one finds
$$
{\rm det}{\bf u}=y_{0}^{2}+y_{1}^{2}+y_{2}^{2}
+y_{3}^{2}=1
\; \; \; \; .
$$
This is the equation of a unit three-sphere centred at the origin,
which means that $SU(2)$ has three-sphere topology and is hence
simply connected (the $n$-sphere is simply connected for all
$n > 1$). More precisely, $SU(2)$ is homeomorphic to
$S^{3} \subset {\Re}^{4}$. 

Consider now the so-called {\it Pauli matrices}:
$$
\sigma_{x}=\pmatrix{0&1 \cr 1&0 \cr}
\; \; \; \; ,
\eqno (1.A.10)
$$
$$
\sigma_{y}=\pmatrix{0&i \cr -i&0 \cr}
\; \; \; \; ,
\eqno (1.A.11)
$$
$$
\sigma_{z}=\pmatrix{-1&0 \cr 0&1 \cr}
\; \; \; \; .
\eqno (1.A.12)
$$
Every two-dimensional matrix with zero trace, say $\bf h$, can
be expressed as a linear combination of these matrices:
$$
{\bf h}=x\sigma_{x}+y\sigma_{y}+z\sigma_{z}=(r,\sigma)
\; \; \; \; .
\eqno (1.A.13)
$$
Explicitly, one has 
$$
{\bf h}=\pmatrix{-z & x+iy \cr x-iy & z \cr}
\; \; \; \; .
\eqno (1.A.14)
$$
In particular, if $x,y$, and $z$ are real, then $\bf h$ is
Hermitian.  

If one transforms $\bf h$ by an arbitrary unitary matrix 
$\bf u$ with unit determinant, one again obtains a matrix
with zero trace, $\overline {\bf h}={\bf u} {\bf h}
{\bf u}^{\dagger}$. Thus, $\overline {\bf h}$ can also be
written as a linear combination of 
$\sigma_{x},\sigma_{y},\sigma_{z}$:
$$
{\overline {\bf h}}={\bf u}{\bf h}{\bf u}^{\dagger}
={\bf u}(r,\sigma){\bf u}^{\dagger}
=x'\sigma_{x}+y'\sigma_{y}+z' \sigma_{z}=(r',\sigma)
\; \; \; \; ,
\eqno (1.A.15)
$$
$$
\pmatrix{a&b \cr -b^{*}&a^{*} \cr}
\pmatrix{-z& x+iy \cr x-iy & z \cr}
\pmatrix{a^{*}& -b \cr b^{*} & a \cr}
= \pmatrix{-z' & x'+iy' \cr x'-iy' & z' \cr}
\; \; \; \; .
\eqno (1.A.16)
$$
Equation (1.A.16) determines $x',y',z'$ as linear
functions of $x,y,z$. The transformation $R_{u}$ which
carries $r=(x,y,z)$ into $R_{u}r=r'=(x',y',z')$ can be
found from Eq. (1.A.16). It is
$$
\eqalignno{
\; & x'={1\over 2}\Bigr(a^{2}+{a^{*}}^{2}-b^{2}
-{b^{*}}^{2}\Bigr)x 
+{i\over 2}\Bigr(a^{2}-{a^{*}}^{2}+b^{2}-{b^{*}}^{2}
\Bigr)y \cr
&+\Bigr(a^{*}b^{*}+ab \Bigr)z
\; \; \; \; ,
&(1.A.17)\cr}
$$
$$
\eqalignno{
\; & y'={i\over 2}\Bigr({a^{*}}^{2}-a^{2}+b^{2}
-{b^{*}}^{2}\Bigr)x
+{1\over 2}\Bigr(a^{2}+{a^{*}}^{2}+b^{2}
+{b^{*}}^{2}\Bigr)y \cr
&+i \Bigr(a^{*}b^{*}-ab \Bigr)z
\; \; \; \; ,
&(1.A.18)\cr}
$$
$$
z'=-(a^{*}b+ab^{*})x+i(a^{*}b-ab^{*})y
+(aa^{*}-bb^{*})z
\; \; \; \; .
\eqno (1.A.19)
$$
The particular form of the matrix $R_{u}$ does not matter; it
is important only that
$$
x'^{2}+y'^{2}+z'^{2}=x^{2}+y^{2}+z^{2}
\; \; \; \; ,
\eqno (1.A.20)
$$
since the determinants of $\overline {\bf h}$ and 
$\bf h$ are equal ($\bf u$ being an element of $SU(2)$). 
According to the analysis in (ii), this
implies that the transformation $R_{u}$ must be complex
orthogonal. Such a property can also be seen directly from
(1.A.17)--(1.A.19).

Moreover, $\overline {\bf h}$ is Hermitian if $\bf h$ is.
In other words, $r'=(x',y',z')$ is real if $r=(x,y,z)$ is
real. This implies, by virtue of (i), that $R_{u}$ is
pure real, as can also be seen directly from 
(1.A.17)--(1.A.19). Thus, $R_{u}$ is a rotation: every 
two-dimensional unitary matrix $\bf u$ of unit determinant
corresponds to a three-dimensional rotation $R_{u}$; the
correspondence is given by (1.A.15) or (1.A.16).

It should be stressed that the determinant of $R_{u}$ is
$+1$, since as $\bf u$ is changed continuously into a unit
matrix, $R_{u}$ goes continuously into the three-dimensional
unit matrix. If its determinant were $-1$ at the beginning
of this process, it would have to make the jump to $+1$.
This is impossible, since the function ``det'' is continuous,
and hence the matrices with negative determinant cannot be
connected to the identity of the group. 
As a corollary of these properties, $R_{u}$ is a pure rotation 
for all $\bf u$.

The above correspondence is such that the product 
$\bf {qu}$ of two unitary matrices $\bf q$ and $\bf u$
corresponds to the product $R_{qu}=R_{q} \cdot R_{u}$ of
the corresponding rotations. According to (1.A.15), applied
to $\bf q$ instead of $\bf u$,
$$
{\bf q}(r,\sigma){\bf q}^{\dagger}=\Bigr(R_{q}r,\sigma\Bigr)
\; \; \; \; ,
\eqno (1.A.21)
$$
and upon transformation with $\bf u$ this yields
$$
{\bf uq}(r,\sigma){\bf q}^{\dagger}{\bf u}^{\dagger}
={\bf u}(R_{q}r,\sigma){\bf u}^{\dagger}
=(R_{u}R_{q}r,\sigma)=(R_{uq}r,\sigma)
\; \; \; \; ,
\eqno (1.A.22)
$$
using (1.A.15) again, with $R_{q}r$ replacing $r$ and
${\bf uq}$ replacing $\bf u$. Thus, a homomorphism exists
between the group of two-dimensional unitary matrices of
determinant +1 (the ``unitary group") and three-dimensional
rotations; the correspondence is given by (1.A.15) or
(1.A.17)--(1.A.19). Recall, by the way, that a homomorphism
of two groups, say $G_{1}$ and $G_{2}$, is a map 
$\phi: G_{1} \rightarrow G_{2}$ such that
$$
\phi(g_{1} \cdot g_{2})=\phi(g_{1}) \; \phi(g_{2})
\; \; \; \; ,
$$
$$
\phi \Bigr({\rm e}_{G_{1}}\Bigr)
=\phi \Bigr({\rm e}_{G_{2}}\Bigr)
\; \; \; \; ,
$$
$$
\phi(g^{-1})=(\phi(g))^{-1}
\; \; \; \; .
$$
However, we note that so far we have
not shown that the homomorphism exists between the 
two-dimensional unitary group and the {\it whole} pure rotation
group. That would imply that $R_{u}$ ranges over all
rotations as $\bf u$ covers the entire unitary group. This
will be proved shortly. It should also be noticed that the
homomorphism {\it is not an isomorphism}, since more than
one unitary matrix corresponds to the same rotation
(see below).

We first assume that $\bf u$ is a diagonal matrix, say
${\bf u}_{1}(\alpha)$ (i.e. we set $b=0$, and, for 
convenience, we write $a=e^{-{i\over 2}\alpha}$). Then
$\mid a^{2} \mid =1$ and $\alpha$ is real:
$$
{\bf u}_{1}(\alpha)=\pmatrix{e^{-{i\over 2}\alpha} & 0 \cr
0 & e^{{i\over 2}\alpha} \cr}
\; \; \; \; .
\eqno (1.A.23)
$$
From (1.A.17)--(1.A.19) one can see that the corresponding
rotation:
$$
R_{u_{1}}=\pmatrix{\cos \alpha & \sin \alpha & 0 \cr
-\sin \alpha & \cos \alpha & 0 \cr
0 & 0 & 1 \cr}
\eqno (1.A.24)
$$
is a rotation about $Z$ through an angle $\alpha$. We next
assume that $\bf u$ is real:
$$
{\bf u}_{2}(\beta)=\pmatrix{\cos {\beta \over 2} &
- \sin {\beta \over 2} \cr
\sin {\beta \over 2} & \cos {\beta \over 2} \cr}
\; \; \; \; .
\eqno (1.A.25)
$$
From (1.A.17)--(1.A.19) the corresponding rotation is
found to be
$$
R_{u_{2}}=\pmatrix{\cos \beta & 0 & - \sin \beta \cr
0 & 1 & 0 \cr 
\sin \beta & 0 & \cos \beta \cr}
\; \; \; \; ,
\eqno (1.A.26)
$$
i.e. a rotation about $Y$ through an angle $\beta$. The 
product of the three unitary matrices 
${\bf u}_{1}(\alpha){\bf u}_{2}(\beta){\bf u}_{1}(\gamma)$
corresponds to the product of a rotation about $Z$ through
an angle $\gamma$, about $Y$ through $\beta$, and about $Z$
through $\alpha$, in other words, to a rotation with Euler
angles $\alpha, \beta, \gamma$. It follows from this that 
the correspondence defined in (1.A.15) not only specified a
three-dimensional rotation for every two-dimensional unitary matrix,
but also {\it at least one unitary matrix} for every pure
rotation. Specifically, the matrix
$$
\eqalignno{
\; & \pmatrix{e^{-{i\over 2}\alpha}&0 \cr 
0 & e^{{i\over 2}\alpha} \cr}
\pmatrix{\cos {\beta \over 2} & - \sin {\beta \over 2} \cr
\sin {\beta \over 2} & \cos {\beta \over 2} \cr}
\pmatrix{e^{-{i\over 2}\gamma} & 0 \cr 
0 & e^{{i\over 2}\gamma} \cr} \cr
&=\pmatrix{e^{-{i\over 2}\alpha} \cos {\beta \over 2}
e^{-{i\over 2}\gamma} 
& -e^{-{i\over 2}\alpha} \sin {\beta \over 2}
e^{{i\over 2}\gamma} \cr
e^{{i\over 2}\alpha} \sin {\beta \over 2}
e^{-{i\over 2}\gamma} &
e^{{i\over 2}\alpha} \cos {\beta \over 2} 
e^{{i\over 2}\gamma} \cr}
&(1.A.27)\cr}
$$
corresponds to the rotation $\left \{ \alpha \beta
\gamma \right \}$. Thus, the homomorphism is in fact
a homomorphism of the unitary group onto the {\it whole}
three-dimensional pure rotation group.

The question remains of the {\it multiplicity} of the
homomorphism, i.e. how many unitary matrices {\bf u}
correspond to the same rotation. For this purpose, it is
sufficient to check how many unitary matrices 
${\bf u}_{0}$ correspond to the identity of the rotation 
group, i.e. to the transformation $x'=x,y'=y,z'=z$. For
these particular ${\bf u}_{0}$'s, the identity
${\bf u}_{0}{\bf h}{\bf u}_{0}^{\dagger}={\bf h}$ should
hold for all $\bf h$; this can only be the case when 
${\bf u}_{0}$ is a constant matrix: $b=0$ and 
$a=a^{*}$, ${\bf u}_{0}= \pm \II$ (since ${\mid a \mid}^{2}
+{\mid b \mid}^{2}=1$). Thus, the two unitary matrices
$+ \II$ and $-\II$, and {\it only these}, correspond to the
identity of the rotation group. These two elements form an
invariant subgroup of the unitary group, and those elements
(and only those) which are in the same coset of the
invariant subgroup, i.e. {\bf u} and $-{\bf u}$, correspond
to the same rotation. Indeed, that $\bf u$ and $-{\bf u}$
actually do correspond to the same rotation can be seen
directly from (1.A.15) or (1.A.17)--(1.A.19).
 
Alternatively, one can simply note that only the 
half-Euler-angles occur in (1.A.27). The Euler angles are
determined by a rotation only up to a multiple of $2\pi$;
the half angles, only up to a multiple of $\pi$. This implies
that the trigonometric functions in (1.A.27) are determined
only up to a sign.

A very important result has been thus obtained: there exists
a two-to-one homomorphism of the group of two-dimensional unitary
matrices with determinant $1$ {\it onto} the three-dimensional pure
rotation group: there is a one-to-one correspondence between
{\it pairs} of unitary matrices ${\bf u}$ and $-{\bf u}$ and
rotations $R_{u}$ in such a way that, from ${\bf uq}={\bf t}$
it also follows that $R_{u}R_{q}=R_{t}$; conversely, from
$R_{u}R_{q}=R_{t}$, one has that ${\bf uq}=\pm {\bf t}$. If the
unitary matrix $\bf u$ is known, the corresponding rotation
is best obtained from (1.A.17)--(1.A.19). Conversely, the 
unitary matrix for a rotation $\left \{ \alpha \beta \gamma
\right \}$ is best found from (1.A.27) (Wigner 1959).  
\vskip 100cm
\centerline {\it CHAPTER TWO}
\vskip 1cm
\centerline {\bf DIFFERENTIAL OPERATORS ON MANIFOLDS}
\vskip 1cm
\noindent
{\bf Abstract.} Clifford algebras play a key role in the
definition of the Dirac operator. Thus, the chapter begins
with a definition of the Clifford algebra of a vector 
space equipped with a non-degenerate quadratic form.
Clifford groups and Pin groups are also briefly introduced.
The following sections are devoted to the Laplace operator,
the signature operator and an intrinsic definition of
ellipticity and index. The chapter ends with a brief
review of properties of the Dirac operator, first in
Euclidean space and then on compact, oriented,
even-dimensional Riemannian manifolds.
\vskip 100cm
\centerline {\bf 2.1 Clifford Algebras}
\vskip 1cm
Let $V$ be a real vector space equipped with an inner product
$\langle \; , \; \rangle$, 
defined by a non-degenerate quadratic form $Q$
of signature $(p,q)$. Let $T(V)$ be the tensor algebra of $V$
and consider the ideal $\cal I$ in $T(V)$ generated by
$x \otimes x +Q(x)$. By definition, $\cal I$ consists of sums
of terms of the kind 
$a \otimes \Bigr \{x \otimes x + Q(x) \Bigr \} 
\otimes b$, $x \in V, a,b \in T(V)$. The quotient space
$$
Cl(V) \equiv Cl(V,Q) \equiv T(V)/{\cal I}
\eqno (2.1.1)
$$
is the {\it Clifford algebra} of the vector space $V$ equipped with
the quadratic form $Q$. The product induced by the tensor product
in $T(V)$ is known as Clifford multiplication or the Clifford
product and is denoted by $x \cdot y$, for $x,y \in Cl(V)$.
The dimension of $Cl(V)$ is $2^{n}$ if dim$(V)=n$. A basis for
$Cl(V)$ is given by the scalar 1 and the products
$$
e_{i_{1}} \cdot e_{i_{2}} \cdot e_{i_{n}}
\; \; \; \; 
i_{1}<...<i_{n}
\; \; \; \; ,
$$
where $\Bigr \{e_{1},...,e_{n}\Bigr \}$ is an orthonormal
basis for $V$. Moreover, the products satisfy (cf. (1.1.13))
$$
e_{i} \cdot e_{j}+e_{j} \cdot e_{i}=0
\; \; \; \; 
i \not = j
\; \; \; \; ,
\eqno (2.1.2)
$$
$$
e_{i} \cdot e_{i}=-2 \langle e_{i},e_{i} \rangle
\; \; \; \; 
i=1,...,n
\; \; \; \; .
\eqno (2.1.3)
$$
As a vector space, $Cl(V)$ is isomorphic to $\Lambda^{*}(V)$,
the Grassmann algebra, with
$$
e_{i_{1}} ... e_{i_{n}} \longrightarrow
e_{i_{1}} \wedge ... \wedge e_{i_{n}}
\; \; \; \; .
$$
There are two natural {\it involutions} on $Cl(V)$. The first,
denoted by $\alpha:Cl(V) \rightarrow Cl(V)$, is induced by the
involution $x \rightarrow -x$ defined on $V$, which extends to
an automorphism of $Cl(V)$. The eigenspace of $\alpha$ with
eigenvalue $+1$ consists of the even elements of $Cl(V)$, and the
eigenspace of $\alpha$ of eigenvalue $-1$ consists of the odd
elements of $Cl(V)$.

The second involution is a map $x \rightarrow x^{t}$, induced
on generators by
$$
\Bigr(e_{i_{1}} ... e_{i_{p}}\Bigr)^{t}
=e_{i_{p}} ... e_{i_{1}}
\; \; \; \; ,
$$
where $e_{i}$ are basis elements of $V$. Moreover, we define
$x \rightarrow {\overline x}$, a third involution of
$Cl(V)$, by ${\overline x} \equiv \alpha(x^{t})$. 

One then defines $Cl^{*}(V)$ to be the group of invertible 
elements of $Cl(V)$, and the {\it Clifford group} $\Gamma(V)$ is
the subgroup of $Cl^{*}(V)$ defined by
$$
\Gamma(V) \equiv \biggr \{x \in Cl^{*}(V):
y \in V \Rightarrow \alpha(x)yx^{-1} \in V \biggr \}
\; \; \; \; .
\eqno (2.1.4)
$$
One can show that the map $\rho : V \rightarrow V$
given by $\rho(x)y=\alpha(x)yx^{-1}$ is an isometry of $V$
with respect to the quadratic form $Q$. The map 
$x \rightarrow \| x \| \equiv 
x{\overline x}$ is the square-norm map,
and enables one to define a remarkable subgroup of
the Clifford group, i.e. (Ward and Wells 1990)
$$
{\rm Pin}(V) \equiv \biggr \{x \in \Gamma(V):
\| x \| =1 \biggr \}
\; \; \; \; .
\eqno (2.1.5)
$$
\vskip 1cm
\centerline {\bf 2.2 Exterior Differentiation and 
Hodge-Laplace Operator}
\vskip 1cm
Let $X$ be an $n$-dimensional $C^{\infty}$ manifold and
let $\Omega^{q}(X)$ denote the space of $C^{\infty}$ 
exterior differential forms of degree $q$ on $X$. One then
has a naturally occurring differential operator
$d: \Omega^{q}(X) \rightarrow \Omega^{q+1}(X)$ which extends
the differential of a function. The sequence (Atiyah 1975a)
$$
0 \rightarrow \Omega^{0}(X) \rightarrow \Omega^{1}(X)
\rightarrow ... \rightarrow \Omega^{n}(X) \rightarrow 0
$$
is a complex in that $d^{2}=0$, and the main theorem of
de Rham states that the cohomology of this complex is 
isomorphic to the cohomology of $X$, with real coefficients.
In other words, if $\gamma_{1},...,\gamma_{N}$ is a basis for
the homology in dimension $q$, then a $q$-form $\omega$ 
exists with $d\omega=0$ having prescribed periods
$\int_{\omega} \gamma_{i}$, and $\omega$ is unique modulo
the addition of forms $d\theta$. 
In particular, if $X={\Re}^{n}$,
the corresponding $\Omega^{q}(X)$ can be identified with
$C^{\infty}$ maps ${\Re}^{n} \rightarrow 
\Lambda^{q}({{\Re}^{n}}^{*})$,
and $d$ becomes a constant-coefficient operator whose
Fourier transform is exterior multiplication by $i\xi$,
for $\xi \in ({\Re}^{n})^{*}$.

If $X$ is compact oriented, and a Riemannian metric is
given on $X$, one can define a positive-definite inner 
product on $\Omega^{q}(X)$ by
$$
\langle u,v \rangle \equiv \int_{X}u_{\wedge} *v
\; \; \; \; ,
\eqno (2.2.1)
$$
where $*: \Omega^{q}(X) \rightarrow \Omega^{n-q}(X)$
is the isomorphism given by the metric. 
By virtue of the identity
$$
d \Bigr(u_{\wedge} *v\Bigr)
=du_{\wedge} *v+(-1)^{q}u_{\wedge}d*v
\; \; \; \; ,
\eqno (2.2.2)
$$
where $u \in \Omega^{q}, v \in \Omega^{q-1}$, one gets
$$
\int_{X}du_{\wedge} *v +(-1)^{q}\int_{X}u_{\wedge}
d*v=0
\; \; \; \; .
\eqno (2.2.3)
$$
This implies
$$
\langle du,v \rangle =(-1)^{q+1} \langle u,{*}^{-1}d*v \rangle
\; \; \; \; .
\eqno (2.2.4)
$$
Thus, the adjoint $d^{*}$ of $d$ on $\Omega^{q}$
is given by
$$
d^{*}=(-1)^{q+1}{*}^{-1}d^{*}
=\epsilon \; { }^{*}d^{*}
\; \; \; \; ,
\eqno (2.2.5)
$$
where $\epsilon$ may take the values $\pm 1$ depending on
$n,q$. 

The corresponding Hodge-Laplace operator on $\Omega^{q}$
is therefore defined as
$$
\bigtriangleup \equiv dd^{*}+d^{*}d
\; \; \; \; ,
\eqno (2.2.6)
$$
and the harmonic forms ${\cal H}^{q}$ are the solutions
of $\bigtriangleup u=0$. The main theorem of Hodge theory
is that ${\cal H}^{q}$ is isomorphic to the $q$-th de Rham
group and hence to the $q$-th cohomology group of $X$.
\vskip 1cm
\centerline {\bf 2.3 Index of Elliptic Operators and
Signature Operator}
\vskip 1cm
It is convenient to consider the operator
$d+d^{*}$ acting on $\Omega^{*}(X) \equiv \oplus_{q}
\Omega^{q}(X)$. Since $d^{2}=(d^{*})^{2}=0$, one has
$\bigtriangleup=(d+d^{*})^{2}$ and the harmonic forms are
also the solutions of $(d+d^{*})u=0$. If we consider 
$d+d^{*}$ as an operator $\Omega^{{\rm ev}} \rightarrow
\Omega^{{\rm odd}}$, its null-space is
$\oplus {\cal H}^{2q}$, while the null-space of its adjoint
is $\oplus {\cal H}^{2q+1}$. Hence its {\it index} is the
Euler characteristic of $X$. It is now appropriate to define
elliptic differential operators and their index.

Let us denote again by $X$ a compact oriented smooth manifold,
and by $E,F$ two smooth complex vector bundles over $X$. We
consider linear differential operators
$$
D:\Gamma(E) \rightarrow \Gamma(F)
\; \; \; \; ,
\eqno (2.3.1)
$$
i.e. linear operators defined on the spaces of smooth
sections and expressible locally by a matrix of partial
derivatives. Note that the extra generality involved in
considering vector bundles presents no serious difficulties
and it is quite essential in a systematic treatment on
manifolds, since all geometrically interesting operators
operate on vector bundles.

Let $T^{*}(X)$ denote the cotangent vector bundle of $X$,
$S(X)$ the unit sphere-bundle in $T^{*}(X)$ (relative to
some Riemannian metric), $\pi: S(X) \rightarrow X$ the
projection. Then, associated with $D$ there is a vector-bundle
homomorphism
$$
\sigma(D):\pi^{*}E \rightarrow \pi^{*}F
\; \; \; \; ,
\eqno (2.3.2)
$$
which is called the symbol of $D$. In terms of local coordinates,
$\sigma(D)$ is obtained from $D$ replacing 
${\partial \over \partial x_{j}}$ by $i\xi_{j}$ in the
highest-order terms of $D$ ($\xi_{j}$ is the $j$th coordinate
in the cotangent bundle). 
By definition, $D$ is {\it elliptic} if $\sigma(D)$
is an {\it isomorphism}. Of course, this implies that
the complex vector bundles $E,F$ have the same dimension.

One of the basic properties of elliptic operators is that
${\rm Ker} \; D$ (i.e. the null-space) and ${\rm Coker} \; D
\equiv \Gamma(F)/D\Gamma(E)$ are both finite-dimensional. The
{\it index} $\gamma(D)$ is defined by (Atiyah and Singer 1963)
$$
\gamma(D) \equiv {\rm dim} \; {\rm Ker} \; D
- {\rm dim} \; {\rm Coker} \; D
\; \; \; \; .
\eqno (2.3.3)
$$
If $D^{*}:\Gamma(F) \rightarrow \Gamma(E)$ denotes the
formal adjoint of $D$, relative to metrics in $E,F,X$,
then $D^{*}$ is also elliptic and
$$
{\rm Coker} \; D \cong {\rm Ker} \; D^{*}
\; \; \; \; ,
\eqno (2.3.4)
$$
so that 
$$
\gamma(D)={\rm dim} \; {\rm Ker} \; D
-{\rm dim} \; {\rm Ker} \; D^{*}
\; \; \; \; .
\eqno (2.3.5)
$$

Getting back to the definition of symbol, we find it
helpful for the reader to say that, for the exterior
derivative $d$, its symbol is exterior multiplication by
$i \xi$, for $\bigtriangleup$ it is $-{\| \xi \|}^{2}$,
and for $d+d^{*}$ it is $iA_{\xi}$, where $A_{\xi}$ is
Clifford multiplication by $\xi$. 

For a given $\alpha \in \Lambda^{p}({\Re}^{2l})
\otimes_{{\Re}}C$ (recall that $\Lambda^{q}(V)$ denotes the
exterior powers of the vector space $V$), 
let us denote by $\tau$ the involution
$$
\tau(\alpha) \equiv i^{p(p-1)+l} \; { }^{*}\alpha
\; \; \; \; .
\eqno (2.3.6)
$$
If $n=2l$, $\tau$ can be expressed as $i^{l}\omega$, where
$\omega$ denotes Clifford multiplication by the volume
form ${ }^{*}1$. Remarkably, $(d+d^{*})$ and $\tau$
anti-commute. In the Clifford algebra one has 
$\xi \omega =-\omega \xi$ for $\xi \in T^{*}$. Thus, if
$\Omega^{\pm}$ denote the $\pm1$-eigenspaces of $\tau$,
$(d+d^{*})$ maps $\Omega^{+}$ into $\Omega^{-}$. The restricted
operator
$$
d+d^{*}:\Omega_{+} \rightarrow \Omega_{-}
\; \; \; \; ,
\eqno (2.3.7)
$$
is called the {\it signature operator} and denoted by $A$.
If $l=2k$, the index of this operator is equal to
${\rm dim} \; {\cal H}_{+}-{\rm dim} \; {\cal H}_{-}$,
where ${\cal H}_{\pm}$ are the spaces of harmonic forms in
$\Omega_{\pm}$. If $q \not = l$, the space 
${\cal H}^{q} \oplus {\cal H}^{n-q}$ is stable under $\tau$,
which just interchanges the two factors. Hence one gets a
vanishing contribution to the index from such a space. For
$q=l=2k$, however, one has $\tau(\alpha)=i^{l(l-1)+l}
{ }^{*}\alpha={ }^{*}\alpha$ for $\alpha \in \Omega^{l}$.
One thus finds
$$
{\rm index} \; A = {\rm dim} \; {\cal H}_{+}^{l}
-{\rm dim} \; {\cal H}_{-}^{l}
\; \; \; \; ,
\eqno (2.3.8)
$$
where ${\cal H}_{\pm}^{l}$ are the $\pm1$-eigenspaces
of $*$ on ${\cal H}^{l}$. Thus, since the inner product in
${\cal H}^{l}$ is given by $\int u_{\wedge}{ }^{*}v$, one
finds
$$
{\rm index} \; A={\rm Sign}(X)
\; \; \; \; ,
\eqno (2.3.9)
$$
where ${\rm Sign}(X)$ is the signature of the quadratic
form on $H^{2k}(X;{\Re})$. Hence the name
for the operator $A$. 

It is now worth recalling that the {\it cup-product}
is a very useful algebraic structure occurring in
cohomology theory. It works as follows. Given $[\omega] \in
H^{p}(M;{\Re})$ and $[\nu] \in H^{q}(M;{\Re})$, 
then one defines the 
cup-product of $[\omega]$ and $[\nu]$, written 
$[\omega] \bigcup [\nu]$, by
$$
[\omega] \bigcup [\nu] \equiv [\omega \wedge \nu]
\; \; \; \; .
\eqno (2.3.10)
$$
The right-hand-side of (2.3.10) is a $(p+q)$-form so that
$[\omega \wedge \nu] \in H^{p+q}(M;{\Re})$. One can check, using
the properties of closed and exact forms, that this is a 
well-defined product of cohomology classes. The cup-product
$\bigcup$ is therefore a map of the form
$$
\bigcup : H^{p}(M;{\Re}) \times H^{q}(M;{\Re}) \rightarrow
H^{p+q}(M;{\Re})
\; \; \; \; .
\eqno (2.3.11)
$$
If the sum of all cohomology groups, $H^{*}(M;{\Re})$, 
is defined by
$$
H^{*}(M;{\Re}) \equiv \oplus_{p>0}H^{p}(M;{\Re})
\; \; \; \; ,
\eqno (2.3.12)
$$
the cup-product has the neater looking form
$$
\bigcup : H^{*}(M;{\Re}) \times H^{*}(M;{\Re})
\rightarrow H^{*}(M;{\Re})
\; \; \; \; .
\eqno (2.3.13)
$$
The product on $H^{*}(M;{\Re})$ defined in (2.3.13) makes 
$H^{*}(M;{\Re})$ into a ring. It can happen that two spaces
$M$ and $N$ have the same cohomology groups and yet are
not topologically the same. This can be proved by evaluating
the cup-product $\bigcup$ for 
$H^{*}(M;{\Re})$ and $H^{*}(N;{\Re})$,
and showing that the resulting rings are different. An example
is provided by choosing $M \equiv S^{2} \times S^{4}$
and $N \equiv CP^{3}$.
\vskip 1cm
\centerline {\bf 2.4 Dirac Operator}
\vskip 1cm
We begin our analysis by considering Euclidean space
${\Re}^{2k}$. The Clifford algebra ${\widetilde C}_{2k}$ 
of section 2.1 is a full matrix algebra acting on the
$2^{k}$-dimensional spin-space $S$. Moreover, $S$ is
given by the direct sum of $S^{+}$ and $S^{-}$, i.e. the
eigenspaces of $\omega=e_{1}e_{2}...e_{2k}$, and 
$\omega=\pm i^{k}$ on $S^{\pm}$. Let us now denote by
$E_{1},...,E_{2k}$ the linear transformations on $S$
representing $e_{1},...,e_{2k}$. The Dirac operator is
then defined as the first-order differential operator
$$
D \equiv \sum_{i=1}^{2k}E_{i}{\partial \over \partial
x_{i}}
\; \; \; \; ,
\eqno (2.4.1)
$$
and satisfies $D^{2}=-\bigtriangleup \cdot I$, where $I$
is the identity of spin-space. Note that, since the $e_{i}$
anti-commute with $\omega$, the Dirac operator interchanges
the positive and negative spin-spaces, $S^{+}$ and $S^{-}$
(cf. appendix 6.A).
Restriction to $S^{+}$ yields the operator
$$
B : C^{\infty}\Bigr({\Re}^{2k},S^{+}\Bigr) \rightarrow
C^{\infty} \Bigr({\Re}^{2k},S^{-}\Bigr)
\; \; \; \; .
\eqno (2.4.2)
$$
Since $E_{i}^{2}=-1$, in the standard metric the $E_{i}$
are unitary and hence skew-adjoint. Moreover, since
${\partial \over \partial x_{i}}$ is also formally 
skew-adjoint, it follows that $D$ is formally self-adjoint,
i.e. the formal adjoint of $B$ is the restriction of
$D$ to $S^{-}$.

The next step is the global situation of a compact oriented
$2k$-dimensional manifold $X$. First, we assume that a
spin-structure exists on $X$. 
From section 1.3, this means that the  
principal $SO(2k)$ bundle $P$ of $X$, consisting of oriented
orthonormal frames, lifts to a principal Spin($2k$) bundle
$Q$, i.e. the map $Q \rightarrow P$ is a double covering
inducing the standard covering Spin($2k$) $\rightarrow
SO(2k)$ on each fibre. If $S^{\pm}$ are the two half-spin
representations of Spin($2k$), we consider the associated
vector bundles on $X$: $E^{\pm} \equiv Q
\times_{{\rm Spin}(2k)}S^{\pm}$. Sections of these vector
bundles are the spinor fields on $X$. The {\it total} Dirac
operator on $X$ is a differential operator acting on
$E=E^{+} \oplus E^{-}$ and switching factors as above. To
define $D$ we have to use the Levi-Civita connection on $P$,
which lifts to one on $Q$. This enables one to define the
covariant derivative
$$
\nabla : C^{\infty}\Bigr(X,E\Bigr) \rightarrow C^{\infty}
\Bigr(X,E \otimes T^{*}\Bigr)
\; \; \; \; ,
\eqno (2.4.3)
$$
and $D$ is defined as the composition of $\nabla$ with
the map $C^{\infty}(X,E \otimes T^{*}\Bigr)
\rightarrow C^{\infty}(X,E)$ induced by Clifford multiplication.
By using an orthonormal base $e_{i}$ of $T$ at any point one
can write (see (1.4.12))
$$
Ds \equiv \sum_{i=1}^{2k} e_{i} \nabla_{i}s
\; \; \; \; ,
\eqno (2.4.4)
$$
where $\nabla_{i}s$ is the covariant derivative in the
direction $e_{i}$, and $e_{i}( \; )$ denotes Clifford
multiplication. By looking at symbols, $D$ is skew-adjoint.
Hence $D-D^{*}$ is an algebraic invariant of the metric and 
it only involves the first derivatives of $g$. Using normal
coordinates one finds that any such invariant vanishes,
which implies $D=D^{*}$, i.e. $D$ is self-adjoint.

As in Euclidean space, the restriction of $D$ to the
half-spinors $E^{+}$ is denoted by
$$
B: C^{\infty}\Bigr(X,E^{+}\Bigr) \rightarrow
C^{\infty} \Bigr(X,E^{-}\Bigr)
\; \; \; \; .
\eqno (2.4.5)
$$
The index of the restricted Dirac operator is given by
$$
{\rm index} \; B={\rm dim} \; {\cal H}^{+}
-{\rm dim} \; {\cal H}^{-}
\; \; \; \; ,
\eqno (2.4.6)
$$
where ${\cal H}^{\pm}$ are the spaces of solutions of
$Du=0$, for $u$ a section of $E^{\pm}$. Since $D$ is
elliptic and self-adjoint, these spaces are also the solutions
of $D^{2}u=0$, and are called {\it harmonic spinors}.
\vskip 1cm
\centerline {\bf 2.5 Some Outstanding Problems}
\vskip 1cm
In the mathematical and physical applications of elliptic
operators, one studies a mathematical framework given
by (complex) vector bundles over a compact Riemannian
manifold, say $M$, where $M$ may or may not have a boundary.
The latter case is simpler and came first in the (historical)
development of index theory. The investigations in Atiyah
and Singer 1968a, Atiyah and Segal 1968, Atiyah and Singer
1968b, Atiyah and Singer 1971a--b, proved that the index of
an elliptic operator (in general, one deals with elliptic
systems) is obtained by integrating over the sphere-bundle 
of $M$ a suitable composition of the Chern character with 
the Todd class of $M$ (see section 3.1). In other words, a
deep relation exists between the dimensions of the 
null-spaces of an elliptic operator and its adjoint on one
hand, and the topological invariants of the fibre bundles
of the problem on the other hand.

The classical Riemann-Roch problem, which is concerned with
giving a formula for the dimension of the space of meromorphic
functions on a compact Riemann surface, having poles of
orders $\leq \nu_{i}$ at points $P_{i}$, may be viewed as an
index problem. In other words, the solution of the index
problem in general is equivalent to finding an extension of the
Riemann-Roch theorem from the domain of holomorphic function
theory to that of general elliptic systems (Atiyah 1966).
In general, one has to find suitable topological invariants
of the pair $(M,D)$, where $M$ is the base manifold and $D$
is the elliptic operator. Moreover, the explicit formula
for ${\rm index}(D)$ has to be expressed in terms of these
invariants. In the particular case of the Riemann-Roch
theorem, the desired topological invariants turn out to be
the genus, and the degree of $\sum_{i}\nu_{i}P_{i}$.

It is now instructive to gain a qualitative understanding of
why an elliptic operator, say $D$, defines invariants of the
characteristic class type (Atiyah 1966). For this purpose,
let us recall that a homogeneous, constant coefficient,
$N \times N$ matrix of differential operators
$$
P=\left[P_{ij} \biggr({\partial \over \partial x_{1}},...,
{\partial \over \partial x_{n}}\biggr)\right] \; \; ,
\; \; i,j=1,...,N
\; \; \; \; ,
\eqno (2.5.1)
$$
is elliptic if, for $\xi=(\xi_{1},...,\xi_{n}) \not = 0$
(and real) one has
$$
{\rm det} \; P (\xi_{1},...,\xi_{n}) \not = 0
\; \; \; \; .
\eqno (2.5.2)
$$
If this condition holds, $P$ defines a map $\xi \rightarrow
P(\xi)$ of $S^{n-1}$ into $GL(N,C)$, where $S^{n-1}$ is the
unit sphere in ${\Re}^{n}$. This is why the homotopy (and
hence the homology) of $GL(N,C)$ enters into the study of
elliptic operators. Now, according to the Bott periodicity
theorem (Bott 1959, Atiyah and Bott 1964), the homotopy
groups $\pi_{n-1}(GL(N,C))$ are $0$ for $n$ odd, and 
isomorphic to the integers for $n$ even, provided that
$2N \geq n$ (see also section 3.6). Thus, if $n$ is even
and $2N \geq n$, $P$ defines an integer which may be called
its {\it degree}. Such a degree may be evaluated explicitly
as an integral
$$
\int_{S^{n-1}}\omega(P)
\; \; \; \; ,
$$
where $\omega(P)$ is a differential expression in $P$
generalizing the familiar formula ${1\over 2 \pi i}
\int_{S^{1}}{dP\over P}$. The use of these invariants of a
general elliptic operator $D$, jointly with the ordinary
characteristic classes of $M$, can give an explicit formula
for ${\rm index}(D)$, which is indeed similar to that 
occurring in the Riemann-Roch formula. As Atiyah (1966) 
put it, the deeper underlying reason is that, as far as
the topology is concerned, the classical operators are just
as complicated as the most general ones, so that the
Riemann-Roch formula contains relevant information about 
the general case.

If one is interested in the actual proof of the index
theorem (cf. section 3.1),
one can say that, when $M$ is a simple space like a sphere,
Bott's theorem can be used to deform $D$ into a standard
operator, whose index may be computed directly. Moreover,
if $M$ is complicated, one can embed 
$M$ in a sphere $S$, and construct
an elliptic operator, say $D'$, on $S$, such that
$$
{\rm index}(D)={\rm index}(D')
\; \; \; \; .
\eqno (2.5.3)
$$
Thus, reduction to the previous case is possible. However,
even if one starts with a ``nice" operator on $M$ (for
example, if $D$ results from a complex structure), it is
not possible, in general, to get a nice operator $D'$ 
on $S$. Thus, the consideration of operators which are
neither ``nice" nor classical is indeed unavoidable. The
topology of linear groups and the analysis of elliptic
operators share important properties like linearity,
stability under deformation, and finiteness. It was suggested
in Atiyah 1966 that all this lies at the very heart of the
close relation between these branches of mathematics. In the
following chapters, we shall also study index problems for
manifolds with boundary, and their deep link with general
properties of complex powers of elliptic operators, and the
asymptotic expansion of the integrated heat-kernel 
(cf. Piazza 1991, Piazza 1993).
\vskip 100cm
\centerline {\it CHAPTER THREE}
\vskip 1cm
\centerline {\bf INDEX PROBLEMS}
\vskip 1cm
\noindent
{\bf Abstract.} This chapter begins with an outline of index
problems for closed manifolds and for manifolds with boundary, 
and of the relation between index theory and
anomalies in quantum field theory. In particular, the index
of the Dirac operator is evaluated in the presence of gauge
fields and gravitational fields. The result is expressed by
integrating the Chern character and the Dirac genus. 
Interestingly, anomaly calculations turn out to be 
equivalent to the analysis of partition functions in
ordinary quantum mechanics. The chapter ends with an
introductory presentation of Bott periodicity and K-theory,
with the aim of describing why the Dirac operator is the 
most fundamental, in the theory of elliptic operators on
compact Riemannian manifolds.
\vskip 100cm
\centerline {\bf 3.1 Index Problem for Manifolds with Boundary}
\vskip 1cm
Following Atiyah and Bott 1965, this section is devoted to 
some aspects of the extension of
the index formula for closed manifolds, to manifolds with
boundary. The naturally occurring question is then how to
measure the topological implications of elliptic boundary
conditions, since boundary conditions have of course a
definite effect on the index.

For example, let $X$ be the unit disk in the plane, let $Y$
be the boundary of $X$, and let $b$ be a nowhere-vanishing 
vector field on $Y$. Denoting by $D$ the Laplacian on $X$,
we consider the operator
$$
(D,b):C^{\infty}(X) \rightarrow C^{\infty}(X) \oplus
C^{\infty}(Y)
\; \; \; \; .
\eqno (3.1.1)
$$
By definition, $(D,b)$ sends $f$ into $Df \oplus (bf \mid Y)$,
where $bf$ is the directional derivative of $f$ along $b$.
Since $D$ is elliptic and $b$ is non-vanishing, kernel and
cokernel of $(D,b)$ are both finite-dimensional, hence the 
index of the boundary-value problem is finite. One would now
like it to express the index in terms of the topological 
data given by $D$ and the boundary conditions.
The solution of this problem is expressed by a formula 
derived by Vekua (H\"{o}rmander 1963), according to which
$$
{\rm index} \; (D,b)=2 \Bigr(1-{\rm winding}
\; {\rm number} \; {\rm of} \; b \Bigr)
\; \; \; \; .
\eqno (3.1.2)
$$

It is now necessary to recall the index formula for closed
manifolds, and this is here done in the case of vector
bundles. From now on, $X$ is the base manifold, $Y$ its
boundary, $T(X)$ the tangent bundle of $X$, $B=B(X)$
the ball-bundle (consisting of vectors in $T(X)$ of
length $\leq 1$), $S=S(X)$ the sphere-bundle of unit vectors
in $T(X)$. Of course, if $X$ is a closed manifold
(hence without boundary), then $S(X)$ is just the boundary of
$B(X)$, i.e.
$$
\partial B(X)=S(X) 
\; \; \; \; {\rm if} \; \partial X = \emptyset
\; \; \; \; .
\eqno (3.1.3)
$$
However, if $X$ has a boundary $Y$, then
$$
\partial B(X)=S(X) \cup B(X) \mid Y
\; \; \; \; ,
\eqno (3.1.4)
$$
where $B(X) \mid Y$ is the subspace of $B(X)$ lying
over $Y$ under the natural map $\pi: B(X) \rightarrow X$.

We now focus on a system of $k$ linear partial differential
operators, i.e. 
$$
Df_{i} \equiv \sum_{j=1}^{k} A_{ij} f_{j}
\; \; \; \; ,
\eqno (3.1.5)
$$
defined on the $n$-dimensional manifold $X$. As we know from
section 2.3, the symbol of $D$ is a function $\sigma(D)$ on
$T^{*}(X)$ attaching to each cotangent vector $\lambda$ of
$X$, the matrix $\sigma(D: \lambda)$ obtained from the 
highest terms of $A_{ij}$ by replacing ${\partial^{\alpha}
\over \partial x^{\alpha}}$ with $(i \lambda)^{\alpha}$.
Moreover, the system $D$ is {\it elliptic} if and only if
the function $\sigma(D)$ maps $S(X)$ into the group 
$GL(k,C)$ of non-singular $k \times k$ matrices with complex
coefficients. Thus, defining
$$
GL \equiv \lim_{m \to \infty}GL(m,C)
\; \; \; \; ,
\eqno (3.1.6)
$$
the symbol defines a map (cf. appendix 3.A)
$$
\sigma(D):S(X) \rightarrow GL
\; \; \; \; .
\eqno (3.1.7)
$$
Since the index of an elliptic system is invariant under
deformations, on a closed manifold the index of $D$ only
depends on the homotopy class of the map $\sigma(D)$. 

To write down the index formula we are looking for, one 
constructs a definite differential form $ch \equiv
\sum_{i}{ch}^{i}$ on $GL$. Note that, strictly, we
define a differential form ${ch}(m)$ on each
$GL(m,C)$ (see Eqs. (3.B.12) and (3.B.13) for the
Chern character). 
Moreover, by using a universal expression in the
curvature of $X$, one constructs the Todd class of 
$TX$, denoted by $td(X)$, as in (3.B.16). The index
theorem for a closed manifold then yields the index of
$D$ as an integral 
$$
{\rm index} (D)=
\int_{S(X)} \sigma(D)^{*} \; {ch}
\wedge \pi^{*} td(X)
\; \; \; \; .
\eqno (3.1.8)
$$
With our notation, $\sigma(D)^{*}ch$ is the form on $GL$
pulled back to $S(X)$ via $\sigma(D)$. By virtue of 
(3.1.3), the integral (3.1.8) may be re-written as
$$
{\rm index}(D)=
\int_{\partial B(X)} \sigma(D)^{*} \; {ch}
\wedge \pi^{*} td(X)
\; \; \; \; .
\eqno (3.1.9)
$$
In this form the index formula is also meaningful
for a manifold with boundary, {\it provided that} $\sigma(D)$,
originally defined on $S(X)$, is {\it extended} to 
$\partial B(X)$. It is indeed in this extension that the
topological data of a set of elliptic boundary conditions
manifest themselves. Following Atiyah 
and Bott 1965, we may in fact state the following theorem
(cf. Booss and Bleecker 1985):
\vskip 0.3cm
\noindent
{\bf Theorem 3.1.1} A set of elliptic boundary conditions,
$B$, on the elliptic system $D$, defines a definite map
$\sigma(D,B):\partial B(X) \rightarrow GL$, which extends
the map $\sigma(D): S(X) \rightarrow GL$ to the whole of
$\partial B(X)$.
\vskip 0.3cm
\noindent
One thus finds an index theorem formally analogous to the 
original one, in that the index of $D$ subject to the
elliptic boundary condition $B$ is given by
$$
{\rm index}(D,B)=\int_{\partial B(X)}
\sigma(D,B)^{*} {ch} \wedge \pi^{*} td(X)
\; \; \; \; .
\eqno (3.1.10)
$$

One can thus appreciate a key feature of index theorems:
they relate analytic properties of differential operators
on fibre bundles to the topological invariants of such bundles
(i.e. their characteristic classes). As will be shown in
the following sections, the relevance of index theorems for
theoretical physics lies in the possibility to evaluate the
number of zero-modes of differential operators, if one
knows the topology of the fibre bundles under consideration.
\vskip 1cm
\centerline {\bf 3.2 Elliptic Boundary Conditions}
\vskip 1cm
Although this section is (a bit) technical, it is necessary
to include it for the sake of completeness. Relevant examples
of elliptic boundary-value problems (see appendix 3.A) 
will be given in chapters 5--10.

Following Atiyah and Bott 1965, 
let $D$ be a $k \times k$ elliptic system of
differential operators on $X$, and let $r$ denote the order
of $D$. The symbol $\sigma(D)$ is a function on the cotangent
vector bundle $T^{*}(X)$ whose values are $(k \times k)$-matrices,
and its restriction to the unit sphere-bundle $S(X)$ takes
non-singular values. For our purposes, it is now necessary to
consider a system $B$ of boundary operators 
(cf. (5.1.18)) given by an
$l \times k$ matrix with rows $b_{1},...,b_{l}$ of orders
$r_{1},...,r_{l}$ respectively. We now denote by $\sigma(b_{i})$
the symbol of $b_{i}$, and by $\sigma(B)$ the matrix with
$\sigma(b_{i})$ as $i$-th row. At a point of the boundary 
$Y$ of $X$, let $\nu$ be the unit 
inward-pointing normal and let $y$
denote any unit tangent vector to $Y$. One puts
$$
\sigma_{y}(D)(t)=\sigma(D)(y+t\nu)
\; \; \; \; ,
\eqno (3.2.1)
$$
$$
\sigma_{y}(B)(t)=\sigma(B)(y+t\nu)
\; \; \; \; ,
\eqno (3.2.2)
$$
so that $\sigma_{y}(D)$ and $\sigma_{y}(B)$ are polynomials
in $t$. It then makes sense to consider the system of
ordinary linear equations
$$
\sigma_{y}(D)\biggr(-i{d\over dt}\biggr)u=0
\; \; \; \; ,
\eqno (3.2.3)
$$
whose space of solutions is denoted by ${\cal M}_{y}$.
The {\it ellipticity} of the system $D$ means, by definition,
that ${\cal M}_{y}$ can be decomposed as
$$
{\cal M}_{y}={\cal M}_{y}^{+} \oplus
{\cal M}_{y}^{-}
\; \; \; \; ,
\eqno (3.2.4)
$$
where ${\cal M}_{y}^{+}$ consists only of exponential
polynomials involving $e^{i\lambda t}$ with 
${\rm Im}(\lambda)>0$, and ${\cal M}_{y}^{-}$ involves
those with ${\rm Im}(\lambda)<0$. The ellipticity
condition for the system of boundary operators, relative
to the elliptic system of differential operators, is that
the equations (3.2.3) should have a {\it unique} solution 
$u \in {\cal M}_{y}^{+}$ satisfying the boundary
condition (cf. appendix 3.A)
$$
\sigma_{y}(B) \biggr(-i{d\over dt}\biggr)
u {\left.  \right |}_{t=0}=V
\; \; \; \; ,
\eqno (3.2.5)
$$
for any given $V \in {\cal C}^{l}$.

Now a lemma can be proved, according to which a natural
isomorphism of vector spaces exists between $M_{y}^{+}$,
the cokernel of $\sigma_{y}(D)(t)$, and ${\cal M}_{y}^{+}$:
$$
M_{y}^{+} \cong {\cal M}_{y}^{+}
\; \; \; \; .
\eqno (3.2.6)
$$
Hence the elliptic boundary condition yields an isomorphism
$$
\beta_{y}^{+}: M_{y}^{+} \rightarrow {\cal C}^{l}
\; \; \; \; .
\eqno (3.2.7)
$$
The set of all $M_{y}^{+}$, for $y \in S(Y)$, forms a vector
bundle $M^{+}$ over $S(Y)$, and (3.2.7) defines an isomorphism
$\beta^{+}$ of $M^{+}$ with the trivial bundle $S(Y) \times 
{\cal C}^{l}$.

Possible boundary conditions are differential or integro-differential.
If only differential boundary conditions are imposed then the
map $\beta_{y}^{+}$, regarded as a function of $y$, cannot be an
arbitrary continuous function. To obtain {\it all} continuous
functions one has to enlarge the problem and consider 
integro-differential boundary conditions as well. This is,
indeed, an important topological simplification. Thus, an
elliptic problem $(D,B)$ has associated with it $\sigma(D),M^{+},
\beta^{+}$, where $\beta^{+}$ can be any vector-bundle isomorphism
of $M^{+}$ with the trivial bundle.
\vskip 1cm
\centerline {\bf 3.3 Index Theorems and Gauge Fields}
\vskip 1cm
Anomalies in gauge theories, and their relation to the index
theory of the Dirac operator, are a key topic in the
quantization of gauge theories of fundamental interactions.
The aim of this section is to introduce the
reader to ideas and problems in this field of research,
following Atiyah 1984 and then Alvarez-Gaum\'e 1983b.

The index theorem is concerned with any elliptic differential
operator, but for physical applications one only needs the
special case of the Dirac operator. From chapter 1, we know
that this is globally defined on any Riemannian manifold
$(M,g)$ provided that $M$ is oriented and has a spin-structure.
If $M$ is compact, the Dirac operator is elliptic, self-adjoint, 
and has a discrete spectrum of eigenvalues. Of particular 
interest is the $0$-eigenspace (or null-space). By definition,
the Dirac operator acts on spinor fields, and it should be
emphasized that there is a basic algebraic difference between 
spinors in even and odd dimensions. In fact in odd dimensions
spinors belong to an irreducible representation of the
spin-group, whereas in even dimensions spinors break up
into two irreducible pieces, here denoted by $S^{+}$ and
$S^{-}$. The Dirac operator interchanges these two, hence it
consists essentially of an operator (cf. section 1.4)
$$
D: S^{+} \rightarrow S^{-}
\; \; \; \; ,
\eqno (3.3.1)
$$
and its adjoint
$$
D^{*}:S^{-} \rightarrow S^{+}
\; \; \; \; .
\eqno (3.3.2)
$$
The dimensions of the null-spaces of $D$ and $D^{*}$ are denoted
by $N^{+}$ and $N^{-}$ respectively, and the index of $D$
(cf. Eqs. (2.3.3)--(2.3.5)) is defined by
$$
{\rm index}(D) \equiv N^{+}-N^{-}
\; \; \; \; .
\eqno (3.3.3)
$$
Note that, although a formal symmetry exists between
positive and negative spinors (in fact they are interchanged
by reversing the orientation of $M$), the numbers
$N^{+}$ and $N^{-}$ need not be equal, due to topological
effects. Hence the origin of the {\it chiral anomaly}
(cf. Rennie 1990).

The index theorem provides an explicit formula for
${\rm index}(D)$ in terms of topological invariants 
of $M$. Moreover, by using the Riemannian metric $g$
and its curvature tensor $R(g)$, one can write an explicit
integral formula
$$
{\rm index}(D)=\int_{M}\Omega(R(g))
\; \; \; \; ,
\eqno (3.3.4)
$$
where $\Omega$ is a formal expression obtained purely
algebraically from $R(g)$. For example, when $M$ is
four-dimensional one finds
$$
\Omega={{\rm Tr} \; \omega^{2} \over 96 \pi^{2}}
\; \; \; \; ,
\eqno (3.3.5)
$$
where $\omega$ is regarded as a matrix of two-forms.
Interestingly, the index formula (3.3.4) is, from the
physical point of view, purely gravitational since it
only involves the metric $g$. Moreover, the index vanishes
for the sphere or torus, and one needs more complicated
manifolds to exhibit a non-vanishing index. However, 
(3.3.4) can be generalized to gauge theories. This means 
that one is given a complex vector bundle $V$ over $M$ with a
unitary connection $A$. If the fibre of $V$ is isomorphic 
with ${\cal C}^{N}$, then $A$ is a $U(N)$ gauge field over $M$.
The {\it extended Dirac operator} now acts on vector-bundle-valued
spinors
$$
D_{A}:S^{+} \otimes V \rightarrow S^{-} \otimes V
\; \; \; \; ,
\eqno (3.3.6)
$$
and one defines again the index of $D_{A}$  by a formula
like (3.3.3). The index formula (3.3.4) is then replaced
by
$$
{\rm index}\Bigr(D_{A}\Bigr)=\int_{M}
\Omega(R(g),F(A))
\; \; \; \; ,
\eqno (3.3.7)
$$
where $F(A)$ is the curvature of $A$, and $\Omega(R,F)$ is a
certain algebraic expression in $R$ and $F$. For example,
if $M$ is four-dimensional, (3.3.5) is generalized by
$$
\Omega={{\rm Tr} \; \omega^{2}\over 96 \pi^{2}}
-{{\rm Tr} \; F^{2} \over 8\pi^{2}}
\; \; \; \; .
\eqno (3.3.8)
$$

We now find it instructive, instead of giving a brief
outline of a wide range of results (see, for example,
Rennie 1990 and Gilkey 1995), to focus on a specific but
highly non-trivial example. Following Alvarez-Gaum\'e
1983b, we deal (again) with the Dirac equation in the 
presence of external gauge and gravitational fields. The
motivations of our investigation lie in the deep link 
between supersymmetry and the Atiyah-Singer index theorem.
Indeed, if one considers supersymmetric quantum mechanics,
which may be viewed as a (0+1)-dimensional field theory,
one finds that this theory has $N$ conserved charges, say
$Q_{i}$, which anticommute with the fermion number operator
$(-1)^{F}$, and which satisfy the supersymmetry algebra
(for all $i,j=1,...,N$):
$$
\left \{ Q_{i}, Q_{j}^{*} \right \}
=2 \delta_{ij} H
\; \; \; \; ,
\eqno (3.3.9)
$$
$$
\left \{ Q_{i}, (-1)^{F} \right \} =0
\; \; \; \; ,
\eqno (3.3.10)
$$
$$
\left \{ Q_{i}, Q_{j} \right \} =0
\; \; \; \; ,
\eqno (3.3.11)
$$
where $H$ is the Hamiltonian of supersymmetric quantum
mechanics. If $Q, Q^{*}$ are any of the $N$ supersymmetric
charges, the operator
$$
S \equiv {1\over \sqrt{2}} (Q+Q^{*})
\eqno (3.3.12)
$$
is Hermitian and satisfies $S^{2}=H$. Moreover, if 
$\mid E \rangle$ is an arbitrary eigenstate of $H$:
$$
H \mid E \rangle = E \mid E \rangle
\; \; \; \; ,
\eqno (3.3.13)
$$
then $S \mid E \rangle $ is another state with the same
energy. Thus, if $\mid E \rangle $ is a bosonic state,
$S \mid E \rangle $ is fermionic, and the other way 
around, and non-zero energy states in the spectrum appear
in Fermi-Bose pairs. This in turn implies that
${\rm Tr}(-1)^{F} e^{-\beta H}$ receives contributions 
only from zero-energy states. Further to this, the work in
Witten 1982 has shown that ${\rm Tr}(-1)^{F}e^{-\beta H}$ is
a {\it topological invariant} of the quantum theory. 
Bearing in mind that bosonic zero-energy states, say
$\mid B \rangle$, are, by definition, solutions of 
the equation
$$
Q \mid B \rangle =0
\; \; \; \; ,
\eqno (3.3.14)
$$
and fermionic zero-modes solve instead the equation
$$
Q^{*} \mid F \rangle =0
\; \; \; \; ,
\eqno (3.3.15)
$$
one finds that
$$
{\rm Index}(Q)={\rm dim} \; {\rm Ker} \; Q
-{\rm dim} \; {\rm Ker} \; Q^{*}
={\rm Tr}\left[(-1)^{F}e^{-\beta H}\right]
\; \; \; \; .
\eqno (3.3.16)
$$
The problem of finding the index of $Q$ is therefore 
reduced to the evaluation of the trace on the right-hand
side of Eq. (3.3.16) in the $\beta \rightarrow 0$ limit.
The idea of Alvarez-Gaum\'e (1983a--b) was to perform this
calculation with the help of a path-integral representation
of the partition function, with periodic boundary 
conditions for fermionic fields:
$$
{\rm Tr}\left[(-1)^{F}e^{-\beta H}\right]=\int_{\Omega_{\beta}}
d\phi(t) \; d\psi(t) \; {\rm exp} - \int_{0}^{\beta}
L_{E}(t) dt
\; \; \; \; ,
\eqno (3.3.17)
$$
where $\phi$ and $\psi$ are required to belong to the
class $\Omega_{\beta}$ of fermionic fields, say $\chi$,
such that
$$
\chi(0)=\chi(\beta)
\; \; \; \; .
\eqno (3.3.18)
$$
The next non-trivial step is to use the Lagrangian of the
supersymmetric non-linear $\sigma$-model obtained by
dimensional reduction from $1+1$ to $0+1$ dimensions:
$$
\eqalignno{
L&={1\over 2} g_{ij}(\phi){\dot \phi}^{i} {\dot \phi}^{j}
+{i\over 2} g_{jk} \psi_{\alpha}^{j} \left[{d\over dt}
\psi_{\alpha}^{k}+ \Gamma_{\; ml}^{k} \; {\dot \phi}^{m}
\; \psi_{\alpha}^{l} \right] \cr
&+{1\over 4}R_{ijkl} \; \psi_{1}^{i} \; \psi_{1}^{j}
\; \psi_{2}^{k} \; \psi_{2}^{l}
\; \; \; \; .
&(3.3.19)\cr}
$$
With this notation, $g_{ij}$ is the metric on the manifold
$M$, $\Gamma_{\; jk}^{i}$ are the Christoffel symbols,
$R_{\; jkl}^{i}$ is the Riemann tensor of $M$, 
$\psi_{\alpha}^{i}(t)$ are real anticommuting fermionic fields
($\alpha=1,2$).

Since our ultimate goal is the analysis of the Dirac equation
in the presence of gauge fields, it is now appropriate to
consider a gauge group, say $G$, acting on $M$, with gauge
connection $A_{i}^{\alpha}(\phi)$ and gauge curvature
$$
F_{\; ij}^{\alpha}(\phi)=\partial_{i}A_{j}^{\alpha}
-\partial_{j}A_{i}^{\alpha}+h f_{\; \beta \gamma}^{\alpha}
\; A_{i}^{\beta} \; A_{j}^{\gamma}
\; \; \; \; ,
\eqno (3.3.20)
$$
where $h$ is the gauge coupling constant and Greek indices
range from 1 through ${\rm dim}(G)$. In the explicit expression
of the first-order derivative resulting from the Dirac 
operator, one has now to include both the effects of 
$A_{i}^{\alpha}$ and the contribution of the spin-connection,
say $\omega_{\; ib}^{a}$. In terms of the vierbein 
$e_{\; i}^{a}$ (cf. section 1.2), and of its dual 
$E_{\; a}^{i}$:
$$
E_{\; a}^{i} \; e_{\; j}^{a}=\delta_{\; j}^{i}
\; \; \; \; ,
\eqno (3.3.21)
$$
one has 
$$
\omega_{\; ib}^{a}=-E_{\; b}^{k} \Bigr(\partial_{i}
e_{\; k}^{a} - \Gamma_{\; ik}^{l} \; e_{\; l}^{a} \Bigr)
\; \; \; \; ,
\eqno (3.3.22)
$$
and the eigenvalue problem for the Dirac operator reads
$$
i \gamma^{k} \Bigr(\partial_{k}+{1\over 2}
\omega_{kab} \sigma^{ab}+ih A_{k}^{\alpha} T^{\alpha}
\Bigr)_{AB} \Bigr(\psi_{\lambda}\Bigr)_{B}
=\lambda \Bigr(\psi_{\lambda}\Bigr)_{A}
\; \; \; \; .
\eqno (3.3.23)
$$
In Eq. (3.3.23), one has (cf. (1.2.10))
$\sigma^{ab} \equiv {1\over 4} \Bigr[\gamma^{a},
\gamma^{b}\Bigr]$,
and $\Bigr(T^{\alpha}\Bigr)_{AB}$ are the generators of the
representation of $G$, with indices $A,B$ ranging from 1
through ${\rm dim}(T)$. The reader should be aware that our
$a,b$ indices, here used to follow the notation in
Alvarez-Gaum\'e 1983b, correspond to the hatted indices 
of section 1.2.

We are eventually interested in the one-dimensional analogue
of Eq. (3.3.23). For this purpose, we consider a pair of
fermionic annihilation and creation operators, say
$C_{A},C_{A}^{*}$:
$$
\left \{ C_{A}, C_{B} \right \}
= \left \{ C_{A}^{*}, C_{B}^{*} \right \}=0
\; \; \; \; ,
\eqno (3.3.24)
$$
$$
\left \{ C_{A}^{*},C_{B} \right \}=\delta_{AB}
\; \; \; \; .
\eqno (3.3.25)
$$
They lead to the eigenvalue equation
$$
i \gamma^{k} \Bigr(\partial_{k}+{1\over 2}
\omega_{kab} \sigma^{ab}+ i h A_{k}^{\alpha}
C^{*} T C \Bigr) \mid \psi \rangle 
= \lambda \mid \psi \rangle
\; \; \; \; .
\eqno (3.3.26)
$$
Interestingly, if $\mid \psi, \lambda \rangle$ is an
eigenfunction of the Dirac operator with eigenvalue 
$\lambda \not = 0$, then $\gamma_{5} \mid \psi , \lambda
\rangle$ turns out to be the eigenfunction belonging to
the eigenvalue $- \lambda$. Thus, since
(hereafter, ${\cal D} \equiv i \gamma^{k} D_{k}$)
$\left[{\cal D}^{2},
\gamma_{5} \right]=0$,
one finds (see (3.3.16))
$$
{\rm Tr}\left[ \gamma_{5} e^{-\beta {\cal D}^{2}}\right]
=n_{E=0}(\gamma_{5}=1)-n_{E=0}(\gamma_{5}=-1)
\; \; \; \; ,
\eqno (3.3.27)
$$
where $n_{E=0}(\gamma_{5}=1)$ and $n_{E=0}(\gamma_{5}=-1)$
denote the number of zero-eigenvalues of $\cal D$ with 
$\gamma_{5}=1$ and $\gamma_{5}=-1$, respectively. On 
writing this equation, we have recognized that 
$(-1)^{F}=\gamma_{5}$ in four dimensions. Now the Euclidean
Lagrangian is inserted into the path-integral formula
(3.3.17), with the understanding that functional integration 
is restricted to the space of one-particle states for the
$(C,C^{*})$ fermions. This restriction is indeed necessary to
ensure that the index of $\cal D$ is evaluated only in the
representation $T^{\alpha}$ of the gauge group $G$. The
boundary conditions are now (3.3.18) for $\phi$ and $\psi$,
jointly with antiperiodic boundary conditions for fermionic
operators:
$$
C(0)=-C(\beta)
\; \; \; \; , \; \; \; \; 
C^{*}(0)=-C^{*}(\beta)
\; \; \; \; .
\eqno (3.3.28)
$$
Moreover, the background-field method is used, writing
(recall that $\psi^{a} \equiv e_{\; i}^{a} \; \psi^{i}$)
$$
\phi^{i}=\phi_{0}^{i}+\xi^{i}
\; \; \; \; , \; \; \; \; 
\psi^{a}=\psi_{0}^{a}+\eta^{a}
\; \; \; \; .
\eqno (3.3.29)
$$
The underlying idea is that, in the $\beta \rightarrow 0$
limit, the functional integral is dominated by time-independent
field configurations $\phi_{0}^{i}$ and $\psi_{0}^{a}$. The
terms $\xi^{i}$ and $\eta^{a}$ are then the fluctuations 
around the constant background values, according to 
(3.3.29). Omitting a few details,
which can be found in Alvarez-Gaum\'e 1983b and references
therein, one obtains
$$
\eqalignno{
{\rm index}(\cal D)&=(i / 2\pi)^{n} \int {\rm dvol}
\int d\psi_{0} \left({\rm Tr} \exp \Bigr(-{i\over 2}
\psi_{0}^{a} \psi_{0}^{b} F_{\; ab}^{\alpha} T^{\alpha}
\Bigr) \right) \cr
& \times \prod_{l=1}^{n}{(i x_{l}/2) \over 
\sinh (i x_{l}/2)} \; \; \; \; ,
&(3.3.30)\cr}
$$
where $(2\pi)^{-n}$ results from the Feynman measure for
constant modes, $(i)^{n}$ is due to integration over constant
real-valued fermionic configurations, and the $x_{l}$ are the
eigenvalues of the matrix ${1\over 2}R_{abcd} \; \psi_{0}^{c}
\; \psi_{0}^{d}$. Note that, considering the Riemann curvature
two-form
$$
R_{ab} \equiv {1\over 2} R_{abcd} \; e^{c} \wedge e^{d}
\; \; \; \; ,
\eqno (3.3.31)
$$
and the gauge curvature two-form
$$
F \equiv {h\over 2} F_{\; ab}^{\alpha} \; T^{\alpha}
\; e^{a} \wedge e^{b}
\; \; \; \; ,
\eqno (3.3.32)
$$
one can form the polynomials (cf. appendix 3.B)
$$
ch(F) \equiv {\rm Tr} \; e^{F/ 2\pi}
\; \; \; \; ,
\eqno (3.3.33)
$$
$$
{\widehat A}(M) \equiv \prod_{\alpha}
{(\omega_{\alpha}/ 4\pi)\over \sinh (\omega_{\alpha}/4\pi)}
\; \; \; \; .
\eqno (3.3.34)
$$
In the definition (3.3.33), the trace is taken over the
representation $T^{\alpha}$ of $G$, and in (3.3.34)
the $\omega_{\alpha}$ are the eigenvalues of the 
antisymmetric matrix (3.3.31). According to a standard
notation, $ch(F)$ is the Chern character of the principal
bundle associated to the gauge field, and ${\widehat A}(M)$
is the Dirac genus of $M$. They make it possible to
re-express the result (3.3.30) in the neat form (cf. (3.1.8))
$$
{\rm index}({\cal D})=\int_{M} ch(F){\widehat A}(M)
\; \; \; \; .
\eqno (3.3.35)
$$ 
In particular, if $M$ is four-dimensional, Eq. (3.3.35) 
leads to (cf. (3.3.7))
$$
{\rm index}({\cal D})={({\rm dim}T)\over 192 \pi^{2}}
\int {\rm Tr} R \wedge R
+{1\over 8 \pi^{2}} \int {\rm Tr} F \wedge F
\; \; \; \; .
\eqno (3.3.36)
$$
As Alvarez-Gaum\'e (1983b) pointed out, this calculation
has great relevance for the theory of anomalies in quantum
field theory. Indeed, anomaly calculations involve usually
traces of the form
$\sum_{n} \psi_{n}^{\dagger}(x) \; L \; \psi_{n}(x)$,
where $L$ is a differential or algebraic operator, and the
$\psi_{n}'s$ are eigenfunctions of the Dirac operator in the
presence of external fields (either gravitational or gauge
fields). The analysis leading to (3.3.36) has shown that
such traces can be turned into functional integrals for
one-dimensional field theories. In other words, anomaly
calculations are equivalent to the analysis of partition
functions in quantum mechanics (Alvarez-Gaum\'e 1983b).

Indeed, another definition of index is also available in the
literature. This is motivated by the analysis of a family
$D$ of elliptic operators $D_{x}$, parametrized by the
points $x$ of a compact space, say $X$. The corresponding 
index is then defined by (Rennie 1990)
$$
{\rm index}(D) \equiv {\rm Ker}(D)-{\rm Coker}(D)
\; \; \; \; ,
\eqno (3.3.37)
$$
where ${\rm Ker}(D)$ denotes the {\it family} of vector
spaces ${\rm Ker}(D_{x})$, and similarly for 
${\rm Coker}(D)$. The index defined in (3.3.37) turns out
to be an element of the $K$-theory of $X$ (see section 3.7).
An outstanding open problem is to prove that the
supersymmetric functional integral admits a $K$-theoretic
interpretation. Moreover, it remains to be seen whether
supersymmetry can be applied to prove the index theorem
for families (cf. Atiyah and Singer 1971a). The author is
indebted to Professor L. Alvarez-Gaum\'e for correspondence
about these issues.
\vskip 1cm
\centerline {\bf 3.4 Index of Two-Parameter Families}
\vskip 1cm
A very important property of the index is that it is
unchanged by perturbation of the operator, hence it can
be given by a topological formula. This is why, when we
vary the gauge field $A$ continuously, the index of the
extended Dirac operator $D_{A}$ does not change. It is 
however possible to extract more topological information from 
a continuous family of operators, and we here focus on the
two-parameter case (Atiyah 1984). 

For this purpose we consider a two-dimensional surface $X$
compact, connected and oriented, and suppose a fibre bundle
$Y$ over $X$ is given with fibre $M$. The twist involved in 
making a non-trivial bundle is an essential topological
feature. By giving $Y$ a Riemannian metric, one gets metrics
$g_{x}$ on all fibres $M_{x}$, hence one has metrics on
$M$ parametrized by $X$. The corresponding family of Dirac
operators on $M$ is denoted by $D_{x}$. We can now define
two topological invariants, i.e. ${\rm index}(D_{x})$ and
the {\it degree} of the family of Dirac operators. To define
such a degree, we restrict ourselves to the particular case
when the index vanishes.

Since ${\rm index}(D_{x})=0$ by hypothesis, the generic
situation (one can achieve this by perturbing the metric 
on $Y$) is that $D_{x}$ is invertible $\forall x \in X$
except at a finite number of points $x_{1},...,x_{k}$.
Moreover, a multiplicity $\nu_{i}$ can be assigned to
each $x_{i}$, which is generically $\pm 1$. To obtain
this one can, in the neighbourhood of each $x_{i}$, replace
$D_{x}$ by a finite matrix $T_{x}$. Zeros of ${\rm det}(T_{x})$
have multiplicity $\nu_{i}$ at $x_{i}$. We are actually dealing
with the phase-variation of ${\rm det}(T_{x})$ as $x$
traverses positively a small circle centred at $x_{i}$.

The degree $\nu$ of the family of Dirac operators $D_{x}$
is defined by
$$
\nu \equiv \sum_{i=1}^{k}\nu_{i}
\; \; \; \; .
\eqno (3.4.1)
$$
This is a topological invariant of the family in that it is
invariant under perturbation. Hence it only depends on the
topology of the fibre bundle $Y$ and vanishes for the 
product $M \times X$. In the case of gauge fields, one can
fix the metric $g$ on $X$ and take $Y= M \times X$, but we
also take a vector bundle $V$ on $Y$ with a connection.
This gives a family $A_{x}$ of gauge fields, and the
corresponding Dirac family has again a degree which now
depends on the topology of $V$. Such a degree vanishes if
the vector bundle $V$ comes from a bundle on $M$.
\vskip 1cm
\centerline {\bf 3.5 Determinants of Dirac Operators}
\vskip 1cm
For fermionic fields one has to define determinants of
Dirac operators, and these are not positive-definite by
virtue of their first-order nature. Consider as before 
an even-dimensional Riemannian manifold $(M,g)$ with Dirac
operators $D_{g}:S^{+} \rightarrow S^{-}$, depending on the
metric $g$. We look for a regularized complex-valued 
determinant ${\rm det}(D_{g})$ which should have the
following properties (Atiyah 1984):

(1) ${\rm det}(D_{g})$ is a differentiable function
of $g$.

(2) ${\rm det}(D_{g})$ is gauge-invariant, i.e.
${\rm det}\Bigr(D_{f_{(g)}}\Bigr)={\rm det}(D_{g})$ for any
diffeomorphism $f$ of $M$.

(3) ${\rm det}(D_{g})=0$ to first-order if and only if
$D_{g}^{*} \; D_{g}$ has exactly one zero-eigenvalue.
\vskip 0.3cm
\noindent
The third property is the characteristic property of determinants
in finite dimensions and it is a minimum requirement for
any regularized determinant. Note also that, since
$D_{g}^{*} \; D_{g}$ is a Laplace-type operator, there is no
difficulty in defining its determinant, so that one can also
define $\left | {\rm det}(D_{g}) \right |$. The problem is to
define the {\it phase} of ${\rm det}(D_{g})$ (cf. section 6
of Martellini and Reina 1985).

Suppose now that one can find a fibration over a surface $X$ with
fibre $M$, so that the Dirac operators $D_{x}$ all have
vanishing index whereas the {\it degree} of the family does
not vanish. This implies that there is no function ${\rm det}(D_{g})$
with the 3 properties just listed. Suppose in fact that such a
${\rm det}(D_{g})$ exists. Then (1) and (2) imply that 
$h(x) \equiv {\rm det} \Bigr(D_{g_{x}}\Bigr)$ is a differentiable
complex-valued function on $X$. Moreover, (3) implies that 
$\nu$ is the number of zeros of $h$ counted with multiplicities
and signs, i.e. that $\nu$ is 
the degree of $h:X \rightarrow {\cal C}$.
But topological arguments may be used to show that the degree
of $h$ has to vanish, hence $\nu=0$, contradicting the
assumption.

We should now bear in mind that the local multiplicities 
$\nu_{i}$ at points $x_{i}$ where $D_{x}$ is not invertible, 
are defined as local phase variations obtained by using an
eigenvalue cut-off. Remarkably, if it were possible to use
such a cut-off consistently at all points, the total phase
variation $\nu$ would have to vanish. Thus, a fibration with
non-vanishing degree implies a behaviour of the eigenvalues
which prevents a good cut-off regularization. Such a remark is
not of purely academic interest, since two-dimensional examples
can be found, involving surfaces $M$ and $X$ of fairly high
genus, where all these things happen. What is essentially
involved is the variation in the conformal structure of $M$
(cf. section 5.3).

Other examples of non-vanishing degree, relevant for gauge theory,
are obtained when $M$ is the $2n$-sphere $S^{2n}$, $X=S^{2}$
and the vector bundle $V$ is $(n+1)$-dimensional, so that the
gauge group $G$ is $U(n+1)$ or $SU(n+1)$. 
If one represents $S^{2}$ as ${\Re}^{2} 
\cup \infty$ then, removing the
point at infinity, one obtains a two-parameter family of gauge 
fields on $M$ parametrized by ${\Re}^{2}$. Since this family is
defined by a bundle over $M \times S^{2}$, gauge fields are all
gauge-equivalent as one goes to $\infty$ in ${\Re}^{2}$. What are
we really up to ? By taking a 
large circle in ${\Re}^{2}$, one gets
a closed path in the group ${\cal G}$ of gauge transformations
of the bundle on $M$. The degree $\nu$ is essentially a
homomorphism $\pi_{1}({\cal G}) \rightarrow Z$, and if
$\nu$ does not vanish, the variation in phase of
${\rm det}(D_{g})$ going round this path in ${\cal G}$ proves
the lack of gauge-invariance. If $M=S^{2n}$ one has
$$
\pi_{1}({\cal G})=\pi_{2n+1}(G)
\; \; \; \; ,
\eqno (3.5.1)
$$
$$
\pi_{2n+1} \Bigr(SU(n+1)\Bigr)=Z
\; \; \; \; ,
\eqno (3.5.2)
$$
and the isomorphism is given by $\nu$, by virtue of the general
index theorem for families (Atiyah and Singer 1971a). 
\vskip 1cm
\centerline {\bf 3.6 Bott Periodicity}
\vskip 1cm
A crucial and naturally occurring problem in all our 
investigations is why the Dirac operator is so important
in index theory. The answer is provided by the so-called
Bott periodicity theorem. To understand the key features,
let us consider an elliptic operator on ${\Re}^{n}$ with
constant coefficients, say $P$, which acts on 
${\cal C}^{m}$-valued functions. The symbol of $P$, denoted
by $\sigma_{P}(\xi)$, provides a map
$$
\sigma_{P}: S^{n-1} \rightarrow GL(m,{\cal C})
\; \; \; \; .
$$
For example, the restricted Dirac operator on ${\Re}^{2k}$,
say $B$, yields the map
$$
\sigma_{B}: S^{2k-1} \rightarrow GL(2^{k-1},{\cal C})
\; \; \; \; .
$$
The Bott periodicity theorem states that this is a stable
generator, in that
\vskip 0.3cm
\noindent
(i) The map
$$
\pi_{i}(GL(m,{\cal C})) \rightarrow 
\pi_{i}(GL(m+1,{\cal C}))
$$
is an isomorphism for $m > {i\over 2}$;
\vskip 0.3cm
\noindent
(ii) The limit group $\pi_{i}(GL(\infty))$ 
is $0$ if $i$ is even, and coincides with $Z$, generated
by $\sigma_{B}$, if $i$ is odd.
\vskip 0.3cm
\noindent
We now find it helpful to state the Bott theorem in an
even simpler form. In other words, one studies the
homotopy groups of certain Lie groups, and one finds that
they are periodic, in that (Rennie 1990)
$$
\pi_{n}(U(N))=0 \; \; {\rm if} \; \; n \; \; 
{\rm is} \; \; {\rm even} \; \; {\rm and} \; \; 
n < 2N \; \; \; \; ,
\eqno (3.6.1)
$$
$$
\pi_{n}(U(N))=Z \; \; {\rm if} \; \; n \; \; 
{\rm is} \; \; {\rm odd} \; \; {\rm and} \; \;
n > 2N \; \; \; \; .
\eqno (3.6.2)
$$
Moreover, in terms of the $N \rightarrow \infty$ limit,
one has the periodicity
$$
\pi_{n}(U(\infty))=\pi_{n+2}(U(\infty))
\; \; \; \; ,
\eqno (3.6.3)
$$
$$
\pi_{n}(O(\infty))=\pi_{n+8}(O(\infty))
\; \; \; \; .
\eqno (3.6.4)
$$

A global version of the Bott theorem, resulting from the
local form, is also available. According to the global
Bott theorem, the symbol of the Dirac operator on a 
compact manifold is a generator, in a suitable sense, of
all elliptic symbols. More precisely, let $X$ be a 
compact, oriented, $2k$-dimensional manifold endowed with
a spin-structure (see section 1.3). We know from section
2.3 that an elliptic differential operator
$$
P: C^{\infty}(X,V) \rightarrow C^{\infty}(X,W)
$$
has a symbol, say $\sigma_{P}$, which is an isomorphism
$$
\pi^{*}V \rightarrow \pi^{*}W
\; \; \; \; ,
$$
where $\pi: S(X) \rightarrow X$ is the projection map 
of the unit sphere-bundle. A {\it trivial symbol} is
a symbol for which $V=W$ and $\sigma$ is the identity 
map. Two symbols are said to be {\it equivalent} if
they become homotopic after adding trivial symbols.

The global Bott theorem can be then stated by saying
that every symbol is equivalent to the symbol of 
$B_{V}$, for some $V$, where $B_{V}$ is the restricted
Dirac operator, extended so as to act on the tensor
product $E^{+} \otimes V$. For index problems, the
interest of this property lies in the fact that two
elliptic operators whose symbols are equivalent turn
out to have the same index. Thus, if one is able to
find the index of $B_{V}$ for all $V$, one can indeed
obtain an index formula for all elliptic operators on $X$.
\vskip 1cm
\centerline {\bf 3.7 K-Theory}
\vskip 1cm 
The global Bott theorem and the global theory of the Dirac
operator find their natural formulation within the
framework of $K$-theory. The formalism is as abstract as
powerful, and some key steps are as follows (Atiyah 1967).

We are interested in complex vector bundles over a compact
manifold $X$. The set $Vect(X)$ of isomorphism classes  
of vector bundles over $X$ has the structure of an Abelian
semi-group (the same holds if $X$ is any space), where the
additive structure is defined by direct sum. If $A$ is
{\it any} Abelian semi-group (recall that a semi-group is
a set of elements $\left \{ a,b,c,... \right \}$ with a
binary operation of product, say $ab$, such that 
$(ab)c=a(bc)$), one can associate to $A$ an
Abelian group, $K(A)$, with the following property: there
is a semi-group homomorphism:
$$
\alpha : A \rightarrow K(A)
\eqno (3.7.1)
$$
such that, if $G$ is {\it any} group, and if
$$
\gamma : A \rightarrow G
\eqno (3.7.2)
$$
is {\it any} semi-group homomorphism, there is a {\it unique}
homomorphism
$$
\kappa : K(A) \rightarrow G
\eqno (3.7.3)
$$
such that
$$
\gamma= \kappa \; \alpha
\; \; \; \; .
\eqno (3.7.4)
$$
If such a $K(A)$ exists, it must be unique. Note that the
order of factors on the right-hand side of Eq. (3.7.4) is
of the right sort: first, $\alpha$ maps $A$ into $K(A)$, then
$\kappa$ maps $K(A)$ into $G$. Altogether, this is a map of
$A$ into $G$. What is non-trivial is the uniqueness property
of the homomorphism.

To define the group $K(A)$, one starts from $F(A)$, the
free Abelian group generated by the elements of $A$. Second,
one considers $E(A)$, the subgroup of $F(A)$ generated by
those elements of the form 
$$
a+a'-(a \oplus a')
\; \; \; \; ,
$$
where $\oplus$ is the addition in $A$, and $a,a' \in A$. 
Then
$$
K(A) \equiv F(A)/E(A)
\eqno (3.7.5)
$$
is the desired Abelian group. If $A$ possesses a multiplication
which is distributive over the addition of $A$, then $K(A)$
is a ring.

Now if $X$ is a space, one writes
$$
K(X) \equiv K(Vect(X))
\eqno (3.7.6)
$$
for the ring $K(Vect(X))$. Moreover, if $E \in Vect(X)$, one
writes $[E]$ for the image of $E$ in $K(X)$. One can prove that
every element of $K(X)$ is of the form $[E]-[F]$, where
$E,F$ are bundles over $X$. Let now $T$ be a bundle such that
$F \oplus T$ is trivial. On writing ${\underline n}$ for the 
trivial bundle of dimension $n$, this means that
$$
F \oplus T = {\underline n}
\; \; \; \; .
\eqno (3.7.7)
$$
One can thus obtain a very useful formula for a generic
element of $K(X)$:
$$
[E]-[F]=[E]+[T]-([F]+[T])
=[E \oplus T]-[{\underline n}]
\; \; \; \; .
\eqno (3.7.8)
$$
In other words, every element of $K(X)$ is of the form
$[H]-[{\underline n}]$. If two bundles become equivalent
when a suitable trivial bundle is added to each of them,
the bundles are said to be {\it stably equivalent}. Thus,
$[E]=[F]$ if and only if $E$ and $F$ are stably equivalent

For a space with base point $x_{0}$, one defines
$$
{\widetilde K}(X) \equiv {\rm Ker} \left \{
K(X) \rightarrow K(x_{0}) \right \}
\; \; \; \; .
\eqno (3.7.9)
$$
Since $K({\rm point}) \cong Z$, one has the isomorphism
$$
K(X) \cong {\widetilde K}(X) \oplus Z
\; \; \; \; .
\eqno (3.7.10)
$$
For a closed sub-space of $X$, say $Y$, one defines
$$
K(X,Y) \equiv {\widetilde K}(X/Y)
\; \; \; \; ,
\eqno (3.7.11)
$$
where $X/Y$ is obtained by collapsing $Y$ to a point,
taken as the base point. For a locally compact space,
one sets
$$
K(X) \equiv {\widetilde K}(X^{+})
\; \; \; \; ,
\eqno (3.7.12)
$$
where $X^{+}$ is the one-point compactification of $X$.
Another elementary lemma can then be proved:
\vskip 0.3cm
\noindent
If $Y \subset X$ is a compact pair, then
$$
K(X,Y) \rightarrow K(X) \rightarrow K(Y)
$$
is an exact sequence.

The tensor product of vector bundles induces a ring structure
on $K(X)$, and $K(Y,X)$ becomes a module on $K(X)$. The
identity map
$$
S^{1} \rightarrow U(1) \subset GL(1,{\cal C})
$$
defines a line-bundle, say $H$, on $S^{2}={\cal C}^{+}$.
Let $b \in K({\cal C})$ be the class of $H-1$. The local
form of the Bott periodicity theorem is therefore
(Atiyah 1975a):
\vskip 0.3cm
\noindent
{\bf Theorem 3.7.1} Multiplication by $b$ provides an
isomorphism
$$
\beta: K(X) \rightarrow K({\cal C} \times X)
\; \; \; \; .
$$
On taking $X={\cal C}^{n}$ and applying induction for
$n \geq 0$, one finds that $K({\cal C}^{n}) \cong Z$.
Moreover, the symbol of the restricted Dirac operator, 
$B$, is eventually identified with the element $b^{n}$.

In the analysis of global properties, one first proves
that an elliptic symbol $\sigma$ defines an element
$[\sigma]$ of $K(T^{*}X)$. The global Bott theorem is then:
\vskip 0.3cm
\noindent
{\bf Theorem 3.7.2} For an even-dimensional spin-manifold,
multiplication by $\sigma_{B}$ provides the isomorphism
$$
K(X) \rightarrow K(T^{*}X)
\; \; \; \; .
$$
In particular, if the manifold $X$ is parallelizable,
theorem 3.7.2 reduces to theorem 3.7.1. In the general
case, one takes a covering of $X$, on which $T^{*}X$
is trivial, and one applies induction by using
the long exact sequence which extends
$$
K(X,Y) \rightarrow K(X) \rightarrow K(Y)
\; \; \; \; .
$$
This long exact sequence involves higher-order groups,
denoted by $K^{-n}(X)$, defined, for $n \geq 0$,
by means of
$$
K^{-n}(X) \equiv K({\Re}^{n} \times X)
\; \; \; \; .
\eqno (3.7.13)
$$

One may wonder what can be said about manifolds 
which are not spin-manifolds. Indeed,
if $X$ is not a spin-manifold one cannot define a Dirac
operator. One can, however, build a signature operator
(see section 2.3), say $A$. Locally, the symbol of $A$
is equivalent to a multiple of the symbol of $B$
(Atiyah 1975a):
$$
\sigma_{A}=2^{k} \; \sigma_{B}
\; \; \; \; .
\eqno (3.7.14)
$$
\vskip 1cm
\centerline {\bf 3.A Appendix}
\vskip 1cm
An elliptic boundary-value problem consists of the pair
$(P,B)$, where, for a given vector bundle $V$ over a
Riemannian manifold $M$, $P$ is a differential operator
of order $d$ on $V$, and $B$ is a boundary operator (see
(5.1.18)). The {\it strong ellipticity} of $(P,B)$ is
defined in terms of the eigenvalue equation for the leading
symbol of $P$, jointly with an asymptotic condition on the
solution of this equation. For us to be able to define these
concepts, it is necessary to start from the structure which
makes it possible to split the tangent bundle of $M$ in the
form of a direct sum (Gilkey 1995)
$$
T(M)=T(\partial M) \oplus T([0,\delta))
\; \; \; \; .
\eqno (3.A.1)
$$
Indeed, following Gilkey 1995, one can use the inward 
geodesic flow to identify a neighbourhood of $\partial M$
with the {\it collar}
$$
{\cal C}: {\partial M} \times [0,i(M))
\; \; \; \; ,
\eqno (3.A.2)
$$
$i(M)$ being the injectivity radius. If $(y_{1},...,y_{m-1})$ 
are local coordinates on the boundary, and if $x_{m}$ is the
geodesic distance to the boundary, a system of local coordinates
on the collar is then $(y_{1},...,y_{m-1},x_{m})$. The normal
derivatives $D_{m}^{k} \Bigr(f \mid_{{\cal C}}\Bigr)$ are
well defined after identifying $V \mid_{{\cal C}}$ with
$V \times [0,i(M))$. We now need to define $d$-graded vector
bundles ($d$ being the order of $P$) and their auxiliary
bundles. For this purpose, following Gilkey 1995, we say
that a $d$-graded vector bundle is a vector bundle, say $U$,
endowed with a decomposition into $d$ sub-bundles
$$
U=U_{0} \oplus ... \oplus U_{d-1}
\; \; \; \; .
\eqno (3.A.3)
$$
Moreover, we consider $W$, defined by
$$
W \equiv [V]_{\partial M} \oplus ... \oplus
[V]_{\partial M}
\; \; \; \; ,
\eqno (3.A.4)
$$
i.e. the $d$-graded vector bundle for boundary data, and
define (Gilkey 1995)
$$
f_{i} \equiv \Bigr[D_{\; m}^{i} (f \mid_{{\cal C}})
\Bigr]_{\partial M}
\; \; \; \; .
\eqno (3.A.5)
$$
The boundary data map is then a map
$$
{\overline \gamma}: C^{\infty}(V) \rightarrow C^{\infty}(W)
$$
such that
$$
{\overline \gamma}(f) \equiv (f_{0},...,f_{d-1})
\; \; \; \; .
\eqno (3.A.6)
$$
Further to this, one has to take into account an {\it auxiliary}
$d$-graded vector bundle over the boundary, say $Y$, whose
dimension satisfies
$$
{\rm dim}(Y)={d\over 2}{\rm dim}(V)
\; \; \; \; .
\eqno (3.A.7)
$$
If $B: C^{\infty}(W) \rightarrow C^{\infty}(Y)$ is a tangential
differential operator defined on $\partial M$, one can 
decompose it as follows:
$$
B \equiv B_{ij}
\; \; \; \; ,
\eqno (3.A.8a)
$$
$$
B_{ij}: C^{\infty}(W_{i}) \rightarrow C^{\infty}(Y_{j})
\; \; \; \; ,
\eqno (3.A.8b)
$$
with ${\rm ord}(B_{ij}) \leq j-i$. Sections of 
$C^{\infty}(W_{i})$ arise from taking the normal derivative
of order $i$.

All these geometric objects make it possible to define the
{\it d-graded leading symbol} of $B$ as follows (Gilkey 1995):
$$
\sigma_{g}(B)_{ij}(y,\zeta) \equiv \sigma_{L}(B_{ij})(y,\zeta)
\; \; {\rm if} \; \; {\rm ord}(B_{ij})=j-i
\; \; \; \; ,
\eqno (3.A.9a)
$$
$$
\sigma_{g}(B)_{ij}(y,\zeta) \equiv 0 \; \; {\rm if} \; \; 
{\rm ord}(B_{ij}) < j-i
\; \; \; \; .
\eqno (3.A.9b)
$$
With the notation described so far, the boundary condition is
expressed by the equation
$$
B \; {\overline \gamma} \; f=0
\; \; \; \; .
\eqno (3.A.10)
$$

To define {\it strong ellipticity} of the pair $(P,B)$ one
assumes that $P$ is a differential operator on $V$ of order $d$,
with elliptic leading symbol, say $p_{d}$. If $\cal K$ is a cone
containing $0$ and contained in $\bf C$, and such that, for
$\xi \not = 0$,
$$
{\rm Spec}(p_{d}(x,\xi)) \subset {\cal K}^{c}
\; \; \; \; ,
\eqno (3.A.11)
$$
one studies, on the boundary, the equation
$$
p_{d}(y,0,\zeta,D_{r})f(r)=\lambda \; f(r)
\; \; \; \; ,
\eqno (3.A.12)
$$
subject to the asymptotic condition
$$
\lim_{r \to \infty}f(r)=0
\; \; \; \; ,
\eqno (3.A.13)
$$
and under the assumption that $(0,0) \not = (\zeta,\lambda)
\in T^{*}(\partial M) \times {\cal K}$. By definition, the
pair $(P,B)$ is said to be {\it strongly elliptic} with respect
to the cone $\cal K$ if, for any $w$ belonging to the auxiliary
$d$-graded vector bundle $Y$, there exists a {\it unique} 
solution of the problem described by Eqs. (3.A.12) and (3.A.13),
with (Gilkey 1995)
$$
\sigma_{g}(B)(y,\zeta){\overline \gamma}(f)=w
\; \; \; \; .
\eqno (3.A.14)
$$
The cone $\cal K$ is normally assumed to coincide with the set
of complex numbers minus ${\Re}^{+}$ (or minus $\Re$ and
${\Re}^{+}$). 

For example, if $P$ is an operator of Laplace type on
$C^{\infty}(V)$, one finds (Gilkey 1995)
$$
p_{d}(y,0,\zeta,D_{r})f(r)=-\partial_{r}^{2}
+{\mid \zeta \mid}^{2}
\; \; \; \; ,
\eqno (3.A.15)
$$
and hence the solutions of Eq. (3.A.12) read
$$
f(r)=w_{0}e^{\mu r}+w_{1}e^{-\mu r}
\; \; \; \; ,
\eqno (3.A.16)
$$
where $\mu \equiv \sqrt{{\mid \zeta \mid}^{2}-\lambda}$.
Bearing in mind that ${\rm Re}(\mu)>0$, the asymptotic
condition (3.A.13) picks out the solution of the form
$$
{\widetilde f}(r)=w_{1} e^{-\mu r}
\; \; \; \; .
\eqno (3.A.17)
$$
These properties are useful in the course of proving that
$P$ is strongly elliptic with respect to the cone 
${\bf C}-{\Re}^{+}$, when the boundary conditions are a
mixture of Dirichlet and Robin conditions (see our Eq.
(5.4.2) and Gilkey 1995). A thorough treatment of elliptic
boundary-value problems, with emphasis on the analytic
approach, may be found in Grubb 1996.
\vskip 1cm
\centerline {\bf 3.B Appendix}
\vskip 1cm
This appendix describes briefly some concepts in homotopy
theory and in the theory of characteristic classes. It is
intended to help the readers who have already attended 
introductory courses, but do not have enough time to read
comprehensive monographs or long review papers.
\vskip 0.3cm
\noindent
(i) A very important concept in homotopy theory is that
of an exact sequence. A sequence of groups and homomorphisms
$$
... {\mathrel{\mathop{\longrightarrow}^{f_{i-1}}}}H_{i}
{\mathrel{\mathop{\longrightarrow}^{f_{i}}}}H_{i+1}
{\mathrel{\mathop{\longrightarrow}^{f_{i+1}}}}H_{i+2}...
$$
is called an {\it exact sequence} if, for all $i$,
$$
{\rm Im} \; f_{i-1}={\rm Ker} \; f_{i}
\; \; \; \; .
\eqno (3.B.1)
$$
A {\it short} exact sequence is a five-term sequence with
trivial end groups, i.e.
$$
0 \longrightarrow H 
{\mathrel{\mathop{\longrightarrow}^{f}}}H' 
{\mathrel{\mathop{\longrightarrow}^{f'}}}H'' 
\longrightarrow 0
\; \; \; \; .
\eqno (3.B.2)
$$
\vskip 0.3cm
\noindent
(ii) If $m$ is a complex $k \times k$ matrix, and $Q(m)$ is
a polynomial in the components of $m$, then $Q(m)$ is called
a {\it characteristic polynomial} if 
$$
Q(m)=Q(h^{-1} \; m \; h)
\; \; \; \; 
\forall h \in GL(k,{\cal C})
\; \; \; \; .
\eqno (3.B.3)
$$
The polynomial $Q(m)$ is a symmetric function of the
eigenvalues $\left \{ \lambda_{1},...,\lambda_{k}
\right \}$ of $m$. The $j$-th symmetric polynomial,
say $T_{j}(\lambda)$, reads
$$
T_{j}(\lambda)=\sum_{i_{1}<i_{2}<...i_{j}}
\lambda_{i_{1}} \; \lambda_{i_{2}} 
... \lambda_{i_{j}}
\; \; \; \; ,
\eqno (3.B.4)
$$
and $Q(m)$ is a polynomial in the $T_{j}(\lambda)$:
$$
Q(m)=\alpha + \beta T_{1}(\lambda)+\gamma T_{2}(\lambda)
+\delta [T_{1}(\lambda)]^{2}+...
\; \; \; \; .
\eqno (3.B.5)
$$
If, in $Q(m)$, one replaces $m$ by the curvature 
two-form $\Omega$, one finds that $Q(\Omega)$ is closed.

The {\it total Chern form} $c(\Omega)$ of a complex vector
bundle $E$ over $M$, with $GL(k,{\cal C})$ transition
functions and connection $\omega$, is defined in terms
of the characteristic polynomial ${\rm det}(I+m)$ for
$\Omega$:
$$
c(\Omega)={\rm det}\left(I+{i\over 2\pi}\Omega \right)
=\sum_{k=0}^{\infty}c_{k}(\Omega)
\; \; \; \; ,
\eqno (3.B.6)
$$
where the various Chern forms in (3.B.6) read
$$
c_{0}(\Omega)=1
\; \; \; \; ,
\eqno (3.B.7)
$$
$$
c_{1}(\Omega)={i\over 2\pi}{\rm Tr}\Omega
\; \; \; \; ,
\eqno (3.B.8)
$$
$$
c_{2}(\Omega)={1\over 8\pi^{2}}
\Bigr[{\rm Tr}\Omega \wedge \Omega
-{\rm Tr}\Omega \wedge {\rm Tr} \Omega \Bigr]
\; \; \; \; ,
\eqno (3.B.9)
$$
plus infinitely many other formulae that can be derived
from the expansion of the determinant in (3.B.6).
Since $Q(\Omega)$ is closed, one finds that any homogeneous
polynomial in the expansion of a characteristic polynomial
is closed:
$$
d \; c_{j}(\Omega)=0
\; \; \; \; .
\eqno (3.B.10)
$$
This means that the Chern forms, $c_{j}(\Omega)$, define
cohomology classes belonging to $H^{2j}(M)$.
\vskip 0.3cm
\noindent
(iii) The {\it Chern numbers} of a fibre bundle are the
numbers found by integrating characteristic polynomials
over the manifold. For example, one has
$$
C_{2}(E) \equiv \int_{M} c_{2}(\Omega)
\; \; \; \; .
\eqno (3.B.11)
$$
\vskip 0.3cm
\noindent
(iv) {\it Pontrjagin classes}, say $p_{k}(E)$, and
{\it Pontrjagin numbers}, are the counterpart, for
{\it real} vector bundles, of Chern classes and Chern
numbers for complex vector bundles.
\vskip 0.3cm
\noindent
(v) In the course of studying index theorems, one needs 
certain combinations of characteristic classes. The first
example is provided by the {\it Chern character},
defined by the invariant polynomial
$$
ch(E) \equiv {\rm Tr} \exp (im / 2\pi)
=\sum_{l=0}^{\infty}{1\over l!}
{\rm Tr}(im / 2\pi)^{l}
\; \; \; \; .
\eqno (3.B.12)
$$
A deep and simple relation exists between the Chern
character and Chern classes, i.e.
$$
ch(E)=\sum_{l=1}^{k}\exp (i \Omega_{l} / 2\pi)
=k+c_{1}(E)+{1\over 2}\Bigr(c_{1}^{2}-2c_{2}\Bigr)(E)
+... \; \; \; \; .
\eqno (3.B.13)
$$
\vskip 0.3cm
\noindent
(vi) Yet another deep concept is the one of {\it genus},
i.e. a combination of characteristic classes that
satisfies the Whitney sum property:
$$
f(E \oplus E')=f(E)f(E')
\; \; \; \; .
\eqno (3.B.14)
$$
It is now convenient to set
$$
k_{l} \equiv {i\over 2\pi} \Omega_{l}
\; \; \; \; .
\eqno (3.B.15)
$$
It is then possible to express the {\it Todd class}
as follows:
$$
td(E) \equiv \prod_{l=1}^{k}{x_{l}\over 1-\exp(-x_{l})}
=1+{1\over 2}c_{1}(E)+{1\over 12}\Bigr(c_{1}^{2}+c_{2}
\Bigr)(E)+...
\; \; \; \; .
\eqno (3.B.16)
$$
Other relevant examples of genera are given by the
{\it Hirzebruch L-polynomial}, i.e.
$$
L(E) \equiv \prod_{l}{x_{l}\over \tanh (x_{l})}
\; \; \; \; ,
\eqno (3.B.17)
$$
and the $\hat A$ polynomial
$$
{\hat A}(E) \equiv \prod_{l}{{x_{l}/2} \over \sinh(x_{l}/2)}
=1-{1\over 24}p_{1}+{1\over 5760}
\Bigr(7 p_{1}^{2}-4 p_{2} \Bigr)+...
\; \; \; \; .
\eqno (3.B.18)
$$
\vskip 0.3cm
\noindent
(vii) A further important set of characteristic classes
consists of the Stiefel-Whitney classes. Unlike all 
characteristic classes described previously, they cannot
be represented by differential forms in terms of
curvature, and are not integral cohomology classes
(Milnor and Stasheff 1974, Rennie 1990). 
By definition, Stiefel-Whitney
classes are the $Z_{2}$ cohomology classes of a {\it real}
bundle, say $E$, over $M$, with k-dimensional fibre:
$$
w_{i} \in H^{i}(M;Z_{2})
\; \; , \; \; i=1,...,n-1
\; \; \; \; .
\eqno (3.B.19)
$$
The {\it total Stiefel-Whitney class} is defined by
$$
w(E) \equiv 1+\sum_{l=1}^{n} w_{l}
\; \; \; \; .
\eqno (3.B.20)
$$
The vanishing of the first Stiefel-Whitney 
class $w_{1}(TM)$ provides a
necessary and sufficient condition for the orientability
of $M$. The vanishing of the second Stiefel-Whitney class
$w_{2}(TM)$ is instead a necessary and sufficient condition
for the existence of spin-structures (see sections 1.2,
1.3 and Milnor 1963).
\vskip 100cm
\centerline {\it CHAPTER FOUR}
\vskip 1cm
\centerline {\bf SPECTRAL ASYMMETRY}
\vskip 1cm
\noindent
{\bf Abstract.} The investigations in spectral asymmetry
and Riemannian geometry by Atiyah, Patodi and Singer
begin by studying, for a Riemannian manifold, say $M$,
the relation between the signature of the quadratic form
on $H^{2}(M;{\Re})$ and the integral over $M$ of the first
Pontrjagin class. It turns out that, if $M$ has a boundary,
the desired relation involves also the value
at the origin of the $\eta$-function obtained from the
eigenvalues of a first-order differential operator. The result is
indeed an example of index theorem for non-local boundary
conditions. The second part of the chapter performs a
detailed $\eta(0)$ calculation with non-local boundary
conditions. Last, the $\eta(0)$ value is obtained for a
first-order operator with periodic boundary conditions.
\vskip 100cm
\centerline {\bf 4.1 Spectral Asymmetry and Riemannian Geometry}
\vskip 1cm
If $M$ is an oriented, $n$- (even) dimensional
Riemannian manifold, the Gauss-Bonnet-Chern theorem
provides a first relevant example of a deep formula relating
cohomological invariants with curvature. In intrinsic
language, the Euler number $\chi$ of $M$, defined as an
alternating sum of Betti numbers:
$$
\chi \equiv \sum_{p=0}^{n}(-1)^{p}B_{p}
\; \; \; \; ,
\eqno (4.1.1)
$$
reads (Chern 1944--45, Greub et al. 1973, Dowker and Schofield 1990)
$$
\chi=\int_{M} \Lambda 
+ \int_{\partial M} X^{*} \Pi
\; \; \; \; ,
\eqno (4.1.2a)
$$
where the vector field $X$ can 
be chosen as any extension of the normal
vector field $\bf n$ on the boundary $\partial M$. Moreover,
$\Pi$ is an $(n-1)$-form on the unit tangent bundle of $M$,
the sphere-bundle $S(M)$, and $\Lambda$ is an $n$-form such
that $\pi^{*}\Lambda$ is the Pfaffian of the curvature 
matrix, where $\pi^{*}$ maps the cohomology of $M$ into that
of $S(M)$. For example, in the two-dimensional case 
Eq. (4.1.2a) reduces to
$$
\chi={1\over 4 \pi}\int_{M}R
+{1\over 2\pi}\int_{\partial M}{\rm Tr}K
\; \; \; \; ,
\eqno (4.1.2b)
$$
where $R$ is the trace of the Ricci tensor, and $K$ is the
second fundamental form of the boundary of $M$ (see
section 5.2).

When $M$ is four-dimensional, 
in addition to (4.1.2a) there is
another formula which relates cohomological invariants with
curvature. In fact, it is known that the signature (i.e.
number of positive eigenvalues minus number of negative 
eigenvalues) of the quadratic form on $H^{2}(M;{\Re})$ 
given by the cup-product (see section 2.3 and Atiyah 1975a)
is expressed, for manifolds without boundary, by 
$$
{\rm sign}(M)={1\over 3} \int_{M}p_{1}
\; \; \; \; .
\eqno (4.1.3)
$$
In (4.1.3), $p_{1}$ is the differential four-form which
represents the first Pontrjagin class, and is equal
to $(2\pi)^{-2} \; {\rm Tr}(R^{2})$, where $R$ is the curvature
matrix. However, (4.1.3) does not hold in general for
manifolds with boundary, so that one has
$$
{\rm sign}(M)-{1\over 3}\int_{M}p_{1}=f(Y) \not = 0
\; \; \; \; ,
\eqno (4.1.4)
$$
where $Y \equiv \partial M$. Thus, if $M'$ is another manifold
with the same boundary, i.e. such that $Y = \partial M'$, one
has
$$
{\rm sign}(M)-{1\over 3}\int_{M}p_{1}
={\rm sign}(M')-{1\over 3}\int_{M'}p_{1}'
\; \; \; \; .
\eqno (4.1.5)
$$
Hence one is looking for a continuous function $f$ of the
metric on $Y$ such that $f(-Y)=-f(Y)$. Atiyah et al. (1975)
were able to prove that $f(Y)$ is a spectral invariant,
evaluated as follows. One looks at the Laplace operator 
$\triangle$ acting on forms as well as on scalar functions.
This operator $\triangle$ is the square of the self-adjoint
first-order operator $B \equiv \pm \Bigr(d*-*d\Bigr)$, where
$d$ is the exterior-derivative operator, and $*$ is the
Hodge-star operator mapping $p$-forms to $(l-p)$-forms in
$l$ dimensions. Thus, if $\lambda$ is an eigenvalue of $B$, 
the eigenvalues of $\triangle$ are of the form $\lambda^{2}$.
However, the eigenvalues of $B$ can be both negative and positive.
One takes this property into account by defining the
$\eta$-function
$$
\eta(s) \equiv \sum_{\lambda \not = 0}d(\lambda)
({\rm sign}(\lambda))
{\mid \lambda \mid}^{-s}
\; \; \; \; ,
\eqno (4.1.6)
$$
where $d(\lambda)$ is the multiplicity of the eigenvalue
$\lambda$. Note that, since $B$ involves the $*$ operator,
in reversing the orientation of the boundary $Y$ we change
$B$ into $-B$, and hence $\eta(s)$ into $-\eta(s)$. The main
result of Atiyah et al., in its simplest form, states therefore
that (Atiyah et al. 1975, Atiyah 1975b)
$$
f(Y)={1\over 2}\eta(0)
\; \; \; \; .
\eqno (4.1.7)
$$
Now, for a manifold $M$ with boundary $Y$, if one tries to
set up an elliptic boundary-value problem for the signature
operator of section 2.3, one finds that there is no local boundary
condition for this operator. For global boundary conditions,
however, expressed by the vanishing of a given integral
evaluated on $Y$, one has a good elliptic theory and a finite
index. Thus, one has to consider the theorem expressed by
(4.1.7) within the framework of index theorems for global
boundary conditions. Atiyah et al. (1975) were also able
to derive the relation between the index of the Dirac operator
on $M$ with a global boundary condition and $\eta(0)$,
where $\eta$ is the $\eta$-function of the Dirac operator
on the boundary of $M$ (cf. (6.1.4) and (6.1.5)).
\vskip 1cm
\centerline {\bf 4.2 $\eta(0)$ Calculation}
\vskip 1cm
Following Atiyah et al. 1975,
we now evaluate $\eta(0)$ in a specific example.
Let $Y$ be a closed manifold, $E$ a vector bundle over $Y$
and $A: C^{\infty}(Y,E) \rightarrow C^{\infty}(Y,E)$ a
self-adjoint, elliptic, first-order differential operator. 
By virtue of this hypothesis, $A$ has a discrete spectrum
with real eigenvalues $\lambda$ and eigenfunctions
$\phi_{\lambda}$. Let $P$ denote the projection of
$C^{\infty}(Y,E)$ onto the space spanned by the $\phi_{\lambda}$
for $\lambda \geq 0$. We now form the product 
$Y \times {\Re}^{+}$ of $Y$ with the half-line $u \geq 0$ and
consider the operator
$$
D \equiv {\partial \over \partial u}+A
\; \; \; \; ,
\eqno (4.2.1)
$$
acting on sections $f(y,u)$ of $E$ lifted to $Y \times {\Re}^{+}$
(still denoted by $E$). Clearly $D$ is elliptic and its
{\it formal} adjoint is
$$
D^{*} \equiv -{\partial \over \partial u}+A
\; \; \; \; .
\eqno (4.2.2)
$$
The following boundary condition is imposed for $D$:
$$
Pf(\cdot,0)=0
\; \; \; \; .
\eqno (4.2.3)
$$
This is a {\it global} condition for the boundary value
$f(\cdot,0)$ in that it is equivalent to
$$
\int_{Y}\Bigr(f(y,0),\phi_{\lambda}(y)\Bigr)=0
\; \; \; \; {\rm for} \; {\rm all} \; 
\lambda \geq 0
\; \; \; \; .
\eqno (4.2.4)
$$
Of course, the adjoint boundary condition to (4.2.3) is
$$
(1-P)f(\cdot,0)=0
\; \; \; \; .
\eqno (4.2.5)
$$
The naturally occurring second-order self-adjoint operators
obtained from $D$ are
$$
\triangle_{1} \equiv {\cal D}^{*}{\cal D}
\; \; \; \; ,
\eqno (4.2.6)
$$
$$
\triangle_{2} \equiv {\cal D}{\cal D}^{*}
\; \; \; \; ,
\eqno (4.2.7)
$$
where ${\cal D}$ is the closure of the operator $D$ on 
$L^{2}$ with domain given by (4.2.3). For $t >0$, one can then
consider the bounded operators $e^{-t \triangle_{1}}$ and
$e^{-t \triangle_{2}}$. The explicit kernels of these operators
will be given in terms of the eigenfunctions $\phi_{\lambda}$ of
$A$. For this purpose, consider first $\triangle_{1}$, i.e. the
operator given by
$$
-{\partial^{2}\over \partial u^{2}}+A^{2}
$$
with the boundary condition
$$
Pf(\cdot,0)=0
\; \; \; \; ,
\eqno (4.2.8a)
$$
and
$$
(1-P){\left({\partial f \over \partial u}+Af \right)}_{u=0}
=0
\; \; \; \; .
\eqno (4.2.8b)
$$
Expansion in terms of the $\phi_{\lambda}$, so that
$f(y,u)=\sum_{\lambda} f_{\lambda}(u)\phi_{\lambda}(y)$, shows
that for each $\lambda$ one has to study the operator
$$
-{d^{2}\over du^{2}}+\lambda^{2}
$$
on $u \geq 0$ with the boundary conditions
$$
f_{\lambda}(0)=0
\; \; \; \; {\rm if} \; \; \; \;
\lambda \geq 0
\; \; \; \; ,
\eqno (4.2.9)
$$
$$
{\left({df_{\lambda}\over du}+\lambda f_{\lambda}\right)}_{u=0}
=0 \; \; \; \; {\rm if} \; \; \; \; 
\lambda <0
\; \; \; \; .
\eqno (4.2.10)
$$
It should be stressed, once more, that (4.2.9) and (4.2.10)
are non-local, since they rely on the separation of the
spectrum of $A$ into its positive and negative part
(cf. section 6.1).

The fundamental solution for 
$$
{\partial \over \partial t}-{\partial^{2}\over \partial u^{2}}
+\lambda^{2}
$$
with the boundary condition (4.2.9) is found to be
$$
W_{A}=
{e^{-\lambda^{2}t}\over \sqrt{4\pi t}}
\left [{\rm exp}\left({-(u-v)^{2}\over 4t}\right)
-{\rm exp}\left({-(u+v)^{2}\over 4t}\right) \right]
\; \; \; \; .
\eqno (4.2.11)
$$
By contrast, when the boundary condition (4.2.10) is
imposed, the use of the Laplace transform
leads to (Carslaw and Jaeger 1959)
$$
\eqalignno{
W_{B}&=
{e^{-\lambda^{2}t}\over \sqrt{4\pi t}}
\left [{\rm exp}\left({-(u-v)^{2}\over 4t}\right)
+{\rm exp}\left({-(u+v)^{2}\over 4t}\right) \right] \cr
&+\lambda e^{-\lambda(u+v)}\; {\rm erfc}
\left[{(u+v)\over 2\sqrt{t}}-\lambda \sqrt{t}\right]
\; \; \; \; ,
&(4.2.12)\cr}
$$
where ${\rm erfc}$ is the (complementary) error function
defined by
$$
{\rm erfc}(x) \equiv {2\over \sqrt{\pi}}
\int_{x}^{\infty}e^{-\xi^{2}} \; d\xi
\; \; \; \; .
\eqno (4.2.13)
$$
Thus, the kernel $K_{1}$ of $e^{-t\triangle_{1}}$ at a point
$(t;y,u;z,v)$ is obtained as
$$
K_{1}(t;y,u;z,v)
=\sum_{\lambda}W_{A}
\phi_{\lambda}(y)
{\overline {\phi_{\lambda}(z)}} 
\; \; \; \; {\rm if} \; \lambda \geq 0
\; \; \; \; ,
\eqno (4.2.14)
$$
$$
K_{1}(t;y,u;z,v)
=\sum_{\lambda}W_{B}
\phi_{\lambda}(y)
{\overline {\phi_{\lambda}(z)}} 
\; \; \; \; {\rm if} \; \lambda <0
\; \; \; \; .
\eqno (4.2.15)
$$

For the operator $\triangle_{2}$, the boundary conditions
for each $\lambda$ are
$$
f_{\lambda}(0)=0
\; \; \; \; {\rm if} \; \; \; \;
\lambda < 0
\; \; \; \; ,
\eqno (4.2.16)
$$
$$
{\left(-{df_{\lambda}\over du}+\lambda f_{\lambda}\right)}_{u=0}
=0 \; \; \; \; {\rm if} \; \; \; \; 
\lambda \geq 0
\; \; \; \; .
\eqno (4.2.17)
$$
The fundamental solution for
$$
{\partial \over \partial t}-{\partial^{2}\over \partial u^{2}}
+\lambda^{2}
$$
subject to (4.2.16) is ${\widetilde W}_{A}=W_{A}$, while for
the boundary conditions (4.2.17) one finds 
${\widetilde W}_{B}=W_{B}(-\lambda)$. Moreover, by virtue
of the inequality
$$
\int_{x}^{\infty}e^{-\xi^{2}} \; d\xi < e^{-x^{2}}
\; \; \; \; ,
\eqno (4.2.18)
$$
$W_{A}$ and $W_{B}$ are both bounded by
$$
F_{\lambda}(t;u,v) \equiv
\left[{e^{-\lambda^{2}t} \over \sqrt{\pi t}}
+{2 \mid \lambda \mid \over \sqrt{\pi}}e^{-\lambda^{2}t}
\right] \;
{\rm exp}\left({-(u-v)^{2}\over 4t}\right)
\; \; \; \; .
\eqno (4.2.19)
$$
If one now multiplies and divides by $\sqrt{t}$ the second
term in square brackets in (4.2.19), application of the
inequality
$$
x \leq e^{x^{2}/2}
\; \; \; \; ,
\eqno (4.2.20)
$$
to the resulting term $\mid \lambda \mid \sqrt{t}$ shows that
the kernel $K_{1}(t;y,u;z,v)$ of $e^{-t \triangle_{1}}$ is
bounded by
$$
\eqalignno{
G(t;y,u;z,v) & \equiv
{3\over 2 \sqrt{\pi t}}
\left[\sum_{\lambda}e^{-\lambda^{2}t/2}
\biggr({\mid \phi_{\lambda}(y) \mid}^{2} \right . \cr
& \left . +{\mid \phi_{\lambda}(z) \mid}^{2}\biggr)\right]
{\rm exp}\left({-(u-v)^{2}\over 4t}\right)
\; \; \; \; .
&(4.2.21)\cr}
$$
Moreover, since the kernel of $e^{-t \triangle_{2}}$ on the
diagonal of $Y \times Y$ is bounded by 
$C t^{-n/2}$, one finds that the kernels of $e^{-t \triangle_{1}}$
and $e^{-t \triangle_{2}}$ are exponentially small in $t$ as
$t \rightarrow 0^{+}$ for $u \not = v$, in that they are bounded by
$$
C \; t^{-(n+1)/2} \; 
{\rm exp}\left({-(u-v)^{2}\over 4t}\right)
\; \; \; \; ,
$$
where $C$ is some constant, as $t \rightarrow 0^{+}$.

Thus, the contribution outside the diagonal is asymptotically
negligible, so that we are mainly interested in the contribution
from the diagonal. For this purpose, we study $K(t,y,u)$,
i.e. the kernel of $e^{-t \triangle_{1}}-e^{-t \triangle_{2}}$ at
the point $(y,u;y,u)$ of $(Y \times {\Re}^{+}) \times
(Y \times {\Re}^{+})$. Defining ${\rm sign}(\lambda)
\equiv +1 \; \forall
\lambda \geq 0$, ${\rm sign}(\lambda)\equiv -1 
\; \forall \lambda <0$, 
one thus finds
$$ 
\eqalignno{
K(t,y,u) & \equiv \Bigr[W_{A}(t;y,u;y,u)-{\widetilde W}_{B}
(t;y,u;y,u)\Bigr] 
+\Bigr[W_{B}(t;y,u;y,u)-{\widetilde W}_{A}(t;y,u;y,u)\Bigr]\cr
&=\sum_{\lambda \geq 0}{\mid \phi_{\lambda}(y)\mid}^{2}
\left[{e^{-\lambda^{2}t} \over \sqrt{4\pi t}}
\left(1-e^{-u^{2}/t}\right) 
-{e^{-\lambda^{2}t} \over \sqrt{4\pi t}}
\left(1+e^{-u^{2}/t}\right) \right. \cr
&\left. -(-\lambda)e^{2\lambda u}
{\rm erfc} \; \left({u\over \sqrt{t}}+\lambda \sqrt{t}\right)
\right]\cr
&+\sum_{\lambda<0}{\mid \phi_{\lambda}(y)\mid}^{2}
\left[{e^{-\lambda^{2}t}\over \sqrt{4\pi t}}
\left(1+e^{-u^{2}/t}\right) 
+\lambda e^{-2\lambda u}{\rm erfc}
\left({u\over \sqrt{t}}-\lambda \sqrt{t}\right) \right. \cr
&\left. -{e^{-\lambda^{2}t}\over \sqrt{4 \pi t}}
\left(1-e^{-u^{2}/t}\right)\right]\cr
&=\sum_{\lambda}
{\mid \phi_{\lambda}(y)\mid}^{2}
{\rm sign}(\lambda)
\left[-{e^{-\lambda^{2}t}e^{-u^{2}/t}\over \sqrt{\pi t}}
+\mid \lambda \mid e^{2\mid \lambda \mid u}
{\rm erfc} \left({u\over \sqrt{t}}+\mid \lambda \mid 
\sqrt{t}\right)\right]\cr
&=\sum_{\lambda}
{\mid \phi_{\lambda}(y)\mid}^{2}
{\rm sign}(\lambda){\partial \over \partial u}
\left[{1\over 2}e^{2 \mid \lambda \mid u}
{\rm erfc} \left({u\over \sqrt{t}}+\mid \lambda \mid 
\sqrt{t}\right)\right]
\; \; \; \; .
&(4.2.22)\cr}
$$
Thus, integration on $Y \times {\Re}^{+}$ and elementary rules
for taking limits yield
$$
K(t) \equiv \int_{0}^{\infty}\int_{Y}K(t,y,u)dy \; du
=-{1\over 2}\sum_{\lambda}{\rm sign}(\lambda) \;
{\rm erfc}(\mid \lambda \mid \sqrt{t})
\; \; \; \; ,
\eqno (4.2.23)
$$
which implies (on differentiating the error function
and using the signature of eigenvalues)
$$
K'(t)={1\over \sqrt{4\pi t}}\sum_{\lambda}\lambda
e^{-\lambda^{2}t}
\; \; \; \; .
\eqno (4.2.24)
$$
It is now necessary to derive the limiting behaviour of the
(integrated) kernel (4.2.23) as $t \rightarrow \infty$
and as $t \rightarrow 0^{+}$. Indeed, denoting by $h$ the
degeneracy of the zero-eigenvalue, one finds
$$
\lim_{t \to \infty}K(t)=-{h\over 2}
\; \; \; \; ,
\eqno (4.2.25)
$$
whereas, as $t \rightarrow 0^{+}$, the following bound holds:
$$
\mid K(t) \mid \leq {1\over 2}\sum_{\lambda}
{\rm erfc}(\mid \lambda \mid \sqrt{t})
\leq {1\over \sqrt{\pi}}\sum_{\lambda}e^{-\lambda^{2}t}
< C \; t^{-n/2}
\; \; \; \; ,
\eqno (4.2.26)
$$
where $C$ is a constant. Moreover, the result (4.2.25) may be
supplemented by saying that $K(t)+{h\over 2}$ tends to $0$
exponentially as $t \rightarrow \infty$. Thus, combining 
(4.2.25), (4.2.26) and this property, one finds that, for
$Re(s)$ sufficiently large, the integral
$$
I(s) \equiv \int_{0}^{\infty}\Bigr(K(t)+{h\over 2}\Bigr)
t^{s-1} \; dt
\; \; \; \; ,
\eqno (4.2.27)
$$
converges. Integration by parts, definition of the $\Gamma$
function
$$
\Gamma(z) \equiv \int_{0}^{\infty}t^{z-1}e^{-t} \; dt
=k^{z}\int_{0}^{\infty}t^{z-1}e^{-kt} \; dt
\; \; \; \; ,
\eqno (4.2.28)
$$
and careful consideration of positive and negative eigenvalues
with their signatures then lead to
$$
I(s)
=-{\Gamma \left(s+{1\over 2}\right) \over 2s\sqrt{\pi}}
\sum_{\lambda}{{\rm sign}(\lambda)\over {{\mid \lambda \mid}^{2s}}}
=-{\Gamma \left(s+{1\over 2}\right) \over 2s\sqrt{\pi}}
\eta(2s)
\; \; \; \; .
\eqno (4.2.29)
$$
The following analysis relies entirely on the {\it assumption}
than an asymptotic expansion of the integrated kernel $K(t)$
exists as $t \rightarrow 0^{+}$.  
By writing such an expansion in the form
$$
K(t) \sim \sum_{k \geq -m} a_{k}t^{k/2}
\; \; \; \; ,
\eqno (4.2.30)
$$
equations (4.2.27)-(4.2.30) yield (on splitting the integral
(4.2.27) into an integral from $0$ to $1$ plus an integral
from $1$ to $\infty$)
$$
\eta(2s) \sim -{2s \sqrt{\pi} \over 
\Gamma \left(s+{1\over 2}\right)}
\left[{h\over 2s}+
\sum_{k=-m}^{N}{a_{k}\over \left({k\over 2}+s\right)}
+\theta_{N}(s)\right]
\; \; \; \; .
\eqno (4.2.31)
$$
This is the analytic continuation of $\eta(2s)$ to the
whole $s$-plane. Hence one finds
$$
\eta(0)=-\Bigr(2a_{0}+h\Bigr)
\; \; \; \; .
\eqno (4.2.32)
$$

Regularity at the origin of the $\eta$-function is an 
important property in the theory of elliptic operators
on manifolds. What one can prove (Atiyah et al. 
1976, Gilkey 1995) is
that the analytic continuation of the $\eta$-function of
a given elliptic operator $A$:
$$
\eta_{A}(s) \equiv {\rm Tr} 
\left[A {\mid A \mid}^{-s-1}\right]
\; \; \; \; ,
\eqno (4.2.33)
$$
is a meromorphic function which is regular at $s=0$. 
Interestingly, such a property is stable under homotopy,
i.e. under a smooth variation of $A$. More precisely, one
considers a $C^{\infty}$ one-parameter family $A_{u}$ of
elliptic operators, and one proves that the residue 
$R(A_{u})$ at $s=0$ of the corresponding $\eta$-functions
is constant, i.e.
$$
{d\over du}R(A_{u})=0
\; \; \; \; .
\eqno (4.2.34)
$$
Thus, the general formula for the analytic continuation
of the $\eta$-function of $A$ reduces to (cf. Atiyah
et al. 1976)
$$
\eta(s)=\sum_{k=-m}^{N} {b_{k}\over (s+{k\over n})}
+\phi_{N}(s) \; \; , \; \; 
k \not = 0
\; \; \; \; ,
\eqno (4.2.35)
$$
where $m$ is the dimension of the (compact) Riemannian
manifold, $n$ is the order of $A$, $\phi_{N}$ is
holomorphic in the half-plane ${\rm Re}(s)> -{N\over n}$,
and is $C^{\infty}$.
\vskip 10cm
\centerline {\bf 4.3 A Further Example}
\vskip 1cm
It is instructive to describe a simpler example of $\eta(0)$
calculation for first-order differential operators. For
this purpose, we consider the first-order operator
$$
A \equiv i {d\over dx}+t
\; \; \; \; ,
\eqno (4.3.1)
$$
where $t$ is a real parameter lying in the open interval
$]0,1[$, and $x$ is an angular coordinate on the circle. 
The boundary conditions on the eigenfunctions of $A$,
denoted by $f$, are periodicity of period $2\pi$:
$$
f(x)=f(x+2\pi)
\; \; \; \; .
\eqno (4.3.2)
$$
Thus, since the eigenvalue equation for $A$:
$$
i{df\over dx}+tf=\lambda \; f
\eqno (4.3.3)
$$
is solved by ($f_{0}$ being $f(x=0)$)
$$
f=f_{0} \; e^{-i(\lambda-t)x}
\; \; \; \; ,
\eqno (4.3.4)
$$
one finds, by virtue of (4.3.2), the eigenvalue condition
$$
e^{-i(\lambda -t)2\pi}=1
\; \; \; \; .
\eqno (4.3.5)
$$
Equation (4.3.5) is solved by
$$
\lambda=t \pm n
\; \; \; \; \forall n=0,1,2,...
\; \; \; \; .
\eqno (4.3.6)
$$
One can now form the corresponding $\eta$-function:
$$
\eta_{t}(s)=\sum_{r=0}^{\infty}(r+t)^{-s}
-\sum_{r=1}^{\infty}(r-t)^{-s}
\; \; \; \; ,
\eqno (4.3.7)
$$
which can be re-arranged as follows:
$$
\eqalignno{
\eta_{t}(s)&= t^{-s}+\sum_{r=1}^{\infty}
{\Bigr[(r-t)^{s}-(r+t)^{s}\Bigr] \over
(r+t)^{s} (r-t)^{s}} \cr
&= t^{-s} -2st \sum_{r=1}^{\infty}r^{-(s+1)}
+s \Sigma_{1}
\; \; \; \; .
&(4.3.8)\cr}
$$
In Eq. (4.3.8), $\Sigma_{1}$ is a series which is absolutely
convergent in the neighbourhood of $s=0$. One thus finds
$$
\eta_{t}(0)=1-2t \lim_{s \to 0} s \zeta_{R}(s+1)
=1-2t
\; \; \; \; ,
\eqno (4.3.9)
$$
where $\zeta_{R}$ is the Riemann $\zeta$-function,
defined as
$$
\zeta_{R}(s) \equiv \sum_{m=1}^{\infty} m^{-s}
\; \; \; \; .
\eqno (4.3.10)
$$
The result (4.3.9) shows that, unless $t$ takes the value
${1\over 2}$, there will be a ``spectral asymmetry" 
expressed by a non-vanishing value of $\eta_{t}(0)$.

It has been our choice to give a very elementary introduction
to the subject of the $\eta$-function. The reader who is
interested in advanced topics is referred to Berline et al.
1992, Branson and Gilkey 1992a--b, 
Booss-Bavnbek and Wojciechowski 1993, Falomir et al. 1996.
Further work on the Dirac operator and its eigenvalues can
be found in Hortacsu et al. 1980, Atiyah and Singer 1984, 
Vafa and Witten 1984, Atiyah 1985, Polychronakos 1987,
Connes 1995, Camporesi and Higuchi 1996, Landi and Rovelli
1996, Carow-Watamura and Watamura 1997.
\vskip 100cm
\centerline {\it CHAPTER FIVE}
\vskip 1cm
\centerline {\bf SPECTRAL GEOMETRY WITH OPERATORS}
\centerline {\bf OF LAPLACE TYPE}
\vskip 1cm
\noindent
{\bf Abstract.} Spectral geometry with operators of
Laplace type is introduced by 
discussing the inverse problem in the theory of vibrating
membranes. This means that a given spectrum of eigenvalues
is given, and one would like to determine uniquely
the shape of the vibrating object from the asymptotic expansion
of the integrated heat kernel. It turns out that, in general,
it is not possible to tell whether the membrane is convex, 
or smooth, or simply connected, but results of a limited
nature can be obtained. These determine, for example,
the volume and the surface area of the body. Starting from
these examples, the very existence of the asymptotic
expansion of the integrated heat hernel is discussed, relying
on the seminal paper by Greiner. A more careful analysis of
the boundary-value problem is then performed, and the recent
results on the asymptotics of the Laplacian on a manifold 
with boundary are presented in detail. For this purpose,
one studies second-order elliptic operators with leading 
symbol given by the metric. The behaviour of the differential
operator, boundary operator and heat-kernel coefficients 
under conformal rescalings of the background metric leads 
to a set of algebraic equations which determine completely
the heat-kernel asymptotics. Such property holds whenever
one studies boundary conditions of Dirichlet or Robin type,
or a mixture of the two. The chapter ends by describing
the heat-equation approach to index theorems, and the link
between heat equation and $\zeta$-function.
\vskip 100cm
\centerline {\bf 5.1 On Hearing the Shape of a Drum}
\vskip 1cm
A classical problem in spectral geometry, which is used here
to introduce the topic, consists in the attempt to deduce the
shape of a drum from the knowledge of its spectrum of eigenvalues,
say ${\lambda_{n}}$. Some progress is possible
on establishing the leading terms of the asymptotic expansion
of the trace function (cf. section 5.2)
$$
\Theta(t)=\sum_{n=1}^{\infty}e^{-\lambda_{n}t}
\; \; \; \; ,
\eqno (5.1.1)
$$
for small positive values of $t$. In particular, if one studies
a simply connected membrane $\Omega$ bounded by a smooth convex
plane $\Gamma$, for which the displacement satisfies the wave
equation
$$
\bigtriangleup \phi={\partial^{2}\phi \over \partial t^{2}}
\; \; \; \; ,
\eqno (5.1.2)
$$
and Dirichlet boundary conditions on $\Gamma$:
$$
\Bigr[\phi \Bigr]_{\Gamma}=0
\; \; \; \; ,
\eqno (5.1.3)
$$
one finds
$$
\Theta(t) \sim {\mid \Omega \mid \over 4 \pi t}
-{L\over 8 \sqrt{\pi t}}+{1\over 6}
\; \; \; \; .
\eqno (5.1.4)
$$
In (5.1.4), $\mid \Omega \mid$ is the area of $\Omega$,
$L$ is the length of $\Gamma$, and the constant 
${1\over 6}$ results from integration of the curvature 
of the boundary. Moreover, if $\Omega$ has a finite number
of smooth convex holes, such a term should be replaced by
${1\over 6}(1-r)$, where $r$ is the number of holes. The
two basic problems in the heat-equation approach to drums
and vibrating membranes are as follows (Kac 1966, 
Stewartson and Waechter 1971, Waechter 1972):
\vskip 0.3cm
\noindent
(i) Given a set $\left \{ \lambda_{n} \right \}$, can a
corresponding shape be found ? If so, is it unique ?
\vskip 0.3cm
\noindent
(ii) Let $\left \{ \lambda_{n} \right \}$ be a given
spectrum of eigenvalues. Can the shape be uniquely 
determined from the asymptotic expansion of $\Theta(t)$
for small positive $t$ ?
\vskip 0.3cm
\noindent
In Stewartson and Waechter 1971, the authors obtained the
asymptotic expansion of $\Theta(t)$ for a membrane bounded
by a smooth boundary. They found that, when the restriction
of convexity is relaxed, and corners and cusps are 
permitted, the shape can be determined precisely if the
membrane is circular; otherwise the asymptotic expansion
of $\Theta(t)$ as $t \rightarrow 0^{+}$ determines the
area, the length of the perimeter and the existence of
outward pointing cusps. Thus, in general, {\it it is
not possible} to tell whether the membrane is convex, 
or smooth, or simply connected. Whether it is possible 
to do so from the complete asymptotic expansion of
$\Theta(t)$ remained, and remains, an open problem.

In the following sections we will give a rigorous 
description of heat-kernel methods, but here we can
complete the outline of the Green-function approach, 
following again Stewartson and Waechter (1971). 
Indeed, if $G(r,r';t)$ is the Green's function of the
diffusion (or heat) equation
$$
\left({\partial \over \partial t}-\bigtriangleup 
\right)\phi=0
\; \; \; \; ,
\eqno (5.1.5)
$$
subject to the Dirichlet condition
$$
G(r,r';t)=0 \; \; {\rm if} \; \; r \in \Gamma
\; \; \; \; ,
\eqno (5.1.6)
$$
and behaving as a Dirac delta: $\delta(r-r')$, as
$t \rightarrow 0^{+}$, the trace function $\Theta(t)$ is
defined by the equation
$$
\Theta(t) \equiv \int \int_{\Omega} G(r,r;t) d\Omega
\; \; \; \; .
\eqno (5.1.7)
$$
Thus, if one writes
$$
G(r,r';t)=\sum_{n=1}^{\infty}e^{-\lambda_{n}t}
\phi_{n}(r)\phi_{n}(r')
\; \; \; \; ,
\eqno (5.1.8)
$$
where $\lambda_{n}$ are the eigenvalues of the operator
$-\bigtriangleup$, and $\phi_{n}$ are the normalized
eigenfunctions, the result (5.1.1) follows on setting $r=r'$
and integrating over $\Omega$. In many applications it 
is useful to consider the split 
$$
G(r,r';t)=G_{0}(r,r';t)+\chi(r,r';t)
\; \; \; \; ,
\eqno (5.1.9)
$$
where (setting $R \equiv \mid r-r' \mid$)
$$
G_{0}(r,r';t)={1\over 4\pi t}e^{-R^{2}/4t}
\eqno (5.1.10)
$$
is the Green's function for the infinite plane, and
$\chi(r,r';t)$ is the compensating solution of the heat
equation (5.1.5), which satisfies the appropriate 
boundary condition on $\Gamma$.

In Waechter 1972, the author extended the
early investigation of the inverse eigenvalue problem for
vibrating membranes to three or more dimensions. Thus, he
considered the boundary-value problem
$$
-\bigtriangleup \phi = \lambda \; \phi
\; \; {\rm in} \; \; \Omega
\; \; \; \; ,
\eqno (5.1.11)
$$
$$
\phi=0 \; \; {\rm on} \; \; S
\; \; \; \; ,
\eqno (5.1.12)
$$
where $\Omega$ is a closed convex region or body in
the $n$-dimensional Euclidean space $E^{n}$, and $S$
is the bounding surface of $\Omega$. Once more, the
problem was to determine the precise shape of $\Omega$,
on being given the spectrum of eigenvalues $\lambda_{n}$
and the heat equation (5.1.5), whose Green's function
satisfies the boundary condition
$$
G(r,r';t)=0 \; \; {\rm if} \; \; r \in S
\; \; \; \; .
\eqno (5.1.13)
$$
The method was always to use the trace function
$$
\Theta(t) \equiv \int \int \int_{\Omega} 
G(r,r;t) d\Omega
\eqno (5.1.14)
$$
to determine the leading terms of the asymptotic expansion
of $\Theta(t)$ for small positive $t$. Interestingly, even
in the higher-dimensional problem, results of a limited
nature can be obtained. For example, the first six terms of
the expansion of $\Theta(t)$ for a sphere were determined,
and for a smooth convex body Waechter found the first four terms: 
$$
\Theta(t) \sim {V\over (4\pi t)^{3/2}}-{S\over 16 \pi t}
+{M\over 6 \pi \sqrt{4\pi t}}
+{1\over 512 \pi} \int \int_{S}(\kappa_{1}-\kappa_{2})^{2}dS
+{\rm O}(\sqrt{t}) \; .
\eqno (5.1.15)
$$
With the notation in Waechter 1972, $V$ and $S$
are the volume and surface area of the body $\Omega$,
respectively, $M$ is the surface integral of mean curvature:
$$
M \equiv \int \int {1\over 2}(\kappa_{1}+\kappa_{2})dS
\; \; \; \; ,
\eqno (5.1.16)
$$
while $\kappa_{1}$ and $\kappa_{2}$ are the principal curvatures
at the surface element $dS$ of the body surface.

So far, nothing has been said about the actual {\it existence}
of the asymptotic expansion of the trace function. The proof is
a very important result due to Greiner (1971). Although we can
only refer the reader to the original paper by Greiner for the
detailed proof, one can describe, however, the mathematical
framework, since this will help to achieve a smooth transition 
towards the following sections. We can thus say that Greiner
studied a compact $m$-dimensional $C^{\infty}$ Riemannian
manifold, say $M$, with $C^{\infty}$ boundary, say 
$\partial M$. The boundary-value problem consists of the pair
$(P,B)$, where $P$ is an elliptic differential operator of
order $2n$:
$$
P: C^{\infty}(V_{M}) \rightarrow C^{\infty}(V_{M})
\; \; \; \; ,
\eqno (5.1.17)
$$
with $V_{M}$ a $C^{\infty}$ (complex) vector bundle over $M$,
and is such that $P+{\partial \over \partial t}$ is parabolic. 
One then says that $P$ is $p$-elliptic. An example is the
operator in round brackets in Eq. (5.1.5), with $P$ being
equal to $-\bigtriangleup$. Moreover, the {\it boundary operator}
is a map
$$
B: C^{\infty}(V_{M}) \rightarrow 
C^{\infty}(G_{\partial M})
\; \; \; \; ,
\eqno (5.1.18)
$$
where $G_{\partial M}$ is a $C^{\infty}$ (complex) vector
bundle over $\partial M$. To obtain the Green's {\it kernel}
$G(x,y;t)$ of $e^{-tP}$, Greiner constructed an operator $C$
with kernel $C(x,y;t)$, compensating for the boundary, 
such that
$$
G(x,y;t)=H(x,y;t)-C(x,y;t)
\; \; \; \; ,
\eqno (5.1.19)
$$
where $H(x,y;t)$ is the kernel of $e^{-tP}$ for a manifold
without boundary. Greiner's result states therefore that,
for the integrated (heat) kernel
$$
G(t) \equiv \int_{M}{\rm Tr} G(x,x;t) dx
={\rm Tr}_{L^{2}}(e^{-tP})
\eqno (5.1.20)
$$
an asymptotic expansion exists, as $t \rightarrow 0^{+}$,
in the form (cf. Greiner 1971)
$$
G(t) \sim t^{-m/2n} \Bigr[G_{0}+G_{1}t^{1/2n}+...
+G_{k}t^{k/2n}+...\Bigr]
\; \; \; \; ,
\eqno (5.1.21)
$$
where 
$$
G_{0}=H_{0}
\; \; \; \; ,
\eqno (5.1.22)
$$
$$
G_{k}=H_{k}-C_{k} \; \; \; \; , \; \; \; \;  
k=1,2,...
\; \; \; \; ,
\eqno (5.1.23)
$$
$$
C_{k}=\int_{\partial M}C_{k}(x')dx'
\; \; \; \; .
\eqno (5.1.24)
$$
Note that the trace in the integral (5.1.20) is the fibre
trace, and that in the mathematics literature one also says
that $G(x,y;t)$ is the Green's kernel for the boundary-value
problem $\Bigr(P+{\partial \over \partial t},B \Bigr)$. The
following sections are devoted to a precise characterization
of the $G_{k}$ coefficients occurring in the asymptotic 
expansion (5.1.21).
\vskip 1cm
\centerline {\bf 5.2 Asymptotics of the Laplacian on a
Manifold with Boundary}
\vskip 1cm
Following Branson and Gilkey 1990, we are interested in a
second-order differential operator, say $P$, with leading
symbol given by the metric tensor on a compact $m$-dimensional
Riemannian manifold $M$ with boundary $\partial M$. Denoting
by $\nabla$ the connection on the vector bundle $V=V_{M}$,
our assumption implies that $P$, called an operator of
Laplace type, reads
$$
P=-g^{ab}\nabla_{a}\nabla_{b}-E
\; \; \; \; ,
\eqno (5.2.1)
$$
where $E$ is an endomorphism of $V$. The heat equation for
the operator $P$ is (cf. (5.1.5))
$$
\left({\partial \over \partial t}+P \right)F(x,x';t)=0
\; \; \; \; .
\eqno (5.2.2)
$$
By definition, the {\it heat kernel} (cf. section 5.1) is the
solution of Eq. (5.2.2) subject to the boundary condition
$$
\Bigr[{\cal B}F(x,x';t)\Bigr]_{\partial M}=0
\; \; \; \; ,
\eqno (5.2.3)
$$
jointly with the (initial) condition
$$
\lim_{t \to 0^{+}} \int_{M}F(x,x';t) \rho(x')dx'
=\rho(x)
\; \; \; \; ,
\eqno (5.2.4)
$$
which is a rigorous mathematical expression for the Dirac
delta behaviour as $t \rightarrow 0^{+}$ (cf. section 5.1).
The heat kernel can be written as (cf. (5.1.8))
$$
F(x,x';t)=\sum_{(n)}\varphi_{(n)}(x) \varphi_{(n)}(x')
e^{-\lambda_{(n)}t}
\; \; \; \; ,
\eqno (5.2.5)
$$
where $\left \{ \varphi_{(n)}(x) \right \}$ is a complete
orthonormal set of eigenfunctions with eigenvalues 
$\lambda_{(n)}$. The index $n$ is enclosed in round brackets,
to emphasize that, in general, a finite collection of
integer labels occurs therein. 

Since, by construction, the heat kernel behaves as a
distribution in the neighbourhood of the boundary, it is
convenient to introduce a smooth function, say 
$f \in C^{\infty}(M)$, and consider a slight generalization
of the trace function (or integrated heat kernel) of section
5.1, i.e. ${\rm Tr}_{L^{2}} \Bigr(f e^{-tP} \Bigr)$. It is
precisely the consideration of $f$ that makes it possible  
to recover the distributional behaviour of the heat kernel
near $\partial M$. A key idea is therefore to work with
arbitrary $f$, and then set $f=1$ only when all coefficients
in the asymptotic expansion
$$
{\rm Tr}_{L^{2}}\Bigr(f e^{-t P} \Bigr) \equiv
\int_{M}{\rm Tr}\Bigr[f F(x,x;t)\Bigr]dx
\sim (4\pi t)^{-m/2} \sum_{n=0}^{\infty}
t^{n/2} a_{n/2}(f,P)
\eqno (5.2.6)
$$
have been evaluated. The term $F(x,x;t)$ is called the
{\it heat-kernel diagonal}. By virtue of Greiner's result,
the coefficients $a_{n/2}(f,P)$, which are said to describe
the {\it asymptotics}, are obtained by integrating {\it local}
formulae. More precisely, they admit a split into integrals
over $M$ (interior terms) and over $\partial M$ (boundary
terms). In such formulae, the integrands are linear combinations
of all geometric invariants of the appropriate dimension
(see below)
which result from the Riemann curvature $R_{\; \; bcd}^{a}$
of the background, the extrinsic curvature of the boundary,
the differential operator $P$ (through the endomorphism 
$E$), and the boundary operator $\cal B$ (through the
endomorphisms, or projection operators, or more general
matrices occurring in it). With our notation, the indices
$a,b,...$ range from 1 through $m$ and index a local
orthonormal frame for the tangent bundle of $M$, $TM$, while
the indices $i,j,...$ range from 1 through $m-1$ and index
the orthonormal frame for the tangent bundle of the 
boundary, $T(\partial M)$. The boundary is defined by the
equations
$$
\partial M: \; \; \; \; \; \; \; \; 
y^{a}=y^{a}(x)
\; \; \; \; ,
\eqno (5.2.7)
$$
in terms of the functions $y^{a}(x)$, $x^{i}$ being the
coordinates on $\partial M$, and the $y^{a}$ those on $M$.
Thus, the {\it intrinsic} metric, $\gamma_{ij}$, on the
boundary hypersurface $\partial M$, is given in terms of the
metric $g_{ab}$ on $M$ by (Eisenhart 1926, Dowker and
Schofield 1989)
$$
\gamma_{ij}=g_{ab} \; y_{\; ,i}^{a} \; y_{\; ,j}^{b}
\; \; \; \; .
\eqno (5.2.8)
$$
On inverting this equation one finds (here, $n^{a}=N^{a}$
is the inward-pointing normal)
$$
g^{ab}=q^{ab}+n^{a}n^{b}
\; \; \; \; ,
\eqno (5.2.9)
$$
where 
$$
q^{ab}=y_{\; ,i}^{a} \; y_{\; ,j}^{b} \; \gamma^{ij}
\; \; \; \; .
\eqno (5.2.10)
$$
The tensor $q^{ab}$ is equivalent to $\gamma^{ij}$ and
may be viewed as the {\it induced metric} on 
$\partial M$, in its contravariant form. The tensor 
$q_{\; \; b}^{a}$ is a projection operator, in that
$$
q_{\; \; b}^{a} \; q_{\; \; c}^{b}=q_{\; \; c}^{a}
\; \; \; \; ,
\eqno (5.2.11)
$$
$$
q_{\; \; b}^{a} \; n^{b}=0
\; \; \; \; .
\eqno (5.2.12)
$$
The extrinsic-curvature tensor $K_{ab}$ (or second fundamental
form of $\partial M$) is defined by the projection of the
covariant derivative of an {\it extension} of the outward,
normal vector field $-n$:
$$
K_{ab} \equiv -n_{c;d} \; q_{\; \; a}^{c} 
\; q_{\; \; b}^{d}
\; \; \; \; ,
\eqno (5.2.13)
$$
and is symmetric if the metric-compatible connection on
$M$ is torsion-free. Only its spatial components, $K_{ij}$,
are non-vanishing. 

The semicolon $;$ denotes multiple covariant differentiation
with respect to the Levi-Civita connection $\nabla_{M}$ 
of $M$, while the stroke $\mid$ denotes multiple covariant
differentiation tangentially with respect to the 
Levi-Civita connection $\nabla_{\partial M}$  
of the boundary. When sections of bundles built from $V$
are involved, the semicolon means
$$
\nabla_{M} \otimes \II + \II \otimes \nabla
\; \; \; \; ,
$$
and the stroke means 
$$
\nabla_{\partial M} \otimes \II
+ \II \otimes \nabla
\; \; \; \; .
$$
The curvature of the connection $\nabla$ on $V$ is denoted
by $\Omega$.

When Dirichlet or Robin boundary conditions are imposed on
sections of $V$:
$$
[\phi]_{\partial M}=0
\; \; \; \; ,
\eqno (5.2.14)
$$
or
$$
\Bigr[(n^{a}\nabla_{a}+S)\phi \Bigr]_{\partial M}=0
\; \; \; \; ,
\eqno (5.2.15)
$$
the asymptotics in (5.2.6) is expressed through some
{\it universal constants}  
$$
\left \{ \alpha_{i},b_{i},
c_{i}, d_{i}, e_{i} \right \}
$$ 
such that (here
$R_{ab} \equiv R_{\; \; abc}^{c}$ is the Ricci tensor,
$R \equiv R_{\; \; a}^{a}$, and $\cstok{\ } \equiv
\nabla^{a}\nabla_{a}=g^{ab}\nabla_{a}\nabla_{b}$)
$$
a_{0}(f,P)=\int_{M}{\rm Tr}(f)
\; \; \; \; ,
\eqno (5.2.16)
$$
$$
a_{1/2}(f,P)=\delta (4\pi)^{1/2}
\int_{\partial M}{\rm Tr}(f)
\; \; \; \; ,
\eqno (5.2.17)
$$
$$
\eqalignno{
\; & a_{1}(f,P)={1\over 6}\int_{M}{\rm Tr}\Bigr[\alpha_{1}
fE+\alpha_{2}fR \Bigr] \cr
&+{1\over 6}\int_{\partial M}{\rm Tr} \Bigr[
b_{0}f({\rm tr}K)+b_{1}f_{;N}+b_{2}fS \Bigr]
\; \; \; \; , 
&(5.2.18)\cr}
$$
$$
\eqalignno{
\; & a_{3/2}(f,P)={\delta \over 96}(4\pi)^{1/2}
\int_{\partial M}{\rm Tr} \Bigr[f(c_{0}E+c_{1}R
+c_{2} R_{\; \; NiN}^{i} \cr
&+c_{3}({\rm tr}K)^{2}+c_{4}K_{ij}K^{ij}
+c_{7}S ({\rm tr}K)+c_{8}S^{2} \Bigr) \cr
&+f_{;N}\Bigr(c_{5}({\rm tr}K)+c_{9}S \Bigr)
+c_{6} f_{;NN} \Bigr]
\; \; \; \; ,
&(5.2.19)\cr}
$$
$$
\eqalignno{
\; & a_{2}(f,P)={1\over 360} \int_{M}{\rm Tr}
\biggr[f \Bigr(\alpha_{3} \cstok{\ }E +\alpha_{4}RE
+\alpha_{5} E^{2} + \alpha_{6} \cstok{\ }R
+\alpha_{7} R^{2} \cr
&+\alpha_{8} R_{ab}R^{ab}+\alpha_{9}R_{abcd}R^{abcd}
+\alpha_{10} \Omega_{ab} \Omega^{ab} \Bigr) \biggr]\cr
&+{1\over 360} \int_{\partial M}{\rm Tr} \biggr[
f \Bigr(d_{1} E_{;N}+d_{2}R_{;N}
+d_{3} ({\rm tr}K)_{\mid i}^{\; \; \; \mid i}
+d_{4} K_{ij}^{\; \; \; \mid ij} \cr
&+d_{5} E ({\rm tr}K)+d_{6}R ({\rm tr}K)
+d_{7} R_{\; \; NiN}^{i} ({\rm tr} K)
+d_{8} R_{iNjN} K^{ij} \cr
&+d_{9} R_{\; \; imj}^{m} K^{ij}
+d_{10}({\rm tr}K)^{3}+d_{11}K_{ij}K^{ij}({\rm tr}K) \cr
&+d_{12}K_{i}^{\; \; j} \; K_{j}^{\; \; l} 
\; K_{l}^{\; \; i}
+d_{13} \Omega_{iN;}^{\; \; \; \; \; i}
+d_{14} SE + d_{15} SR \cr
&+d_{16} S R_{\; \; NiN}^{i}
+d_{17} S ({\rm tr}K)^{2}+d_{18}S K_{ij} K^{ij} \cr
&+d_{19} S^{2} ({\rm tr}K)+d_{20} S^{3}
+d_{21} S_{\mid i}^{\; \; \; \mid i} \Bigr)
+f_{;N} \Bigr(e_{1}E + e_{2} R \cr
&+e_{3} R_{\; \; NiN}^{i}+e_{4}({\rm tr}K)^{2}
+e_{5}K_{ij}K^{ij}+e_{8}S ({\rm tr}K)+e_{9}S^{2}\Bigr)\cr
&+f_{;NN} \Bigr(e_{6} ({\rm tr}K)+e_{10}S \Bigr)
+e_{7} f_{;a \; \; N}^{\; \; \; a} \biggr]
\; \; \; \; .
&(5.2.20)\cr}
$$
These formulae may seem to be very complicated, but there
is indeed a systematic way to write them down and then
compute the universal constants. To begin, note that,
if $k$ is odd, $a_{k/2}(f,P)$ receives contributions from
boundary terms only, whereas both interior terms and 
boundary terms contribute to $a_{k/2}(f,P)$, if $k$ is
even and positive. In the $a_{1}$ coefficient, the 
integrand in the interior term must be linear in the 
curvature, and hence it can only be a linear combination
of the trace of the Ricci tensor, and of the endomorphism
$E$ in the differential operator. In the $a_{2}$ 
coefficient, the integrand in the interior term must be
quadratic in the curvature, and hence one needs a linear
combination of the eight geometric invariants
(cf. Schwinger 1951, DeWitt 1965)
$$
\cstok{\ }E \; , \; 
RE \; , \; E^{2} \; , \;
\cstok{\ }R \; , \;
R^{2} \; , \; R_{ab}R^{ab} \; , \;
R_{abcd}R^{abcd} \; , \; \Omega_{ab}\Omega^{ab} \; .
$$

In the $a_{1}$ coefficient, the integrand in the boundary
term is a local expression given by a linear combination
of all invariants linear in the extrinsic curvature:
${\rm tr}K, S$ and $f_{;N}$. In the $a_{3/2}$ coefficient,
the integrand in the boundary term must be quadratic in
the extrinsic curvature. Thus, bearing in mind the
Gauss-Codazzi equations, one finds the general result
(5.2.19). Last, in the $a_{2}$ coefficient, the integrand
in the boundary term must be cubic in the extrinsic 
curvature. This leads to the boundary integral in (5.2.20),
bearing in mind that $f_{;N}$ is linear in $K_{ij}$,
while $f_{;NN}$ is quadratic in $K_{ij}$.

Note that the interior invariants are built universally and
polynomially from the metric tensor, its inverse, and the
covariant derivatives of $R_{\; bcd}^{a}, \Omega_{ab}$
and $E$. By virtue of Weyl's work on the invariants of the
orthogonal group (Weyl 1946, Branson and Gilkey 1990,
Gilkey 1995), these polynomials can be formed using only
tensor products and contraction of tensor arguments. Here,
the structure group is $O(m)$. However, when a boundary occurs,
the boundary structure group is $O(m-1)$. Weyl's theorem is
used again to construct invariants as in the previous 
equations (Branson and Gilkey 1990).
\vskip 1cm
\centerline {\bf 5.3 Functorial Method}
\vskip 1cm
Let $T$ be a map which carries finite-dimensional vector
spaces into finite-dimensional vector spaces. Thus, to
every vector space $V$ one has an associated vector space
$T(V)$. The map $T$ is said to be a {\it continuous functor}
if, for all $V$ and $W$, the map
$$
T: Hom(V,W) \longrightarrow Hom (T(V),T(W))
$$
is continuous (Atiyah 1967).

In the theory of heat-kernels, the functorial method is nothing
but the analysis of heat-equation asymptotics with respect to
conformal variations. Indeed, the behaviour of classical and
quantum field theories under conformal rescalings of the metric:
$$
{\widehat g}_{ab}=\Omega^{2} \; g_{ab}
\; \; \; \; ,
\eqno (5.3.1)
$$
with $\Omega$ a smooth function, is at the heart of many deep
properties: light-cone structure, conformal curvature (i.e.
the Weyl tensor), conformal-infinity techniques, massless
free-field equations, twistor equation, twistor spaces, 
Hodge-star operator in four dimensions, conformal anomalies
(Penrose and Rindler 1986, Ward and Wells 1990, Esposito 
1994a, Esposito 1995, Esposito et al. 1997). In the 
functorial method, one chooses $\Omega$ in the form
$$
\Omega=e^{\varepsilon f}
\; \; \; \; ,
\eqno (5.3.2)
$$
where $\varepsilon$ is a real-valued parameter, and
$f \in C^{\infty}(M)$ is the smooth function considered
in section 5.2. One then deals with a one-parameter 
family of differential operators
$$
P(\varepsilon)=e^{-2 \varepsilon f} \; P(0)
\; \; \; \; ,
\eqno (5.3.3)
$$
boundary operators
$$
{\cal B}(\varepsilon)=e^{-\varepsilon f} \; {\cal B}(0)
\; \; \; \; ,
\eqno (5.3.4)
$$
connections $\nabla^{\varepsilon}$ on $V$,
endomorphisms $E(\varepsilon)$ of $V$, and metrics
$$
g_{ab}(\varepsilon)=e^{2 \varepsilon f} \; g_{ab}(0)
\; \; \; \; .
\eqno (5.3.5)
$$
For example, the form (5.3.2) of the conformal factor should
be inserted into the general formulae which describe the
transformation of Christoffel symbols under conformal
rescalings:
$$
{\widehat \Gamma}_{\; bc}^{a}=\Gamma_{\; bc}^{a}
+ \Omega^{-1} \Bigr(\delta_{\; b}^{a} \; \Omega_{,c}
+\delta_{\; c}^{a} \; \Omega_{,b}
-g_{bc}g^{ad} \Omega_{,d} \Bigr)
\; \; \; \; .
\eqno (5.3.6)
$$
This makes it possible to obtain the conformal variation
formulae for the Riemann tensor $R_{\; bcd}^{a}$ and for
all tensors involving the effect of Christoffel symbols.
For the extrinsic-curvature tensor defined in Eq. (5.2.13)
one finds
$$
{\widehat K}_{ab}=\Omega K_{ab}-n_{a}\nabla_{b}\Omega
+g_{ab} \nabla_{(n)}\Omega
\; \; \; \; ,
\eqno (5.3.7)
$$
which implies
$$
K_{ij}(\varepsilon)=e^{\varepsilon f} \; K_{ij}
-\varepsilon \; g_{ij} \; e^{\varepsilon f} \; f_{;N}
\; \; \; \; .
\eqno (5.3.8)
$$

The application of these methods to heat-kernel asymptotics
relies on the work by Branson et al. (1990) and Branson
and Gilkey (1990). Within this framework, a crucial role is
played by the following ``functorial" formulae
($F$ being another smooth function):
$$
\left[{d\over d\varepsilon}a_{n/2}\Bigr(1,e^{-2\varepsilon f}
P(0)\Bigr)\right]_{\varepsilon=0}=(m-n)a_{n/2}(f,P(0))
\; \; \; \; ,
\eqno (5.3.9)
$$
$$
\left[{d\over d\varepsilon}a_{n/2}\Bigr(1,P(0)
-\varepsilon F \Bigr)\right]_{\varepsilon=0}
=a_{{n/2}-1}(F,P(0))
\; \; \; \; ,
\eqno (5.3.10)
$$
$$
\left[{d\over d\varepsilon}a_{n/2}\Bigr(e^{-2 \varepsilon f}F,
e^{-2 \varepsilon f}P(0)\Bigr)\right]_{\varepsilon=0}=0
\; \; \; \; .
\eqno (5.3.11)
$$
Equation (5.3.11) is obtained when $m=n+2$. These properties
can be proved by (formal) differentiation, as follows.

If the conformal variation of an operator of Laplace type
reads
$$
P(\varepsilon)=e^{-2 \varepsilon f} P(0)
-\varepsilon F
\; \; \; \; ,
\eqno (5.3.12)
$$
one finds
$$
\eqalignno{
\; & \left[{d\over d\varepsilon} {\rm Tr}_{L^{2}}
\Bigr(e^{-t P(\varepsilon)} \Bigr)\right]_{\varepsilon=0}
={\rm Tr}_{L^{2}} \biggr[\Bigr(2tf P(0)+tF \Bigr)
e^{-t P(0)} \biggr] \cr
&=-2t {\partial \over \partial t}{\rm Tr}_{L^{2}}
\biggr(f \; e^{-t P(0)} \biggr)+t {\rm Tr}_{L^{2}}
\biggr(F \; e^{-t P(0)} \biggr)
\; \; \; \; .
&(5.3.13)\cr}
$$
Moreover, by virtue of the asymptotic expansion (5.2.6), 
one has (the numerical factors $(4\pi)^{-m/2}$ are omitted
for simplicity, since they do not affect the form of Eqs.
(5.3.9)--(5.3.11); following Branson and Gilkey 1990 one 
can, instead, absorb such factors into the definition of 
the coefficients $a_{n/2}(f,P)$)
$$
{\partial \over \partial t}{\rm Tr}_{L^{2}}
\biggr(f \; e^{-t P(0)} \biggr) \sim -{1\over 2}
\sum_{n=0}^{\infty}(m-n)t^{{n\over 2}-{m\over 2}-1}
\; a_{n/2}(f,P(0))
\; \; \; \; .
\eqno (5.3.14)
$$
Thus, if $F$ vanishes, Eqs. (5.3.13) and (5.3.14) lead 
to the result (5.3.9). By contrast, if $f$ is set to
zero, one has $P(\varepsilon)=P(0)-\varepsilon F$, which
implies
$$
\eqalignno{
\left[{d\over d \varepsilon}{\rm Tr}_{L^{2}}
\Bigr(e^{-t P(\varepsilon)}\Bigr)\right]_{\varepsilon=0}
& \sim t^{-m/2} \sum_{n=0}^{\infty}t^{{n/2}+1}
\; a_{n/2}(F,P(0)) \cr
&= t^{-m/2} \sum_{l=2}^{\infty}t^{l/2} \;
a_{{l/2}-1}(F,P(0))
\; \; \; \; ,
&(5.3.15)\cr}
$$
which leads in turn to Eq. (5.3.10). Last, to obtain the
result (5.3.11), one considers the two-parameter 
conformal variation
$$
P(\varepsilon,\gamma)=e^{-2 \varepsilon f}P(0)
-\gamma e^{-2 \varepsilon f} F
\; \; \; \; .
\eqno (5.3.16)
$$
Now in Eq. (5.3.9) we first replace $n$ by $n+2$, and then
set $m=n+2$. One then has, from Eq. (5.3.16):
$$
{\partial \over \partial \varepsilon}a_{{n/2}+1}
(1,P(\varepsilon,\gamma))=0
\; \; \; \; .
\eqno (5.3.17)
$$
Equation (5.3.17) can be differentiated with respect to
$\gamma$, i.e. (see (5.3.10))
$$ 
\eqalignno{
0&= {\partial^{2}\over \partial \gamma \partial \varepsilon}
a_{{n/2}+1}(1,P(\varepsilon,\gamma))
={\partial \over \partial \varepsilon}
{\partial \over \partial \gamma}a_{{n/2}+1}
\Bigr(1,e^{-2 \varepsilon f}(P(0)-\gamma F)\Bigr) \cr
&={\partial \over \partial \varepsilon}a_{{n/2}}
\Bigr(e^{-2 \varepsilon f}F, e^{-2 \varepsilon f}P(0)\Bigr)
\; \; \; \; ,
&(5.3.18)\cr}
$$
and hence Eq. (5.3.11) is proved.

To deal with Robin boundary conditions (called Neumann in
Branson and Gilkey 1990) one needs another lemma, which is
proved following again Branson and Gilkey. What we obtain
is indeed a particular case of a more general property, which
is proved in section 4 of Avramidi and Esposito 1997. Our
starting point is $M$, a compact, connected one-dimensional
Riemannian manifold with boundary. In other words, one deals
with the circle or with a closed interval. If
$$
b: C^{\infty}(M) \longrightarrow {\Re}
$$
is a smooth, real-valued function, one can form the
first-order operator
$$
A \equiv {d\over dx}-b
\; \; \; \; ,
\eqno (5.3.19)
$$
and its (formal) adjoint
$$
A^{\dagger} \equiv -{d\over dx}-b
\; \; \; \; .
\eqno (5.3.20)
$$
From these operators, one can form the second-order
operators (cf. sections 5.5 and 5.6)
$$
D_{1} \equiv A^{\dagger} A = -\left[{d^{2}\over dx^{2}}
-b_{x}-b^{2}\right] \; \; \; \; ,
\eqno (5.3.21)
$$
$$
D_{2} \equiv A A^{\dagger}=-\left[{d^{2}\over dx^{2}}
+b_{x}-b^{2} \right] \; \; \; \; ,
\eqno (5.3.22)
$$
where $b_{x} \equiv {db\over dx}$. For $D_{1}$, Dirichlet
boundary conditions are taken, while Robin boundary conditions
are assumed for $D_{2}$. Defining
$$
f_{x} \equiv {df\over dx}
\; \; \; \; , \; \; \; \; 
f_{xx} \equiv {d^{2}f \over dx^{2}}
\; \; \; \; ,
$$
one then finds the result (cf. Branson and Gilkey 1990):
$$
(n-1)\biggr[a_{n/2}(f,D_{1})-a_{n/2}(f,D_{2})\biggr]
=a_{{n/2}-1} \Bigr(f_{xx}+2b f_{x}, D_{1} \Bigr)
\; \; \; \; .
\eqno (5.3.23)
$$
As a first step in the proof of (5.3.23), one takes a spectral
resolution for $D_{1}$, say $\left \{ \theta_{\nu}, \lambda_{\nu}
\right \}$, where $\theta_{\nu}$ is the eigenfunction of $D_{1}$
belonging to the eigenvalue $\lambda_{\nu}$:
$$
D_{1} \; \theta_{\nu}=\lambda_{\nu} \; \theta_{\nu}
\; \; \; \; .
\eqno (5.3.24)
$$
Thus, differentiation with respect to $t$ of the
heat-kernel diagonal:
$$
K(t,x,x,D_{1})=\sum_{\nu}e^{-t \lambda_{\nu}}
\; \theta_{\nu}^{2}(x) \; \; \; \; ,
\eqno (5.3.25)
$$
yields
$$
{\partial \over \partial t}K(t,x,x,D_{1})
=-\sum_{\nu}\lambda_{\nu} e^{-t \lambda_{\nu}}
\; \theta_{\nu}^{2}(x)=-\sum_{\nu}e^{-t \lambda_{\nu}}
(D_{1} \; \theta_{\nu})\theta_{\nu}
\; \; \; \; .
\eqno (5.3.26)
$$
Moreover, for any $\lambda_{\nu} \not = 0$, the set
$$
\left \{ {A \theta_{\nu}\over \sqrt{\lambda_{\nu}}},
\lambda_{\nu} \right \}
$$
provides a spectral resolution of $D_{2}$ on
${\rm Ker}(D_{2})^{\perp}$, and one finds
$$
\eqalignno{
\; & {\partial \over \partial t}K(t,x,x,D_{2})
=-\sum_{\lambda_{\nu} \not = 0} \lambda_{\nu}
e^{-t \lambda_{\nu}} \; \theta_{\nu}^{2}(x) \cr
&= -\sum_{\lambda_{\nu} \not = 0} e^{-t \lambda_{\nu}}
\Bigr(\sqrt{\lambda_{\nu}} \; \theta_{\nu} \Bigr)
\Bigr(\sqrt{\lambda_{\nu}} \; \theta_{\nu} \Bigr)
=-\sum_{\lambda_{\nu} \not = 0} e^{-t \lambda_{\nu}}
(A \theta_{\nu}) (A \theta_{\nu}) \; .
&(5.3.27)\cr}
$$
Bearing in mind that $A \theta_{\nu}=0$ if $\lambda_{\nu}=0$,
summation may be performed over all values of $\nu$, to find
$$
\eqalignno{
\; & 2 {\partial \over \partial t}\Bigr[K(t,x,x,D_{1})
-K(t,x,x,D_{2})\Bigr] \cr
&=2 \sum_{\nu}e^{-t \lambda_{\nu}}\biggr[\Bigr(
\theta_{\nu}'' \theta_{\nu}-b' \theta_{\nu}^{2}
-b^{2} \theta_{\nu}^{2}\Bigr)+(\theta_{\nu}'-b \theta_{\nu})
(\theta_{\nu}'-b \theta_{\nu})\biggr] \cr
&= 2 \sum_{\nu}e^{-t \lambda_{\nu}}\biggr[\theta_{\nu}''
\theta_{\nu}-b' \theta_{\nu}^{2}+(\theta_{\nu}')^{2}
-2b \theta_{\nu}' \theta_{\nu} \biggr]
\; \; \; \; .
&(5.3.28)\cr}
$$
On the other hand, differentiation with respect to $x$ yields
$$
\left({\partial \over \partial x}-2b \right)K(t,x,x,D_{1})
=\sum_{\nu}e^{-t \lambda_{\nu}}\Bigr(2 \theta_{\nu}
\theta_{\nu}' -2b \theta_{\nu}^{2} \Bigr)
\; \; \; \; ,
\eqno (5.3.29)
$$
which implies
$$
f {\partial \over \partial x} \left({\partial \over \partial x}
-2b \right)K(t,x,x,D_{1})=2f {\partial \over \partial t}
\Bigr[K(t,x,x,D_{1})-K(t,x,x,D_{2})\Bigr]
\; \; \; \; .
\eqno (5.3.30)
$$
We now integrate this formula over $M$ and use the boundary
conditions described before, jointly with integration by parts.
All boundary terms are found to vanish, so that
$$
\eqalignno{
\; & \int_{M}f {\partial \over \partial x}\left(
{\partial \over \partial x}-2b \right)K(t,x,x,D_{1})dx
=\int_{M} \left({\partial^{2}f \over \partial x^{2}}
+2b {\partial f \over \partial x}\right)
K(t,x,x,D_{1})dx \cr
&=\int_{M}2f {\partial \over \partial t} \Bigr[
K(t,x,x,D_{1})-K(t,x,x,D_{2})\Bigr]
\; \; \; \; .
&(5.3.31)\cr}
$$
Bearing in mind the standard notation for heat-kernel
traces, Eq. (5.3.31) may be re-expressed as
$$
\eqalignno{
\; & 2 {\partial \over \partial t}\left[
{\rm Tr}_{L^{2}}\Bigr(f e^{-t D_{1}} \Bigr)
-{\rm Tr}_{L^{2}} \Bigr(f e^{-t D_{2}} \Bigr)\right] \cr
&={\rm Tr}_{L^{2}} \left[\Bigr(f_{xx}+2b f_{x}\Bigr)
e^{-t D_{1}} \right] \; \; \; \; ,
&(5.3.32)\cr}
$$
which leads to Eq. (5.3.23) by virtue of the asymptotic
expansion (5.2.6). 

The algorithm resulting from Eq. (5.3.10) is sufficient to
determine almost all interior terms in heat-kernel asymptotics.
To appreciate this, notice that one is dealing with conformal
variations which only affect the endomorphism of the operator
$P$ in (5.2.1). For example, on setting $n=2$ in Eq. (5.3.10),
one ends up by studying (the tilde symbol is now used for
interior terms)
$$
{\tilde a}_{1}(1,P) \equiv {1\over 6} \int_{M}{\rm Tr}
\Bigr[\alpha_{1}E+\alpha_{2}R \Bigr]
={\tilde a}_{1}(E,R)
\; \; \; \; ,
\eqno (5.3.33)
$$
which implies
$$
\eqalignno{
\; & {\tilde a}_{1}(1,P(0)-\varepsilon F)
={\tilde a}_{1}(E,R)-{\tilde a}_{1}(E-\varepsilon F,R) \cr
&={1\over 6} \int_{M}{\rm Tr} \biggr[\alpha_{1}
(E-(E-\varepsilon F))+\alpha_{2}(R-R)\biggr] \cr
&={1\over 6} \int_{M}{\rm Tr}(\alpha_{1}\varepsilon F)
\; \; \; \; ,
&(5.3.34)\cr}
$$
and hence
$$
\eqalignno{
\; & \left[{d\over d \varepsilon}{\tilde a}_{1}
(1,P(0)-\varepsilon F)\right]_{\varepsilon=0}
={1\over 6} \int_{M} {\rm Tr}(\alpha_{1}F) \cr
&={\tilde a}_{0}(F,P(0))=\int_{M}{\rm Tr}(F)
\; \; \; \; .
&(5.3.35)\cr}
$$
By comparison, Eq. (5.3.35) shows that 
$$
\alpha_{1}=6
\; \; \; \; .
\eqno (5.3.36)
$$
An analogous procedure leads to (see (5.2.20))
$$
\eqalignno{
\; & {\tilde a}_{2}(1,P(0)-\varepsilon F)
={\tilde a}_{2}\Bigr(E,R,{\rm Ric},{\rm Riem},\Omega \Bigr)
-{\tilde a}_{2}\Bigr(E-\varepsilon F,R,{\rm Ric},{\rm Riem},
\Omega \Bigr) \cr
&={1\over 360} \int_{M}{\rm Tr} \Bigr[\alpha_{4}R \varepsilon F
+ \alpha_{5} (E^{2}-(E-\varepsilon F)^{2})\Bigr] \cr
&={1\over 360} \int_{M}{\rm Tr} \biggr[\alpha_{4}R \varepsilon
F + \alpha_{5}\Bigr(-{\varepsilon}^{2}F^{2}
+ 2 \varepsilon E F \Bigr)\biggr]
\; \; \; \; .
&(5.3.37)\cr}
$$
Now one can apply Eq. (5.3.10) when $n=4$, to find
$$
\eqalignno{
\; & \left[{d\over d \varepsilon}{\tilde a}_{2}
(1,P(0)-\varepsilon F) \right]_{\varepsilon=0}
={1\over 360} \int_{M}{\rm Tr}\Bigr[\alpha_{4}FR
+2 \alpha_{5} FE \Bigr] \cr
&={\tilde a}_{1}(F,P(0))={1\over 6}\int_{M}{\rm Tr}
\Bigr[\alpha_{1}FE+\alpha_{2}FR \Bigr]
\; \; \; \; .
&(5.3.38)\cr}
$$
Equating the coefficients of the invariants occurring
in Eq. (5.3.38) one finds
$$
\alpha_{2}={1\over 60}\alpha_{4}
\; \; \; \; ,
\eqno (5.3.39)
$$
$$
\alpha_{5}=30 \alpha_{1}=180 \; \; \; \; .
\eqno (5.3.40)
$$
Further, the consideration of Eq. (5.3.11) when $n=2$ yields
$$
\eqalignno{
\; & \left[{d\over d\varepsilon}{\tilde a}_{1}
\Bigr(e^{-2 \varepsilon f}F, e^{-2 \varepsilon f} P(0)
\Bigr)\right]_{\varepsilon=0}={1\over 6}\int_{M}
\left \{ \biggr[{d\over d\varepsilon} {\rm Tr}
(\alpha_{1}FE)\biggr]_{\varepsilon=0}
+2f {\rm Tr}(\alpha_{1}FE) \right . \cr
& \left . + \left[{d\over d\varepsilon} {\rm Tr}
(\alpha_{2}FR)\right]_{\varepsilon=0}
+ 2f {\rm Tr}(\alpha_{2}FR) \right \}
\; \; \; \; .
&(5.3.41)\cr}
$$
At this stage, we need the conformal variation formulae
$$
\left[{d\over d\varepsilon}E(\varepsilon) 
\right]_{\varepsilon=0}
=-2fE+{1\over 2}(m-2) \cstok{\ } f
\; \; \; \; ,
\eqno (5.3.42)
$$
$$
\left[{d\over d \varepsilon}R(\varepsilon) 
\right]_{\varepsilon=0}
=-2fR-2(m-1)\cstok{\ }f
\; \; \; \; .
\eqno (5.3.43)
$$
Since we are studying the case $m=n+2=4$, we find
$$
\left[{d\over d \varepsilon}{\tilde a}_{1}
\Bigr(e^{-2 \varepsilon f}F, e^{-2 \varepsilon f}P(0)
\Bigr)\right]_{\varepsilon=0}={1\over 6} \int_{M}
{\rm Tr} \Bigr[(\alpha_{1}-6 \alpha_{2})F 
\cstok{\ }f \Bigr]=0
\; \; \; \; ,
\eqno (5.3.44)
$$
which implies
$$
\alpha_{2}={1\over 6} \alpha_{1}=1
\; \; \; \; ,
\eqno (5.3.45)
$$
$$
\alpha_{4}=60 \alpha_{2}=60
\; \; \; \; .
\eqno (5.3.46)
$$
After considering the Laplacian acting on functions for a
product manifold $M=M_{1} \times M_{2}$, the complete set
of coefficients for interior terms can be determined
(Branson and Gilkey 1990):
$$
\alpha_{3}=60 \; , \; \alpha_{6}=12 \; , \;
\alpha_{7}=5 \; , \; \alpha_{8}=-2 \; , \;
\alpha_{9}=2 \; , \; \alpha_{10}=30 
\; \; \; \; .
\eqno (5.3.47)
$$

As far as interior terms are concerned, one has to use
Eqs. (5.3.9) and (5.3.11), jointly with two conformal
variation formulae which provide divergence terms that
play an important role (Branson and Gilkey 1990):
$$
\eqalignno{
\; & \left[{d\over d \varepsilon}a_{1} \Bigr(F,
e^{-2 \varepsilon f} P(0) \Bigr)\right]_{\varepsilon=0}
-(m-2) a_{1}(fF,P(0)) \cr
&={1\over 6}(m-4) \int_{M} {\rm Tr} \Bigr(F
\cstok{\ }f \Bigr) \; \; \; \; ,
&(5.3.48)\cr}
$$
$$
\eqalignno{
\; & \left[{d\over d \varepsilon}a_{2} \Bigr(F,
e^{-2 \varepsilon f}P(0) \Bigr)\right]_{\varepsilon=0}
-(m-4)a_{2} (fF,P(0)) \cr
&={1\over 360}(m-6) \int_{M} F {\rm Tr} \biggr[
6 f_{\; \; \; ;b}^{;b \; \; \; a}
+10 f^{;a} R \cr
&+60 f^{;a}E + 4 f_{;b} R^{ab} \biggr]_{;a}
\; \; \; \; .
&(5.3.49)\cr}
$$
Equations (5.3.48) and (5.3.49) are proved for manifolds
without boundary in Lemma 4.2 of Branson and Gilkey 1990. 
They imply that, for manifolds with boundary, the right-hand
side of Eq. (5.3.48), evaluated for $F=1$, should be added
to the left-hand side of Eq. (5.3.9) when $n=2$. Similarly,
the right-hand side of Eq. (5.3.49), evaluated for $F=1$,
should be added to the left-hand side of Eq. (5.3.9)
when $n=4$. Other useful formulae involving boundary effects
are (Branson and Gilkey 1990)
$$
\int_{M} \Bigr(f_{;b} R^{ab} \Bigr)_{;a}
=\int_{\partial M} \biggr[f_{;j} \Bigr(
K_{\; \; \; \; \mid i}^{ij}-({\rm tr}K)^{\mid j}\Bigr)
+f_{;N} R_{\; \; NiN}^{i} \biggr]
\; \; \; \; ,
\eqno (5.3.50)
$$
$$
f_{;i}^{\; \; \; ;i}=f_{\mid i}^{\; \; \; \mid i}
-({\rm tr}K)f_{;N}
\; \; \; \; ,
\eqno (5.3.51)
$$
$$
\int_{M} \cstok{\ }f= - \int_{\partial M} f_{;N}
\; \; \; \; ,
\eqno (5.3.52)
$$
$$
\int_{\partial M} \cstok{\ } f
=\int_{\partial M} \Bigr[f_{;NN}-({\rm tr}K)f_{;N}\Bigr]
\; \; \; \; .
\eqno (5.3.53)
$$

On taking into account Eqs. (5.3.48)--(5.3.53), the application
of Eq. (5.3.9) when $n=2,3,4$ leads to 18 equations which are
obtained by setting to zero the coefficients multiplying
$$
f_{;N} \; \; ({\rm when} \; n=2)
\; \; \; \; ,
$$
$$
f_{;NN} \; \; , \; \; f_{;N}({\rm tr}K) \; \; , \; \; 
f_{;N}S \; \; ({\rm when} \; n=3)
\; \; \; \; ,
$$
and
$$
f_{;a \; \; \; N}^{\; \; ;a} \; \; , \; \;
f_{;N}E \; \; , \; \; 
f_{;N}R \; \; , \; \; 
f_{;N} R_{\; \; NiN}^{i} \; \; \; \; ,
$$
$$
f_{;NN} ({\rm tr}K) \; \; , \; \;
f_{;N}({\rm tr}K)^{2} \; \; , \; \;
f_{;N}K_{ij}K^{ij} \; \; , \; \; 
f_{\mid i}({\rm tr}K)^{\mid i} \; \; \; \; ,
$$
$$
f_{\mid i}K_{\; \; \; \mid j}^{ij} \; \; , \; \;
f_{\mid i} \Omega_{\; N}^{i} \; \; , \; \; 
f_{;N}S ({\rm tr}K) \; \; , \; \; 
f_{;N}S^{2} \; \; \; \; ,
$$
$$
f_{;NN}S \; \; , \; \; 
f_{\mid i}^{\; \; \; \mid i} \; S
\; \; ({\rm when} \; n=4)
\; \; \; \; .
$$
The integrals of these 18 terms have a deep geometric nature
in that they form a basis for the integral invariants. The
resulting 18 equations are
$$
-b_{0}(m-1)-b_{1}(m-2)+{1\over 2}b_{2}(m-2)-(m-4)=0
\; \; \; \; ,
\eqno (5.3.54)
$$
$$
{1\over 2}c_{0}(m-2)-2c_{1}(m-1)+c_{2}(m-1)-c_{6}(m-3)=0
\; \; \; \; ,
\eqno (5.3.55)
$$
$$
\eqalignno{
\; & -{1\over 2}c_{0}(m-2)+2c_{1}(m-1)-c_{2}-2c_{3}(m-1) \cr
&-2c_{4}-c_{5}(m-3)+{1\over 2}c_{7}(m-2)=0
\; \; \; \; ,
&(5.3.56)\cr}
$$
$$
-c_{7}(m-1)+c_{8}(m-2)-c_{9}(m-3)=0
\; \; \; \; ,
\eqno (5.3.57)
$$
$$
-6(m-6)+{1\over 2}d_{1}(m-2)-2d_{2}(m-1)-e_{7}(m-4)=0
\; \; \; \; ,
\eqno (5.3.58)
$$
$$
-60(m-6)-2d_{1}-d_{5}(m-1)+{1\over 2}d_{14}(m-2)
-e_{1}(m-4)=0
\; \; \; \; ,
\eqno (5.3.59)
$$
$$
\eqalignno{
\; & -10(m-6)-2d_{2}-d_{6}(m-1)\cr 
&+d_{9}+{1\over 2}d_{15}(m-2)
-e_{2}(m-4)=0
\; \; \; \; ,
&(5.3.60)\cr}
$$
$$
\eqalignno{
\; & -d_{7}(m-1)-d_{8}+2d_{9}-e_{3}(m-4)\cr 
&+{1\over 2}d_{16}(m-2)
+4(m-6)=0
\; \; \; \; ,
&(5.3.61)\cr}
$$
$$
{1\over 2}d_{5}(m-2)-2d_{6}(m-1)+d_{7}(m-1)+d_{8}
-e_{6}(m-4)=0
\; \; \; \; ,
\eqno (5.3.62)
$$
$$ 
\eqalignno{
\; & -{1\over 2}d_{5}(m-2)+2d_{6}(m-1)-d_{7}-d_{9}
-3d_{10}(m-1) \cr
&-2d_{11}+{1\over 2}d_{17}(m-2)-e_{4}(m-4)=0
\; \; \; \; ,
&(5.3.63)\cr}
$$
$$
\eqalignno{
\; & -d_{8}-d_{9}(m-3)-d_{11}(m-1)-3d_{12}\cr 
&+{1\over 2}
d_{18}(m-2)-e_{5}(m-4)=0
\; \; \; \; ,
&(5.3.64)\cr}
$$
$$
d_{3}(m-4)-{1\over 2}d_{5}(m-2)+2d_{6}(m-1)
-d_{7}-d_{9}-4(m-6)=0
\; \; \; \; ,
\eqno (5.3.65)
$$
$$
d_{4}(m-4)-d_{8}-d_{9}(m-3)+4(m-6)=0
\; \; \; \; ,
\eqno (5.3.66)
$$
$$
(m-4)d_{13}=0
\; \; \; \; ,
\eqno (5.3.67)
$$
$$ 
\eqalignno{
\; & -{1\over 2}d_{14}(m-2)+2d_{15}(m-1)-d_{16}
-2d_{17}(m-1) \cr
&-2d_{18}+d_{19}(m-2)-e_{8}(m-4)=0
\; \; \; \; ,
&(5.3.68)\cr}
$$
$$
-d_{19}(m-1)+{3\over 2}d_{20}(m-2)-e_{9}(m-4)=0
\; \; \; \; ,
\eqno (5.3.69)
$$
$$
{1\over 2}d_{14}(m-2)-2d_{15}(m-1)+d_{16}(m-1)
-e_{10}(m-4)=0
\; \; \; \; ,
\eqno (5.3.70)
$$
$$
{1\over 2}d_{14}(m-2)-2d_{15}(m-1)+d_{16}
-d_{21}(m-4)=0 
\; \; \; \; .
\eqno (5.3.71)
$$
This set of algebraic equations among universal constants holds
independently of the choice of Dirichlet or Robin boundary
conditions. Another set of equations which hold for either 
Dirichlet or Robin boundary conditions is obtained by applying
Eq. (5.3.11) when $n=3,4$. One then obtains 5 equations which
result from setting to zero the coefficients multiplying
$$
F_{;N} f_{;N} \; \; ({\rm when} \; n=3)
\; \; \; \; ,
$$
$$
f_{;N}F_{;NN} \; \; , \; \;
f_{;NN} F_{;N} \; \; , \; \; 
f_{;N} F_{;N} ({\rm tr}K) \; \; , \; \; 
f_{;N} F_{;N} S \; \; ({\rm when} \; n=4)
\; \; \; \; .
$$
The explicit form of these equations is (Branson and
Gilkey 1990)
$$
-4 c_{5}-5c_{6}+{3\over 2}c_{9}=0
\; \; \; \; ,
\eqno (5.3.72)
$$
$$
-5e_{6}-4e_{7}+2e_{10}=0
\; \; \; \; ,
\eqno (5.3.73)
$$
$$
2e_{1}-10e_{2}+5e_{3}-2e_{7}=0
\; \; \; \; ,
\eqno (5.3.74)
$$
$$
-2e_{1}+10e_{2}-e_{3}-10e_{4}-2e_{5}-5e_{6}+6e_{7}
+2e_{8}=0
\; \; \; \; ,
\eqno (5.3.75)
$$
$$
-5e_{8}+4e_{9}-5e_{10}=0
\; \; \; \; .
\eqno (5.3.76)
$$

Last, one has to use the Lemma expressed by Eq. (5.3.23)
when $n=2,3,4$, bearing in mind that
$$
E_{1} \equiv E(D_{1})=-b_{x}-b^{2}
\; \; \; \; ,
\eqno (5.3.77)
$$
$$
E_{2} \equiv E(D_{2})=b_{x}-b^{2}
\; \; \; \; .
\eqno (5.3.78)
$$
For example, when $n=2$, one finds
$$
a_{1}(f,D_{1})-a_{1}(f,D_{2})-a_{0}\Bigr(f_{xx}
+2bf_{x},D_{1} \Bigr)=0
\; \; \; \; ,
\eqno (5.3.79)
$$
which implies (with Dirichlet for $D_{1}$ and Robin
for $D_{2}$)
$$
\int_{M}\Bigr[6f(E_{1}-E_{2})-6f_{xx}-12S f_{x} \Bigr]
+\int_{\partial M}\Bigr[-b_{2}fS-(3+b_{1})f_{;N}\Bigr]=0
\; \; \; \; .
\eqno (5.3.80)
$$
The integrand of the interior term in Eq. (5.3.80) may be
re-expressed as a total divergence, and hence one gets
$$
\int_{\partial M} \Bigr[(12-b_{2})fS+(3-b_{1})f_{;N}\Bigr]=0
\; \; \; \; ,
\eqno (5.3.81)
$$
which leads to
$$
b_{1}=3 \; \; , \; \; b_{2}=12
\; \; \; \; .
\eqno (5.3.82)
$$
Further details concern only the repeated application of all
these algorithms, and hence we refer the reader to
Branson and Gilkey 1990.
\vskip 1cm
\centerline {\bf 5.4 Mixed Boundary Conditions}
\vskip 1cm
Mixed boundary conditions occur naturally in the theory of
fermionic fields, gauge fields and gravitation, in that some
components of the field obey one set of boundary conditions,
and the remaining part of the field obeys a complementary set
of boundary conditions (Esposito et al. 1997). 
Here, we focus on some mathematical
aspects of the problem, whereas more difficult problems are
studied in chapters 6, 9 and 10. The framework of our
investigation consists, as in section 5.3, of a compact
Riemannian manifold, say $M$, with smooth boundary 
$\partial M$. Given a vector bundle $V$ over $M$, we assume
that $V$ can be decomposed as the direct sum
$$
V=V_{n} \oplus V_{d}
\; \; \; \; ,
\eqno (5.4.1)
$$
near $\partial M$. The corresponding projection operators
are denoted by $\Pi_{n}$ and $\Pi_{d}$, respectively. On
$V_{n}$ one takes Neumann boundary conditions modified by some
endomorphism, say $S$, of $V_{n}$ (see (5.2.15)), while
Dirichlet boundary conditions hold on $V_{d}$. The (total)
boundary operator reads therefore (Gilkey 1995)
$$
{\cal B}f \equiv \biggr[\Bigr(\Pi_{n}f \Bigr)_{;N}
+S \Pi_{n}f \biggr]_{\partial M} \oplus
\Bigr[ \Pi_{d}f \Bigr]_{\partial M}
\; \; \; \; .
\eqno (5.4.2)
$$
On defining
$$
\psi \equiv \Pi_{n}-\Pi_{d}
\; \; \; \; ,
\eqno (5.4.3)
$$
seven new universal constants are found to contribute to
heat-kernel asymptotics for an operator of Laplace type,
say $P$. In other words, the linear combination of
projectors considered in Eq. (5.4.3) gives rise to seven
new invariants in the calculation of heat-kernel coefficients
up to $a_{2}$: one contributes to 
${\tilde a}_{3/2}(f,P,{\cal B})$, whereas 
the other six contribute to 
${\tilde a}_{2}(f,P,{\cal B})$ (of course, the
number of invariants is continuously increasing as one 
considers higher-order heat-kernel coefficients). The 
dependence on the boundary operator is emphasized by 
including it explicitly into the arguments of heat-kernel
coefficients. One can thus write the general formulae
(cf. (5.2.19) and (5.2.20))
$$
{\tilde a}_{3/2}(f,P,{\cal B})={\delta \over 96}
(4\pi)^{1/2} \int_{\partial M}{\rm Tr} \Bigr[
\beta_{1} f \psi_{\mid i} \; \psi^{\mid i} \Bigr]
+a_{3/2}(f,P,{\cal B})
\; \; \; \; ,
\eqno (5.4.4)
$$
$$
\eqalignno{
{\tilde a}_{2}(f,P,{\cal B})&={1\over 360}
\int_{\partial M}{\rm Tr} \Bigr[\beta_{2}f \psi \psi_{\mid i}
\; \Omega_{\; N}^{i}+\beta_{3}f \psi_{\mid i} \; 
\psi^{\mid i} ({\rm tr}K) \cr
&+\beta_{4}f \psi_{\mid i} \; \psi_{\mid j} K^{ij}
+\beta_{5}f \psi_{\mid i} \; \psi^{\mid i} S \cr
&+ \beta_{6} f_{;N} \psi_{\mid i} \; \psi^{\mid i}
+\beta_{7} f \psi_{\mid i} \; \Omega_{\; N}^{i} \Bigr]
+a_{2}(f,P,{\cal B}) \; \; \; \; ,
&(5.4.5)\cr}
$$
where $a_{3/2}(f,P,{\cal B})$ is formally analogous to
Eq. (5.2.19), but with some universal constants replaced by
linear functions of $\psi, \Pi_{n}, \Pi_{d}$, and similarly
for $a_{2}(f,P,{\cal B})$ and Eq. (5.2.20). The work in
Branson and Gilkey 1992b, Vassilevich 1995a and Gilkey 1995 has
fixed the following values for the universal constants 
$\left \{ \beta_{i} \right \}$ occurring in Eqs. (5.4.4)
and (5.4.5):
$$
\beta_{1}=-12 \; \; , \; \; \beta_{2}=60 \; \; , \; \;
\beta_{3}=-12 \; \; , \; \; \beta_{4}=-24 \; \; \; \; ,
\eqno (5.4.6)
$$
$$
\beta_{5}=-120 \; \; , \; \; 
\beta_{6}=-18 \; \; , \; \; 
\beta_{7}=0 \; \; \; \; .
\eqno (5.4.7)
$$
To obtain this result, it is crucial to bear in mind that
the correct functorial formula for the endomorphism $S$ is
$$
\left[{d\over d \varepsilon}S(\varepsilon) 
\right]_{\varepsilon=0}
=-fS +{1\over 2}(m-2)f_{;N} \Pi_{n}
\; \; \; \; .
\eqno (5.4.8)
$$
This result was first obtained in Vassilevich 1995a, where the
author pointed out that $\Pi_{n}$ should be included, since
the variation of $S$ should compensate the one of $\omega_{n}$
only on the subspace $V_{n}$. The unfortunate omission of
$\Pi_{n}$ led to incorrect results in physical applications, 
which were later corrected in Moss and Poletti 1994, hence
confirming the analytic results in D'Eath and Esposito 1991a
and Esposito et al. 1994a.

There is indeed a very rich literature on the topics in
spectral geometry studied in this chapter and in the previous
one. In particular, further to the literature cited so far,
we would like to recommend the work in Gilkey 1975, 
Kennedy 1978, B\'erard 1986, Amsterdamski et al. 1989,
Avramidi 1989, Moss and Dowker 1989, Avramidi 1990a--b, 
Avramidi 1991, McAvity and Osborn 1991a--b, Fursaev 1994,
Alexandrov and Vassilevich 1996, Falomir et al. 1996,
Branson et al. 1997.
\vskip 1cm
\centerline {\bf 5.5 Heat Equation and Index Theorem}
\vskip 1cm
Heat-kernel asymptotics makes it possible to obtain a
deep and elegant formula for the index of elliptic
operators. The problem is studied by many authors
(e.g. Atiyah and Patodi 1973, Gilkey 1995). Here, we follow
the outline given in Atiyah 1975a.

Let $P$ be an elliptic differential operator on a compact
manifold $X$ without boundary, and let $P^{\dagger}$ be
its adjoint. One can then consider the two self-adjoint
operators $P^{\dagger}P$ and $PP^{\dagger}$. If $\phi$ is
any eigenfunction of $P^{\dagger}P$ with eigenvalue 
$\lambda$:
$$
P^{\dagger}P \; \phi = \lambda \; \phi
\; \; \; \; ,
\eqno (5.5.1)
$$
one has, acting with $P$ on the left:
$$
(PP^{\dagger})P \; \phi = \lambda \; P \phi
\; \; \; \; .
\eqno (5.5.2)
$$
This means that $P \phi$ is an eigenfunction of
$PP^{\dagger}$ with eigenvalue $\lambda$, provided
that $\lambda$ does not vanish. Conversely, if $\Phi$
is any eigenfunction of $PP^{\dagger}$ with eigenvalue
$\lambda$:
$$
PP^{\dagger} \; \Phi= \lambda \; \Phi
\; \; \; \; ,
\eqno (5.5.3)
$$
one finds
$$
(P^{\dagger}P)P^{\dagger} \; \Phi
= \lambda \; P^{\dagger}\Phi
\; \; \; \; ,
\eqno (5.5.4)
$$
i.e. $P^{\dagger}\Phi$ is an eigenfunction of
$P^{\dagger}P$ with eigenvalue $\lambda$, provided that
$\lambda \not = 0$. In other words, the operators 
$P^{\dagger}P$ and $PP^{\dagger}$ have the same non-zero
eigenvalues.

The next step is the analysis of the operators
$e^{-t P^{\dagger}P}$ and $e^{-t P P^{\dagger}}$. They are
fundamental solutions of the corresponding heat equations
(cf. Eq. (5.2.2)). For positive values of $t$, these have
$C^{\infty}$ kernels and hence are of trace class (i.e.
they have finite norm). Their $L^{2}$ traces read
$$
{\rm Tr}_{L^{2}}\Bigr(e^{-t P^{\dagger}P}\Bigr)
=\sum_{\lambda}e^{-t \lambda}
\; \; \; \; ,
\eqno (5.5.5)
$$
$$
{\rm Tr}_{L^{2}}\Bigr(e^{-tPP^{\dagger}}\Bigr)
=\sum_{\mu}e^{-t \mu}
\; \; \; \; ,
\eqno (5.5.6)
$$
where the summations run over the respective spectra.
Thus, bearing in mind that the non-vanishing $\lambda$
coincide with the non-vanishing $\mu$, while $\lambda=0$
corresponds to the null-space of $P$, and $\mu=0$ 
corresponds to the null-space of $P^{\dagger}$, one finds
$$
{\rm index}(P)={\rm Tr}_{L^{2}}\Bigr(e^{-t P^{\dagger}P}
\Bigr)-{\rm Tr}_{L^{2}}\Bigr(e^{-tPP^{\dagger}}\Bigr)
\; \; \; \; .
\eqno (5.5.7)
$$
On the other hand, it is well known from section 5.2,
following the work in Minakshisundaram and Pleijel 1949
and Greiner 1971, that for any elliptic self-adjoint 
differential operator $A$ of order $n$ with non-negative
spectrum, the integrated heat-kernel has an asymptotic
expansion ($m$ being the dimension of $X$)
$$
{\rm Tr}_{L^{2}}e^{-tA} \sim \sum_{k=-m}^{\infty}
a_{k} \; t^{k/n}
\; \; \; \; .
\eqno (5.5.8)
$$
The coefficients $a_{k}$ are obtained by integrating over
$X$ a linear combination of local invariants (see
(5.2.16)--(5.2.20)):
$$
a_{k}=\int_{X} \alpha_{k}
\; \; \; \; .
\eqno (5.5.9)
$$
In our case, Eqs. (5.5.7)--(5.5.9) lead to a formula of
the kind
$$
{\rm index}(P)=\int_{X}(\alpha_{0}-\beta_{0})
\; \; \; \; ,
\eqno (5.5.10)
$$
where $\alpha_{0}$ and $\beta_{0}$ refer to the operators
$P^{\dagger}P$ and $PP^{\dagger}$, respectively. If the
manifold $X$ has a boundary, boundary terms will contribute
to the index. As we know from sections 3.6 and 3.7, it is
sufficient to refer to the index of the Dirac operator,
say $\cal D$, which takes the general form
$$
{\rm index}({\cal D})=\int_{X}(\alpha_{0}-\beta_{0})
+ \int_{\partial X}(\gamma_{0}-\delta_{0})
\; \; \; \; ,
\eqno (5.5.11)
$$
where $\gamma_{0}$ and $\delta_{0}$ are linear combinations
of local invariants on the boundary.
\vskip 1cm
\centerline {\bf 5.6 Heat-Kernel and $\zeta$-Function}
\vskip 1cm
If $\cal A$ is an elliptic, self-adjoint, positive-definite
operator, the spectral theorem makes it possible to define
its complex power ${\cal A}^{-s}$, where $s$ is any complex
number (Seeley 1967). The $L^{2}$ trace of the complex power of 
$\cal A$ is, by definition, its $\zeta$-function:
$$
\zeta_{{\cal A}}(s) \equiv {\rm Tr}_{L^{2}} {\cal A}^{-s}
\; \; \; \; .
\eqno (5.6.1a)
$$
In the literature, Eq. (5.6.1a) is frequently re-expressed
in the form
$$
\zeta_{{\cal A}}(s)=\sum_{{\lambda > 0}}\lambda^{-s}
\; \; \; \; ,
\eqno (5.6.1b)
$$
where $\lambda$ runs over the (discrete) eigenvalues of
$\cal A$, counted with their multiplicity. As it stands,
this infinite sum converges if ${\rm Re}(s)$ is greater
than a lower limit depending on the dimension $m$ of the
Riemannian manifold under consideration, and on the order
$n$ of the operator $\cal A$. Interestingly, 
$\zeta_{{\cal A}}(s)$ can be analytically continued as a
meromorphic function on the whole $s$-plane (Seeley 1967,
Atiyah et al. 1976, Esposito et al. 1997), which is
regular at $s=0$. Moreover, a deep link exists between
the $\zeta$-function of $\cal A$ and its integrated 
heat-kernel. This is expressed by the so-called inverse
Mellin transform (Hawking 1977, Gilkey 1995):
$$
\zeta_{{\cal A}}(s)={1\over \Gamma(s)}
\int_{0}^{\infty}t^{s-1} \; {\rm Tr}_{L^{2}}
\Bigr(e^{-t {\cal A}}\Bigr) dt
\; \; \; \; ,
\eqno (5.6.2)
$$
where $\Gamma$ is the $\Gamma$-function, defined in (4.2.28).
If $\cal A$ is a second-order operator, the identity (5.6.2),
jointly with the asymptotic expansion 
$$
{\rm Tr}_{L^{2}} \Bigr(e^{-t {\cal A}}\Bigr) \sim
\sum_{k=0}^{\infty}A_{k/2} t^{(k-m)/2}
\; \; \; \; ,
\eqno (5.6.3)
$$
implies that $\zeta(0)$ takes the form
$$
\zeta_{{\cal A}}(0)=A_{m/2}=\int_{X}Q_{m/2}
+\int_{{\partial X}}S_{m/2}
\; \; \; \; ,
\eqno (5.6.4)
$$
where $Q_{m/2}$ and $S_{m/2}$ are linear combinations
of geometric invariants. Their expression can be derived,
for a given value of $m$, from the general method described
in sections 5.2 and 5.3. In quantum field theory, Eq. 
(5.6.4) is found to express the conformal anomaly or the
one-loop divergence (Hawking 1977, Esposito et al. 1997).

Moreover, for a given elliptic operator, say $D$, one can
define the operators (cf. section 5.5)
$$
\bigtriangleup_{0} \equiv 1+D^{\dagger}D
\; \; \; \; ,
\eqno (5.6.5)
$$
$$
\bigtriangleup_{1} \equiv 1+DD^{\dagger}
\; \; \; \; ,
\eqno (5.6.6)
$$
and the $\zeta$-functions
$$
\zeta_{i}(s) \equiv {\rm Tr}_{L^{2}} 
{\bigtriangleup}_{i}^{-s}
\; \; \; \; i=1,2
\; \; \; \; .
\eqno (5.6.7)
$$
It is then possible to prove that (Atiyah 1966)
$$
{\rm index}(D)=\zeta_{0}(s)-\zeta_{1}(s)
\; \; \; \; .
\eqno (5.6.8)
$$
Remarkably, this formula holds {\it for all values} of $s$.
In particular, it is more convenient to choose $s=0$ for
the explicit calculation. The equations (5.5.11),
(5.6.4) and (5.6.8) show the deep link between index theory,
heat-kernels and $\zeta$-functions.
\vskip 100cm
\centerline {\it CHAPTER SIX}
\vskip 1cm
\centerline {\bf NON-LOCAL BOUNDARY CONDITIONS FOR}
\centerline {\bf MASSLESS SPIN-${1\over 2}$ FIELDS}
\vskip 1cm
\noindent
{\bf Abstract.} This chapter studies the one-loop 
approximation for a massless spin-${1\over 2}$
field on a flat four-dimensional Euclidean background 
bounded by two concentric three-spheres, when non-local boundary
conditions of the spectral type are imposed. The use of
$\zeta$-function regularization shows that the conformal
anomaly vanishes, as in the case when the same field
is subject to local boundary conditions involving projectors.
A similar analysis of non-local boundary conditions can be
performed for massless supergravity models on manifolds
with boundary, to study their one-loop properties.
Moreover, it is shown that the proof of 
self-adjointness for the boundary-value problem can be
obtained by means of the limit-point criterion. 
\vskip 100cm
\centerline {\bf 6.1 Introduction}
\vskip 1cm
The quantum theory of fermionic fields can be expressed,
following the ideas of Feynman, in terms of amplitudes
of going from suitable fermionic data on a spacelike surface
${\cal S}_{I}$, say, to fermionic data on a spacelike
surface ${\cal S}_{F}$. To make sure that the quantum 
boundary-value problem is well posed, one has actually to
consider the Euclidean formulation, where the boundary
three-surfaces, $\Sigma_{I}$ and $\Sigma_{F}$, say, may be
regarded as (compact) Riemannian three-manifolds bounding a
Riemannian four-manifold. In the case of massless spin-1/2
fields, which are the object of our investigation, one thus
deals with transition amplitudes
$$
{\cal A}[{\rm boundary \; data}]=\int e^{-I_{E}}
{\cal D}\psi \; {\cal D}{\widetilde \psi}
\; \; \; \; ,
\eqno (6.1.1)
$$
where $I_{E}$ is the Euclidean action functional, and
the integration is over all massless spin-1/2 fields 
matching the boundary data on $\Sigma_{I}$ and $\Sigma_{F}$. 
The path-integral representation of the quantum
amplitude (6.1.1) is then obtained with the help of Berezin
integration rules, and one has a choice of non-local 
(D'Eath and Esposito 1991b) or local (D'Eath and
Esposito 1991a) boundary conditions. The mathematical foundations
of the former lie in the theory of spectral asymmetry and
Riemannian geometry (Atiyah et al. 1975, Atiyah et al. 1976),
and their formulation can be described
as follows. In two-component spinor 
notation (see appendix 6.A), a massless
spin-1/2 field in a four-manifold with positive-definite 
metric is represented by a pair
$\Bigr(\psi^{A},{\widetilde \psi}_{A'}\Bigr)$ of independent
spinor fields, not related by any spinor conjugation. Suppose
now that $\psi^{A}$ and ${\widetilde \psi}^{A'}$ are expanded
on a family of concentric three-spheres as
$$
\psi^{A}={1\over 2\pi}\tau^{-{3\over 2}}\sum_{n=0}^{\infty}
\sum_{p,q=1}^{(n+1)(n+2)}\alpha_{n}^{pq}
\Bigr[m_{np}(\tau)\rho^{nqA}+{\widetilde r}_{np}(\tau)
{\overline \sigma}^{nqA}\Bigr]
\; \; \; \; ,
\eqno (6.1.2)
$$
$$
{\widetilde \psi}^{A'}={1\over 2\pi}\tau^{-{3\over 2}}
\sum_{n=0}^{\infty}\sum_{p,q=1}^{(n+1)(n+2)}
\alpha_{n}^{pq}\Bigr[{\widetilde m}_{np}(\tau)
{\overline \rho}^{nqA'}+r_{np}(\tau)\sigma^{nqA'}\Bigr]
\; \; \; \; .
\eqno (6.1.3)
$$
With a standard notation,
$\tau$ is the Euclidean-time coordinate which
plays the role of a radial coordinate, and the block-diagonal
matrices $\alpha_{n}^{pq}$ and the $\rho$- and 
$\sigma$-harmonics are described in detail in 
D'Eath and Halliwell 1987.
One can now check that the harmonics $\rho^{nqA}$ 
have positive eigenvalues for the intrinsic
three-dimensional Dirac operator on $S^{3}$:
$$
{\cal D}_{AB} \equiv {_{e}n_{AB'}} \; e_{B}^{\; \; B'j}
\; { }^{(3)}D_{j}
\; \; \; \; , 
\eqno (6.1.4)
$$ 
and similarly for the harmonics $\sigma^{nqA'}$ and the
Dirac operator
$$
{\cal D}_{A'B'} \equiv {_{e}n_{BA'}} \; e_{B'}^{\; \; \; Bj}
\; { }^{(3)}D_{j} 
\; \; \; \; .
\eqno (6.1.5)
$$ 
With our notation, ${_{e}n_{AB'}}$ is the Euclidean normal
to the boundary, $e_{B}^{\; \; B'j}$ are the spatial 
components of the two-spinor version of the tetrad, and
${ }^{(3)}D_{j}$ denotes three-dimensional covariant
differentiation on $S^{3}$. 
By contrast, the harmonics ${\overline \sigma}^{nqA}$ and
${\overline \rho}^{nqA'}$ have negative eigenvalues for
the operators (6.1.4) and (6.1.5), respectively.

The so-called {\it spectral} boundary conditions
rely therefore on a non-local operation, i.e. the separation of
the spectrum of a first-order elliptic operator 
(our (6.1.4) and (6.1.5))
into a positive and a negative part. They require that half of
the spin-1/2 field should vanish on $\Sigma_{F}$, where this half
is given by those modes $m_{np}(\tau)$ and $r_{np}(\tau)$ which
multiply harmonics having positive eigenvalues 
for (6.1.4) and (6.1.5), respectively.
The remaining half of the field should vanish on $\Sigma_{I}$, and
is given by those modes ${\widetilde r}_{np}(\tau)$ and
${\widetilde m}_{np}(\tau)$ which multiply harmonics having
negative eigenvalues for 
(6.1.4) and (6.1.5), respectively. One thus writes 
$$
\Bigr[\psi^{A}_{(+)}\Bigr]_{\Sigma_{F}}=0
\Longrightarrow \Bigr[m_{np}\Bigr]_{\Sigma_{F}}=0
\; \; \; \; ,
\eqno (6.1.6)
$$
$$
\Bigr[{\widetilde \psi}^{A'}_{(+)}\Bigr]_{\Sigma_{F}}=0
\Longrightarrow \Bigr[r_{np}\Bigr]_{\Sigma_{F}}=0
\; \; \; \; ,
\eqno (6.1.7)
$$
and
$$
\Bigr[\psi^{A}_{(-)}\Bigr]_{\Sigma_{I}}=0 
\Longrightarrow 
\Bigr[{\widetilde r}_{np}\Bigr]_{\Sigma_{I}}=0
\; \; \; \; ,
\eqno (6.1.8)
$$
$$
\Bigr[{\widetilde \psi}^{A'}_{(-)}\Bigr]_{\Sigma_{I}}=0
\Longrightarrow
\Bigr[{\widetilde m}_{np}\Bigr]_{\Sigma_{I}}=0 
\; \; \; \; .
\eqno (6.1.9)
$$
Massless spin-1/2 fields are here studied since they provide
an interesting example of conformally invariant field
theory for which the spectral boundary conditions (6.1.6)--(6.1.9)
occur naturally already at the classical level 
(D'Eath and Halliwell 1987).

Section 6.2 is devoted to the evaluation of the
$\zeta(0)$ value resulting from the boundary conditions
(6.1.6)--(6.1.9). This yields the one-loop divergence of the
quantum amplitude, and coincides with the conformaly anomaly
in our model. Essential self-adjointness is proved
in section 6.3.
\vskip 1cm
\centerline {\bf 6.2 $\zeta(0)$ Value with 
Non-Local Boundary Conditions}
\vskip 1cm
As shown in D'Eath and Esposito 1991a-b, Esposito 1994a, 
the modes occurring in the expansions
(6.1.2) and (6.1.3) obey a coupled set of equations, i.e.
$$
\left({d\over d\tau}-{\Bigr(n+{3\over 2}\Bigr)\over \tau}
\right)x_{np}=E_{np} \; {\widetilde x}_{np}
\; \; \; \; ,
\eqno (6.2.1)
$$
$$
\left(-{d\over d\tau}-{\Bigr(n+{3\over 2}\Bigr)\over \tau}
\right){\widetilde x}_{np}=E_{np} \; x_{np}
\; \; \; \; ,
\eqno (6.2.2)
$$
where $x_{np}$ denotes $m_{np}$ or $r_{np}$, and
${\widetilde x}_{np}$ denotes ${\widetilde m}_{np}$ or
${\widetilde r}_{np}$. Setting $E_{np}=M$ for simplicity
of notation one thus finds, for all $n \geq 0$, the
solutions of (6.2.1) and (6.2.2) in the form
$$
m_{np}(\tau)=\beta_{1,n}\sqrt{\tau}I_{n+1}(M\tau)
+\beta_{2,n}\sqrt{\tau}K_{n+1}(M\tau)
\; \; \; \; ,
\eqno (6.2.3)
$$
$$
r_{np}(\tau)=\beta_{1,n}\sqrt{\tau}I_{n+1}(M\tau)
+\beta_{2,n}\sqrt{\tau}K_{n+1}(M\tau)
\; \; \; \; ,
\eqno (6.2.4)
$$
$$
{\widetilde m}_{np}(\tau)=\beta_{1,n}\sqrt{\tau}I_{n+2}(M\tau)
-\beta_{2,n}\sqrt{\tau}K_{n+2}(M\tau)
\; \; \; \; ,
\eqno (6.2.5)
$$
$$
{\widetilde r}_{np}(\tau)=\beta_{1,n}\sqrt{\tau}I_{n+2}(M\tau)
-\beta_{2,n}\sqrt{\tau}K_{n+2}(M\tau)
\; \; \; \; ,
\eqno (6.2.6)
$$
where $\beta_{1,n}$ and $\beta_{2,n}$ are some constants.
The insertion of (6.2.3)--(6.2.6) into 
the boundary conditions (6.1.6)--(6.1.9)
leads to the equations (hereafter $b$ and $a$ are the
radii of the two concentric three-sphere boundaries, with $b>a$,
and we define $\beta_{n} \equiv \beta_{2,n}/\beta_{1,n}$)
$$
I_{n+1}(Mb)+\beta_{n}K_{n+1}(Mb)=0
\; \; \; \; ,
\eqno (6.2.7)
$$
for $m_{np}$ and $r_{np}$ modes, and 
$$
I_{n+2}(Ma)-\beta_{n}K_{n+2}(Ma)=0
\; \; \; \; ,
\eqno (6.2.8) 
$$
for ${\widetilde m}_{np}$ and ${\widetilde r}_{np}$ modes,
with the same value of $M$. One thus finds two equivalent
formulae for $\beta_{n}$:
$$
\beta_{n}=-{I_{n+1}(Mb)\over K_{n+1}(Mb)}
={I_{n+2}(Ma)\over K_{n+2}(Ma)}
\; \; \; \; ,
\eqno (6.2.9)
$$
which lead to the eigenvalue condition
$$
I_{n+1}(Mb)K_{n+2}(Ma)+I_{n+2}(Ma)K_{n+1}(Mb)=0 
\; \; \; \; .
\eqno (6.2.10)
$$
The full degeneracy is $2(n+1)(n+2)$, for all $n \geq 0$,
since each set of modes contributes to (6.2.7) and (6.2.8) with
degeneracy $(n+1)(n+2)$ (D'Eath and Esposito 1991b).

We can now apply $\zeta$-function regularization to evaluate the
resulting conformal anomaly, following the algorithm developed 
in Barvinsky et al. 1992 and applied several 
times in the recent literature (Esposito et al. 1997).
The basic properties are as follows. Let us denote by $f_{n}$ the
function occurring in the equation obeyed by the eigenvalues by 
virtue of boundary conditions, after taking out fake roots (e.g.
$x=0$ is a fake root of order $\nu$ of the Bessel function
$I_{\nu}(x)$). Let $d(n)$ be the degeneracy of the eigenvalues
parametrized by the integer $n$. One can then define the function
$$
I(M^{2},s) \equiv \sum_{n=n_{0}}^{\infty}
d(n)n^{-2s}\log f_{n}(M^{2})
\; \; \; \; ,
\eqno (6.2.11)
$$
and the work in Barvinsky et al. 1992 
shows that such a function admits an
analytic continuation to the complex-$s$ plane as a meromorphic
function with a simple pole at $s=0$, in the form
$$
``I(M^{2},s)"={I_{\rm pole}(M^{2})\over s}+I^{R}(M^{2})
+{\rm O}(s) 
\; \; \; \; .
\eqno (6.2.12)
$$
The function $I_{\rm pole}(M^{2})$ is the residue at $s=0$,
and makes it possible to obtain the $\zeta(0)$ value as
$$
\zeta(0)=I_{\rm log}+I_{\rm pole}(M^{2}=\infty)
-I_{\rm pole}(M^{2}=0)
\; \; \; \; ,
\eqno (6.2.13)
$$
where $I_{\rm log}$ is the coefficient of the $\log(M)$ term
in $I^{R}$ as $M \rightarrow \infty$. The contributions
$I_{\rm log}$ and $I_{\rm pole}(\infty)$ are obtained from the
uniform asymptotic expansions of basis functions as 
$M \rightarrow \infty$ and their order $n \rightarrow \infty$
(Olver 1954), while
$I_{\rm pole}(0)$ is obtained by taking the
$M \rightarrow 0$ limit of the eigenvalue condition, and then
studying the asymptotics as $n \rightarrow \infty$. 
More precisely, $I_{\rm pole}(\infty)$ coincides with the
coefficient of ${1\over n}$ in the expansion as 
$n \rightarrow \infty$ of
$$
{1\over 2}d(n)\log \Bigr[\rho_{\infty}(n)\Bigr]
\; \; \; \; ,
$$
where $\rho_{\infty}(n)$ is the $n$-dependent term in the
eigenvalue condition as $M \rightarrow \infty$ and
$n \rightarrow \infty$. The $I_{\rm pole}(0)$ value is
instead obtained as the coefficient of ${1\over n}$ in
the expansion as $n \rightarrow \infty$ of
$$
{1\over 2}d(n) \log \Bigr[\rho_{0}(n)\Bigr]
\; \; \; \; ,
$$
where $\rho_{0}(n)$ is the $n$-dependent term in the
eigenvalue condition as $M \rightarrow 0$ and
$n \rightarrow \infty$ (Barvinsky et al. 1992,
Kamenshchik and Mishakov 1992).

In our problem, using the limiting form of Bessel functions 
when the argument tends to zero, one finds that the
left-hand side of Eq. (6.2.10) is proportional to $M^{-1}$ as
$M \rightarrow 0$. Hence one has to multiply by $M$ to get
rid of fake roots. Moreover, in the uniform asymptotic
expansion of Bessel functions as $M \rightarrow \infty$ and
$n \rightarrow \infty$, both $I$ and $K$ functions contribute
a ${1\over \sqrt{M}}$ factor. These properties imply that
$I_{\rm log}$ vanishes:
$$
I_{\rm log}={1\over 2}\sum_{l=1}^{\infty}2l(l+1)
\Bigr(1-{1\over 2}-{1\over 2}\Bigr)=0 
\; \; \; \; .
\eqno (6.2.14)
$$
Moreover, 
$$
I_{\rm pole}(\infty)=0
\; \; \; \; ,
\eqno (6.2.15)
$$
since there is no $n$-dependent coefficient in 
the uniform asymptotic expansion 
of Eq. (6.2.10). Last, one finds
$$
I_{\rm pole}(0)=0
\; \; \; \; ,
\eqno (6.2.16)
$$
since the limiting form of Eq. (6.2.10) as $M \rightarrow 0$ 
and $n \rightarrow \infty$ is
$$
{2\over Ma}(b/a)^{n+1} 
\; \; \; \; .
$$
The results (6.2.14)--(6.2.16), jointly with the general formula
(6.2.13), lead to a vanishing value of the one-loop divergence:
$$
\zeta(0)=0
\; \; \; \; .
\eqno (6.2.17)
$$

Our detailed calculation shows that, 
in flat Euclidean four-space, the conformal 
anomaly for a massless spin-1/2 field subject to non-local
boundary conditions of the spectral type on two concentric
three-spheres vanishes, as in the case when the same
field is subject to the local boundary conditions
$$
\sqrt{2} \; {_{e}n_{A}^{\; \; A'}} \; \psi^{A}
= \pm {\widetilde \psi}^{A'} 
\; \; \; \; {\rm on} \; \Sigma_{I} \; {\rm and}
\; \Sigma_{F} 
\; \; \; \; .
\eqno (6.2.18)
$$
If Eq. (6.2.18) holds and the spin-1/2 field is massless, the work
in Kamenshchik and Mishakov 1994 shows in fact that $\zeta(0)=0$.

Backgrounds given by flat Euclidean four-space bounded by two
concentric three-spheres are not the ones occurring in the 
Hartle-Hawking proposal for quantum cosmology, where the
initial three-surface $\Sigma_{I}$ shrinks to a point 
(Hartle and Hawking 1983, Hawking 1984).
Nevertheless, they are relevant for the quantization 
programme of gauge fields and gravitation in the presence
of boundaries (Esposito et al. 1997). 
In particular, similar techniques
have been used in section 5 of 
Esposito and Kamenshchik 1996 to study a two-boundary
problem for simple supergravity subject to spectral boundary
conditions in the axial gauge. One then finds the eigenvalue
condition
$$
I_{n+2}(Mb)K_{n+3}(Ma)+I_{n+3}(Ma)K_{n+2}(Mb)=0
\; \; \; \; ,
\eqno (6.2.19)
$$
for all $n \geq 0$. The analysis of Eq. (6.2.19) along the same
lines of what we have done for Eq. (6.2.10)
shows that three-dimensional transverse-traceless gravitino
modes yield a vanishing contribution to $\zeta(0)$, unlike
three-dimensional transverse-traceless 
modes for gravitons, which instead
contribute $-5$ to $\zeta(0)$ (Esposito et al. 1994b). 

Thus, the results in Esposito and Kamenshchik 1996 seem to
show that, at least in finite regions bounded by one three-sphere
or two concentric three-spheres, simple supergravity is not
one-loop finite in the presence of boundaries. Of course,
more work is in order to check this property, and then
compare it with the finiteness of scattering problems suggested
in D'Eath 1996. Further 
progress is thus likely to occur by virtue of
the fertile interplay of geometric and analytic techniques 
in the investigation of heat-kernel asymptotics and (one-loop)
quantum cosmology.
\vskip 1cm
\centerline {\bf 6.3 Self-Adjointness}
\vskip 1cm
This section is devoted to the mathematical foundations
of the one-loop analysis described so far. We shall see 
that the boundary conditions
are enough to determine a real and positive
spectrum for the squared Dirac operator (from which the
$\zeta(0)$ value can be evaluated as we just did). 

We here consider the portion of flat Euclidean four-space bounded
by a three-sphere of radius $a$. On inserting the expansions
(6.1.2) and (6.1.3) into the massless spin-1/2 action
$$
I_{E}={i\over 2}\int_{M}\biggr[{\widetilde \psi}^{A'}
\Bigr(\nabla_{AA'}\psi^{A}\Bigr)
-\Bigr(\nabla_{AA'} \; {\widetilde \psi}^{A'}\Bigr)
\psi^{A}\biggr]\sqrt{{\rm det} \; g} \; d^{4}x 
\; \; \; \; ,
\eqno (6.3.1)
$$
and studying the spin-1/2 eigenvalue equations, one
finds that the modes obey the second-order differential 
equations 
$$
P_{n}{\widetilde m}_{np}=P_{n}{\widetilde m}_{n,p+1}
=P_{n}{\widetilde r}_{np}=P_{n}{\widetilde r}_{n,p+1}=0
\; \; \; \; ,
\eqno (6.3.2)
$$
$$
Q_{n}r_{np}=Q_{n}r_{n,p+1}=Q_{n}m_{np}=Q_{n}m_{n,p+1}=0
\; \; \; \; ,
\eqno (6.3.3)
$$
where 
$$
P_{n} \equiv {d^{2}\over d\tau^{2}}
+\left[E_{n}^{2}-{((n+2)^{2}-{1\over 4})\over \tau^{2}}\right]
\; \; \; \; ,
\eqno (6.3.4)
$$
$$
Q_{n} \equiv {d^{2}\over d\tau^{2}}
+\left[E_{n}^{2}-{((n+1)^{2}-{1\over 4})\over \tau^{2}}\right]
\; \; \; \; ,
\eqno (6.3.5)
$$
and $E_{n}$ are the eigenvalues of the mode-by-mode form of the
Dirac operator (Esposito 1994a).

The spectral boundary conditions (6.1.6)--(6.1.9) imply
that, in our case,
$$
m_{np}(a)=0
\; \; \; \; ,
\eqno (6.3.6)
$$
$$
r_{np}(a)=0 
\; \; \; \; .
\eqno (6.3.7)
$$
Thus, one studies the one-dimensional operators $Q_{n}$
defined in (6.3.5), and the eigenmodes are requested to be
regular at the origin, and to obey (6.3.6) and (6.3.7) on
$S^{3}$. For this purpose, it is convenient to consider,
for all $n =0,1,2,...$, the differential operators
$$
{\widetilde Q}_{n} \equiv -{d^{2}\over d\tau^{2}}
+{((n+1)^{2}-{1\over 4})\over \tau^{2}}
\; \; \; \; ,
\eqno (6.3.8)
$$
with domain given by the functions $u$ in $AC^{2}[0,a]$
such that $u(a)=0$.
These are particular cases of a large class of operators
considered in the literature. They can be studied by using
the following definitions and theorems from 
Reed and Simon 1975 (cf. Chernoff 1977, Weidmann 1980):
\vskip 0.3cm
\noindent
{\it Definition 6.3.1.} The function $V$ is in the  
{\it limit circle} case at zero if for some, and therefore 
all $\lambda$, {\it all} solutions of the equation
$$
-\varphi''(x)+V(x)\varphi(x)=\lambda \varphi(x)
\eqno (6.3.9)
$$
are square integrable at zero.
\vskip 0.3cm
\noindent
{\it Definition 6.3.2.} If $V(x)$ is not in the limit circle
case at zero, it is said to be in the {\it limit point}
case at zero.
\vskip 0.3cm
\noindent
{\it Theorem 6.3.1.} Let $V$ be continuous and {\it positive}
near zero. If $V(x) \geq {3\over 4}x^{-2}$ near zero, then
$\cal O$ is in the limit point case at zero. 
\vskip 0.3cm
\noindent
When $V(x)$ takes the form ${c\over x^{2}}$ for $c>0$, 
theorem 6.3.1 can be proved as follows 
(Reed and Simon 1975). The equation 
$$
-\varphi''(x)+{c\over x^{2}}\varphi(x)=0
\eqno (6.3.10)
$$
admits solutions of the form $x^{\alpha}$, where
$\alpha$ takes the values
$$
\alpha_{1}={1\over 2}+{1\over 2}\sqrt{1+4c}
\; \; \; \; ,
\eqno (6.3.11)
$$
$$
\alpha_{2}={1\over 2}-{1\over 2}\sqrt{1+4c}
\; \; \; \; .
\eqno (6.3.12)
$$
When $\alpha=\alpha_{1}$ the solution is obviously 
square integrable at zero. However, when 
$\alpha=\alpha_{2}$, the solution is square integrable
at zero if and only if $\alpha_{2}>-{1\over 2}$,
which implies $c<{3\over 4}$. By virtue of definitions
6.3.1 and 6.3.2, this means that $V(x)$ is in the limit point 
at zero if and only if $c \geq {3\over 4}$. 

The operator ${\widetilde Q}_{n}$ obeys the limit point condition
at $\tau=0$, as we have just proved, and the limit circle condition
at $\tau=a$. These properties, jointly with the homogeneous 
Dirichlet condition at $\tau=a$, are sufficient to obtain a
self-adjoint boundary-value problem.
The calculation of conformal anomalies
for massless spin-${1\over 2}$ fields on flat Euclidean 
manifolds with boundary, initiated 
in D'Eath and Esposito 1991a--b and confirmed in
Kamenshchik and Mishakov 1992--1993,
has led over the last few years to a substantial
improvement of the understanding of one-loop divergences in
quantum field theory. Various independent techniques, which 
often rely on the structures applied in the spin-${1\over 2}$
analysis, have been used to deal with the whole set of
perturbative modes for Euclidean Maxwell theory and Euclidean
quantum gravity (Esposito 1994a--b; Esposito et al. 
1994a--b, 1995a--b; Esposito et al. 1997).
Nevertheless, a systematic analysis
aimed at explaining why the various $\zeta(0)$ values are
well defined was still lacking in the literature.

Our investigation, which relies in part on Esposito et al. 1996, 
represents the first step in this
direction, in the case of the squared Dirac operator with the
spectral boundary conditions (6.1.6)--(6.1.9).
Note that such non-local boundary conditions play a crucial
role in our proof, and that the
extension to the local boundary conditions studied in
D'Eath and Esposito 1991a (see section 10.3)
remains the main open problem in our
investigation. For this purpose, the calculation of deficiency
indices (Reed and Simon 1975) appears 
more appropriate. This would complete the
analysis in Esposito 1994a, Esposito 1995, 
where the existence of self-adjoint 
extensions is proved by showing that a 
linear, anti-involutory operator
$F$ exists which is norm-preserving, commutes with the Dirac
operator $\cal D$ and maps the domain of $\cal D$ into itself.

Moreover, it remains to be seen how to apply similar techniques to
the analysis of higher-spin fields. These are gauge fields and
gravitation, which obey a complicated set of mixed boundary
conditions (Esposito et al. 1997). 
The mixed nature of the boundary conditions
results from the request of their invariance under infinitesimal
gauge transformations. In particular, for the gravitational
field, at least five different sets of mixed boundary conditions
have been proposed in the literature (Barvinsky 1987, Luckock
and Moss 1989, Luckock 1991,
Esposito and Kamenshchik 1995, Marachevsky and Vassilevich
1996, Moss and Silva 1997).
If this investigation could be completed,
it would put on solid ground the current work on trace anomalies
and one-loop divergences
on manifolds with boundary, and it would add evidence 
in favour of quantum cosmology being at the very heart of many
exciting developments in quantum field theory, analysis and
differential geometry (Esposito et al. 1997).
\vskip 1cm
\centerline {\bf 6.A Appendix}
\vskip 1cm
Two-component spinor calculus is a powerful tool for
studying classical field theories in four-dimensional
space-time models. Within this framework,
the basic object is spin-space,
a two-dimensional complex vector space $S$ with a
symplectic form $\varepsilon$, i.e. an antisymmetric
complex bilinear form. Unprimed spinor indices 
$A,B,...$ take the values $0,1$ whereas primed spinor
indices $A',B',...$ take the values $0',1'$ since there
are actually two such spaces: unprimed spin-space
$(S,\varepsilon)$ and primed spin-space $(S',\varepsilon')$.
The whole two-spinor calculus in {\it Lorentzian}
four-manifolds relies on three fundamental properties
(Penrose and Rindler 1984, Ward and Wells 1990)

(i) The isomorphism between $\Bigr(S,\varepsilon_{AB}\Bigr)$ and
its dual $\Bigr(S^{*},\varepsilon^{AB}\Bigr)$. This is provided
by the symplectic form $\varepsilon$, which raises and
lowers indices according to the rules
$$
\varepsilon^{AB} \; \varphi_{B}=\varphi^{A} \; \in \; S
\; \; \; \; ,
\eqno (6.A.1)
$$
$$
\varphi^{B} \; \varepsilon_{BA}=\varphi_{A} \; \in \; S^{*}
\; \; \; \; .
\eqno (6.A.2)
$$
Thus, since
$$
\varepsilon_{AB}=\varepsilon^{AB}=\pmatrix {0&1\cr -1&0 \cr}
\; \; \; \; ,
\eqno (6.A.3)
$$
one finds in components $\varphi^{0}=\varphi_{1},
\varphi^{1}=-\varphi_{0}$. 

Similarly, one has the
isomorphism $\Bigr(S',\varepsilon_{A'B'}\Bigr)
\cong \Bigr((S')^{*},\varepsilon^{A'B'}\Bigr)$, which implies
$$
\varepsilon^{A'B'} \; \varphi_{B'}=\varphi^{A'} \; \in \; S'
\; \; \; \; ,
\eqno (6.A.4)
$$
$$
\varphi^{B'} \; \varepsilon_{B'A'}=\varphi_{A'} \; \in 
\; (S')^{*}
\; \; \; \; ,
\eqno (6.A.5)
$$
where
$$
\varepsilon_{A'B'}=\varepsilon^{A'B'}=\pmatrix
{0'&1'\cr -1'&0'\cr}
\; \; \; \; .
\eqno (6.A.6)
$$

(ii) The anti-isomorphism between $\Bigr(S,\varepsilon_{AB}\Bigr)$
and $\Bigr(S',\varepsilon_{A'B'}\Bigr)$, called complex conjugation,
and denoted by an overbar. According to a standard convention,
one has
$$
{\overline {\psi^{A}}} \equiv {\overline \psi}^{A'}
\; \in \; S'
\; \; \; \; ,
\eqno (6.A.7)
$$
$$
{\overline {\psi^{A'}}} \equiv {\overline \psi}^{A}
\; \in \; S
\; \; \; \; .
\eqno (6.A.8)
$$
Thus, complex conjugation maps elements of a spin-space to
elements of the {\it complementary} spin-space. Hence we
say that it is an anti-isomorphism (Stewart 1991).
In components, if $w^{A}$ is viewed as
$w^{A}=\pmatrix {\alpha \cr \beta \cr}$, the action of (6.A.7)
leads to
$$
{\overline {w^{A}}} \equiv {\overline w}^{A'}
\equiv \pmatrix {{\overline \alpha} \cr {\overline \beta}\cr}
\; \; \; \; ,
\eqno (6.A.9)
$$
whereas, if $z^{A'}=\pmatrix {\gamma \cr \delta \cr}$, then
(6.A.8) leads to
$$
{\overline {z^{A'}}} \equiv {\overline z}^{A}
=\pmatrix {{\overline \gamma}\cr {\overline \delta}\cr}
\; \; \; \; .
\eqno (6.A.10)
$$
With our notation, $\overline \alpha$ denotes complex
conjugation of the function $\alpha$, and so on. Note that
the symplectic structure is preserved by complex conjugation,
since ${\overline \varepsilon}_{A'B'}=\varepsilon_{A'B'}$.

(iii) The isomorphism between the tangent space $T$ at a
point of space-time and the tensor product of the 
unprimed spin-space $\Bigr(S,\varepsilon_{AB}\Bigr)$ and the
primed spin-space $\Bigr(S',\varepsilon_{A'B'}\Bigr)$:
$$
T \cong \Bigr(S,\varepsilon_{AB}\Bigr) \otimes
\Bigr(S',\varepsilon_{A'B'}\Bigr)
\; \; \; \; .
\eqno (6.A.11)
$$
The Infeld-van der
Waerden symbols $\sigma_{\; \; AA'}^{a}$ and
$\sigma_{a}^{\; \; AA'}$ express this isomorphism, and the
correspondence between a vector $v^{a}$ and a spinor 
$v^{AA'}$ is given by
$$
v^{AA'} \equiv v^{a} \; \sigma_{a}^{\; \; AA'}
\; \; \; \; ,
\eqno (6.A.12)
$$
$$
v^{a} \equiv v^{AA'} \; \sigma_{\; \; AA'}^{a}
\; \; \; \; .
\eqno (6.A.13)
$$
These mixed spinor-tensor symbols obey the identities
(Ward and Wells 1990)
$$
{\overline \sigma}_{a}^{\; \; AA'}=\sigma_{a}^{\; \; AA'}
\; \; \; \; ,
\eqno (6.A.14)
$$
$$
\sigma_{a}^{\; \; AA'} \; \sigma_{\; \; AA'}^{b}
=\delta_{a}^{\; \; b}
\; \; \; \; ,
\eqno (6.A.15)
$$
$$
\sigma_{a}^{\; \; AA'} \; \sigma_{\; \; BB'}^{a}
=\varepsilon_{B}^{\; \; A} \; \varepsilon_{B'}^{\; \; \; A'}
\; \; \; \; ,
\eqno (6.A.16)
$$
$$
\sigma_{[a}^{\; \; AA'} \; \sigma_{b]A}^{\; \; \; \; \; B'}
=-{i\over 2} \; \varepsilon_{abcd} \; \sigma^{cAA'}
\; \sigma_{\; \; A}^{d \; \; B'}
\; \; \; \; .
\eqno (6.A.17)
$$
Similarly, a one-form $\omega_{a}$ has a spinor equivalent
$$
\omega_{AA'} \equiv \omega_{a} \; \sigma_{\; \; AA'}^{a}
\; \; \; \; ,
\eqno (6.A.18)
$$
whereas the spinor equivalent of the metric is
$$
\eta_{ab} \; \sigma_{\; \; AA'}^{a}
\; \sigma_{\; \; BB'}^{b} \equiv
\varepsilon_{AB} \; \varepsilon_{A'B'}
\; \; \; \; .
\eqno (6.A.19)
$$
In particular, in Minkowski space-time Eqs. (6.A.12)
and (6.A.17) enable one to write down a coordinate system
in $2 \times 2$ matrix form
$$
x^{AA'}={1\over \sqrt{2}}
\pmatrix {{x^{0}+x^{3}}&{x^{1}-ix^{2}}\cr
{x^{1}+ix^{2}}&{x^{0}-x^{3}}\cr}
\; \; \; \; .
\eqno (6.A.20)
$$

In the Lorentzian-signature case, the Maxwell 
two-form $F \equiv F_{ab}dx^{a} \wedge dx^{b}$ can be
written spinorially (Ward and Wells 1990) as
$$
F_{AA'BB'}={1\over 2}\Bigr(F_{AA'BB'}-F_{BB'AA'}\Bigr)
=\varphi_{AB} \; \varepsilon_{A'B'}
+\varphi_{A'B'} \; \varepsilon_{AB}
\; \; \; \; ,
\eqno (6.A.21)
$$
where
$$
\varphi_{AB} \equiv {1\over 2} 
F_{AC'B}^{\; \; \; \; \; \; \; \; \; C'}
=\varphi_{(AB)}
\; \; \; \; ,
\eqno (6.A.22)
$$
$$
\varphi_{A'B'} \equiv {1\over 2}
F_{CB' \; \; A'}^{\; \; \; \; \; \; C}
=\varphi_{(A'B')}
\; \; \; \; .
\eqno (6.A.23)
$$
These formulae are obtained by applying the identity
$$
T_{AB}-T_{BA}=\varepsilon_{AB} \; T_{C}^{\; \; C}
\eqno (6.A.24)
$$
to express ${1\over 2}\Bigr(F_{AA'BB'}-F_{AB'BA'}\Bigr)$
and ${1\over 2}\Bigr(F_{AB'BA'}-F_{BB'AA'}\Bigr)$.
Note also that round brackets $(AB)$ denote (as usual)
symmetrization over the spinor indices $A$ and $B$, and that
the antisymmetric part of $\varphi_{AB}$ vanishes by virtue
of the antisymmetry of $F_{ab}$, since (Ward and Wells 1990)
$\varphi_{[AB]}={1\over 4}\varepsilon_{AB} \;
F_{CC'}^{\; \; \; \; \; CC'}={1\over 2}\varepsilon_{AB}
\; \eta^{cd} \; F_{cd}=0$. Last, but not least, in the 
Lorentzian case
$$
{\overline {\varphi_{AB}}} \equiv {\overline \varphi}_{A'B'}
=\varphi_{A'B'}
\; \; \; \; .
\eqno (6.A.25)
$$
The symmetric spinor fields $\varphi_{AB}$ and
$\varphi_{A'B'}$ are the anti-self-dual and self-dual parts
of the gauge curvature two-form, respectively.

Similarly, the Weyl curvature $C_{\; \; bcd}^{a}$, i.e. the
part of the Riemann curvature tensor invariant under conformal
rescalings of the metric, may be expressed spinorially,
omitting soldering forms (see below) 
for simplicity of notation, as
$$
C_{abcd}=\psi_{ABCD} \; \varepsilon_{A'B'} \;
\varepsilon_{C'D'}
+{\overline \psi}_{A'B'C'D'} \;
\varepsilon_{AB} \; \varepsilon_{CD}
\; \; \; \; .
\eqno (6.A.26)
$$
It should be emphasized that it is exactly the omission of
soldering forms that makes it possible to achieve the
most transparent translation of tensor fields into 
their spinor equivalent.

In canonical gravity, two-component spinors
lead to a considerable simplification of calculations. Denoting
by $n^{\mu}$ the future-pointing unit timelike normal to a
spacelike three-surface, its spinor version obeys the relations
$$
n_{AA'} \; e_{\; \; \; \; \; i}^{AA'}=0
\; \; \; \; ,
\eqno (6.A.27)
$$
$$
n_{AA'} \; n^{AA'}=1
\; \; \; \; ,
\eqno (6.A.28)
$$
where $e_{\; \; \; \; \; \mu}^{AA'} \equiv e_{\; \; \mu}^{a}
\; \sigma_{a}^{\; \; AA'}$ is the two-spinor version of the tetrad,
often referred to in the literature as soldering form (Ashtekar 1988).
Denoting by $h$ the induced metric on the three-surface, other
useful relations are (D'Eath 1984, Esposito 1994a)
$$
h_{ij}=-e_{AA'i} \; e_{\; \; \; \; \; j}^{AA'}
\; \; \; \; ,
\eqno (6.A.29)
$$
$$
e_{\; \; \; \; \; 0}^{AA'}=N \; n^{AA'}
+N^{i} \; e_{\; \; \; \; \; i}^{AA'}
\; \; \; \; ,
\eqno (6.A.30)
$$
$$
n_{AA'} \; n^{BA'}={1\over 2} \varepsilon_{A}^{\; \; B}
\; \; \; \; ,
\eqno (6.A.31)
$$
$$
n_{AA'} \; n^{AB'}={1\over 2} \varepsilon_{A'}^{\; \; \; B'}
\; \; \; \; ,
\eqno (6.A.32)
$$
$$
n_{[EB'} \; n_{A]A'}={1\over 4}\varepsilon_{EA} \;
\varepsilon_{B'A'}
\; \; \; \; ,
\eqno (6.A.33)
$$
$$
e_{AA'j} \; e_{\; \; \; \; \; k}^{AB'}
=-{1\over 2}h_{jk} \; \varepsilon_{A'}^{\; \; \; B'}
-i \varepsilon_{jkl}\sqrt{{\rm det} \; h} \;
n_{AA'} \; e^{AB'l}
\; \; \; \; .
\eqno (6.A.34)
$$
In Eq. (6.A.30), $N$ and $N^{i}$ are the lapse and shift
functions respectively (Esposito 1994a).

To obtain the space-time curvature, we first need to define
the spinor covariant derivative $\nabla_{AA'}$. 
If $\theta,\phi,\psi$ are spinor fields, $\nabla_{AA'}$
is a map such that (Penrose and Rindler 1984, Stewart 1991)
\vskip 0.3cm
\noindent
(1) $\nabla_{AA'}(\theta+\phi)=\nabla_{AA'}\theta
+\nabla_{AA'}\phi$ (i.e. linearity).
\vskip 0.3cm
\noindent
(2) $\nabla_{AA'}(\theta \psi)=\Bigr(\nabla_{AA'}\theta\Bigr)\psi
+\theta \Bigr(\nabla_{AA'}\psi\Bigr)$ (i.e. Leibniz rule).
\vskip 0.3cm
\noindent
(3) $\psi=\nabla_{AA'}\theta$ implies
${\overline \psi}=\nabla_{AA'}{\overline \theta}$
(i.e. reality condition).
\vskip 0.3cm
\noindent
(4) $\nabla_{AA'}\varepsilon_{BC}=\nabla_{AA'}\varepsilon^{BC}=0$,
i.e. the symplectic form may be used to raise or lower indices
within spinor expressions acted upon by $\nabla_{AA'}$, in
addition to the usual metricity condition
$\nabla g=0$, which involves instead the product of two
$\varepsilon$-symbols. 
\vskip 0.3cm
\noindent
(5) $\nabla_{AA'}$ commutes with any index substitution
not involving $A,A'$.
\vskip 0.3cm
\noindent
(6) For any function $f$, one finds
$\Bigr(\nabla_{a}\nabla_{b}-\nabla_{b}\nabla_{a}\Bigr)f
=2S_{ab}^{\; \; \; c} \; \nabla_{c}f$, where 
$S_{ab}^{\; \; \; c}$ is the torsion tensor.
\vskip 0.3cm
\noindent
(7) For any derivation $D$ acting on spinor fields, a spinor
field $\xi^{AA'}$ exists such that $D \psi=\xi^{AA'}
\; \nabla_{AA'} \psi, \forall \psi$.
\vskip 0.3cm
\noindent
As proved in Penrose and Rindler 1984, such a spinor covariant
derivative exists and is unique.

If Lorentzian space-time is replaced by a complex or
real Riemannian four-manifold, an important modification
should be made, since the anti-isomorphism between
unprimed and primed spin-space no longer exists. This
means that primed spinors can no longer be regarded as
complex conjugates of unprimed spinors, or viceversa,
as in (6.A.7) and (6.A.8). In particular, Eqs. (6.A.21)
and (6.A.26) should be re-written as
$$
F_{AA'BB'}=\varphi_{AB} \; \varepsilon_{A'B'}
+{\widetilde \varphi}_{A'B'} \; \varepsilon_{AB}
\; \; \; \; ,
\eqno (6.A.35)
$$
$$
C_{abcd}=\psi_{ABCD} \; \varepsilon_{A'B'}
\; \varepsilon_{C'D'}
+{\widetilde \psi}_{A'B'C'D'} \; 
\varepsilon_{AB} \; \varepsilon_{CD}
\; \; \; \; .
\eqno (6.A.36)
$$
With our notation, $\varphi_{AB},{\widetilde \varphi}_{A'B'}$,
as well as $\psi_{ABCD},{\widetilde \psi}_{A'B'C'D'}$
are {\it completely independent} symmetric spinor fields,
not related by any conjugation.

Indeed, a conjugation can still be defined in the real
Riemannian case, but it no longer 
relates $\Bigr(S,\varepsilon_{AB}\Bigr)$
to $\Bigr(S',\varepsilon_{A'B'}\Bigr)$. It is instead an
anti-involutory operation which maps elements of a spin-space
(either unprimed or primed) to elements of the {\it same}
spin-space. By anti-involutory we mean that, when applied twice
to a spinor with an odd number of indices, 
it yields the same spinor with the opposite
sign, i.e. its square is minus the identity, whereas the square
of complex conjugation as defined in 
(6.A.9) and (6.A.10) equals
the identity. Following Woodhouse 1985 and Esposito 1994a,
Euclidean conjugation, denoted by a {\it dagger}, is defined as follows:
$$
{\Bigr(w^{A}\Bigr)}^{\dagger} \equiv 
\pmatrix {{\overline \beta}\cr -{\overline \alpha}\cr}
\; \; \; \; ,
\eqno (6.A.37)
$$
$$
{\Bigr(z^{A'}\Bigr)}^{\dagger} \equiv
\pmatrix {-{\overline \delta}\cr {\overline \gamma}\cr}
\; \; \; \; .
\eqno (6.A.38)
$$
This means that, in flat Euclidean four-space, a unit
$2 \times 2$ matrix $\delta_{BA'}$ exists such that
$$
{\Bigr(w^{A}\Bigr)}^{\dagger} \equiv
\varepsilon^{AB} \; \delta_{BA'} \;
{\overline w}^{A'}
\; \; \; \; .
\eqno (6.A.39)
$$
We are here using the freedom to regard $w^{A}$ either as an
$SL(2,C)$ spinor for which complex conjugation can be defined,
or as an $SU(2)$ spinor for which Euclidean conjugation is
instead available. The soldering forms for $SU(2)$ spinors
only involve spinor indices of the same spin-space, i.e.
${\widetilde e}_{i}^{\; \; AB}$ and 
${\widetilde e}_{i}^{\; \; A'B'}$
(cf. Ashtekar 1991). More precisely, denoting by $E_{a}^{i}$
a real {\it triad}, where $i=1,2,3$, and by 
$\tau_{\; \; A}^{a \; \; \; B}$ the three Pauli matrices
obeying the identity
$$
\tau_{\; \; A}^{a \; \; \; B} \;
\tau_{\; \; B}^{b \; \; \; D}
=i \; \varepsilon^{abc} \; \tau_{cA}^{\; \; \; \; D}
+\delta^{ab} \; \delta_{A}^{\; \; D}
\; \; \; \; ,
\eqno (6.A.40)
$$
the $SU(2)$ soldering forms are defined by 
$$
{\widetilde e}_{\; \; A}^{j \; \; \; B} \equiv
-{i \over \sqrt{2}} \; E_{a}^{j} \; \tau_{\; \; A}^{a \; \; \; B}
\; \; \; \; .
\eqno (6.A.41)
$$
Note that our conventions differ from the ones in Ashtekar 1991,
i.e. we use ${\widetilde e}$ 
instead of $\sigma$, and $a,b$ for Pauli-matrix
indices, $i,j$ for tangent-space indices on a three-manifold $\Sigma$,
to agree with our previous notation. The soldering form in (6.A.41)
provides an isomorphism between the three-real-dimensional tangent
space at each point of $\Sigma$, and the three-real-dimensional
vector space of $2 \times 2$ trace-free Hermitian matrices.
The Riemannian three-metric on $\Sigma$ is then given by
$$
h^{ij}=-{\widetilde e}_{\; \; A}^{i \; \; \; B} \; 
{\widetilde e}_{\; \; B}^{j \; \; \; A}
\; \; \; \; .
\eqno (6.A.42)
$$
\vskip 100cm
\centerline {\it CHAPTER SEVEN}
\vskip 1cm
\centerline {\bf MASSLESS SPIN-${3\over 2}$ POTENTIALS} 
\vskip 1cm
\noindent
{\bf Abstract.} This chapter studies the two-spinor form of the
Rarita-Schwinger potentials subject to local boundary conditions
compatible with local supersymmetry.
The massless Rarita-Schwinger field equations are studied in 
four-real-dimensional Riemannian backgrounds with boundary. Gauge
transformations on the potentials are shown to be compatible
with the field equations providing the background is Ricci-flat,
in agreement with previous results in the literature.
However, the preservation of boundary conditions under such
gauge transformations leads to a restriction of the gauge freedom.
The recent construction by Penrose
of a second set of potentials which supplement the 
Rarita-Schwinger potentials is then applied.
The equations for these potentials, jointly with
the boundary conditions, 
imply that the background four-geometry is further
restricted to be totally flat. The analysis of other gauge
transformations confirms that, in the massless case, the
only admissible class of Riemannian backgrounds with
boundary is totally flat.
\vskip 100cm
\centerline {\bf 7.1 Introduction}
\vskip 1cm
Over the last few years, much effort has been made to
study locally supersymmetric boundary conditions in perturbative
quantum cosmology (Luckock and Moss 1989, Luckock 1991,
D'Eath and Esposito 1991a, Esposito 1994a-b, Kamenshchik
and Mishakov 1993, Kamenshchik and Mishakov 1994, Esposito
et al. 1997). The aim of 
this chapter is to perform a complete 
analysis of the corresponding {\it classical} elliptic boundary-value
problems. Indeed, in Esposito and Pollifrone 1994 
it was shown that one possible set of local
boundary conditions for a massless field of spin $s$, 
involving field strengths $\phi_{A...L}$ and ${\widetilde \phi}_{A'...L'}$ 
and the Euclidean normal ${_{e}n^{AA'}}$ to the boundary:
$$
2^{s} \; {_{e}n^{AA'}} ... {_{e}n^{LL'}} \; \phi_{A...L}
= \pm \; {\widetilde \phi}^{A'...L'} 
\; \; \; \; {\rm at} \; \; \; \; {\partial M} 
\; \; \; \; ,
\eqno (7.1.1)
$$
can only be imposed in flat Euclidean backgrounds with boundary.
However, such boundary conditions [motivated
by supergravity theories in anti-de Sitter 
space-time (Breitenlohner and Freedman 1982, Hawking 1983), where
(7.1.1) is essential to obtain a well-defined quantum theory
after taking the covering space of anti-de Sitter]
do not make it possible to relate bosonic
and fermionic fields through the action of 
complementary projection operators at the boundary 
(Luckock 1991). For this purpose, one has
to impose another set of local and supersymmetric boundary
conditions, first proposed in Luckock and Moss 1989. 
These are in general {\it mixed}, and 
involve in particular Dirichlet conditions for the transverse modes of
the vector potential of electromagnetism, a mixture of Dirichlet and
Neumann conditions for scalar fields, and local boundary conditions for
the spin-${1\over 2}$ field and the spin-${3\over 2}$ potential. Using
two-component spinor notation for supergravity 
(D'Eath 1984, Esposito 1995, D'Eath 1996), the
spin-${3\over 2}$ boundary conditions 
relevant for quantum cosmology take the form (Esposito 1994a)
$$
\sqrt{2} \; {_{e}n_{A}^{\; \; A'}} \;
\psi_{\; \; i}^{A}= \pm  
{\widetilde \psi}_{\; \; i}^{A'} 
\; \; \; \; {\rm at} \; \; \; \; \partial M 
\; \; \; \; .
\eqno (7.1.2)
$$
With our notation, 
$\Bigr(\psi_{\; \; i}^{A},{\widetilde \psi}_{\; \; i}^{A'}\Bigr)$
are the {\it independent} (i.e. not related by any conjugation) 
spatial components (hence $i=1,2,3$) of the spinor-valued 
one-forms appearing in the action 
functional of Euclidean supergravity (D'Eath 1996).

In the light of the results outlined so far, a naturally occurring
question is whether an analysis motivated by the one in 
Esposito and Pollifrone 1994 can be used to derive 
restrictions on the classical boundary-value problem corresponding
to (7.1.2). Such a question is of crucial importance for at least
two reasons:
\vskip 0.3cm
\noindent
(i) In the absence of boundaries, extended supergravity theories
are naturally formulated on curved backgrounds with a cosmological
constant (Breitenlohner and Freedman 1982, Hawking 1983).  
Thus, if a local theory in terms of 
spin-${3\over 2}$ potentials and in the presence of boundaries can
only be studied in flat Euclidean four-space, this result would 
make it impossible to consider 
the most interesting supergravity models when a four-manifold 
with boundaries occurs.
\vskip 0.3cm
\noindent
(ii) One of the main problems of the twistor programme for general
relativity lies in the impossibility of achieving a twistorial
reconstruction of (complex) vacuum space-times which are not
right-flat (i.e. such that the Ricci spinor $R_{AA'BB'}$ and 
the self-dual Weyl spinor ${\widetilde \psi}_{A'B'C'D'}$ vanish).
To overcome this difficulty, Penrose has proposed a new definition
of twistors as charges for massless spin-${3\over 2}$ fields in
Ricci-flat Riemannian manifolds (Penrose 1994, Esposito 1995).
However, since gravitino potentials have been studied also
in backgrounds which are not 
Ricci-flat, one is led to
ask whether the recent Penrose formalism can be applied to the
analysis of a larger class of Riemannian four-manifolds with boundary.

For this purpose, we introduce in section 7.2 the Rarita-Schwinger
potentials with their gauge transformations in Riemannian
background four-geometries. Section 7.3
derives compatibility conditions from the gauge transformations of
section 7.2, and from the boundary conditions (7.1.2). 
Section 7.4 is devoted to the potentials which
supplement the Rarita-Schwinger potentials in 
Ricci-flat backgrounds. Section 7.5 studies other
sets of gauge transformations.
Concluding remarks and open problems are presented in section 7.6.
Relevant details about the two-spinor form of Rarita-Schwinger equations
are given in the appendix.
\vskip 1cm
\centerline {\bf 7.2 Rarita-Schwinger Potentials}
\vskip 1cm
For the reasons described in the introduction, we are here interested
in the independent spatial components $\Bigr(\psi_{\; \; i}^{A},
{\widetilde \psi}_{\; \; i}^{A'}\Bigr)$ of the gravitino field
in Riemannian backgrounds. In terms of the spatial components
$e_{AB'i}$ of the tetrad, and of spinor fields, they can be
expressed as (Aichelburg and Urbantke 1981, D'Eath 1984,
Penrose 1991)
$$
\psi_{A \; i}= \Gamma_{\; \; AB}^{C'}
\; e_{\; \; C'i}^{B} 
\; \; \; \; ,
\eqno (7.2.1)
$$
$$
{\widetilde \psi}_{A' \; i}= 
\gamma_{\; \; A'B'}^{C} \; 
e_{C \; \; \; i}^{\; \; B'} 
\; \; \; \; .
\eqno (7.2.2)
$$
A first important difference with respect to the Dirac form of
the potentials studied in Esposito and Pollifrone
1994 is that the spinor fields 
$\Gamma_{\; \; AB}^{C'}$ and 
$\gamma_{\; \; A'B'}^{C}$ are no longer symmetric in
the second and third index. From now on, they will be 
referred to as spin-${3\over 2}$ potentials.
They obey the differential equations 
(see appendix and cf. Rarita and Schwinger 1941, 
Aichelburg and Urbantke 1981, Penrose 1991)
$$
\varepsilon^{B'C'} \; \nabla_{A(A'} \; \gamma_{\; \; B')C'}^{A}
=-3 \Lambda  \; {\widetilde \alpha}_{A'} 
\; \; \; \; ,
\eqno (7.2.3)
$$
$$
\nabla^{B'(B} \; \gamma_{\; \; \; B'C'}^{A)}
=\Phi_{\; \; \; \; \; \; \; \; C'}^{ABL'}
\; {\widetilde \alpha}_{L'} 
\; \; \; \; ,
\eqno (7.2.4)
$$
$$
\varepsilon^{BC} \; \nabla_{A'(A} \; \Gamma_{\; \; B)C}^{A'}
=-3\Lambda \; \alpha_{A} 
\; \; \; \; ,
\eqno (7.2.5)
$$
$$
\nabla^{B(B'} \; \Gamma_{\; \; \; BC}^{A')}
={\widetilde \Phi}_{\; \; \; \; \; \; \; \; \; C}^{A'B'L}
\; \alpha_{L} 
\; \; \; \; ,
\eqno (7.2.6)
$$
where $\nabla_{AB'}$ is the spinor covariant derivative corresponding
to the curved connection $\nabla$ of the background,
the spinors $\Phi_{\; \; \; \; \; C'D'}^{AB}$ and
${\widetilde \Phi}_{\; \; \; \; \; \; \; CD}^{A'B'}$ 
correspond to the trace-free part of the Ricci tensor, the
scalar $\Lambda$ corresponds to the
scalar curvature $R=24\Lambda$ of the background,
and $\alpha_{A},{\widetilde \alpha}_{A'}$ are a pair of
independent spinor fields, corresponding to the Majorana
field in the Lorentzian regime.
Moreover, the potentials are subject to the gauge transformations
(cf. section 7.5)
$$
{\widehat \gamma}_{\; \; B'C'}^{A}
\equiv \gamma_{\; \; B'C'}^{A}
+\nabla_{\; \; B'}^{A} \; \lambda_{C'} 
\; \; \; \; ,
\eqno (7.2.7)
$$
$$
{\widehat \Gamma}_{\; \; BC}^{A'}
\equiv \Gamma_{\; \; BC}^{A'}
+\nabla_{\; \; B}^{A'} \; \nu_{C} 
\; \; \; \; .
\eqno (7.2.8)
$$
A second important difference with respect 
to the Dirac potentials  
is that the spinor fields $\nu_{B}$ and $\lambda_{B'}$ are no
longer taken to be solutions of the Weyl equation.
They should be freely specifiable (see section 7.3).
\vskip 1cm
\centerline {\bf 7.3 Compatibility Conditions}
\vskip 1cm
Our task is now to derive compatibility conditions, by requiring
that the field equations (7.2.3)--(7.2.6) should also be satisfied by the
gauge-transformed potentials appearing on the left-hand side of
Eqs. (7.2.7)--(7.2.8). For this purpose, after defining the
operators
$$
\cstok{\ }_{AB} \equiv \nabla_{M'(A} 
\; \nabla_{B)}^{\; \; \; M'} 
\; \; \; \; ,
\eqno (7.3.1)
$$
$$
\cstok{\ }_{A'B'} \equiv \nabla_{F(A'} 
\; \nabla_{B')}^{\; \; \; \; F} 
\; \; \; \; ,
\eqno (7.3.2)
$$
we need the standard identity
$
\Omega_{[AB]} = {1\over 2} \varepsilon_{AB} \; \Omega_{C}^{\; \; C}
$
and the spinor Ricci identities 
$$
\cstok{\ }_{AB} \; \nu_{C}=\psi_{ABCD} \; \nu^{D}
-2 \Lambda \; \nu_{(A} \; \varepsilon_{B)C} 
\; \; \; \; ,
\eqno (7.3.3)
$$
$$
\cstok{\ }_{A'B'} \lambda_{C'}=
{\widetilde \psi}_{A'B'C'D'} \; \lambda^{D'}
-2 \Lambda \; \lambda_{(A'} \;
\varepsilon_{B')C'} 
\; \; \; \; ,
\eqno (7.3.4)
$$
$$
\cstok{\ }^{AB} \; \lambda_{B'}=\Phi_{\; \; \; \; M'B'}^{AB}
\; \lambda^{M'} 
\; \; \; \; ,
\eqno (7.3.5)
$$
$$
\cstok{\ }^{A'B'} \; \nu_{B}
={\widetilde \Phi}_{\; \; \; \; \; \; MB}^{A'B'}
\; \nu^{M} 
\; \; \; \; .
\eqno (7.3.6)
$$
Of course, ${\widetilde \psi}_{A'B'C'D'}$ and $\psi_{ABCD}$ are
the self-dual and anti-self-dual Weyl spinors respectively.

Thus, on using the Eqs. (7.2.3)--(7.2.8) and (7.3.1)--(7.3.6),
the basic rules of two-spinor calculus (Penrose and Rindler 
1986, Ward and Wells 1990, Stewart 1991, Esposito 1995)
lead to the compatibility equations 
$$
3 \Lambda \; \lambda_{A'}=0 
\; \; \; \; ,
\eqno (7.3.7)
$$
$$
\Phi_{\; \; \; \; M'}^{AB \; \; \; C'} \; \lambda^{M'}=0 
\; \; \; \; ,
\eqno (7.3.8)
$$
$$
3\Lambda \; \nu_{A}=0 
\; \; \; \; ,
\eqno (7.3.9)
$$
$$
{\widetilde \Phi}_{\; \; \; \; \; \; M}^{A'B' \; \; C}
\; \nu^{M}=0 
\; \; \; \; .
\eqno (7.3.10)
$$
Non-trivial solutions of (7.3.7)--(7.3.10) only exist if 
the scalar curvature and the trace-free part of the Ricci
tensor vanish. Hence the gauge transformations 
(7.2.7) and (7.2.8)
lead to spinor fields $\nu_{A}$ and $\lambda_{A'}$ which are
freely specifiable {\it inside} Ricci-flat backgrounds,
while the boundary conditions (7.1.2) are preserved under the
action of (7.2.7) and (7.2.8) provided that the following 
conditions hold at the boundary:
$$
\sqrt{2} \; {_{e}n_{A}^{\; \; A'}} \;
\Bigr(\nabla^{AC'} \; \nu^{B}\Bigr)e_{BC'i}
=\pm \Bigr(\nabla^{CA'} 
\lambda^{B'}\Bigr) e_{CB'i} 
\; \; \; \; {\rm at} \; \; \; \; \partial M 
\; \; \; \; .
\eqno (7.3.11)
$$
\vskip 1cm
\centerline {\bf 7.4 Second Set of 
Potentials in Ricci-Flat Backgrounds}
\vskip 1cm
As shown by Penrose (1994), in a Ricci-flat manifold the 
Rarita-Schwinger potentials may be supplemented by secondary
potentials. Here we use such a construction in its local form.
For this purpose, we introduce
a second set of potentials for spin ${3\over 2}$ by
requiring that locally (see Penrose 1994)
$$
\gamma_{A'B'}^{\; \; \; \; \; \; \; C} \equiv  \nabla_{BB'} \; 
\rho_{A'}^{\; \; \; CB} 
\; \; \; \; .
\eqno (7.4.1)
$$
Of course, special attention should be payed to the index ordering
in (7.4.1), since the spin-${3\over 2}$ potentials are not
symmetric. On inserting (7.4.1) into (7.2.3), a repeated use
of symmetrizations and anti-symmetrizations leads to the equation
(hereafter $\cstok{\ } \equiv \nabla_{CF'} \nabla^{CF'}$)
$$ \eqalignno{
\; & \varepsilon_{FL} \; \nabla_{AA'} \;
\nabla^{B'(F} \; \rho_{B'}^{\; \; \; A)L}
+{1\over 2} \nabla_{\; \; A'}^{A} \;
\nabla^{B'M} \; \rho_{B'(AM)} \cr
&+ \cstok{\ }_{AM} \; \rho_{A'}^{\; \; \; (AM)}
+{3\over 8} \cstok{\ } \rho_{A'}
= 0 \; \; \; \; ,
&(7.4.2)\cr}
$$
where, following Penrose 1994, we have defined
$$
\rho_{A'} \equiv \rho_{A' C}^{\; \; \; \; \; C} 
\; \; \; \; ,
\eqno (7.4.3)
$$
and we bear in mind that our background has to be Ricci-flat.
Thus, if the following equation holds (cf. Penrose 1994):
$$
\nabla^{B'(F} \; \rho_{B'}^{\; \; \; A)L}=0 
\; \; \; \; ,
\eqno (7.4.4)
$$
one finds
$$
\nabla^{B'M} \; \rho_{B'(AM)}={3\over 2} \; 
\nabla_{A}^{\; \; F'} \; \rho_{F'} \; \; \; \; ,
\eqno (7.4.5)
$$
and hence Eq. (7.4.2) may be cast in the form
$$
\cstok{\ }_{AM} \; \rho_{A'}^{\; \; \; (AM)}=0 
\; \; \; \; .
\eqno (7.4.6)
$$
On the other hand,
a very useful identity resulting from Eq. (4.9.13)
of Penrose and Rindler 1984 enables one to show that
$$
\cstok{\ }_{AM} \; \rho_{A'}^{\; \; \; (AM)}
=-\Phi_{AMA'}^{\; \; \; \; \; \; \; \; \; \; L'} \; 
\rho_{L'}^{\; \; \; (AM)} 
\; \; \; \; .
\eqno (7.4.7)
$$
Hence Eq. (7.4.6) reduces to an identity by virtue of
Ricci-flatness. Moreover, we have to insert (7.4.1) into the
field equation (7.2.4) for $\gamma$-potentials. By virtue of
Eq. (7.4.4) and of the identities (cf. Penrose and Rindler 1984)
$$
\cstok{\ }^{BM} \; \rho_{B' \; \; M}^{\; \; \; A}
=-\psi^{ABLM} \; \rho_{(LM)B'}
-\Phi_{\; \; \; \; \; B'}^{BM \; \; \; D'}
\; \rho_{\; \; MD'}^{A}
+4\Lambda \; \rho_{\; \; \; \; \; \; \; B'}^{(AB)} \; ,
\eqno (7.4.8)
$$
$$
\cstok{\ }^{B'F'} \; \rho_{B'}^{\; \; \; (AB)}
=3 \Lambda \; \rho^{(AB)F'}
+{\widetilde \Phi}_{\; \; \; \; \; \; \; L}^{B'F' \; \; \; A}
\; \rho_{\; \; \; \; \; \; \; B'}^{(LB)}
+{\widetilde \Phi}_{\; \; \; \; \; \; \; \; \; \; L}^{B'F'B}
\; \rho_{\; \; \; \; \; \; \; B'}^{(AL)} \; ,
\eqno (7.4.9)
$$
this leads to the equation
$$ 
\psi^{ABLM} \; \rho_{(LM)C'}=0 
\; \; \; \; ,
\eqno (7.4.10)
$$
where we have again used the Ricci-flatness condition.

Of course, potentials supplementing the $\Gamma$-potentials
may also be constructed locally. On defining (cf. (7.4.1))
$$
\Gamma_{AB}^{\; \; \; \; \; C'} \equiv
\nabla_{B'B} \; \theta_{A}^{\; \; C'B'} 
\; \; \; \; ,
\eqno (7.4.11)
$$
$$
\theta_{A} \equiv \theta_{AC'}^{\; \; \; \; \; C'} 
\; \; \; \; ,
\eqno (7.4.12)
$$
and requiring that (Penrose 1994, Esposito 1995)
$$
\nabla^{B(F'} \; \theta_{B}^{\; \; A')L'}=0 
\; \; \; \; ,
\eqno (7.4.13)
$$
one finds
$$
\nabla^{BM'} \; \theta_{B(A'M')}={3\over 2}
\nabla_{A'}^{\; \; \; F} \; \theta_{F} 
\; \; \; \; ,
\eqno (7.4.14)
$$
and a similar calculation yields an identity and the equation
$$ 
{\widetilde \psi}^{A'B'L'M'}
\; \theta_{(L'M')C}=0 
\; \; \; \; .
\eqno (7.4.15)
$$
Note that Eqs. (7.4.10) and (7.4.15) relate explicitly the
second set of potentials to 
the curvature of the background. This inconsistency
is avoided if one of the following conditions holds:
\vskip 0.3cm
\noindent
(i) The whole conformal curvature of the background vanishes.
\vskip 0.3cm
\noindent
(ii) $\psi^{ABLM}$ and $\theta_{(L'M')C}$, or
${\widetilde \psi}^{A'B'L'M'}$ and 
$\rho_{(LM)C'}$, vanish.
\vskip 0.3cm
\noindent
(iii) The symmetric parts of the 
$\rho$- and $\theta$-potentials vanish.
\vskip 0.3cm
\noindent
In the first case one finds that the only admissible 
background is again flat Euclidean four-space with boundary,
as in Esposito and Pollifrone 1994. By contrast, in the other cases, 
left-flat, right-flat or Ricci-flat backgrounds are still
admissible, provided that the $\rho$- and
$\theta$-potentials take the form
$$
\rho_{A'}^{\; \; \; CB}=\varepsilon^{CB} \; 
{\widetilde \alpha}_{A'} 
\; \; \; \; ,
\eqno (7.4.16)
$$
$$
\theta_{A}^{\; \; C'B'}=\varepsilon^{C'B'} \;
\alpha_{A} 
\; \; \; \; ,
\eqno (7.4.17)
$$
where $\alpha_{A}$ and ${\widetilde \alpha}_{A'}$ solve
the Weyl equations
$$
\nabla^{AA'} \; \alpha_{A}=0 
\; \; \; \; ,
\eqno (7.4.18)
$$
$$
\nabla^{AA'} \; {\widetilde \alpha}_{A'}=0 
\; \; \; \; .
\eqno (7.4.19)
$$
Eqs. (7.4.16)--(7.4.19) ensure also the validity of Eqs.
(7.4.4), (7.4.13), jointly with (7.A.6)
and (7.A.7) of the appendix.

However, if one requires the preservation of Eqs. (7.4.4)
and (7.4.13) under the following gauge transformations for 
$\rho$- and $\theta$-potentials (the order of the indices $AL$, $A'L'$
is of crucial importance):
$$
{\widehat \rho}_{B'}^{\; \; \; AL} \equiv
\rho_{B'}^{\; \; \; AL}+\nabla_{B'}^{\; \; \; A}
\; \mu^{L} 
\; \; \; \; ,
\eqno (7.4.20)
$$
$$
{\widehat \theta}_{B}^{\; \; A'L'} \equiv
\theta_{B}^{\; \; A'L'}+\nabla_{B}^{\; \; A'}
\; \sigma^{L'} 
\; \; \; \; ,
\eqno (7.4.21)
$$
one finds compatibility conditions in Ricci-flat backgrounds
of the form
$$
\psi_{AFLD} \; \mu^{D}=0 
\; \; \; \; ,
\eqno (7.4.22)
$$
$$
{\widetilde \psi}_{A'F'L'D'} \; \sigma^{D'}=0 
\; \; \; \; .
\eqno (7.4.23)
$$
Thus, to ensure {\it unrestricted} gauge freedom (except at
the boundary) for the second set of potentials, one is forced
to work with flat Euclidean backgrounds. The boundary
conditions (7.1.2) play a role in this respect, since they make
it necessary to consider both $\psi_{i}^{A}$ and
${\widetilde \psi}_{i}^{A'}$, and hence both
$\rho_{B'}^{\; \; \; AL}$ and $\theta_{B}^{\; \; A'L'}$.
Otherwise, one might use Eq. (7.4.22) to set to zero the 
anti-self-dual Weyl spinor only, {\it or} Eq. (7.4.23) to set to
zero the self-dual Weyl spinor only, so that self-dual
(left-flat) or anti-self-dual (right-flat) Riemannian 
backgrounds with boundary would survive.
\vskip 1cm
\centerline {\bf 7.5 Other Gauge Transformations}
\vskip 1cm
In the massless case, flat Euclidean backgrounds with
boundary are really the only possible choice for
spin-${3\over 2}$ potentials with a gauge freedom. 
To prove this,
we have also investigated an alternative set of gauge
transformations for spin-${3\over 2}$ potentials, written in
the form (cf. (7.2.7) and (7.2.8))
$$
{\widehat \gamma}_{\; \; B'C'}^{A} \equiv
\gamma_{\; \; B'C'}^{A}+\nabla_{\; \; C'}^{A} 
\; \lambda_{B'} 
\; \; \; \; ,
\eqno (7.5.1)
$$
$$
{\widehat \Gamma}_{\; \; \; BC}^{A'} \equiv
\Gamma_{\; \; \; BC}^{A'}+\nabla_{\; \; \; C}^{A'}
\; \nu_{B} 
\; \; \; \; .
\eqno (7.5.2)
$$
These gauge transformations {\it do not} correspond to the
usual formulation of the Rarita-Schwinger system, but we
will see that they can be interpreted in terms of
familiar physical concepts.

On imposing that the field equations (7.2.3)--(7.2.6) should be
preserved under the action of (7.5.1) and (7.5.2), and setting to
zero the trace-free part of the Ricci spinor (since it is
inconsistent to have gauge fields $\lambda_{B'}$ and
$\nu_{B}$ which depend explicitly 
on the curvature of the background) one finds compatibility
conditions in the form of differential equations, i.e. 
(cf. Esposito 1995)
$$
\cstok{\ }\lambda_{B'}=-2\Lambda \; \lambda_{B'} 
\; \; \; \; ,
\eqno (7.5.3)
$$
$$
\nabla^{(A(B'} \; \nabla^{C')B)} \lambda_{B'}=0 
\; \; \; \; ,
\eqno (7.5.4)
$$
$$
\cstok{\ }\nu_{B}=-2\Lambda \; \nu_{B} 
\; \; \; \; ,
\eqno (7.5.5)
$$
$$
\nabla^{(A'(B} \; \nabla^{C)B')} \; \nu_{B}=0 
\; \; \; \; .
\eqno (7.5.6)
$$
In a flat Riemannian four-manifold with flat connection $D$,
covariant derivatives commute and $\Lambda=0$. Hence it is
possible to express $\lambda_{B'}$ and $\nu_{B}$ as
solutions of the Weyl equations
$$
D^{AB'} \; \lambda_{B'}=0 
\; \; \; \; ,
\eqno (7.5.7)
$$
$$
D^{BA'} \; \nu_{B}=0 
\; \; \; \; ,
\eqno (7.5.8)
$$
which agree with the flat-space version of (7.5.3)--(7.5.6).
The boundary conditions (7.1.2) are then preserved under the
action of (7.5.1) and (7.5.2) if $\nu_{B}$ and $\lambda_{B'}$
obey the boundary conditions (cf. (7.3.11))
$$
\sqrt{2} \; {_{e}n_{A}^{\; \; A'}}
\Bigr(D^{BC'} \; \nu^{A}\Bigr)e_{BC'i}
=\pm \Bigr(D^{CB'} \; \lambda^{A'}\Bigr)
e_{CB'i} \; \; \; \; {\rm at} \; \; \; \; 
{\partial M} 
\; \; \; \; .
\eqno (7.5.9)
$$

In the curved case, on defining
$$
\phi^{A} \equiv \nabla^{AA'} \; \lambda_{A'} 
\; \; \; \; ,
\eqno (7.5.10)
$$
$$
{\widetilde \phi}^{A'} \equiv \nabla^{AA'} \; \nu_{A} 
\; \; \; \; ,
\eqno (7.5.11)
$$
equations (7.5.4) and (7.5.6) imply that 
these spinor fields solve the equations (cf. Esposito 1995)
$$
\nabla_{C'}^{\; \; \; (A} \; \phi^{B)}=0 
\; \; \; \; ,
\eqno (7.5.12)
$$
$$
\nabla_{C}^{\; \; (A'} \; {\widetilde \phi}^{B')}=0 
\; \; \; \; .
\eqno (7.5.13)
$$
Moreover, Eqs. (7.5.3), (7.5.5) and the spinor Ricci identities
imply that
$$
\nabla_{AB'} \; \phi^{A}=2\Lambda \; \lambda_{B'} 
\; \; \; \; ,
\eqno (7.5.14)
$$
$$
\nabla_{BA'} \; {\widetilde \phi}^{A'}=2\Lambda \; \nu_{B} 
\; \; \; \; .
\eqno (7.5.15)
$$
Remarkably, the Eqs. (7.5.12) and (7.5.13) are the twistor
equations (Penrose and Rindler 1986) in Riemannian four-geometries. 
The consistency conditions for the existence of non-trivial
solutions of such equations in curved four-manifolds 
are given by (Penrose and Rindler 1986)
$$
\psi_{ABCD}=0 
\; \; \; \; ,
\eqno (7.5.16)
$$
and
$$
{\widetilde \psi}_{A'B'C'D'}=0 
\; \; \; \; ,
\eqno (7.5.17)
$$
respectively, unless one regards $\phi^{B}$ as a four-fold
principal spinor of $\psi_{ABCD}$, and
${\widetilde \phi}^{B'}$ as a four-fold principal spinor
of ${\widetilde \psi}_{A'B'C'D'}$.

Further consistency conditions for our problem are derived
by acting with covariant differentiation on the twistor
equation, i.e.
$$
\nabla_{A'}^{\; \; \; C} \; \nabla^{AA'} \; \phi^{B}
+\nabla_{A'}^{\; \; \; C} \; \nabla^{BA'} \; \phi^{A}=0 
\; \; \; \; .
\eqno (7.5.18)
$$
While the complete symmetrization in $ABC$ yields Eq. (7.5.16),
the use of Eq. (7.5.18), jointly with the spinor Ricci identities
of section 7.3, yields
$$
\cstok{\ }\phi^{B}=2\Lambda \; \phi^{B} 
\; \; \; \; ,
\eqno (7.5.19)
$$
and an analogous equation is found for ${\widetilde \phi}^{B'}$.
Thus, since Eq. (7.5.12) implies 
$$
\nabla_{C'}^{\; \; \; A} \; \phi^{B}=\varepsilon^{AB}
\; \pi_{C'} 
\; \; \; \; ,
\eqno (7.5.20)
$$
we may obtain from (7.5.20) the equation 
$$
\nabla^{BA'} \; \pi_{A'}=2\Lambda \; \phi^{B} 
\; \; \; \; ,
\eqno (7.5.21)
$$
by virtue of the spinor Ricci identities and of Eq. (7.5.19).
On the other hand, in the light of (7.5.20), Eq. (7.5.14)
leads to
$$
\nabla_{AB'} \; \phi^{A}=2\pi_{B'}
=2\Lambda \; \lambda_{B'} 
\; \; \; \; .
\eqno (7.5.22)
$$
Hence $\pi_{A'}=\Lambda \; \lambda_{A'}$, and the 
definition (7.5.10) yields
$$
\nabla^{BA'} \; \pi_{A'}=\Lambda \; \phi^{B} 
\; \; \; \; .
\eqno (7.5.23)
$$
By comparison of Eqs. (7.5.21) and (7.5.23), one gets the
equation $\Lambda \; \phi^{B}=0$. If $\Lambda \not = 0$,
this implies that $\phi^{B}$, $\pi_{B'}$ and
$\lambda_{B'}$ have to vanish, and there is no gauge
freedom fou our model. This inconsistency is avoided
if and only if $\Lambda=0$, and the corresponding 
background is forced to be totally flat, since we have
already set to zero the trace-free part of the Ricci 
spinor and the whole conformal curvature. The same argument
applies to ${\widetilde \phi}^{B'}$ and the gauge field
$\nu_{B}$. The present analysis corrects the statements
made in section 8.8 of Esposito 1995, where it was not realized 
that, in our massless model, a non-vanishing cosmological
constant is incompatible with a gauge freedom for the
spin-${3\over 2}$ potential. More precisely, if one sets
$\Lambda=0$ from the beginning in Eqs. (7.5.3) and (7.5.5), the
system (7.5.3)--(7.5.6) admits solutions of the Weyl equation in
Ricci-flat manifolds. These backgrounds are further
restricted to be totally flat on considering the Eqs.
(7.4.10) and (7.4.15) for an arbitrary 
form of the $\rho$- and $\theta$-potentials.
As already pointed out at the end of section 7.4,
the boundary conditions (7.1.2) play a role, since otherwise
one might focus on right-flat or left-flat Riemannian
backgrounds with boundary.

Yet other gauge transformations can be studied (e.g. the
ones involving gauge fields $\lambda_{B'}$ and $\nu_{B}$
which solve the twistor equations), but they are all
incompatible with a non-vanishing cosmological 
constant in the massless case.
\vskip 1cm
\centerline {\bf 7.6 Concluding Remarks} 
\vskip 1cm
The consideration of boundary conditions is essential
for obtaining a well-defined formulation of physical theories
in quantum cosmology (Hartle and Hawking 1983, 
Hawking 1984). In particular, one-loop
quantum cosmology (Esposito 1994a, Esposito et al. 1997)
makes it necessary to study 
spin-${3\over 2}$ potentials 
about four-dimensional Riemannian backgrounds with
boundary. The corresponding classical analysis has been performed
in our chapter in the massless case, to supersede the analysis  
appearing in Esposito and Pollifrone 1994. Our results are as follows.

First, the gauge transformations (7.2.7)
and (7.2.8) are compatible with the
massless Rarita-Schwinger equations provided that the 
background four-geometry is Ricci-flat
(Deser and Zumino 1976). However, the presence
of a boundary restricts the gauge freedom, since the boundary
conditions (7.1.2) are preserved under 
the action of (7.2.7) and (7.2.8)
only if the boundary conditions (7.3.11) hold.

Second, the Penrose construction of 
a second set of potentials in Ricci-flat
four-manifolds shows that the admissible backgrounds may be
further restricted to be totally flat, or left-flat, or 
right-flat, unless these potentials take the special 
form (7.4.16) and (7.4.17). Hence the potentials
supplementing the Rarita-Schwinger potentials have a very clear 
physical meaning in Ricci-flat four-geometries with boundary:
they are related to the spinor fields $\Bigr(\alpha_{A},
{\widetilde \alpha}_{A'}\Bigr)$ corresponding to the Majorana
field in the Lorentzian version of Eqs. (7.2.3)--(7.2.6). (One 
should bear in mind that, in real Riemannian 
four-manifolds, the only admissible
spinor conjugation is Euclidean conjugation, which is anti-involutory
on spinor fields with an odd number 
of indices (Woodhouse 1985, Esposito 1995). Hence no Majorana
field can be defined in real Riemannian four-geometries.)

Third, to ensure unrestricted gauge freedom for the $\rho$- and
$\theta$-potentials, one is forced to work with flat Euclidean
backgrounds, when the boundary conditions (7.1.2) are imposed.
Thus, the very restrictive results obtained 
in Esposito and Pollifrone 1994
for massless Dirac potentials with the boundary conditions
(7.1.1) are indeed confirmed also for massless Rarita-Schwinger 
potentials subject to the supersymmetric boundary conditions
(7.1.2). Interestingly, a formalism originally
motivated by twistor theory (Penrose and Rindler 1986,
Ward and Wells 1990, Esposito 1995) 
has been applied to classical
boundary-value problems relevant for one-loop quantum cosmology.

Fourth, the gauge transformations 
(7.5.1) and (7.5.2) with non-trivial
gauge fields are compatible with the field equations 
(7.2.3)--(7.2.6) if and only if the background is totally flat.
The corresponding gauge fields solve the Weyl equations
(7.5.7) and (7.5.8), subject to the boundary conditions (7.5.9).
Indeed, it is well known that the Rarita-Schwinger 
description of a massless spin-${3\over 2}$ field is
equivalent to the Dirac description in a special choice of
gauge (Penrose 1994). In such a gauge, the spinor fields 
$\lambda_{B'}$ and $\nu_{B}$ solve the Weyl equations,
and this is exactly what we find in section 7.5 on choosing
the gauge transformations (7.5.1) and (7.5.2).

A non-vanishing cosmological constant can be consistently
studied when a {\it massive} spin-${3\over 2}$ potential
is studied (Townsend 1977). For this purpose, one has to replace the
spinor covariant derivative $\nabla_{AA'}$ in the field
equations (7.2.3)--(7.2.6) by a new spinor covariant derivative
$S_{AA'}$ which reduces to $\nabla_{AA'}$ when $\Lambda=0$.
In the language of $\gamma$-matrices, one has
(cf. Townsend 1977)
$$
S_{\mu} \equiv \nabla_{\mu}+f(\Lambda)\gamma_{\mu} 
\; \; \; \; ,
\eqno (7.6.1)
$$
where $f(\Lambda)$ vanishes at $\Lambda=0$, and $\gamma_{\mu}$
are the $\gamma$-matrices. The following chapter studies the
reformulation of sections 7.2--7.5 
in terms of the definition (7.6.1).

Moreover, other interesting problems are found to arise:
\vskip 0.3cm
\noindent
(i) Can one relate Eqs. (7.4.4) and (7.4.13) to the theory of
integrability conditions relevant for massless fields in
curved backgrounds (see Penrose 1994 and our appendix) ?
What happens when such equations do not hold ?
\vskip 0.3cm
\noindent
(ii) Is there an underlying global theory of Rarita-Schwinger
potentials ? In the affirmative case, what are the key features
of the global theory ? 
\vskip 0.3cm
\noindent
(iii) Can one reconstruct the Riemannian four-geometry from the
twistor space in Ricci-flat or conformally flat backgrounds with
boundary, or from whatever is going to replace twistor
space ?

Thus, the results and problems presented in our chapter seem to add
evidence in favour of a deep link existing between twistor
geometry, quantum cosmology and modern field theory.
\vskip 1cm
\centerline {\bf 7.A Appendix}
\vskip 1cm
Following Aichelburg and Urbantke 1981, 
one can locally express the $\Gamma$-potentials 
of (7.2.1) as (cf. (7.4.11))
$$
\Gamma_{\; \; BB'}^{A} \equiv \nabla_{BB'} \; \alpha^{A} 
\; \; \; \; .
\eqno (7.A.1)
$$
Thus, acting with $\nabla_{CC'}$ on both sides of (7.A.1), 
symmetrizing over $C'B'$
and using the spinor Ricci identity (7.3.6), one finds
$$
\nabla_{C(C'} \; \Gamma_{\; \; \; \; \; B')}^{AC}
={\widetilde \Phi}_{B'C'L}^{\; \; \; \; \; \; \; \; \; \; A} \;
\alpha^{L} \; \; \; \; .
\eqno (7.A.2)
$$
Moreover, acting with $\nabla_{C}^{\; \; C'}$ 
on both sides of (7.A.1),
putting $B'=C'$ (with contraction over this index), and using the
spinor Ricci identity (7.3.3) leads to
$$
\varepsilon^{AB} \; \nabla_{(C}^{\; \; \; C'} \; 
\Gamma_{\mid A \mid B)C'}=-3\Lambda \; \alpha_{C} 
\; \; \; \; .
\eqno (7.A.3)
$$
Eqs. (7.A.1)--(7.A.3) rely on the conventions 
in Aichelburg and Urbantke 1981. However, to
achieve agreement with the conventions in Penrose 1994
and in our book, Eqs. (7.2.3)--(7.2.6) are obtained
by defining (for the effect of
torsion terms, see comments following Eq. (21) in 
Aichelburg and Urbantke 1981)
$$
\Gamma_{B \; \; \; \; B'}^{\; \; \; A}
\equiv \nabla_{BB'} \; \alpha^{A} 
\; \; \; \; ,
\eqno (7.A.4)     
$$
$$
\gamma_{A' \; \; \; \; C}^{\; \; \; B'}
\equiv \nabla_{CA'} \; {\widetilde \alpha}^{B'} 
\; \; \; \; .
\eqno (7.A.5)
$$
On requiring that (7.A.5) and (7.4.1) should agree,
one finds by comparison that 
$$
\nabla_{BB'} \; \rho_{A'}^{\; \; \; (CB)}
=2 \nabla_{\; \; [A'}^{C} \;
{\widetilde \alpha}_{B']} 
\; \; \; \; ,
\eqno (7.A.6)
$$
which is obviously satisfied if $\rho_{A'}^{\; \; \; (CB)}=0$
and ${\widetilde \alpha}_{B'}$ obeys the Weyl equation (7.4.19).
Similarly, by comparison of (7.A.4) and (7.4.11) one finds
$$
\nabla_{B'B} \; \theta_{A}^{\; \; (C'B')}
=2 \nabla_{\; \; \; [A}^{C'} \; \alpha_{B]} 
\; \; \; \; ,
\eqno (7.A.7)
$$
which is satisfied if Eqs. (7.4.17) and (7.4.18) hold.

In the original approach by Penrose (1994), one describes
Rarita-Schwinger potentials in flat space-time in terms of a
rank-three vector bundle with local coordinates 
$\Bigr(\eta_{A},\zeta \Bigr)$, and an operator $\Omega_{AA'}$
whose action is defined by
$$
\Omega_{AA'}(\eta_{B},\zeta) \equiv
\Bigr({\cal D}_{AA'}\eta_{B},{\cal D}_{AA'}\zeta
-\eta^{C}\rho_{A'AC}\Bigr) 
\; \; \; \; ,
\eqno (7.A.8)
$$
$\cal D$ being the flat Levi-Civita connection of
Minkowski space-time. The gauge transformations are then
$$
\Bigr({\widehat \eta}_{B},{\widehat \zeta}\Bigr) 
\equiv \Bigr(\eta_{B},\zeta+\eta_{A}\xi^{A}\Bigr) 
\; \; \; \; ,
\eqno (7.A.9)
$$
$$
{\widehat \rho}_{A'AB} \equiv \rho_{A'AB}
+{\cal D}_{AA'}\xi_{B} 
\; \; \; \; .
\eqno (7.A.10)
$$
For the operator defined in (7.A.8), the integrability
condition on $\beta$-planes turns out to be
$$
{\cal D}^{A'(A} \; \rho_{A'}^{\; \; \; B)C}=0 
\; \; \; \; .
\eqno (7.A.11)
$$
\vskip 100cm
\centerline {\it CHAPTER EIGHT}
\vskip 1cm
\centerline {\bf MASSIVE SPIN-${3\over 2}$ POTENTIALS}
\vskip 1cm
\noindent
{\bf Abstract}. The two-component spinor form of massive
spin-${3\over 2}$ potentials in conformally flat Einstein
four-manifolds is studied. Following earlier work in the
literature, a non-vanishing cosmological constant makes it
necessary to introduce a supercovariant derivative operator. 
The analysis of supergauge transformations of potentials
for spin ${3\over 2}$ shows that the gauge freedom for massive 
spin-${3\over 2}$ potentials is generated by solutions of the
supertwistor equations. The supercovariant form of a partial
connection on a non-linear bundle is then obtained, and
the basic equation of massive secondary potentials is shown
to be the integrability condition on super $\beta$-surfaces of a
differential operator on a vector bundle of rank three.
\vskip 100cm
\centerline {\bf 8.1 Introduction}
\vskip 1cm
The local theory of spin-${3\over 2}$ potentials in real Riemannian
four-geometries is receiving careful consideration in the current
literature. As we know from chapter seven,
there are a number of deep motivations for this analysis. In
Minkowski space-time, twistors arise naturally as charges for massless
spin-${3\over 2}$ fields (Penrose 1991, Esposito 1995). 
In Ricci-flat four-manifolds (Deser and Zumino 1976) such fields are
well defined (Ricci-flatness being a necessary and sufficient consistency
condition), and a suitable generalization of the concept of twistors 
would make it possible to reconstruct solutions of the vacuum
Einstein equations from the resulting twistor space. In
extended supergravity theories, however, it is necessary to consider
massive spin-${3\over 2}$ fields in Riemannian backgrounds.
For this purpose, a careful spinorial analysis of the 
problem is in order.

We have thus studied massive spin-${3\over 2}$ potentials in 
four-manifolds with non-vanishing cosmological constant, considering the
supercovariant derivative compatible with a non-vanishing scalar
curvature. This is the content of section 8.2. Section 8.3 studies the
gauge freedom of the second kind, which is generated by a particular
type of twistors, i.e. the Euclidean Killing spinors. Section 8.4
studies the preservation of spin-${3\over 2}$ field equations under
the supergauge transformations of 
Rarita-Schwinger potentials. Section 8.5 studies the 
second set of potentials for spin ${3\over 2}$ in the massive
case. In section 8.6, a partial 
superconnection acting on a bundle 
with non-linear fibres is introduced. 
Section 8.7 studies 
the action of a superconnection 
on a vector bundle of rank three, 
and the corresponding integrability condition on 
super $\beta$-surfaces is derived.
Results and open problems are described in section 8.8. 
\vskip 10cm
\centerline {\bf 8.2 The Superconnection}
\vskip 1cm
In the massless case, the two-spinor form of the 
Rarita-Schwinger equations is the one given in chapter 7,
where $\nabla_{AA'}$ is the spinor covariant derivative
corresponding to the connection $\nabla$ of the background.
In the massive case, however, the appropriate connection,
hereafter denoted by $S$, has an additional term which couples
to the cosmological constant $\lambda=6\Lambda$.
In the language of
$\gamma$-matrices, the new covariant derivative $S_{\mu}$ to be
inserted {\it in the field equations} 
(Townsend 1977) takes the form 
$$
S_{\mu} \equiv \nabla_{\mu}+f(\Lambda)\gamma_{\mu} 
\; \; \; \; ,
\eqno (8.2.1)
$$
where $f(\Lambda)$ vanishes at $\Lambda=0$, and $\gamma_{\mu}$
are the curved-space $\gamma$-matrices. 
Since, following Esposito and Pollifrone 1996, 
we are interested in the two-spinor formulation of the problem,
we have to bear in mind the action of $\gamma$-matrices on any
spinor $\varphi \equiv \Bigr(\beta^{C},
{\widetilde \beta}_{C'}\Bigr)$. Note that unprimed and primed
spin-spaces are no longer (anti-)isomorphic in the case of
positive-definite four-metrics, since there is no complex
conjugation which turns primed spinors into unprimed spinors,
or the other way around (Penrose and Rindler 1986,
Esposito 1995). Hence $\beta^{C}$ and
${\widetilde \beta}_{C'}$ are totally unrelated. With this
understanding, we write the supergauge transformations for
massive spin-${3\over 2}$ potentials in the form
(cf. (7.2.7) and (7.2.8))
$$
{\widehat \gamma}_{\; \; B'C'}^{A} \equiv
\gamma_{\; \; B'C'}^{A}+S_{\; \; B'}^{A} 
\; \lambda_{C'} 
\; \; \; \; ,
\eqno (8.2.2)
$$
$$
{\widehat \Gamma}_{\; \; \; BC}^{A'} \equiv
\Gamma_{\; \; \; BC}^{A'}+S_{\; \; \; B}^{A'} \; \nu_{C}
\; \; \; \; ,
\eqno (8.2.3)
$$
where the action of $S_{AA'}$ on the gauge fields  
$\Bigr(\nu^{B},\lambda_{B'}\Bigr)$ is defined by (cf. (8.2.1))
$$
S_{AA'} \; \nu_{B} \equiv \nabla_{AA'} \; \nu_{B}
+f_{1}(\Lambda)\varepsilon_{AB} \; \lambda_{A'} 
\; \; \; \; ,
\eqno (8.2.4)
$$
$$
S_{AA'} \; \lambda_{B'} \equiv \nabla_{AA'} \; \lambda_{B'}
+f_{2}(\Lambda) \varepsilon_{A'B'} \; \nu_{A} 
\; \; \; \; .
\eqno (8.2.5)
$$
With our notation, $R=24 \Lambda$ is the scalar curvature,
$f_{1}$ and $f_{2}$ are two functions which vanish at
$\Lambda=0$, whose form will be determined later by a
geometric analysis. 

The action of $S_{AA'}$ on a many-index
spinor $T_{B'...F'}^{A...L}$ can be obtained by expanding
such a $T$ as a sum of products of spin-vectors, 
i.e. (Penrose and Rindler 1984)
$$
T_{B'...F'}^{A...L}=\sum_{i} \alpha_{(i)}^{A} ...
\beta_{(i)}^{L} \; \gamma_{B'}^{(i)} 
... \delta_{F'}^{(i)} 
\; \; \; \; ,
\eqno (8.2.6)
$$
and then applying the Leibniz rule and the definitions 
(8.2.4) and (8.2.5), where $\alpha_{(i)}^{A}$ has an independent
partner ${\widetilde \alpha}_{(i)}^{A'}$, ... ,
$\gamma_{B'}^{(i)}$ has an independent partner
${\widetilde \gamma}_{B}^{(i)}$, ... , and so on. 
Thus, one has for example
$$ \eqalignno{
\Bigr(S_{AA'}-\nabla_{AA'}\Bigr) 
\; T_{BCE'}&=\sum_{i}\Bigr[f_{1} \varepsilon_{AB}
\; {\widetilde \alpha}_{A'}^{(i)} \; \beta_{C}^{(i)}
\; \gamma_{E'}^{(i)}
+ f_{1} \varepsilon_{AC} \; \alpha_{B}^{(i)} \; 
{\widetilde \beta}_{A'}^{(i)} \;
\gamma_{E'}^{(i)} \cr
&+ f_{2} \varepsilon_{A'E'} \; \alpha_{B}^{(i)} \; \beta_{C}^{(i)}
\; {\widetilde \gamma}_{A}^{(i)}\Bigr] 
\; \; \; \; . 
&(8.2.7)\cr}
$$

A further requirement is that $S_{AA'}$ should annihilate
the curved $\varepsilon$-spinors. Hence in our analysis we
always assume that
$$
S_{AA'} \; \varepsilon_{BC}=0 
\; \; \; \; ,
\eqno (8.2.8)
$$
$$
S_{AA'} \; \varepsilon_{B'C'}=0 
\; \; \; \; .
\eqno (8.2.9)
$$
In the light of the definitions and assumptions presented
so far, one can write the Rarita-Schwinger equations
with non-vanishing cosmological constant
$\lambda=6\Lambda$, i.e. 
$$
\varepsilon^{B'C'} \; S_{A(A'} \;
\gamma_{\; \; B')C'}^{A}=\Lambda 
\; {\widetilde F}_{A'} 
\; \; \; \; ,
\eqno (8.2.10)
$$
$$
S^{B'(B} \; \gamma_{\; \; \; B'C'}^{A)}=0 
\; \; \; \; ,
\eqno (8.2.11)
$$
$$
\varepsilon^{BC} \; S_{A'(A} \; \Gamma_{\; \; \; B)C}^{A'}
=\Lambda \; F_{A} 
\; \; \; \; ,
\eqno (8.2.12)
$$
$$
S^{B(B'} \; \Gamma_{\; \; \; \; BC}^{A')}=0 
\; \; \; \; .
\eqno (8.2.13)
$$
With our notation, $F_{A}$ and ${\widetilde F}_{A'}$ are
spinor fields proportional to the traces of the second set
of potentials for spin ${3\over 2}$. These will be studied in
section 8.5.
\vskip 10cm
\centerline {\bf 8.3 Gauge Freedom of the Second Kind}
\vskip 1cm
The gauge freedom of the second kind is the one which does
not affect the potentials after a gauge 
transformation. This 
requirement corresponds to the case analyzed in Siklos 1985, 
where it is pointed out that, while the Lagrangian of
$N=1$ supergravity is invariant under gauge transformations 
with arbitrary spinor fields $\Bigr(\nu^{A},\lambda_{A'}\Bigr)$,
the actual {\it solutions} are only invariant if the 
supercovariant derivatives (8.2.4) and (8.2.5) vanish.

On setting to zero $S_{AA'} \; \nu_{B}$ and 
$S_{AA'} \; \lambda_{B'}$, one gets a coupled set of
equations which are the Euclidean version of the
Killing-spinor equation (Siklos 1985), i.e.
$$
\nabla_{\; \; \; B}^{A'} \; \nu_{C}=-f_{1}(\Lambda)
\lambda^{A'} \; \varepsilon_{BC} 
\; \; \; \; ,
\eqno (8.3.1)
$$
$$
\nabla_{\; \; B'}^{A} \; \lambda_{C'}=-f_{2}(\Lambda)
\nu^{A} \; \varepsilon_{B'C'} 
\; \; \; \; .
\eqno (8.3.2)
$$
What is peculiar of Eqs. (8.3.1) and (8.3.2) is that their
right-hand sides involve spinor fields which are,
themselves, solutions of the twistor equation. Hence one
deals with a special type of twistors, which do not exist
in a generic curved manifold. Equation (8.3.1) can
be solved for $\lambda^{A'}$ as
$$
\lambda_{C'}={1\over 2f_{1}(\Lambda)}\nabla_{C'}^{\; \; \; B}
\; \nu_{B} 
\; \; \; \; .
\eqno (8.3.3)
$$
The insertion of (8.3.3) into Eq. (8.3.2) and the use of spinor
Ricci identities (see (7.3.3)--(7.3.6)) 
yields the second-order equation
$$
\cstok{\ }\nu_{A}
+(6\Lambda+8f_{1}f_{2})\nu_{A}=0 
\; \; \; \; .
\eqno (8.3.4)
$$
On the other hand, Eq. (8.3.1) implies the twistor equation
$$
\nabla_{\; \; \; (B}^{A'} \; \nu_{C)}=0 
\; \; \; \; .
\eqno (8.3.5)
$$
Covariant differentiation of Eq. (8.3.5), jointly with spinor Ricci
identities, leads to
$$
\cstok{\ }\nu_{A}-2\Lambda \nu_{A}=0 
\; \; \; \; .
\eqno (8.3.6)
$$
By comparison of Eqs. (8.3.4) and (8.3.6) one finds the condition
$f_{1}f_{2}=-\Lambda$. The integrability condition of Eq. (8.3.5)
is given by (Penrose and Rindler 1986)
$$
\psi_{ABCD} \; \nu^{D}=0 
\; \; \; \; .
\eqno (8.3.7)
$$
This means that our manifold is conformally left-flat, unless
$\nu^{D}$ is a four-fold principal spinor of the
anti-self-dual Weyl spinor. The latter possibility is here
ruled out, {\it to avoid having gauge fields related explicitly
to the curvature of the background}.

The condition $f_{1}f_{2}=-\Lambda$ is also obtained by 
comparison of first-order equations, since for example
$$
\nabla^{AA'} \; \nu_{A}=2f_{1}\lambda^{A'}
=-2{\Lambda \over f_{2}} \lambda^{A'} 
\; \; \; \; .
\eqno (8.3.8)
$$
The first equality in (8.3.8) results from Eq. (8.3.1), while the
second one is obtained since the twistor equations also
imply that (see Eq. (8.3.2))
$$
\nabla^{AA'} \Bigr(-f_{2}\nu_{A}\Bigr)
=2\Lambda \; \lambda^{A'} 
\; \; \; \; .
\eqno (8.3.9)
$$
Entirely analogous results are obtained on considering the
twistor equation resulting from Eq. (8.3.2), i.e.
$$
\nabla_{\; \; (B'}^{A} \; \lambda_{C')}=0 
\; \; \; \; .
\eqno (8.3.10)
$$
The integrability condition of Eq. (8.3.10) is
$$
{\widetilde \psi}_{A'B'C'D'} \; \lambda^{D'}=0 
\; \; \; \; .
\eqno (8.3.11)
$$
Since our gauge fields cannot be four-fold principal
spinors of the Weyl spinor (cf. Lewandowski 1991), Eqs. (8.3.7) and 
(8.3.11) imply that our background geometry is conformally flat.
\vskip 10cm
\centerline {\bf 8.4 Compatibility Conditions}
\vskip 1cm
We now require that the field equations (8.2.10)--(8.2.13) should
be preserved under the action of the supergauge transformations
(8.2.2) and (8.2.3). This is the procedure one follows in the massless
case, and is a milder requirement with respect to the analysis
of section 8.3.

If $\nu^{B}$ and $\lambda_{B'}$ are twistors, but not necessarily
Killing spinors, they obey the Eqs. (8.3.5) and (8.3.10), which
imply that, for some independent spinor fields $\pi^{A}$ and
${\widetilde \pi}^{A'}$, one has
$$
\nabla_{\; \; \; B}^{A'} \; \nu_{C}
=\varepsilon_{BC} \; {\widetilde \pi}^{A'} 
\; \; \; \; ,
\eqno (8.4.1)
$$
$$
\nabla_{\; \; B'}^{A} \; \lambda_{C'}
=\varepsilon_{B'C'} \; \pi^{A} 
\; \; \; \; .
\eqno (8.4.2)
$$
In the compatibility equations, whenever one has terms of the
kind $S_{AA'} \; \nabla_{\; \; B'}^{A} \; \lambda_{C'}$, it is
therefore more convenient to symmetrize and anti-symmetrize over
$B'$ and $C'$. A repeated use of this algorithm leads to a
considerable simplification of the lengthy calculations. For
example, the preservation condition of Eq. (8.2.10) has the 
general form
$$ 
3f_{2}\Bigr(\nabla_{AA'} \; \nu^{A}+2f_{1}\lambda_{A'}\Bigr)
+\varepsilon^{B'C'}\biggr[S_{AA'}\Bigr(\nabla_{\; \; B'}^{A}
\; \lambda_{C'}\Bigr)+S_{AB'}\Bigr(\nabla_{\; \; A'}^{A}
\; \lambda_{C'}\Bigr)\biggr]=0 \; .
\eqno (8.4.3)
$$
By virtue of Eq. (8.4.2), Eq. (8.4.3) becomes
$$
f_{2}\Bigr(\nabla_{AA'} \; \nu^{A}+2f_{1}\lambda_{A'}\Bigr)
+S_{AA'} \; \pi^{A}=0 
\; \; \; \; .
\eqno (8.4.4)
$$
Following (8.2.4) and (8.2.5), the action of the
supercovariant derivative on $\pi_{A},{\widetilde \pi}_{A'}$
yields
$$
S_{AA'} \; \pi_{B} \equiv \nabla_{AA'} \; \pi_{B}
+f_{1}(\Lambda)\varepsilon_{AB} 
\; {\widetilde \pi}_{A'} 
\; \; \; \; ,
\eqno (8.4.5)
$$
$$
S_{AA'} \; {\widetilde \pi}_{B'} \equiv 
\nabla_{AA'} \; {\widetilde \pi}_{B'}
+f_{2}(\Lambda)\varepsilon_{A'B'} \; \pi_{A} 
\; \; \; \; .
\eqno (8.4.6)
$$
Equations (8.4.4) and (8.4.5), jointly with the equations
$$
\cstok{\ }\lambda_{A'}-2\Lambda 
\; \lambda_{A'}=0 
\; \; \; \; ,
\eqno (8.4.7)
$$
$$
\nabla^{AA'} \; \pi_{A}=2\Lambda \; \lambda^{A'} 
\; \; \; \; ,
\eqno (8.4.8)
$$
which result from Eq. (8.4.2), lead to
$$
(f_{1}+f_{2}){\widetilde \pi}_{A'}
+(f_{1}f_{2}-\Lambda)\lambda_{A'}=0 
\; \; \; \; .
\eqno (8.4.9)
$$
Moreover, the preservation of Eq. (8.2.11) under (8.2.2) leads to
the equation
$$
S^{B'(A} \; \pi^{B)} + f_{2} \nabla^{B'(A} \;
\nu^{B)}=0 
\; \; \; \; ,
\eqno (8.4.10)
$$
which reduces to
$$
\nabla^{B'(A} \; \pi^{B)}=0 
\; \; \; \; ,
\eqno (8.4.11)
$$
by virtue of (8.4.1) and (8.4.5). Note that a supertwistor is
also a twistor, since
$$
S^{B'(A} \; \pi^{B)}=\nabla^{B'(A} \; \pi^{B)} 
\; \; \; \; ,
\eqno (8.4.12)
$$
by virtue of the definition (8.4.5). It is now clear that,
for a gauge freedom generated by twistors, the preservation
of Eqs. (8.2.12) and (8.2.13) under (8.2.3) leads to the compatibility 
equations
$$
(f_{1}+f_{2})\pi_{A}+(f_{1}f_{2}-\Lambda)\nu_{A}=0 
\; \; \; \; ,
\eqno (8.4.13)
$$
$$
\nabla^{B(A'} \; {\widetilde \pi}^{B')}=0 
\; \; \; \; ,
\eqno (8.4.14)
$$
where we have also used the equation 
(see Eqs. (8.3.6) and (8.4.1))
$$
\nabla^{AA'} \; {\widetilde \pi}_{A'}=2\Lambda \; \nu^{A} 
\; \; \; \; .
\eqno (8.4.15)
$$
Note that, if $f_{1}+f_{2} \not = 0$,
one can solve Eqs. (8.4.9) and (8.4.13) as
$$
{\widetilde \pi}_{A'}={(\Lambda-f_{1}f_{2})\over (f_{1}+f_{2})}
\lambda_{A'} 
\; \; \; \; ,
\eqno (8.4.16)
$$
$$
\pi_{A}={(\Lambda-f_{1}f_{2})\over (f_{1}+f_{2})}\nu_{A} 
\; \; \; \; ,
\eqno (8.4.17)
$$
and hence one deals again with Euclidean Killing spinors as
in section 8.3. However, if
$$
f_{1}+f_{2}=0 
\; \; \; \; ,
\eqno (8.4.18)
$$
$$
f_{1}f_{2}-\Lambda=0 
\; \; \; \; ,
\eqno (8.4.19)
$$
the spinor fields ${\widetilde \pi}_{A'}$ and $\lambda_{A'}$
become {\it unrelated}, as well as $\pi_{A}$ and $\nu_{A}$. 
This is a crucial point. Hence one may have 
$f_{1}=\pm \sqrt{-\Lambda}$, $f_{2}=\mp \sqrt{-\Lambda}$, and
one finds a more general structure.

In the generic case, we do not assume that $\nu^{B}$ and
$\lambda_{B'}$ obey any equation. This means that, on the
second line of Eq. (8.4.3), it is more convenient to
express the term in square brackets as 
$2S_{A(A'} \; \nabla_{\; \; B')}^{A} \; \lambda_{C'}$. The
rule (8.2.7) for the action of $S_{AA'}$ on spinors with
many indices leads therefore to the compatibility 
conditions 
$$ 
3f_{2} \Bigr(\nabla_{AA'} \; \nu^{A}+2f_{1}\lambda_{A'}
\Bigr)-6\Lambda \; \lambda_{A'} 
+4f_{1}{\widetilde P}_{(A'B')}^{\; \; \; \; \; \; \; \; B'}
+3f_{2}{\widetilde Q}_{A'}=0 \; \; ,
\eqno (8.4.20)
$$
$$ 
3f_{1} \Bigr(\nabla_{AA'} \; \lambda^{A'}
+2f_{2} \nu_{A}\Bigr)-6\Lambda \; \nu_{A} 
+4f_{2}P_{(AB)}^{\; \; \; \; \; \; B}
+3f_{1}Q_{A}=0 \; \; ,
\eqno (8.4.21)
$$
$$
\Phi_{\; \; \; \; C'D'}^{AB} \; \lambda^{D'}
+f_{2}U_{\; \; \; \; \; \; C'}^{(AB)}
-f_{2}\nabla_{C'}^{\; \; \; (A} \; \nu^{B)}=0 
\; \; \; \; ,
\eqno (8.4.22)
$$
$$
{\widetilde \Phi}_{\; \; \; \; \; \; CD}^{A'B'} \; \nu^{D}
+f_{1}{\widetilde U}_{\; \; \; \; \; \; \; \; C}^{(A'B')}
-f_{1}\nabla_{C}^{\; \; (A'} \; \lambda^{B')}=0 
\; \; \; \; ,
\eqno (8.4.23)
$$
where the detailed form of 
$P,{\widetilde P},Q,{\widetilde Q},U,{\widetilde U}$ 
is not strictly necessary,
but we can say that they do not depend explicitly on the
trace-free part of the Ricci spinor, or on the Weyl spinors.
Note that, in the massless limit $f_{1}=f_{2}=0$, the Eqs.
(8.4.20)--(8.4.23) reduce to the familiar form of compatibility
equations which admit non-trivial solutions only in Ricci-flat
backgrounds.

Our consistency analysis still makes it necessary to set to
zero $\Phi_{\; \; \; \; C'D'}^{AB}$ (and hence 
${\widetilde \Phi}_{\; \; \; \; \; \; CD}^{A'B'}$ 
by reality (Penrose and Rindler 1984)).
The remaining contributions to (8.4.20)--(8.4.23) should then
become algebraic relations by virtue of the twistor equation.
This is confirmed by the analysis of gauge freedom for
the second set of potentials in section 8.5.
\vskip 1cm
\centerline {\bf 8.5 Second Set of Potentials}
\vskip 1cm
According to the prescription of section 8.2, which replaces
$\nabla_{AA'}$ by $S_{AA'}$ in the field
equations (Townsend 1977), we now {\it assume} that the super 
Rarita-Schwinger equations corresponding to (7.4.4) 
and (7.4.13) are (see section 8.7)
$$
S^{B'(F} \; \rho_{B'}^{\; \; \; A)L}=0 
\; \; \; \; ,
\eqno (8.5.1)
$$
$$
S^{B(F'} \; \theta_{B}^{A')L'}=0 
\; \; \; \; ,
\eqno (8.5.2)
$$
where the secondary potentials are subject locally to the
supergauge transformations
$$
{\widehat \rho}_{B'}^{\; \; \; AL} \equiv
\rho_{B'}^{\; \; \; AL}+S_{B'}^{\; \; \; A} \; \mu^{L} 
\; \; \; \; ,
\eqno (8.5.3)
$$
$$
{\widehat \theta}_{B}^{\; \; A'L'} \equiv 
\theta_{B}^{\; \; A'L'}+S_{B}^{\; \; A'} \; \zeta^{L'} 
\; \; \; \; .
\eqno (8.5.4)
$$
The analysis of the gauge freedom of the second kind is 
entirely analogous to the one in section 8.3, since equations
like (8.2.4) and (8.2.5) now apply to $\mu_{L}$ and $\zeta_{L'}$.
Hence we do not repeat this investigation.

A more general gauge freedom of the twistor type relies on
the supertwistor equations (see (8.4.12))
$$
S_{B'}^{\; \; \; (A} \; \mu^{L)}
=\nabla_{B'}^{\; \; \; (A} \; \mu^{L)}=0 
\; \; \; \; ,
\eqno (8.5.5)
$$
$$
S_{B}^{\; \; (A'} \; \zeta^{L')}=\nabla_{B}^{\; \; (A'} \;
\zeta^{L')}=0 
\; \; \; \; .
\eqno (8.5.6)
$$
Thus, on requiring the preservation of the super Rarita-Schwinger
equations (8.5.1) and (8.5.2) under the supergauge transformations
(8.5.3) and (8.5.4), one finds the preservation conditions
$$
S^{B'(F} \; S_{B'}^{\; \; \; A)} \; \mu^{L}=0 
\; \; \; \; ,
\eqno (8.5.7)
$$
$$
S^{B(F'} \; S_{B}^{\; \; A')} \; \zeta^{L'}=0 
\; \; \; \; ,
\eqno (8.5.8)
$$
which lead to
$$
(f_{1}+f_{2})\pi_{F}+(f_{1}f_{2}-\Lambda)\mu_{F}=0 
\; \; \; \; ,
\eqno (8.5.9)
$$
$$
(f_{1}+f_{2}){\widetilde \pi}_{F'}
+(f_{1}f_{2}-\Lambda)\zeta_{F'}=0 
\; \; \; \; .
\eqno (8.5.10)
$$
Hence we can repeat the remarks following Eqs. 
(8.4.16)--(8.4.19). Again, it is essential that $\pi_{F},\mu_{F}$
and ${\widetilde \pi}_{F'},\zeta_{F'}$ may be unrelated if
(8.4.18) and (8.4.19) hold. In the massless case, this is impossible, 
and hence there is no gauge freedom compatible with a
non-vanishing cosmological constant.

If one does not assume the validity of Eqs. (8.5.5) and (8.5.6),
the general preservation equations (8.5.7) and (8.5.8) lead instead
to the compatibility conditions 
$$ 
\eqalignno{
\; & \psi_{\; \; \; \; \; \; D}^{AFL} \; \mu^{D}
-2\Lambda \; \mu^{(A} \; \varepsilon^{F)L}
+2f_{2} \omega^{(AF)L} 
+f_{1} \varepsilon^{L(A} \; T^{F)} \cr
&+f_{1} \varepsilon^{L(A} \; S^{F)B'} \; \zeta_{B'}=0 
\; \; \; \; ,
&(8.5.11)\cr}
$$
$$
\eqalignno{
\; & {\widetilde \psi}_{\; \; \; \; \; \; \; \; \; D'}^{A'F'L'}
\; \zeta^{D'} -2\Lambda \; \zeta^{(A'} \; \varepsilon^{F')L'}
+2f_{1}{\widetilde \omega}^{(A'F')L'} 
+f_{2} \varepsilon^{L'(A'} \; {\widetilde T}^{F')} \cr
&+f_{2} \varepsilon^{L'(A'} \; 
{\widetilde S}^{F')B} \; \mu_{B}=0 
\; \; \; \; .
&(8.5.12)\cr}
$$
If we now combine the compatibility equations (8.4.20)--(8.4.23)
with (8.5.11) and (8.5.12), and require that the gauge fields
$\nu_{A},\lambda_{A'},\mu_{A},\zeta_{A'}$ should not depend
explicitly on the curvature of the background, we find
that the trace-free part of the Ricci spinor has to
vanish, and the Riemannian four-geometry is forced to be
conformally flat, since under our assumptions the equations
$$
\psi_{AFLD} \; \mu^{D}=0 
\; \; \; \; ,
\eqno (8.5.13)
$$
$$
{\widetilde \psi}_{A'F'L'D'} \; \zeta^{D'}=0 
\; \; \; \; ,
\eqno (8.5.14)
$$
force the anti-self-dual and self-dual Weyl spinors to vanish.
Remarkably, Eqs. (8.5.13) and (8.5.14) are just the integrability
conditions for the existence of non-trivial solutions of the
supertwistor equations (8.5.5) and (8.5.6). Hence the spinor fields
$\omega,S,T,{\widetilde \omega},{\widetilde S}$ 
and $\widetilde T$ in
(8.5.11) and (8.5.12) are such that these equations reduce to 
(8.5.9) and (8.5.10). In other words, for massive spin-${3\over 2}$
potentials, the gauge freedom is indeed
generated by solutions of the twistor equations in conformally
flat Einstein four-manifolds.

Last, on inserting the local equations (7.4.1) and (7.4.11) into 
the second half of the Rarita-Schwinger equations,
and then replacing
$\nabla_{AA'}$ by $S_{AA'}$, one finds equations whose
preservation under the supergauge transformations (8.5.3) and (8.5.4)
is again guaranteed if the supertwistor equations 
(8.5.5) and (8.5.6) hold.
\vskip 1cm
\centerline {\bf 8.6 Non-Linear Superconnection}
\vskip 1cm
As a first step in the proof
that Eqs. (8.5.1) and (8.5.2) arise naturally as 
integrability conditions of a suitable connection, 
we introduce a partial superconnection 
$W_{A'}$ (cf. Penrose 1994) acting on unprimed spinor  
fields $\eta_{D}$ defined on the Riemannian background.
 
With our notation
$$
W_{A'} \; \eta_{D} \equiv \eta^{A} \; S_{AA'}
\; \eta_{D} - \eta_{B} \; \eta_{C} \; 
\rho_{A'}^{\; \; \; BC}
\; \eta_{D} 
\; \; \; \; .
\eqno (8.6.1)
$$
Writing
$$ 
W_{A'}=\eta^{A} \; \Omega_{AA'} 
\; \; \; \; , 
\eqno (8.6.2)
$$
where the operator $\Omega_{AA'}$ acts on spinor fields 
$\eta_{D}$, we obtain 
$$
\eta^{A} \; \Omega_{AA'}
=\eta^{A} \; S_{AA'}-\eta_{B} \; \eta_{C} \;
\rho_{A'}^{\; \; \; BC} 
\; \; \; \; .
\eqno (8.6.3)
$$
Following Penrose 1994, we require that $\Omega_{AA'}$ 
should provide a genuine 
superconnection on the spin-bundle, 
so that it acts in any direction.
Thus, from (8.6.3) one can take (cf. Penrose 1994)
$$
\Omega_{AA'} \equiv S_{AA'}-\eta^{C} \; \rho_{A'AC}=
S_{AA'}-\eta^{C} \; \rho_{A'(AC)}
+{1\over 2}\eta_{A} \; \rho_{A'} 
\; \; \; \; .
\eqno (8.6.4)
$$
Note that (8.6.4) makes it necessary to know the trace $\rho_{A'}$, 
while in (8.6.1) only the symmetric part of 
$\rho_{A'}^{\; \; \; BC}$  survives. 
Thus we can see that, independently of the analysis in the 
previous sections, the definition of $\Omega_{AA'}$ picks out 
a potential of the Rarita-Schwinger type (Penrose 1994).
\vskip 1cm
\centerline {\bf 8.7 Integrability Condition}
\vskip 1cm
In section 8.6 we have introduced a  
superconnection $\Omega_{AA'}$ which acts on a bundle
with non-linear fibres, where the term $-\eta^{C} \;
\rho_{A'AC}$ is responsible for the non-linear nature
of $\Omega_{AA'}$ (see (8.6.4)). Following Penrose 1994,
we now pass to a description in terms of a vector bundle 
of rank three. On introducing the local coordinates
$(u_{A},\xi)$, where
$$
u_{A}=\xi \; \eta_{A} 
\; \; \; \; ,
\eqno (8.7.1)
$$
the action of the new operator
${\widetilde \Omega}_{AA'}$ reads (cf. Penrose 1994)
$$
{\widetilde\Omega}_{AA'}(u_{B},\xi) 
\equiv  \Bigr(S_{AA'} \; u_{B}, 
S_{AA'} \; \xi-u^{C} \; \rho_{A'AC}\Bigr) 
\; \; \; \; .
\eqno (8.7.2)
$$
Now we are able to prove that Eqs. (8.5.1) and (8.5.2) are 
integrability conditions.

The super $\beta$-surfaces are totally null two-surfaces 
whose tangent vector has the form $u^{A} \; \pi^{A'}$,
where $\pi^{A'}$ is varying and $u^{A}$ obeys the equation
$$
u^{A} \; S_{AA'} \; u_{B}=0 
\; \; \; \; ,
\eqno (8.7.3)
$$ 
which means that $u^{A}$ is supercovariantly constant
over the surface. On defining
$$
\tau_{A'} \equiv u_{B} \; u_{C} \; \rho_{A'}^{\; \; \; BC} 
\; \; \; \; ,
\eqno (8.7.4)
$$
the condition for ${\widetilde \Omega}_{AA'}$ to be
integrable on super $\beta$-surfaces is (cf. Penrose 1994)
$$
u^{A} \; {\widetilde \Omega}_{AA'} \; \tau^{A'}=
u_{A} \; u_{B} \; u_{C} \; 
S^{A'(A} \; \rho_{A'}^{\; \; \; B)C}= 0 
\; \; \; \; ,
\eqno (8.7.5)
$$
by virtue of the Leibniz rule and of (8.7.2)--(8.7.4).
Equation (8.7.5) implies
$$
S^{A'(A} \; \rho_{A'}^{\; \; \; B)C}=0 
\; \; \; \; ,
\eqno (8.7.6)
$$
which is the Eq. (8.5.1). Similarly, on studying
super $\alpha$-surfaces defined by the equation
$$
{\widetilde u}^{A'} \; S_{AA'} \; {\widetilde u}_{B'}=0 
\; \; \; \; ,
\eqno (8.7.7)
$$
one obtains Eq. (8.5.2). Thus, although Eqs. (8.5.1) and (8.5.2) 
are naturally suggested by the local theory of 
spin-${3\over 2}$ potentials, they have a deeper geometric
origin, as shown.
\vskip 1cm
\centerline {\bf 8.8 Results and Open Problems}
\vskip 1cm
We have given an entirely two-spinor description of massive
spin-${3\over 2}$ potentials in Einstein four-geometries.
Although the supercovariant derivative
(8.2.1) was well known in the literature, following the work
in Townsend 1977, and its Lorentzian version was already applied in
Perry 1984 and Siklos 1985, 
the systematic analysis of spin-${3\over 2}$
potentials with the local form of their supergauge 
transformations was not yet available in the literature, to
the best of our knowledge, before the work in
Esposito and Pollifrone 1996.

Our first result is the two-spinor
proof that, for massive spin-${3\over 2}$ potentials, the
gauge freedom is generated by solutions of
the supertwistor equations in conformally flat Einstein
four-manifolds. Moreover, we have shown that the first-order
equations (8.5.1) and (8.5.2),
whose consideration is suggested by the local
theory of massive spin-${3\over 2}$ potentials, admit a
deeper geometric interpretation as integrability
conditions on super $\beta$- and super $\alpha$-surfaces 
of a connection on a rank-three vector bundle.
This result generalizes the analysis of massless
spin-${3\over 2}$ fields appearing in chapter seven.
One now has to find explicit solutions
of the Eqs. (8.2.10)--(8.2.13), and the supercovariant form
of $\beta$-surfaces studied in our chapter deserves 
a more careful consideration. 
Hence we hope that our work can lead to a better understanding
of twistor geometry and consistent supergravity theories
in four dimensions.
\vskip 100cm
\centerline {\it CHAPTER NINE}
\vskip 1cm
\centerline {\bf LOCAL BOUNDARY CONDITIONS IN}
\centerline {\bf QUANTUM SUPERGRAVITY}
\vskip 1cm
\noindent
{\bf Abstract.} When quantum supergravity is studied on
manifolds with boundary, one may consider local boundary conditions
which fix on the initial surface the whole primed part
of tangential components of gravitino perturbations, and fix
on the final surface the whole unprimed part of tangential
components of gravitino perturbations. This chapter studies such
local boundary conditions in a flat Euclidean background 
bounded by two concentric three-spheres. It is shown that, 
as far as three-dimensional 
transverse-traceless perturbations are concerned,
the resulting contribution to $\zeta(0)$ vanishes
when such boundary data are set to zero,
exactly as in the case when non-local boundary conditions of the
spectral type are imposed. These properties may be used to
show that one-loop finiteness of massless supergravity models
is only achieved when two boundary three-surfaces occur, and there 
is no exact cancellation of the contributions of gauge and
ghost modes in the Faddeev-Popov path integral. In these
particular cases, which rely on the use of covariant
gauge-averaging functionals, pure gravity is one-loop 
finite as well.
\vskip 100cm
The problem of a consistent formulation of quantum supergravity
on manifolds with boundary is still receiving careful
consideration in the current literature (Esposito 1994a, 
D'Eath 1996, Esposito and Kamenshchik 1996, 
Esposito et al. 1997). In particular,
many efforts have been produced to understand whether simple
supergravity is one-loop finite 
in the presence of boundaries.
In the analysis of such an issue, the first problem consists,
of course, in a careful choice of boundary conditions. For
massless gravitino potentials, which are the object of our
investigation, these may be non-local of the spectral type
(D'Eath and Esposito 1991b) or local (Esposito 1996).

In the former case the idea is to fix at the boundary half of
the gravitino potential. On the final surface $\Sigma_{F}$ one
can fix those perturbative modes which multiply harmonics 
having positive eigenvalues of the intrinsic three-dimensional
Dirac operator $\cal D$ of the boundary. On the initial surface
one can instead fix those gravitino modes which multiply
harmonics having negative eigenvalues of the intrinsic
three-dimensional Dirac operator of the boundary. What is
non-local in this procedure is the separation of the spectrum
of a first-order elliptic operator (our $\cal D$) into a positive
and a negative part. This leads to a sort of positive- and
negative-frequency split which is typical for scattering
problems (D'Eath 1996), 
but may also be applied to the analysis of 
quantum amplitudes in finite regions (Esposito et al. 1997). 

Our chapter deals instead with the latter choice, i.e. local
boundary conditions for quantum supergravity. By this one 
usually means a formulation where complementary projection
operators act on gravitational and spin-${3\over 2}$
perturbations. Local boundary conditions of this type were
investigated in Luckock and Moss 1989, Luckock 1991, 
and then applied to quantum cosmological backgrounds in
Esposito 1994a--b, Moss and Poletti 1994, Esposito et al.
1994a--b, Esposito et al. 1995a--b, Esposito et al. 1997.

More recently, another choice of local boundary conditions 
for gravitino perturbations has been considered in D'Eath 1996.
Using two-component spinor notation, and referring 
the reader to D'Eath 1984 and D'Eath 1996 
for notation and background material,
we here represent the spin-${3\over 2}$ potential by a pair
of independent spinor-valued one-forms
$\Bigr(\psi_{\mu}^{A},{\widetilde \psi}_{\mu}^{A'}\Bigr)$ in
a Riemannian four-manifold which is taken to be flat
Euclidean four-space bounded by two concentric three-spheres
(Esposito and Kamenshchik 1996). 
Denoting by $S_{I}$ and $S_{F}$ the boundary three-spheres,
with radii $a$ and $b$ respectively (here $b>a$), the local
boundary conditions proposed in D'Eath 1996 read in our case
($i=1,2,3$)
$$
\Bigr[{\widetilde \psi}_{i}^{A'}\Bigr]_{S_{I}}
=F_{i}^{A'} 
\; \; \; \; ,
\eqno (9.1)
$$
$$
\Bigr[\psi_{i}^{A}\Bigr]_{S_{F}}=H_{i}^{A} 
\; \; \; \; ,
\eqno (9.2)
$$
where $F_{i}^{A'}$ and $H_{i}^{A}$ are boundary
data which may or may not satisfy the classical constraint
equations (D'Eath 1996). With the choice (9.1) and (9.2), the whole 
primed part of the tangential components of the spin-${3\over 2}$
potential is fixed on $S_{I}$, and the whole unprimed part
of the tangential components of the spin-${3\over 2}$ potential
is fixed on $S_{F}$.

In a Hamiltonian analysis, $\psi_{0}^{A}$ and 
${\widetilde \psi}_{0}^{A'}$ are 
Lagrange multipliers (D'Eath 1996), and hence
boundary conditions for them look un-natural in a one-loop
calculation, especially if one is interested in reduction
to three-dimensional transverse-traceless (TT) 
gravitino modes (usually regarded
as the {\it physical} part of gravitinos). In a covariant
path-integral analysis, however, one cannot disregard the issue
of boundary conditions on normal components of gravitinos.
We shall thus post-pone the discussion of this point, and we
will focus on the TT sector of the boundary conditions (9.1)
and (9.2).

With the notation of Esposito 1994a, the expansion in harmonics on
three-spheres of the TT part of tangential components of
spin-${3\over 2}$ perturbations reads
$$
\psi_{i}^{A}={\tau^{-{3\over 2}}\over 2\pi}\sum_{n=0}^{\infty}
\sum_{p,q=1}^{(n+1)(n+4)}\alpha_{n}^{pq}
\Bigr[m_{np}(\tau)\beta^{nqACC'}
+{\widetilde r}_{np}(\tau){\overline \mu}^{nqACC'}\Bigr]
e_{CC'i} 
\; \; \; \; ,
\eqno (9.3)
$$
$$
{\widetilde \psi}_{i}^{A'}={\tau^{-{3\over 2}}\over 2\pi}
\sum_{n=0}^{\infty}\sum_{p,q=1}^{(n+1)(n+4)}\alpha_{n}^{pq}
\Bigr[{\widetilde m}_{np}(\tau)
{\overline \beta}^{nqA'C'C}+r_{np}(\tau)
\mu^{nqA'C'C}\Bigr]e_{CC'i} 
\; \; \; \; ,
\eqno (9.4)
$$
where $\beta^{nqACC'} \equiv -\rho^{nq(ACD)}n_{D}^{\; \; C'}$
and $\mu^{nqA'C'C} \equiv -\sigma^{nq(A'C'D')}
n_{\; \; D'}^{C}$. Of course, round brackets denote complete
symmetrization over spinor indices, and $n_{\; \; D'}^{C}$ is
obtained from the Euclidean normal as $n_{\; \; D'}^{C}
=i {_{e}n_{\; \; D'}^{C}}$. Variation of the TT gravitino action
yields, for all integer $n \geq 0$, the following eigenvalue 
equations for gravitino modes:
$$
\biggr({d\over d\tau}-{(n+5/2)\over \tau}\biggr)x_{np}
=E_{np} \; {\widetilde x}_{np} 
\; \; \; \; ,
\eqno (9.5)
$$
$$
\biggr(-{d\over d\tau}-{(n+5/2)\over \tau}\biggr)
{\widetilde x}_{np}=E_{np} \; x_{np} 
\; \; \; \; ,
\eqno (9.6)
$$
where $x_{np}=m_{np}$ and ${\widetilde x}_{np}=
{\widetilde m}_{np}$, or $x_{np}=r_{np}$ and
${\widetilde x}_{np}={\widetilde r}_{np}$. The Eqs. (9.5) and
(9.6) lead to the following basis functions in terms of
modified Bessel functions (hereafter $M \equiv E_{np}$ for
simplicity of notation, while $\beta_{1,n}$ and $\beta_{2,n}$
are some constants):
$$
m_{np}(\tau)=\beta_{1,n}\sqrt{\tau}I_{n+2}(M\tau)
+\beta_{2,n}\sqrt{\tau}K_{n+2}(M\tau) 
\; \; \; \; ,
\eqno (9.7)
$$
$$
{\widetilde m}_{np}(\tau)=\beta_{1,n}\sqrt{\tau}I_{n+3}(M\tau)
-\beta_{2,n}\sqrt{\tau}K_{n+3}(M\tau) 
\; \; \; \; ,
\eqno (9.8)
$$
$$
r_{np}(\tau)=\beta_{1,n}\sqrt{\tau}I_{n+2}(M\tau)
+\beta_{2,n}\sqrt{\tau}K_{n+2}(M\tau) 
\; \; \; \; ,
\eqno (9.9)
$$
$$
{\widetilde r}_{np}(\tau)=\beta_{1,n}\sqrt{\tau}I_{n+3}(M\tau)
-\beta_{2,n}\sqrt{\tau}K_{n+3}(M\tau) 
\; \; \; \; .
\eqno (9.10)
$$
By virtue of (9.1) and (9.2), these modes obey the boundary
conditions
$$
{\widetilde m}_{np}(a)=A_{n} 
\; \; \; \; ,
\eqno (9.11a)
$$
$$
r_{np}(a)=A_{n} 
\; \; \; \; ,
\eqno (9.11b)
$$
$$
m_{np}(b)=B_{n} 
\; \; \; \; ,
\eqno (9.12a)
$$
$$
{\widetilde r}_{np}(b)=B_{n} 
\; \; \; \; ,
\eqno (9.12b)
$$
where $A_{n}$ and $B_{n}$ are constants resulting from the
boundary data $F_{i}^{A'}$ and $H_{i}^{A}$ respectively. The
boundary conditions (9.11) and (9.12) lead therefore, 
if $A_{n}=B_{n}=0$, $\forall n$, to the eigenvalue condition
$$
I_{n+3}(Mr)K_{n+2}(Mr)+I_{n+2}(Mr)K_{n+3}(Mr)=0 
\; \; \; \; ,
\eqno (9.13)
$$
where $r=a$ or $b$. 

We can now apply $\zeta$-function regularization to evaluate 
the resulting TT contribution to the one-loop divergence,
following the algorithm developed in Barvinsky et al. 1992
and already used in section 6.2. The basic steps
are as follows. Let us denote by $f_{l}$ the function occurring
in the equation obeyed by the eigenvalues by virtue of boundary
conditions, after taking out fake roots (e.g. $x=0$ is a fake
root of order $n$ of the Bessel function $I_{n}$). Let $d(l)$ 
be the degeneracy of the eigenvalues parametrized by the integer
$l$. One can then define the function
$$
I(M^{2},s) \equiv \sum_{l=l_{0}}^{\infty} d(l) l^{-2s}
\log f_{l}(M^{2}) 
\; \; \; \; .
\eqno (9.14)
$$
Such a function admits an analytic continuation to the
complex-$s$ plane as a meromorphic function with a simple pole
at $s=0$, in the form
$$
``I(M^{2},s)"={I_{\rm pole}(M^{2})\over s}
+I^{R}(M^{2})+{\rm O}(s) 
\; \; \; \; .
\eqno (9.15)
$$
The function $I_{\rm pole}(M^{2})$ is the residue at $s=0$, 
and makes it possible to obtain the $\zeta(0)$ value as
$$
\zeta(0)=I_{\rm log}+I_{\rm pole}(M^{2}=\infty)
-I_{\rm pole}(M^{2}=0) 
\; \; \; \; ,
\eqno (9.16)
$$
where $I_{\rm log}$ is the coefficient of the $\log M$ term in
$I^{R}$ as $M \rightarrow \infty$. Moreover, $I_{\rm pole}(\infty)$
coincides with the coefficient of ${1\over l}$ in the expansion 
as $l \rightarrow \infty$ of ${1\over 2}d(l) \log[\rho_{\infty}(l)]$,
where $\rho_{\infty}(l)$ is the $l$-dependent term in the eigenvalue
condition as $M \rightarrow \infty$ and $l \rightarrow \infty$. 
The $I_{\rm pole}(0)$ value is instead obtained as the coefficient
of ${1\over l}$ in the expansion as $l \rightarrow \infty$ of
${1\over 2}d(l) \log[\rho_{0}(l)]$, where $\rho_{0}(l)$ is the
$l$-dependent term in the eigenvalue condition as 
$M \rightarrow 0$ and $l \rightarrow \infty$.

In our problem, using the limiting form of Bessel functions when
the argument tends to zero, one finds that the left-hand side 
of Eq. (9.13) is proportional to $M^{-1}$ as $M \rightarrow 0$. Hence
one has to multiply by $M$ to get rid of fake roots. Moreover,
in the uniform asymptotic expansion of Bessel functions as
$M \rightarrow \infty$ and $n 
\rightarrow \infty$ (Olver 1954), both $I$
and $K$ functions contribute a ${1\over \sqrt{M}}$ factor. 
These properties imply that $I_{\rm log}$ vanishes
(hereafter $l \equiv n+1$):
$$
I_{\rm log}={1\over 2}\sum_{l=1}^{\infty}2l(l+3)
\Bigr(1-1/2-1/2 \Bigr)=0 
\; \; \; \; .
\eqno (9.17)
$$
Moreover, $I_{\rm pole}(\infty)$ vanishes since there is no
$l$-dependent coefficient in the uniform asymptotic expansion of
Eq. (9.13) as $M \rightarrow \infty$ and $l \rightarrow \infty$. 
Last, $I_{\rm pole}(0)$ vanishes as well, since the limiting
form of Eq. (9.13) as $M \rightarrow 0$ and $l \rightarrow \infty$
is ${1\over r}M^{-1}$. One thus finds for gravitinos
$$
\zeta_{TT}(0)=0 
\; \; \; \; .
\eqno (9.18)
$$
 
It is now clear that local boundary conditions for 
$\psi_{0}^{A}$ and ${\widetilde \psi}_{0}^{A'}$ along the
same lines of (9.1) and (9.2), i.e. (here $\kappa^{A'}$ and
$\mu^{A}$ are some boundary data)
$$
\Bigr[{\widetilde \psi}_{0}^{A'}\Bigr]_{S_{I}}
=\kappa^{A'} 
\; \; \; \; ,
\eqno (9.19)
$$
$$
\Bigr[\psi_{0}^{A}\Bigr]_{S_{F}}=\mu^{A} 
\; \; \; \; ,
\eqno (9.20)
$$
give again a vanishing contribution to $\zeta(0)$ in this
two-boundary problem if $\kappa^{A'}$ and $\mu^{A}$
are set to zero, since the resulting eigenvalue condition
is analogous to Eq. (9.13) with $n$ replaced by $n-1$. 

A suitable set of local boundary conditions on metric perturbations
$h_{\mu \nu}$ are the ones considered by Luckock, Moss and 
Poletti (Luckock 1991, Moss and Poletti 1994, Moss 1996, Moss
and Silva 1997). In our problem, denoting by $g_{\mu \nu}$ the
background four-metric, they read (Esposito et al. 1997)
$$
\Bigr[h_{ij}\Bigr]_{\partial M}=0 
\; \; \; \; ,
\eqno (9.21)
$$
$$
\biggr[{\partial h_{00}\over \partial \tau}
+{6\over \tau}h_{00}-{\partial \over \partial \tau}
\Bigr(g^{ij}h_{ij}\Bigr)\biggr]_{\partial M}=0 
\; \; \; \; ,
\eqno (9.22)
$$
$$
\Bigr[h_{0i}\Bigr]_{\partial M}=0 
\; \; \; \; ,
\eqno (9.23)
$$
while the ghost one-form is subject to the mixed boundary
conditions 
$$
\Bigr[\varphi_{0}\Bigr]_{\partial M}=0 
\; \; \; \; ,
\eqno (9.24)
$$
$$
\biggr[{\partial \varphi_{i} \over \partial \tau}
-{2\over \tau}\varphi_{i} \biggr]_{\partial M}=0 
\; \; \; \; .
\eqno (9.25)
$$
As first shown in Esposito et al. 1994b, 
graviton TT modes contribute
$$
\zeta_{TT}(0)=-5 
\; \; \; \; ,
\eqno (9.26)
$$
while, using a covariant gauge-averaging functional of the
de Donder type, gauge and ghost modes contribute 
$$
\zeta(0)_{\rm {gauge \; and \; ghost}}=5 
\; \; \; \; .
\eqno (9.27)
$$
One thus finds that the one-loop path integral for the
gravitational sector, including TT, gauge and ghost modes,
gives a vanishing contribution to the one-loop divergence.

What is left are gauge and ghost modes for gravitino
perturbations. In general, their separate values depend on
the gauge-averaging functional being used (only the full 
one-loop divergence should be gauge-independent (Endo 1995)).
However, on general ground, since in our {\it flat} Euclidean
background all possible contributions to $\zeta(0)$ involve 
surface integrals of terms like (cf. section 5.2)
$$
{\rm Tr}(K^{3}) \; , \; ({\rm Tr}K)({\rm Tr}K^{2})
\; , \; ({\rm Tr}K)^{3} 
\; \; \; \; ,
$$
$K$ denoting the extrinsic-curvature tensor of the boundary, 
any dependence on the three-spheres radii $a$ and $b$ 
disappears after integration over $\partial M$. Moreover,
{\it if} the formalism is gauge-independent ax expected, the
one-loop result can only coincide with the one found in Sec. V 
of Esposito and Kamenshchik 1996 in the axial gauge:
$$
\zeta_{3\over 2}(0)=0 
\; \; \; \; .
\eqno (9.28)
$$
The following concluding remarks are now in order:
\vskip 0.3cm
\noindent
(i) The one-loop finiteness suggested by our analysis
does not seem to contradict the results of Esposito
and Kamenshchik 1996. What is
shown therein is instead that, when flat Euclidean 
four-space is bounded by {\it only one} three-sphere, simple
supergravity fails to be one-loop finite (either spectral
boundary conditions with non-covariant gauge, or local boundary
conditions in covariant gauge). This is {\it not} our background.
Moreover, the two-boundary problem of 
Esposito and Kamenshchik 1996 was studied in
a non-covariant gauge of the axial type, when the Faddeev-Popov
path integral is supplemented by the integrability condition
for the eigenvalue equations on graviton and gravitino
perturbations. This quantization was found to pick out TT
modes only, but differs from the scheme proposed in
our chapter (Esposito 1996).
\vskip 0.3cm
\noindent
(ii) The result (9.28) is crucial for one-loop finiteness to
hold (when combined with (9.26) and (9.27)), 
and its explicit proof has not yet been achieved.
\vskip 0.3cm
\noindent
(iii) As a further check of one-loop finiteness (or of its
lack, cf. van Nieuwenhuizen and Vermaseren 1976), 
one should now perform two-boundary calculations of
$\zeta(0)$ when Luckock-Moss-Poletti local boundary conditions
are imposed on gravitino perturbations.
\vskip 0.3cm
\noindent
(iv) Yet another check might be obtained by combining 
Barvinsky boundary conditions for pure 
gravity (Barvinsky 1987) in the
de Donder gauge (these are completely invariant under infinitesimal
diffeomorphisms) with local or non-local boundary conditions 
for gravitinos in covariant gauges (Endo 1995).
\vskip 0.3cm
\noindent
(v) The calculations performed in Avramidi et al. 1996,
Esposito and Kamenshchik 1996, Esposito et al. 1997 and in
our chapter show that, when cancellation of the effects of gauge
and ghost modes is achieved, only the effects of TT modes 
survive, and hence both pure gravity and simple supergravity
are not even one-loop finite. By contrast, if the effects of
gauge and ghost modes do not cancel each other exactly (e.g.
by using covariant gauge-averaging functionals in the 
Faddeev-Popov path integral), then both pure gravity and 
simple supergravity may turn out to be one-loop finite in
the presence of two bounding three-spheres.
\vskip 0.3cm
\noindent
(vi) A deep problem is the relation between the
Hamiltonian analysis in D'Eath 1996, where auxiliary fields
play an important role but ghost fields are not studied,
and the path-integral approach of 
Esposito and Kamenshchik 1996, where ghost
fields are analyzed in detail but auxiliary fields are not
found to affect the one-loop calculation.

All this adds evidence in favour of quantum cosmology being
able to lead to new perspectives in Euclidean quantum gravity
and quantum supergravity (cf. van Nieuwenhuizen 1981,
Esposito et al. 1997).
\vskip 100cm
\centerline {\it CHAPTER TEN}
\vskip 1cm
\centerline {\bf NEW FRONTIERS}
\vskip 1cm
\noindent
{\bf Abstract.} Quarks are spin-${1\over 2}$ particles for which
a Dirac operator can be studied. The local boundary conditions
studied for models of quark confinement coincide with the local
boundary conditions studied, more recently, in one-loop quantum
cosmology. Further developments
lie in the possibility to study quantization schemes in
conformally invariant gauges. This possibility is investigated
in the case of the Eastwood-Singer gauge for vacuum Maxwell
theory on manifolds with boundary. This is part of a more general
scheme, leading to the analysis of conformally covariant
operators. These are also presented, with emphasis on the
Paneitz operator. Last, a class of boundary operators are
described which include the effect of tangential derivatives.
They lead to many new invariants in the general form of
heat-kernel asymptotics for operators of Laplace type. The
consideration of tangential derivatives arises naturally 
within the framework of recent attempts to obtain 
BRST-invariant boundary conditions in quantum field theory. 
However, in Euclidean quantum gravity, it remains unclear
how to write even just the general form of the various 
heat-kernel coefficients.
\vskip 100cm
\centerline {\bf 10.1 Introduction}
\vskip 1cm
So far we have dealt with many aspects of manifolds with
boundary in mathematics and physics. Hence it seems
appropriate to begin the last chapter of our monograph 
with a brief review of the areas of research which provide
the main motivations for similar investigations. They are
as follows.
\vskip 0.3cm
\noindent
{\it (i) Index theorems for manifolds with boundary}. One
wants to understand how to extend index theorems, originally
proved for closed manifolds, to manifolds with boundary
(Atiyah and Singer 1963, Atiyah and Bott 1965, Atiyah 1975b).
A naturally occurring analytic tool is indeed the use of
complex powers of elliptic operators, via $\eta$- and
$\zeta$-functions. 
\vskip 0.3cm
\noindent
{\it (ii) Spectral geometry with mixed boundary conditions}.
Spinor fields, gauge fields and gravitation are subject to
mixed boundary conditions. These may be local or non-local
(Esposito et al. 1997), and reflect the first-order nature
of the Dirac operator, and the symmetries of gauge fields
and gravitation (e.g. invariance under local gauge
transformations, invariance under (infinitesimal) 
diffeomorphisms, BRST symmetry, supersymmetry). The resulting
boundary operators may be (complementary) projectors,
or first-order differential operators. They lead to a
heat-kernel asymptotics which is completely understood
only in the cases involving a mixture of Dirichlet and Robin
boundary conditions (Gilkey 1995, Esposito et al. 1997,
Avramidi and Esposito 1997).
\vskip 0.3cm
\noindent
{\it (iii) Quantum cosmology}. The path-integral approach to
quantum cosmology needs a well-defined prescription for the
``sum over histories" which should define the quantum state
of the universe (Hawking 1979, Hawking 1982, Hartle and
Hawking 1983, Hawking 1984, Gibbons and Hawking 1993, 
Esposito 1994a, Esposito et al. 1997).
\vskip 0.3cm
\noindent
{\it (iv) Quantum field theory on manifolds with boundary}. 
Boundaries play a crucial role to obtain a complete and
well defined mathematical model for the quantization of all
fundamental interactions. Indeed, one is already familiar
with the need to specify boundary data from the analysis of
potential theory, waveguides, and the classical variational
problem (cf. York 1986).
\vskip 0.3cm
\noindent
{\it (v) Boundary counterterms in supergravity}. The presence
of boundaries provides a crucial test of the finiteness
properties of supersymmetric theories of gravitation, e.g.
simple supergravity (D'Eath 1996, Esposito and Kamenshchik
1996, Esposito 1996, Moniz 1996, Esposito et al. 1997).
\vskip 0.3cm
\noindent
{\it (vi) Applications of quantum field theory}. Variations
of zero-point energies of the quantized electromagnetic
field are finite, measurable, and depend on the geometry
of the problem. This is what one learns from the Casimir
effect (Casimir 1948, Sparnay 1958, Boyer 1968, Grib
et al. 1994, Esposito et al. 1997, Lamoreaux 1997).
Moreover, the consideration of boundary effects provides a
useful toy model for the investigation of quark confinement,
as is shown in the following section (Chodos et al. 1974).
\vskip 1cm
\centerline {\bf 10.2 Quark Boundary Conditions}
\vskip 1cm
A non-trivial application of spin-${1\over 2}$ fields to
modern physics consists of the mathematical models for
quark fields. We here focus on the model proposed by Chodos
et al. (1974), which describes elementary particles as
composite systems, with their internal structure being
associated with quark and gluon field variables. Their main
idea was to account for the internal structure with the help
of fields. However, the fields which describe the sub-structure
of the hadron {\it belong only to the sub-structure of a 
particle}, and cannot be extended to all points of space,
unlike the normal situation in field theory. In other words, 
field variables are confined on the subset of points which are
inside of an extended particle. Such a set of points represents
{\it the bag}. Since then, the model has been called M. I. T. bag
model (the authors being research workers of the M. I. T.).
Following the presentation in Johnson 1975, one can say that a
local disturbance of some medium leads to a collective motion
described by fields, which are taken to be the quark-gluon
fields. The localized excitation consists of a particle,
i.e. a hadron.

When the M. I. T. model was first proposed, there was (of 
course) great excitement in the physics community. Whether
or not the model by Chodos et al. can still have an impact on
current developments in particle physics, it provides a first
relevant example of local boundary conditions on fields acted
upon by a Dirac operator, and this is the feature that we want
to emphasize hereafter. For this purpose, we follow again 
Johnson (1975), and we denote by $q_{a}(x)$ the quark field.
This is a Dirac field carrying colour and flavour. Bearing
in mind what we said at the beginning, the quark fields can
only be associated with points {\it inside the hadron}. 
Outside the hadron, $q_{a}(x)$ vanishes by definition. The
local flux of colour and flavour quantum numbers inside the
hadron is
$$
j_{ab}^{\mu}(x) \equiv 
{\overline q}_{a}(x) \; \gamma^{\mu} \; q_{b}(x)
\; \; \; \; .
\eqno (10.2.1)
$$
To avoid losing quantum numbers through the surface, one has
to impose the boundary condition (hereafter, $n^{\mu}$ is the
inward-pointing normal to the boundary)
$$
\Bigr[n_{\mu} \; j_{ab}^{\mu}(x)\Bigr]_{\partial M}=0
\; \; \; \; ,
\eqno (10.2.2a)
$$
where $\partial M$ denotes the surface of the hadron. By
virtue of (10.2.1), the boundary condition (10.2.2a) may be
re-expressed in the form
$$
\Bigr[{\overline q}_{a}(x) \; n_{\mu} \; \gamma^{\mu}
\; q_{b}(x)\Bigr]_{\partial M}=0
\; \; \; \; .
\eqno (10.2.2b)
$$
At this stage we remark that, denoting by $\varepsilon$ a
real-valued function of position on the boundary 
($\varepsilon$ was taken equal to $\pm 1$ in Johnson 1975):
$$
\varepsilon: y \in \partial M \longrightarrow
\varepsilon(y) \in {\Re}
\; \; \; \; ,
$$
if one imposes the local boundary condition
$$
\biggr[\Bigr(i \; n_{\mu} \; \gamma^{\mu}- \varepsilon
\Bigr)q_{a}(x)\biggr]_{\partial M}=0
\; \; \; \; ,
\eqno (10.2.3)
$$
one obtains also the ``conjugate" boundary condition
$$
\biggr[{\overline q}_{a}(x)\Bigr(i \; n_{\mu} \;
\gamma^{\mu}+ \varepsilon \Bigr)\biggr]_{\partial M}=0
\; \; \; \; .
\eqno (10.2.4)
$$
By virtue of Eq. (10.2.3) one now finds
$$
\Bigr[i \; n_{\mu} \; j_{ab}^{\mu}(x)\Bigr]_{\partial M}
=\Bigr[{\overline q}_{a}(x) \; i \; n_{\mu} \; 
\gamma^{\mu} \; q_{b}(x)\Bigr]_{\partial M}
=\Bigr[\varepsilon \; {\overline q}_{a}(x) \;
q_{b}(x) \Bigr]_{\partial M}
\; \; \; \; .
\eqno (10.2.5)
$$
Moreover, by virtue of Eq. (10.2.4) one also has
$$
\Bigr[i \; n_{\mu} \; j_{ab}^{\mu}(x)\Bigr]_{\partial M}
=- \Bigr[\varepsilon \; {\overline q}_{a}(x) \; q_{b}(x)
\Bigr]_{\partial M}
\; \; \; \; .
\eqno (10.2.6)
$$
The comparison of Eqs. (10.2.5) and (10.2.6) leads to
$$
\Bigr[{\overline q}_{a}(x) \; q_{b}(x)\Bigr]_{\partial M}=0
\; \; \; \; ,
\eqno (10.2.7)
$$
which in turn ensures the fulfillment of the condition
(10.2.2a), bearing in mind the definition (10.2.1).

The local boundary conditions (10.2.3) find another relevant
physical application in a completely different framework,
i.e. (one-loop) quantum cosmology. A review of the necessary
steps is given in the following section.
\vskip 1cm
\centerline {\bf 10.3 Quantum Cosmology}
\vskip 1cm
In the early nineties, there was an intensive investigation
of local boundary conditions for the Dirac operator in one-loop
quantum cosmology (e.g. D'Eath and Esposito 1991a, 
Kamenshchik and Mishakov 1993, Esposito 1994a, Moss and
Poletti 1994). The motivations for this programme were
as follows.
\vskip 0.3cm
\noindent
(i) If one studies local boundary conditions for solutions
of the massless free-field equations, following the work of
Breitenlohner and Freedman 1982 and Hawking 1983, one finds 
that there exist solutions of the Euclidean twistor equation
which generate {\it rigid supersymmetry transformations}
among {\it classical solutions} obeying boundary conditions
of the type
$$
\sqrt{2} \; {_{e}n_{A}^{\; \; A'}} \; \phi^{A}
= \pm {\widetilde \phi}^{A'}
\; \; \; \; {\rm on} \; \; \; \; \partial M
\; \; \; \; ,
\eqno (10.3.1)
$$
$$
2 \; {_{e}n_{A}^{\; \; A'}} \;
{_{e}n_{B}^{\; \; B'}} \; \rho^{AB}=\pm
{\widetilde \rho}^{A'B'}
\; \; \; \; {\rm on} \; \; \; \; \partial M
\; \; \; \; .
\eqno (10.3.2)
$$
\vskip 0.3cm
\noindent
(ii) Following instead the work in Luckock and Moss 1989,
Luckock 1991, one finds that {\it local supersymmetry 
transformations} exist in supergravity which fix at the
boundary the spatial components of the tetrad and a projector
acting on spinor-valued one-forms which represent the
gravitino potentials. On extending this scheme to lower-spin
fields in a supergravity multiplet, one obtains the local
boundary conditions
$$
\sqrt{2} \; {_{e}n_{A}^{\; \; A'}} \; \psi^{A}
= \pm {\widetilde \psi}^{A'}
\; \; \; \; {\rm on} \; \; \; \; 
\partial M 
\eqno (10.3.3)
$$
for a massless spin-${1\over 2}$ field which can be expanded
in a complete orthonormal set of eigenfunctions of the Dirac
operator.

In particular, on studying a portion of flat Euclidean 
four-space bounded by a three-sphere of radius $a$ 
(Schleich 1985), the boundary conditions (10.3.3) were found
to imply the eigenvalue condition (D'Eath and Esposito 1991a,
Esposito 1994a)
$$
F(E) \equiv [J_{n+1}(Ea)]^{2}-[J_{n+2}(Ea)]^{2}=0
\; \; \; \; \forall n \geq 0
\; \; \; \; .
\eqno (10.3.4)
$$
Although the boundary conditions (10.3.3) were the
two-component spinor version of the boundary conditions
(10.2.3), it was a non-trivial step, both technically
and conceptually, to undertake the analysis of heat-kernel
asymptotics with this class of local boundary conditions
for the Dirac operator (see Falomir 1996 for the discussion
of possible topological obstructions). Nevertheless, the
result in D'Eath and Esposito 1991a (see also our
appendix 10.A):
$$
\zeta(0)={11\over 360}
\; \; \; \; ,
\eqno (10.3.5)
$$
later confirmed, independently, by several authors
(Kamenshchik and Mishakov 1993, Moss and Poletti 1994,
Dowker 1996, Dowker et al. 1996, Kirsten and Cognola 1996),
provided evidence in favour of local boundary conditions for
the Dirac operator being admissible in a number of 
relevant cases. Note that, by requiring that Eq. (10.3.3)
should be preserved under the action of the Dirac operator,
one obtains the complementary condition (Esposito 1994a)
$$
\left[\biggr({_{e}n^{BB'}} \nabla_{BB'}
+{1\over 2}({\rm Tr}K)\biggr)
\biggr(\sqrt{2} \; {_{e}n_{A}^{\; \; A'}} \; \psi^{A}
\pm {\widetilde \psi}^{A'}\biggr)\right]_{\partial M}=0
\; \; \; \; .
\eqno (10.3.6)
$$
In other words, denoting by $P_{-}$ and $P_{+}$ two
complementary projectors, where
$$
P_{\pm} \equiv {1\over 2} \Bigr(1 \pm i \gamma_{5}
\; \gamma_{\mu} \; n^{\mu} \Bigr)
\; \; \; \; ,
\eqno (10.3.7)
$$
the boundary conditions (10.3.3) and (10.3.6) can be
cast in the form
$$
\Bigr[P_{-}\phi \Bigr]_{\partial M}=0
\; \; \; \; ,
\eqno (10.3.8)
$$
$$
\left[\Bigr(n^{a}\nabla_{a}+{1\over 2}({\rm Tr}K)
\Bigr) P_{+}\phi \right]_{\partial M}=0
\; \; \; \; ,
\eqno (10.3.9)
$$
which is the form of mixed boundary conditions studied
in section 5.4.

Another important result was that, for a massless 
spin-${1\over 2}$ field subject to non-local boundary
conditions of the form described in section 6.1, one finds
again the conformal anomaly (10.3.5) when the same flat
background with boundary is studied. This result was
obtained in D'Eath and Esposito 1991b by using the Laplace
transform of the heat equation and, independently, by
Kamenshchik and Mishakov (1992), who used instead the powerful
analytic technique derived by Barvinsky et al. (1992).
The result was non-trivial because no geometric formulae for
heat-kernel asymptotics with non-local boundary conditions
were available at that time in the literature. Thus, analytic
methods for $\zeta(0)$ calculations were the only way to
investigate such issues.

The current developments in one-loop quantum cosmology deal
instead with mixed boundary conditions for gauge fields and
gravitation, and with different quantization techniques
for such fields. Since many topics are already studied in
detail by Esposito et al. (1994a--b, 1995a--b, 1997), we
prefer to focus, in the following section, on open problems.
As one might expect, the main source of fascinating open
problems lies in the attempts to quantize the gravitational 
field.
\vskip 1cm
\centerline {\bf 10.4 Conformal Gauges and Manifolds
with Boundary}
\vskip 1cm
The recent attempts to quantize Euclidean Maxwell theory in quantum
cosmological backgrounds have led to a detailed investigation of the
quantized Maxwell field in covariant and non-covariant gauges on
manifolds with boundary (Esposito 1994a--b, Esposito and Kamenshchik 1994,
Vassilevich 1995a-b, Esposito et al. 1997). The main emphasis
has been on the use of
analytic or geometric techniques to evaluate the one-loop semiclassical
approximation of the wave function of the universe, when magnetic
or electric boundary conditions are imposed. In the former case
one sets to zero at the boundary the tangential components $A_{k}$
of the potential (the background value of $A_{\mu}$ is taken to
vanish), the real-valued ghost fields $\omega$ and $\psi$
(or, equivalently, a complex-valued ghost zero-form $\varepsilon$),
and the gauge-averaging functional $\Phi(A)$:
$$
\Bigr[A_{k}\Bigr]_{\partial M}=0 
\; \; \; \; ,
\eqno (10.4.1)
$$
$$
[\varepsilon]_{\partial M}=0 
\; \; \; \; ,
\eqno (10.4.2)
$$
$$
\Bigr[\Phi(A)\Bigr]_{\partial M}=0 
\; \; \; \; .
\eqno (10.4.3)
$$
In the electric scheme one sets instead to zero at the boundary
the normal component of $A_{\mu}$, jointly with the normal
derivative of the ghost and the normal derivative of $A_{k}$:
$$
\Bigr[A_{0}\Bigr]_{\partial M}=0 
\; \; \; \; ,
\eqno (10.4.4)
$$
$$
\Bigr[{\partial \varepsilon}/ {\partial n}\Bigr]_{\partial M}=0 
\; \; \; \; ,
\eqno (10.4.5)
$$
$$
\Bigr[{\partial A_{k}}/ {\partial n}\Bigr]_{\partial M}=0 
\; \; \; \; .
\eqno (10.4.6)
$$
The boundary conditions (10.4.1)--(10.4.3) and
(10.4.4)--(10.4.6) are found to be invariant under 
infinitesimal gauge transformations
on $A_{\mu}$, as well as under BRST transformations 
(Esposito et al. 1997). 

On the other hand, the gauge-averaging functionals studied in
Esposito 1994, Esposito and Kamenshchik 1994, Vassilevich 
1995a--b, Esposito et al. 1997,
were not conformally invariant, although a conformally
invariant choice of gauge was already known, at the {\it classical}
level, from the work of Eastwood and Singer (1985). 
It has been therefore our aim
to investigate the {\it quantum} counterpart of the conformally
invariant scheme proposed in 
Eastwood and Singer 1985, to complete the current
work on quantized gauge fields. For this purpose, we have
studied a portion of flat Euclidean 
four-space bounded by three-dimensional
surfaces. The vanishing curvature of 
the four-dimensional background is
helpful to obtain a preliminary understanding of the quantum
operators, which will be shown to have highly nontrivial properties.
In our scheme, all curvature effects result from the boundary only.

In flat Euclidean four-space, the conformally invariant gauge
proposed in Eastwood and Singer 1985 reads (hereafter $b,c=0,1,2,3$)
$$
\nabla_{b}\nabla^{b}\nabla^{c}A_{c}
=\cstok{\ } \nabla^{c}A_{c}=0 
\; \; \; \; .
\eqno (10.4.7)
$$
If the classical potential is subject to an infinitesimal
gauge transformation
$$
{ }^{f}A_{b}=A_{b}+\nabla_{b}f 
\; \; \; \; ,
\eqno (10.4.8)
$$
the gauge condition (10.4.7) is satisfied by ${ }^{f}A_{b}$ if
and only if $f$ obeys the fourth-order equation
$$
{\cstok{\ }}^{2}f=0 
\; \; \; \; ,
\eqno (10.4.9)
$$
where ${\cstok{\ }}^{2}$ is the $\cstok{\ }$ operator composed
with itself, i.e.
${\cstok{\ }}^{2} \equiv g^{ab}g^{cd}\nabla_{a}\nabla_{b}
\nabla_{c}\nabla_{d}$. 

In the quantum theory via path integrals,
however, one performs Gaussian averages over gauge functionals
$\Phi(A)$ which ensure that well-defined Feynman Green's
functions for the $\cal P$ operator on $A_{b}$, and for the
ghost operator, actually exist (DeWitt 1981, Esposito
et al. 1997). This means that the
left-hand side of Eq. (10.4.7) is no longer set to zero. One defines
instead a gauge-averaging functional
$$
\Phi(A) \equiv \cstok{\ } \nabla^{b}A_{b} 
\; \; \; \; ,
\eqno (10.4.10)
$$
and the gauge-averaging term ${\beta \over 2\alpha}
[\Phi(A)]^{2}$, with $\beta$ a dimensionful and 
$\alpha$ a dimensionless parameter,
is added to the Maxwell Lagrangian ${1\over 4}F_{ab}F^{ab}$.
A double integration by parts is then necessary to express 
the Euclidean Lagrangian in the form
${1\over 2}A_{b}{\cal P}^{bc}A_{c}$, where
$$
{\cal P}^{bc} \equiv -g^{bc} \cstok{\ }
+\biggr(1-{\beta \over \alpha}{\cstok{\ }}^{2}\biggr)
\nabla^{b}\nabla^{c} 
\; \; \; \; .
\eqno (10.4.11)
$$
The operator ${\cal P}^{bc}$ is a complicated sixth-order
elliptic operator, and it is unclear how to deal properly with
it for finite values of $\alpha$. However, in the limit
as $\alpha \rightarrow \infty$, it reduces to the following
second-order operator:
$$
P^{bc}=-g^{bc}\cstok{\ }+\nabla^{b}\nabla^{c} 
\; \; \; \; .
\eqno (10.4.12)
$$
This operator remains non-minimal, since the
term $\nabla^{b}\nabla^{c}$ survives.
In this particular case, we still need to specify boundary
conditions on $A_{b}$ and ghost perturbations. For this 
purpose, we put to zero at the boundary the whole set of
$A_{b}$ perturbations:
$$
\Bigr[A_{b}\Bigr]_{\partial M}=0 \; \forall b=0,1,2,3 
\; \; \; \; ,
\eqno (10.4.13)
$$
and we require invariance of Eq. (10.4.13) under infinitesimal gauge
transformations on $A_{b}$. This leads to (hereafter $\tau$
is a radial coordinate (Esposito et al. 1997))
$$
[\varepsilon]_{\partial M}=0 
\; \; \; \; ,
\eqno (10.4.14)
$$
$$
\Bigr[{\partial \varepsilon}/ \partial \tau 
\Bigr]_{\partial M}=0 
\; \; \; \; .
\eqno (10.4.15)
$$
Condition (10.4.14) results from the gauge invariance of the 
Dirichlet condition on $A_{k}$, $\forall k=1,2,3$,
and condition (10.4.15) results from the gauge invariance of the
Dirichlet condition on $A_{0}$. Note that it would be
inconsistent to impose the boundary conditions (10.4.13)--(10.4.15)
when the Lorentz gauge-averaging functional is chosen, since
the corresponding ghost operator is second-order. 

When two boundary three-surfaces occur, 
Eqs. (10.4.14) and (10.4.15) lead to
$$
[\varepsilon]_{\Sigma_{1}}
=[\varepsilon]_{\Sigma_{2}}=0 
\; \; \; \; ,
\eqno (10.4.16)
$$
$$
\Bigr[{\partial \varepsilon}/ \partial \tau 
\Bigr]_{\Sigma_{1}}
=\Bigr[{\partial \varepsilon}/ \partial \tau 
\Bigr]_{\Sigma_{2}}=0 
\; \; \; \; .
\eqno (10.4.17)
$$
When Eq. (10.4.10) is used, and the ghost operator is hence
${\cstok{\ }}^{2}$, the four boundary conditions (10.4.16) and
(10.4.17) provide enough conditions to determine completely
the coefficients $C_{1},...,C_{4}$ in the linear combination
$$
\varepsilon_{(\lambda)}=\sum_{i=1}^{4}C_{i}
\rho_{i_{(\lambda)}} 
\; \; \; \; ,
\eqno (10.4.18)
$$
where $\rho_{1},...,\rho_{4}$ are four linearly independent
solutions of the fourth-order eigenvalue equation
$$
{\cstok{\ }}^{2}\varepsilon_{(\lambda)}=\lambda \;
\varepsilon_{(\lambda)} 
\; \; \; \; .
\eqno (10.4.19)
$$
We therefore find that, when the conformally invariant gauge
functionals (10.4.10) are used, the admissible boundary conditions
differ substantially from the magnetic and electric schemes 
outlined in Eqs. (10.4.1)--(10.4.3) 
and (10.4.4)--(10.4.6), and are conformally invariant
by construction (with the exception of Eq. (10.4.15)). 

Had we set to zero at the boundary 
$A_{k}$ ($k=1,2,3$) and the
functional (10.4.10), we would not have obtained enough boundary
conditions for ghost perturbations, since both choices lead to
Dirichlet conditions on the ghost. The boundary conditions (10.4.13)
are also very important since they ensure the vanishing of all 
boundary terms resulting from integration by parts in the 
Faddeev-Popov action.
In the particular case when
the three-surface $\Sigma_{1}$ shrinks to a point, which is relevant
for (one-loop) quantum cosmology 
(Esposito et al. 1997), the boundary conditions
read (here $\Sigma$ is the bounding three-surface)
$$
\Bigr[A_{b}\Bigr]_{\Sigma}=0 
\; \; \; \; 
\forall b=0,1,2,3 
\; \; \; \; ,
\eqno (10.4.20)
$$
$$
[\varepsilon]_{\Sigma}=0 
\; \; \; \; ,
\eqno (10.4.21)
$$
$$
\Bigr[{\partial \varepsilon}/ 
\partial \tau \Bigr]_{\Sigma}=0 
\; \; \; \; ,
\eqno (10.4.22)
$$
jointly with regularity at $\tau=0$ of $A_{b}, \varepsilon$
and ${\partial \varepsilon \over \partial \tau}$. Many fascinating
problems are now in sight. They are as follows:
\vskip 0.3cm
\noindent
(i) To prove uniqueness of the solution of the classical 
boundary-value problem
$$
{\cstok{\ }}^{2}f=0 
\; \; \; \; ,
\eqno (10.4.23)
$$
$$
[f]_{\Sigma_{1}}= [f]_{\Sigma_{2}}=0 
\; \; \; \; ,
\eqno (10.4.24)
$$
$$
\Bigr[{\partial f} / {\partial \tau} \Bigr]_{\Sigma_{1}}
=\Bigr[{\partial f} / 
{\partial \tau} \Bigr]_{\Sigma_{2}}=0 
\; \; \; \; .
\eqno (10.4.25)
$$
\vskip 0.3cm
\noindent
(ii) To study the quantum theory resulting from the operator
(10.4.11) for finite values of $\alpha$. Interestingly, the
Feynman choice $\alpha=1$ does not get rid of the sixth-order
nature of the operator ${\cal P}^{bc}$.
\vskip 0.3cm
\noindent
(iii) To evaluate the one-loop semiclassical approximation,
at least when ${\cal P}^{bc}$ reduces to the form (10.4.12) 
in the presence of three-sphere boundaries. The ghost operator
is then found to take the form
$$
\eqalignno{
{\cstok{\ }}^{2}&=
{\partial^{4}\over \partial \tau^{4}}
+{6\over \tau}{\partial^{3}\over \partial \tau^{3}}
+{3\over \tau^{2}}{\partial^{2}\over \partial \tau^{2}}
-{3\over \tau^{3}}{\partial \over \partial \tau} \cr 
&+ {2\over \tau^{2}}\biggr({\partial^{2}\over \partial \tau^{2}}
+{1\over \tau}{\partial \over \partial \tau}\biggr)
{\;}_{\mid i}^{\; \; \; \mid i}
+{1\over \tau^{4}}\Bigr(
{\;}_{\mid i}^{\; \; \; \mid i}\Bigr)^{2} 
\; \; \; \; .
&(10.4.26)\cr}
$$
With a standard notation, we denote 
by $\mid$ the operation of covariant 
differentiation tangentially with respect to the 
three-dimensional Levi-Civita connection of the boundary
(see section 5.2).
If one expands the ghost perturbations on a family of
three-spheres centred on the origin as (Esposito 1994a--b)
$$
\varepsilon(x,\tau)=\sum_{n=1}^{\infty}
\varepsilon_{n}(\tau)Q^{(n)}(x) 
\; \; \; \; ,
$$
the operator (10.4.26), jointly with the properties of scalar
harmonics, leads to the eigenvalue equation (cf. Eq. (10.4.19))
$$
\eqalignno{
\; & {d^{4}\varepsilon_{n}\over d\tau^{4}}
+{6\over \tau}{d^{3}\varepsilon_{n}\over d\tau^{3}}
-{(2n^{2}-5)\over \tau^{2}}{d^{2}\varepsilon_{n}
\over d\tau^{2}} \cr 
&- {(2n^{2}+1)\over \tau^{3}}
{d\varepsilon_{n}\over d\tau} 
+\left({(n^{2}-1)^{2}\over \tau^{4}}-\lambda_{n} \right)
\varepsilon_{n}=0 
\; \; \; \; .
&(10.4.27)\cr}
$$
This equation admits a power-series solution in the form
$$
\varepsilon_{n}(\tau)=\tau^{\rho}\sum_{k=0}^{\infty}
b_{n,k}(n,k,\lambda_{n})\tau^{k} 
\; \; \; \; .
\eqno (10.4.28)
$$
The values of $\rho$ are found by solving the fourth-order
algebraic equation
$$
\rho^{4}-2(n^{2}+1)\rho^{2}
+(n^{2}-1)^{2}=0 
\; \; \; \; ,
\eqno (10.4.29)
$$
which has the four real roots $\pm (n \pm 1)$.
Moreover, the only non-vanishing $b_{n,k}$ coefficients are 
of the form $b_{n,4k}$, $\forall k=0,1,2,...$, and are given
by (assuming that $b_{n,0}$ has been fixed)
$$
b_{n,l}={\lambda_{n} \; b_{n,l-4}\over
F(l,n,\rho)} \; , \; \forall l=4,8,12, ... 
\; \; \; \; ,
\eqno (10.4.30)
$$
where we have defined ($\forall k = 0,1,2,...$)
$$
\eqalignno{
F(k,n,\rho)& \equiv (\rho+k)(\rho+k-1) 
(\rho+k-2)(\rho+k-3) \cr
&+ 6(\rho+k)(\rho+k-1)(\rho+k-2) 
-(2n^{2}-5)(\rho+k)(\rho+k-1) \cr 
&- (2n^{2}+1)(\rho+k)
+(n^{2}-1)^{2} 
\; \; \; \; .
&(10.4.31)\cr}
$$
As far as we can see, the solution (10.4.28) can be expressed
in terms of Bessel functions and modified Bessel functions
of order $n$ (cf. section 3.5 of Magnus et al. 1966). 
\vskip 0.3cm
\noindent
(iv) To include the effects of curvature. 
As shown in Eastwood and Singer 1985,
if the background four-geometry is curved, with Riemann tensor
$R_{\; \; bcd}^{a}$, the conformally invariant 
gauge-averaging functional reads (cf. Eq. (10.4.10))
$$
\Phi(A) \equiv  \cstok{\ } \nabla^{b}A_{b} 
+ \nabla_{c} \left[\Bigr(-2R^{bc}+{2\over 3}R g^{bc}\Bigr)
A_{b}\right] 
\; \; \; \; .
\eqno (10.4.32)
$$
It would be interesting to study the (one-loop) quantum theory,
at least when $\alpha \rightarrow \infty$, on curved backgrounds
like $S^{4}$, which is relevant for inflation 
(Esposito et al. 1997), or
$S^{2} \times S^{2}$, which is relevant for the bubbles picture
in Euclidean quantum gravity, as proposed in 
Hawking 1996. 

To our knowledge, the form (10.4.11) of the differential operator
on perturbations of the electromagnetic potential in the
quantum theory, the boundary conditions 
(10.4.13)--(10.4.15), and the
analytic solution (10.4.28)--(10.4.31) for ghost basis functions
are entirely new. Thus, quantization via path integrals in
conformally invariant gauges possesses some new peculiar
properties, which are now under investigation for the
first time. This, in turn,
seems to add evidence in favor of Euclidean quantum
gravity having a deep influence on current developments in
quantum field theory (Esposito et al. 1997).
\vskip 1cm
\centerline {\bf 10.5 Conformally Covariant Operators}
\vskip 1cm
This section, motivated by the previous example of the
conformal gauge for Maxwell theory, is devoted to a brief
review of some key properties of conformally covariant
operators. An important example is provided by the Paneitz
operator. The framework for the introduction of such
operator (Paneitz 1983) is a Riemannian manifold $(M,g)$
of dimension $m \geq 3$ with Riemann curvature $R$,
Ricci curvature $\rho$ and scalar curvature $\tau$. Denoting
by $d$ the exterior derivative, and by $\delta$ its formal
adjoint, the Laplace operator on scalar functions reads
$$
\bigtriangleup \equiv \delta \; d
\; \; \; \; .
\eqno (10.5.1)
$$
Moreover, one can consider (Branson 1996)
$$
J \equiv {\tau \over 2(m-1)}
\; \; \; \; ,
\eqno (10.5.2)
$$
$$
V \equiv {(\rho - J g)\over (m-2)}
\; \; \; \; ,
\eqno (10.5.3)
$$
$$
T \equiv (m-2)J-4V {\cdot}
\; \; \; \; ,
\eqno (10.5.4)
$$
$$
Q \equiv {m\over 2}J^{2}-2{\mid V \mid}^{2}
+ \bigtriangleup J
\; \; \; \; .
\eqno (10.5.5)
$$
The operation $V {\cdot}$ is defined by
$$
\Bigr(V {\cdot} \varphi \Bigr)_{a} \equiv V_{\; a}^{b}
\; \varphi_{b}
\; \; \; \; ,
\eqno (10.5.6)
$$
while ${\mid V \mid}^{2} \equiv V^{cd} V_{cd}$, 
with the indices $a,b,c,d$ 
ranging from 1 through $m$. The Weyl tensor, i.e.
the part of Riemann which is invariant under conformal
rescalings of the metric (and such that all its contractions
vanish) can be therefore expressed as
$$
C_{\; bcd}^{a}=R_{\; bcd}^{a}+V_{bc} \; \delta_{\; d}^{a}
-V_{bd} \; \delta_{\; c}^{a}
+V_{\; d}^{a} \; g_{bc}-V_{\; c}^{a} \; g_{bd}
\; \; \; \; .
\eqno (10.5.7)
$$

The Paneitz operator is, by definition, the following
fourth-order operator:
$$
P \equiv {\bigtriangleup}^{2}+\delta T d
+{(m-4)\over 2}Q
\; \; \; \; .
\eqno (10.5.8)
$$
This operator is {\it conformally covariant} in that, if the
metric is rescaled according to (cf. (5.3.1))
$$
g_{\omega} \equiv e^{2\omega} \; g_{0}
\; \; \; \; ,
\eqno (10.5.9)
$$
with $\omega \in C^{\infty}(M)$, then $P$ rescales as
$$
P_{\omega}=e^{-(m+4)\omega /2} \; P_{0} \; 
\Bigr[e^{(m-4)\omega /2}\Bigr]
\; \; \; \; ,
\eqno (10.5.10)
$$
where square brackets are used to denote a multiplication
operator, $[F]$, for any $F \in C^{\infty}(M)$. Indeed,
there is an analogy with the construction of the
{\it conformal Laplacian}, defined by
$$
Y \equiv \bigtriangleup +{(m-2)\over 2} J
\; \; \; \; ,
\eqno (10.5.11)
$$
which is conformally covariant in that
$$
Y_{\omega}=e^{-(m+2)\omega /2} \; Y_{0} \;
\Bigr[e^{(m-2)\omega /2} \Bigr]
\; \; \; \; .
\eqno (10.5.12)
$$

If $A$ is a formally self-adjoint differential operator
with positive-definite leading symbol, its order can only
be an even number, say $2l > 0$. For a smooth function 
$f \in C^{\infty}(M)$, the $L^{2}$ trace of $f e^{-tA}$ has
an asymptotic expansion as $t \rightarrow 0^{+}$ (cf.
section 5.2)
$$
{\rm Tr}_{L^{2}} \Bigr(f e^{-tA} \Bigr) \sim
\sum_{k=0}^{\infty}t^{(k-m)/2l}
\int_{M} f U_{j}[A] \; dv
\; \; \; \; ,
\eqno (10.5.13)
$$
where $dv$ is the Riemannian measure on $M$, and the
$U_{j}[A]$ are invariants consisting of a polynomial in the
total symbol, whose coefficients are smooth in the leading
symbol (Branson 1996). Since $M$ is here assumed to be a
manifold without boundary, $U_{k}[A]$ vanishes for all odd
values of $k$. The condition of conformal covariance for the
operator $A$ is that, under the conformal rescaling (10.5.9),
$A$ rescales according to the relation
$$
{\overline A}_{\omega}=e^{-b \omega} \; A_{0} \;
[e^{a \omega}] \; \; \; \; ,
\eqno (10.5.14)
$$
for some real parameters $a$ and $b$. 

For the Paneitz operator, the invariant $U_{4}[P]$ is
expressed, after introducing
$$
u(m) \equiv 720(m^{2}-4)(4\pi)^{m/2}
{\Gamma((m-4)/2)\over \Gamma((m-4)/4)}
\; \; \; \; ,
\eqno (10.5.15)
$$
by (Branson 1996)
$$
\eqalignno{
U_{4}[P]&=u^{-1}(m) \left \{ -(m-8)(m+4)(m+2)(m-12)
\biggr[{\bigtriangleup} J +{1\over 2}(m-4)J^{2} \biggr]
\right . \cr
& \left . + (m-8)\Bigr(m^{3}-52m-24 \Bigr)Q
+2(m^{2}-4) {\mid C \mid}^{2} \right \}
\; \; \; \; .
&(10.5.16)\cr}
$$

In general, if $A$ is conformally covariant, or a power 
of such (Eastwood and Rice 1987, Baston and Eastwood 1990,
Graham et al. 1992, Branson 1995), the number of negative
eigenvalues (counted, of course, with their multiplicity),
and the multiplicity of $0$ as an eigenvalue, are conformal
invariants. One can then define the determinant of $A$ by
$$
-\log \mid {\rm det} \; A \mid \equiv \zeta_{A}'(0)
\; \; \; \; ,
\eqno (10.5.17)
$$
with 
$$
{\rm sign} \; {\rm det} \; A = {\rm number} \; {\rm of}
\; \left \{ \lambda_{k}=0 \right \}
\; \; \; \; ,
\eqno (10.5.18)
$$
where $\zeta_{A}(s)$ is the analytic continuation to the
complex plane of 
$$
\sum_{\lambda_{j} \not = 0} {\mid \lambda_{j} \mid}^{-s}
\; \; \; \; .
$$
One can then study ${\rm det} \; A$ as a functional on the
conformal class determined by the rescalings (10.5.9), by
virtue of the relation
$$
\zeta_{A_{\omega}}'(0)-\zeta_{A_{0}}'(0)
=-\log {{\rm det} \; A_{\omega} \over {\rm det} \;
A_{0}} \; \; \; \; .
\eqno (10.5.19)
$$
The problem of finding the extremals of the functional
determinant within a conformal class is now under 
intensive investigation (Branson 1996).

Another important reference on conformal geometry and
conformally covariant operators is the work by Parker
and Rosenberg (1987). More recent developments are described
in the work by Avramidi (1997), where the author has studied
the structure of diagonal singularities of Green functions
of partial differential operators of even order acting on
elements of $C^{\infty}(V,M)$, i.e. smooth sections of a
vector bundle $V$ over a Riemannian 
manifold $M$. In particular, the
work in Avramidi 1997 has investigated operators obtained
by composition of second-order operators of Laplace type
(cf. our section 10.4). Interestingly, singularities of the
corresponding Green functions have been expressed in terms of
heat-kernel coefficients.
\vskip 1cm
\centerline {\bf 10.6 Euclidean Quantum Gravity}
\vskip 1cm
The main physical motivations for studying the problem of a
quantum theory of gravity (as far as the author can see) are
as follows.
\vskip 0.3cm
\noindent
(i) Gravity is responsible for the large-scale structure of
the universe we live in, and Einstein's theory of general
relativity provides a good model of space-time physics, 
satisfying both mathematical elegance and agreement with
observations (Hawking and Ellis 1973, Hawking 1979).
\vskip 0.3cm
\noindent
(ii) The formalism of quantum field theory, despite the lack
of a rigorous mathematical setting, has led to many exciting
developments in the quantum theory of the electromagnetic
field and Yang-Mills fields, leading in turn to new results
in mathematics (Donaldson and Kronheimer 1990, Freed and
Uhlenbeck 1995).
\vskip 0.3cm
\noindent
(iii) The application of perturbative or non-perturbative
methods to the quantization of Einstein's gravity, or of
its supersymmetric version, has shown that formidable 
technical problems always emerge at some stage, possibly
pointing out that all known ideas and methods lead to a
mathematically inconsistent formulation of quantum gravity
(DeWitt 1965, Hawking 1979, van Nieuwenhuizen 1981, 
Ashtekar 1988, Ashtekar 1991, Esposito 1994a, Esposito 1995,
D'Eath 1996, Esposito et al. 1997).

Nevertheless, decades of efforts by thousands of scientists
from all over the world do not seem to have been useless.
For example, one knows that the effective action provides,
in principle, a tool for studying quantum theory as a
theory of small disturbances of the underlying classical 
theory, as well as many non-perturbative properties in
field theory (Jona-Lasinio 1964,
DeWitt 1965, DeWitt 1981, Esposito et al. 
1997 and references therein). The basic object of a 
space-time covariant formulation of quantum gravity may be
viewed as being the path-integral representation of the
$\langle {\rm out} \mid {\rm in} \rangle$ amplitude
(Hawking 1979), which involves the consideration of ghost
fields which reflect the gauge freedom of the classical 
theory (see Esposito et al. 1997 and the many references
therein). In particular, what seems to emerge is that the
consideration of the elliptic boundary-value problems
of quantum gravity sheds new light on the one-loop
semiclassical approximation, which is the ``bridge" in
between the classical world and the as yet unknown (full)
quantum theory (Gibbons and Hawking 1993, Esposito et
al. 1997). We shall thus focus on this part of the quantum
gravity problem, i.e. the boundary conditions on metric
perturbations, when a Riemannian four-manifold with boundary
is considered (this may be a portion of flat Euclidean
four-space, or part of the de Sitter four-sphere, or a 
more general curved background). To begin, let us assume that
spatial components of metric perturbations, say $h_{ij}$,
are set to zero at the boundary: 
$$
\Bigr[h_{ij}\Bigr]_{\partial M}=0
\; \; \; \; .
\eqno (10.6.1)
$$
Of course, this is suggested by what one would do in
linearized theory at the classical level. A basic ingredient
in our analysis is that Eq. (10.6.1) should be preserved
under infinitesimal diffeomorphisms on metric perturbations.
Their action on $h_{ij}$ reads (Esposito et al. 1995b,
Avramidi et al. 1996, Esposito et al. 1997)
$$
{ }^{\varphi}h_{ij}=h_{ij}+\varphi_{(i \mid j)}
+K_{ij} \varphi_{0}
\; \; \; \; ,
\eqno (10.6.2)
$$
where $\varphi_{b}dx^{b}$ is the ghost one-form. It is thus
clear that, if the extrinsic-curvature tensor does not
vanish, a necessary and sufficient condition for the preservation
of the boundary conditions (10.6.1) under the transformations
(10.6.2) is that the following boundary conditions should be
imposed on the normal and tangential components of the 
ghost one-form:
$$
\Bigr[\varphi_{0}\Bigr]_{\partial M}=0
\; \; \; \; ,
\eqno (10.6.3)
$$
$$
\Bigr[\varphi_{i}\Bigr]_{\partial M}=0
\; \; \; \; .
\eqno (10.6.4)
$$
At this stage, the remaining set of boundary conditions on
metric perturbations, whose invariance under infinitesimal
diffeomorphisms 
$$
{ }^{\varphi}h_{ab} \equiv h_{ab}+\nabla_{(a} \;
\varphi_{b)}
\eqno (10.6.5)
$$
is again guaranteed by Eqs. (10.6.3) and (10.6.4), involves
setting to zero at the boundary the gauge-averaging functional,
say $\Phi_{a}(h)$, in the Faddeev-Popov action for Euclidean 
quantum gravity (Esposito et al. 1997). What happens is that,
under the transformations (10.6.5), one finds
$$
\Phi_{a}(h)-\Phi_{a}({ }^{\varphi}h)
={\cal F}_{a}^{\; b} \; \varphi_{b}
\; \; \; \; ,
\eqno (10.6.6)
$$
where ${\cal F}_{a}^{\; b}$ is the ghost operator. If now the
ghost one-form is expanded into a complete set of eigenfunctions
of ${\cal F}_{a}^{\; b}$, say $\left \{ f_{b}^{(\lambda)}
\right \}$, which obey the eigenvalue equations
$$
{\cal F}_{a}^{\; b} \; f_{b}^{(\lambda)}
=\lambda \; f_{a}^{(\lambda)}
\; \; \; \; ,
\eqno (10.6.7)
$$
the boundary conditions
$$
\Bigr[\Phi_{a}(h)\Bigr]_{\partial M}=0
\eqno (10.6.8)
$$
turn out to be invariant under (10.6.5) if and only if
$$
\Bigr[f_{a}^{(\lambda)}\Bigr]_{\partial M}=0 \; \;
\forall a=0,1,2,3
\; \; \; \; ,
\eqno (10.6.9)
$$
which implies in turn that (10.6.3) and (10.6.4) should
hold, bearing in mind that
$$
\varphi_{a}=\sum_{\lambda} C_{\lambda} \; f_{a}^{(\lambda)}
\; \; \; \; ,
\eqno (10.6.10)
$$
where $C_{\lambda}$ are some constant coefficients.

In particular, if a covariant gauge-averaging functional of
the de Donder type is used, i.e.
$$
\Phi_{a}(h) \equiv \nabla^{b} \Bigr(h_{ab}
-{1\over 2}g_{ab} g^{cd}h_{cd} \Bigr)
\; \; \; \; ,
\eqno (10.6.11)
$$
the boundary conditions (10.6.8) lead to normal and tangential
derivatives of the normal components $n^{a}n^{b}h_{ab}$
and $n^{b}h_{ab}$, where $n^{b}$ denotes the inward-pointing
normal to the boundary of $M$. 

In geometric language, one can say that metric perturbations
$h_{ab}$ are sections of the vector bundle $V$ of symmetric
rank-two tensors on $M$. On using the tensor field $q_{ab}$
occurring in Eq. (5.2.9), one can define the projection
operator (Avramidi et al. 1996, Moss and Silva 1997)
$$
\Pi_{ab}^{\; \; \; cd} \equiv q_{\; (a}^{c} \; q_{\; b)}^{d}
\; \; \; \; ,
\eqno (10.6.12)
$$
and the boundary operator corresponding to (10.6.1) and (10.6.8)
may be split as the direct sum (cf. (5.4.2))
$$
{\cal B}={\cal B}_{1} \oplus {\cal B}_{2}
\; \; \; \; ,
\eqno (10.6.13)
$$
where ${\cal B}_{1}$ is proportional to $\Pi_{ab}^{\; \; \; cd}$,
and takes into account the boundary conditions (cf. (10.6.1))
$$
\Bigr[\Pi_{ab}^{\; \; \; cd} \; h_{cd}\Bigr]_{\partial M}=0
\; \; \; \; ,
\eqno (10.6.14)
$$
while ${\cal B}_{2}$ reads (Avramidi et al. 1996, Avramidi
and Esposito 1997)
$$
{\cal B}_{2} \equiv \Bigr(\II - \Pi \Bigr)
\left[H \nabla_{N}+{1\over 2} \Bigr(\Gamma^{i}
{\widehat \nabla}_{i}+{\widehat \nabla}_{i}\Gamma^{i}
\Bigr)+S \right]
\; \; \; \; .
\eqno (10.6.15)
$$
The correct way to interpret ${\widehat \nabla}_{i}\Gamma^{i}$ 
in (10.6.15) is via the Leibniz rule
$$
{\widehat \nabla}_{i}\Gamma^{i} \; h
=\Bigr({\widehat \nabla}_{i}\Gamma^{i}\Bigr)h
+\Gamma^{i}{\widehat \nabla}_{i}h
\; \; \; \; .
$$
With our notation, $H$ is the metric on the bundle $V$, while
$\Gamma^{i}$ and $S$ are endomorphisms of $V$. Their form is
analyzed in detail in Avramidi et al. 1996, Avramidi and 
Esposito 1997. Thus, it is not repeated here, where the emphasis 
is instead put on some qualitative features. First, note that,
if $H$ is replaced by the identity matrix, the operator in
square brackets in (10.6.15) reduces to the form first studied
in McAvity and Osborn 1991b:
$$
{\widetilde {\cal B}} \equiv \nabla_{N}
+{1\over 2}\Bigr(\Gamma^{i}{\widehat \nabla}_{i}
+{\widehat \nabla}_{i}\Gamma^{i}\Bigr)+S
\; \; \; \; .
\eqno (10.6.16)
$$
At this stage, it is already clear that tangential derivatives
lead to a highly non-trivial ``invariance theory" (Gilkey
1995). More precisely, we know from section 5.2, which relies
in turn on Weyl's theory of the invariants of the orthogonal 
group (Weyl 1946, Gilkey 1995), that the integrand in the 
general formulae for interior terms in heat-kernel asymptotics 
is a polynomial which can be found using only tensor products 
and contraction of tensor arguments. Thus, by virtue of 
(10.6.16), it is clear that the integrand in the boundary term
in (5.2.18) should be supplemented by
$$
f \; K_{ij} \; \Gamma^{i} \; \Gamma^{j}
\; \; \; \; ,
$$
since this is the only local invariant which is linear in
the extrinsic curvature and is built from contractions of
$K_{ij}$ with $\Gamma^{i}$. The resulting $a_{1}$ coefficient 
can be denoted, for clarity, by $a_{1}(f,P,{\widetilde
{\cal B}})$, where $P$ is an operator of Laplace type 
(see (5.2.1)).

Similarly, to obtain the form of $a_{3/2}(f,P,{\widetilde
{\cal B}})$ one has to consider, further to the invariants
occurring in (5.2.19), all possible contractions of the
matrices $\Gamma^{i}$ with geometric objects of the form
($K$ being the second fundamental form of the boundary)
$$
fK^{2} \; \; , \; \; fKS \; \; , \; \; 
f{\widehat \nabla}K \; \; , \; \; 
f{\widehat \nabla}S \; \; , \; \; 
fR \; \; , \; \; f \Omega \; \; , \; \;
f_{;N}K \; \; \; \; .
$$
As shown in Avramidi and Esposito 1997, this leads to 11
new invariants in the general formula for 
$a_{3/2}(f,P,{\widetilde {\cal B}})$. The number of such
invariants is rapidly increasing for higher-order
heat-kernel coefficients, and the same method shows
that 68 new invariants contribute to $a_{2}(f,P,{\widetilde
{\cal B}})$ (Avramidi and Esposito 1997). It should be
stressed that this property only holds if $\Gamma^{i}$ is
covariantly constant with respect to the induced Levi-Civita
connection of the boundary (McAvity and Osborn 1991b, 
Avramidi and Esposito 1997):
$$
{\widehat \nabla}_{i}\Gamma^{j}=0
\; \; \; \; ,
\eqno (10.6.17)
$$
and if $\Gamma^{i}$ commutes with both $S$ and $\Gamma^{2}$:
$$
\Bigr[\Gamma^{i},S \Bigr]=0
\; \; \; \; ,
\eqno (10.6.18)
$$
$$
\Bigr[\Gamma^{i},\Gamma^{2} \Bigr]=0
\; \; \; \; .
\eqno (10.6.19)
$$
Under these assumptions, the boundary operator reduces to
$$
{\widehat {\cal B}} \equiv \nabla_{N}
+\Gamma^{i}{\widehat \nabla}_{i}+S
\; \; \; \; .
\eqno (10.6.20)
$$
Moreover, all the invariants contributing to interior terms 
are weighted by {\it universal functions}, rather than the
universal constants of section 5.2. By this we mean functions
which depend on the local coordinates on the boundary, whose
dependence on $\Gamma^{i}$ is realized by means of functions
of $\Gamma^{2} \equiv \Gamma_{i}\Gamma^{i}$. Hence they are
not affected by conformal rescalings of the metric, and their
contribution to conformal variation formulae is purely algebraic.
This point was not appreciated in McAvity and Osborn 1991b, but
is investigated in detail in Avramidi and Esposito 1997. 
Nevertheless, many unsolved problems remain. They are as follows.
\vskip 0.3cm
\noindent
(i) The functorial method is, by itself, unable to determine
all universal functions. Since tangential derivatives occur
in the boundary operator (10.6.20), the universal functions 
obey an involved set of equations.
What is lacking at present is a systematic
algorithm for the generation of all such equations. This
achievement, jointly with the conformal variation formulae in
Avramidi and Esposito 1997, should lead in turn to a complete
understanding of heat-kernel asymptotics with the generalized
boundary operator (10.6.20).
\vskip 0.3cm
\noindent
(ii) Euclidean quantum gravity, however, needs much more,
since, with Barvinsky boundary conditions (Barvinsky 1987,
Esposito et al. 1995b, Avramidi et al. 1996, Esposito et
al. 1997), the boundary operator is expressed by the direct
sum (10.6.13), with ${\cal B}_{2}$ given by Eq. (10.6.15).
As shown in Avramidi and Esposito 1997, the condition
(10.6.19) no longer holds in this case, and hence an infinite
number of universal functions seems to contribute to the
heat-kernel coefficients $a_{k/2}(f,P,{\cal B})$, for all
$k \geq 2$.
\vskip 0.3cm
\noindent
(iii) Barvinsky boundary conditions turn out to be a particular
case of the most general set of BRST-invariant boundary
conditions, derived recently by Moss and Silva (1997). The
heat-kernel asymptotics corresponding to Moss-Silva boundary
conditions remains unknown. In particular, it is unclear
whether the existence of infinitely many universal functions
is a generic property of Euclidean quantum gravity on 
manifolds with boundary.
\vskip 0.3cm
\noindent
(iv) At the mathematical level, there is also the non-trivial
problem of heat-content asymptotics with the boundary operators
(10.6.13), or (10.6.16) or (10.6.20). Within this framework,
for a given smooth vector bundle $V$ over $M$, with dual bundle
$V^{*}$, denoting by $\langle {\cdot}, {\cdot} \rangle$ the
natural pairing between $V$ and $V^{*}$, with $f_{1} \in
C^{\infty}(V)$ and $f_{2} \in C^{\infty}(V^{*})$, one defines
(see (5.2.1))
$$
\gamma(f_{1},f_{2},P,B)(t) \equiv \langle e^{-tP}f_{1},
f_{2} \rangle_{L^{2}}
\; \; \; \; ,
\eqno (10.6.21)
$$
where $B$ is the boundary operator. Following van den Berg
and Gilkey 1994 and Gilkey 1995, one knows that, as
$t \rightarrow 0^{+}$, $\gamma$ has an asymptotic expansion
$$
\gamma(f_{1},f_{2},P,B) \sim \sum_{k=0}^{\infty}
t^{k/2} \; \gamma_{k}(f_{1},f_{2},P,B)
\; \; \; \; .
\eqno (10.6.22)
$$
For Dirichlet or Neumann boundary conditions, there exist local
invariants such that
$$
\gamma_{k}(f_{1},f_{2},P,B)=\gamma_{k}^{(int)}(f_{1},f_{2},P)[M]
+\gamma_{k}^{(bd)}(f_{1},f_{2},P,B)[\partial M]
\; \; \; \; .
\eqno (10.6.23)
$$
This corresponds to the (physical) problem of finding the
asymptotic expansion of the total heat content of $M$ for a
given initial temperature. However, the invariants contributing
to Eq. (10.6.23) when the boundary operator $B$ coincides with
(10.6.13), or (10.6.16), or (10.6.20) or with the Moss-Silva
boundary operators, have not yet been studied.

From the point of view of theoretical physics, these 
investigations remain very valuable, since they show that,
upon viewing quantum field theory as a theory of small
disturbances of the underlying classical theory (DeWitt 1965,
DeWitt 1981, Esposito et al. 1997), the problem of finding 
the first non-trivial corrections to classical field theory
is a problem in invariance theory, and the result may be
expressed in purely geometric terms. Whether this will be
enough to begin a (new) series of substantial achievements in
quantum gravity, is a fascinating problem for the years
to come. 
\vskip 1cm
\centerline {\bf 10.A Appendix}
\vskip 1cm
The structure of the $\zeta(0)$ calculation resulting from
the eigenvalue condition (10.3.4) is so interesting that we
find it appropriate, for completeness, to present a brief
outline in this appendix, although a more extensive 
treatment can be found in D'Eath and Esposito 1991a,
Esposito 1994a, Esposito et al. 1997. 

The function $F$ occurring in Eq. (10.3.4) 
can be expressed in terms of its zeros 
$\mu_{i}$ as ($\gamma$ being a constant)
$$
F(z)=\gamma \; z^{2(n+1)} \prod_{i=1}^{\infty}
\biggr(1-{z^{2}\over \mu_{i}^{2}}\biggr)
\; \; \; \; .
\eqno (10.A.1)
$$
Thus, setting $m \equiv n+2$, one finds 
$$
J_{m-1}^{2}(x)-J_{m}^{2}(x)={J_{m}'}^{2}
+\biggr({m^{2}\over x^{2}}-1\biggr)J_{m}^{2}
+2{m\over x}J_{m} J_{m}' 
\; \; \; \; .
\eqno (10.A.2)
$$
Thus, on making the analytic continuation 
$x \rightarrow ix$ and then defining 
$\alpha_{m} \equiv \sqrt{m^{2}+x^{2}}$, one obtains 
$$ \eqalignno{
\log \biggr[(ix)^{-2(m-1)}\Bigr(J_{m-1}^{2}-J_{m}^{2}\Bigr)
(ix)\biggr] 
& \sim -\log(2\pi) + \log(\alpha_{m})+2\alpha_{m} \cr
&-2m \log(m+\alpha_{m})+\log({\widetilde \Sigma}) \; .
&(10.A.3)\cr}
$$
In the asymptotic expansion (10.A.3), 
$\log({\widetilde \Sigma})$ admits
an asymptotic series in the form
$$
\log ({\widetilde \Sigma}) \sim
\biggr[\log(c_{0})+{A_{1}\over \alpha_{m}}
+{A_{2}\over \alpha_{m}^{2}}
+{A_{3}\over \alpha_{m}^{3}}
+... \biggr]
\; \; \; \; ,
\eqno (10.A.4)
$$
where, on using the Debye polynomials for uniform 
asymptotic expansions of Bessel functions (Olver 1954),
one finds (hereafter, $t \equiv {m\over \alpha_{m}}$)
$$
c_{0}=2(1+t)
\; \; \; \; ,
\eqno (10.A.5)
$$
$$
A_{1}=\sum_{r=0}^{2}k_{1r} t^{r}
\; \; \; \; , \; \; \; \;
A_{2}=\sum_{r=0}^{4}k_{2r}t^{r}
\; \; \; \; , \; \; \; \;
A_{3}=\sum_{r=0}^{6}k_{3r}t^{r}
\; \; \; \; ,
\eqno (10.A.6)
$$
where 
$$
k_{10}=-{1\over 4}
\; \; \; \; k_{11}=0 \; \; \; \; k_{12}={1\over 12}
\; \; \; \; ,
\eqno (10.A.7)
$$
$$
k_{20}=0 
\; \; \; \; k_{21}=-{1\over 8} \; \; \; \; 
k_{22}=k_{23}={1\over 8} \; \; \; \;
k_{24}=-{1\over 8}
\; \; \; \; ,
\eqno (10.A.8)
$$
$$
k_{30}={5\over 192}
\; \; \; \; k_{31}=-{1\over 8} \; \; \; \;
k_{32}={9\over 320} \; \; \; \; 
k_{33}={1\over 2}
\; \; \; \; ,
\eqno (10.A.9)
$$
$$
k_{34}=-{23\over 64} 
\; \; \; \; k_{35}=-{3\over 8} \; \; \; \;
k_{36}={179\over 576}
\; \; \; \; .
\eqno (10.A.10)
$$
The corresponding $\zeta$-function at large $x$ (Moss 1989) 
has a uniform asymptotic expansion given by
$$
\Gamma(3)\zeta(3,x^{2}) \sim W_{\infty}
+\sum_{n=5}^{\infty}{\hat q}_{n} x^{-2-n}
\; \; \; \; ,
\eqno (10.A.11)
$$
where, defining
$$
S_{1}(m,\alpha_{m}(x)) \equiv -\log(\pi)+2\alpha_{m}
\; \; \; \; ,
\eqno (10.A.12)
$$
$$
S_{2}(m,\alpha_{m}(x)) \equiv -(2m-1)\log(m+\alpha_{m})
\; \; \; \; ,
\eqno (10.A.13)
$$
$$
S_{3}(m,\alpha_{m}(x)) \equiv \sum_{r=0}^{2}k_{1r} \; m^{r}
\alpha_{m}^{-r-1}
\; \; \; \; ,
\eqno (10.A.14)
$$
$$
S_{4}(m,\alpha_{m}(x)) \equiv \sum_{r=0}^{4}
k_{2r} \; m^{r} \alpha_{m}^{-r-2}
\; \; \; \; ,
\eqno (10.A.15)
$$
$$
S_{5}(m,\alpha_{m}(x)) \equiv \sum_{r=0}^{6}
k_{3r} \; m^{r} \alpha_{m}^{-r-3}
\; \; \; \; ,
\eqno (10.A.16)
$$
$W_{\infty}$ can be obtained as 
$$
W_{\infty}=\sum_{m=0}^{\infty}\Bigr(m^{2}-m\Bigr)
{\biggr({1\over 2x}{d\over dx}\biggr)}^{3}
\left[\sum_{i=1}^{5}S_{i}(m,\alpha_{m}(x))\right]
\; \; \; \; .
\eqno (10.A.17)
$$
The resulting $\zeta(0)$ value receives contributions from
$S_{2},S_{4}$ and $S_{5}$ only, and is given by 
$$
\zeta(0)=-{1\over 120}+{1\over 24}+{1\over 2}\sum_{r=0}^{4}k_{2r}
-{1\over 2}\sum_{r=0}^{6}k_{3r}={11\over 360}
\; \; \; \; .
\eqno (10.A.18)
$$
Of course, for a massless Dirac field, the full $\zeta(0)$ is
twice the value in (10.A.18): 
$$
\zeta_{{\rm Dirac}}(0)={11\over 180}
\; \; \; \; .
\eqno (10.A.19)
$$
Remarkably, the same $\zeta(0)$ values are found for a massless
spin-1/2 field in a four-sphere background bounded by a 
three-sphere. The detailed derivation of this property can be 
found in Kamenshchik and Mishakov 1993, jointly with the
analysis in the massive case.
\vskip 100cm
\parindent=0pt
\everypar{\hangindent=20pt \hangafter=1}
\centerline {\bf BIBLIOGRAPHY}
\vskip 1cm

[1] Aichelburg P. C. and Urbantke H. K. (1981) Necessary
and Sufficient Conditions for Trivial Solutions in
Supergravity, {\it Gen. Rel. Grav.} {\bf 13}, 817--828.

[2] Alexandrov S. and Vassilevich D. V. (1996) Heat Kernel
for Non-Minimal Operators on a K\"{a}hler Manifold,
{\it J. Math. Phys.} {\bf 37}, 5715--5718.

[3] Alvarez-Gaum\'e L. (1983a) Supersymmetry and The 
Atiyah-Singer Index Theorem, {\it Commun. Math. Phys.}
{\bf 90}, 161--173.

[4] Alvarez-Gaum\'e L. (1983b) A Note on the Atiyah-Singer
Index Theorem, {\it J. Phys.} {\bf A 16}, 4177--4182.

[5] Amsterdamski P. , Berkin A. and O'Connor D. (1989)
$b_{8}$ Hamidew Coefficient for a Scalar Field,
{\it Class. Quantum Grav.} {\bf 6}, 1981--1991.

[6] Ashtekar A. (1988) {\it New Perspectives in Canonical
Gravity} (Naples: Bibliopolis).

[7] Ashtekar A. (1991) {\it Lectures on Non-Perturbative 
Canonical Gravity} (Singapore: World Scientific).

[8] Atiyah M. F. and Singer I. M. (1963) The Index of 
Elliptic Operators on Compact Manifolds, {\it Bull. Amer.
Math. Soc.} {\bf 69}, 422--433.

[9] Atiyah M. F. and Bott R. (1964) On the Periodicity 
Theorem for Complex Vector Bundles, {\it Acta Math.}
{\bf 112}, 229--247.

[10] Atiyah M. F. and Bott R. H. (1965) The Index Problem
for Manifolds with Boundary, in {\it Differential Analysis},
papers presented at the Bombay Colloquium 1964 (Oxford:
Oxford University Press) 175--186.

[11] Atiyah M. F. (1966) Global Aspects of the Theory of
Elliptic Differential Operators, in {\it Proc. Int. Congress
of Mathematicians}, Moscow, 57--64.

[12] Atiyah M. F. (1967) {\it K-Theory} (New York: Benjamin).

[13] Atiyah M. F. and Singer I. M. (1968a) The Index of
Elliptic Operators: I, {\it Ann. of Math.} 
{\bf 87}, 484--530.

[14] Atiyah M. F. and Segal G. B. (1968) The Index of
Elliptic Operators: II, {\it Ann. of Math.} 
{\bf 87}, 531--545.

[15] Atiyah M. F. and Singer I. M. (1968b) The Index of
Elliptic Operators: III, {\it Ann. of Math.}
{\bf 87}, 546--604.

[16] Atiyah M. F. and Singer I. M. (1971a) The Index of
Elliptic Operators: IV, {\it Ann. of Math.}
{\bf 93}, 119--138.

[17] Atiyah M. F. and Singer I. M. (1971b) The Index of
Elliptic Operators: V, {\it Ann. of Math.}
{\bf 93}, 139--149.

[18] Atiyah M. F. and Patodi V. K. (1973) On the Heat
Equation and the Index Theorem, {\it Inventiones Math.}
{\bf 19}, 279--330 [Erratum in {\it Inventiones Math.}
{\bf 28}, 277--280].

[19] Atiyah M. F. (1975a) Classical Groups and Classical
Differential Operators on Manifolds, in {\it Differential
Operators on Manifolds}, CIME, Varenna 1975
(Roma: Edizione Cremonese) 5--48.

[20] Atiyah M. F. (1975b) Eigenvalues and Riemannian Geometry,
in {\it Proc. Int. Conf. on Manifolds and Related Topics
in Topology} (Tokyo: University of Tokyo Press) 5--9.

[21] Atiyah M. F. , Patodi V. K. and Singer I. M. (1975) 
Spectral Asymmetry and Riemannian Geometry I, {\it Math. Proc.
Camb. Phil. Soc.} {\bf 77}, 43--69.

[22] Atiyah M. F. , Patodi V. K. and Singer I. M. (1976)
Spectral Asymmetry and Riemannian Geometry III,
{\it Math. Proc. Camb. Phil. Soc.} {\bf 79}, 71--99. 

[23] Atiyah M. F. (1984) Anomalies and Index Theory, in
{\it Lecture Notes in Physics}, Vol. {\bf 208}
(Berlin: Springer-Verlag) 313--322.

[24] Atiyah M. F. and Singer I. M. (1984) Dirac Operators
Coupled to Vector Potentials, {\it Proc. Nat. Acad. of Sci.
USA} {\bf 81}, 2597--2600.

[25] Atiyah M. F. (1985) Eigenvalues of the Dirac Operator,
in {\it Proc. of 25th Mathematics Arbeitstagung, Bonn
1984, Lecture Notes in Mathematics} (Berlin: Springer-Verlag)
251--260.

[26] Avis S. J. and Isham C. J. (1980) 
Generalized Spin-Structures
on Four-Dimensional Space-Times, {\it Commun. Math. Phys.}
{\bf 72}, 103--118.

[27] Avramidi I. G. (1989) Background Field Calculations in
Quantum Field Theory (Vacuum Polarization), {\it Theor.
Math. Phys.} {\bf 79}, 494--502.

[28] Avramidi I. G. (1990a) The Non-Local Structure of
One-Loop Effective Action via Partial Summation of
Asymptotic Expansion, {\it Phys. Lett.} 
{\bf B 236}, 443--449.

[29] Avramidi I. G. (1990b) The Covariant Technique for
Calculation of the Heat-Kernel Asymptotic Expansion,
{\it Phys. Lett.} {\bf B 238}, 92--97.

[30] Avramidi I. G. (1991) A Covariant Technique for the
Calculation of the One-Loop Effective Action, {\it Nucl.
Phys.} {\bf B 355}, 712--754.

[31] Avramidi I. G. , Esposito G. and Kamenshchik A. Yu. (1996)
Boundary Operators in Euclidean Quantum Gravity,
{\it Class. Quantum Grav.} {\bf 13}, 2361--2373.

[32] Avramidi I. G. and Esposito G. (1997) Heat-Kernel
Asymptotics with Generalized Boundary Conditions
(hep-th/9701018).

[33] Avramidi I. G. (1997) Singularities of Green Functions
of the Products of the Laplace Type Operators
(hep-th/9703005).

[34] Barvinsky A. O. (1987) The Wave Function and the Effective 
Action in Quantum Cosmology: Covariant Loop Expansion,
{\it Phys. Lett.} {\bf B 195}, 344--348.

[35] Barvinsky A. O. , Kamenshchik A. Yu. and Karmazin I. P. (1992)
One-Loop Quantum Cosmology: $\zeta$-Function Technique for the
Hartle-Hawking Wave Function of the Universe,
{\it Ann. Phys.} {\bf 219}, 201--242.

[36] Baston R. J. and Eastwood M. (1990) Invariant Operators,
in {\it Twistors in Mathematics and Physics}, L.M.S. 
Lecture Notes {\bf 156}, eds. T. N. Bailey and R. J.
Baston (Cambridge: Cambridge University Press).

[37] B\'erard P. H. (1986) {\it Spectral Geometry: Direct
and Inverse Problems}, Lecture Notes in Mathematics,
Vol. 1207 (Berlin: Springer-Verlag).

[38] Berline N. , Getzler E. and Vergne M. (1992)
{\it Heat Kernels and Dirac Operators} (Berlin:
Springer-Verlag).

[39] Booss B. and Bleecker D. D. (1985) {\it Topology
and Analysis: The Atiyah-Singer Index Formula and
Gauge-Theoretic Physics} (Berlin: Springer-Verlag).

[40] Booss-Bavnbek B. and Wojciechowski K. P. (1993)
{\it Elliptic Boundary Problems for Dirac Operators}
(Boston: Birkh\"{a}user).

[41] Bott R. (1959) The Stable Homotopy of the Classical
Groups, {\it Ann. of Math.} {\bf 70}, 313--337.

[42] Boyer T. H. (1968) Quantum Electromagnetic Zero-Point
Energy of a Conducting Spherical Shell and the Casimir
Model for a Charged Particle, {\it Phys. Rev.}
{\bf 174}, 1764--1776.

[43] Branson T. P. , Gilkey P. B. and Orsted B. (1990)
Leading Terms in the Heat Invariants, {\it Proc. Amer.
Math. Soc.} {\bf 109}, 437--450.

[44] Branson T. P. and Gilkey P. B. (1990) The Asymptotics of the 
Laplacian on a Manifold with Boundary,
{\it Commun. in Partial Differential
Equations} {\bf 15}, 245--272.

[45] Branson T. P. and Gilkey P. B. (1992a) Residues of
the $\eta$-Function for an Operator of Dirac Type,
{\it J. Funct. Anal.} {\bf 108}, 47--87.

[46] Branson T. P. and Gilkey P. B. (1992b) Residues of
the $\eta$-Function for an Operator of Dirac Type with
Local Boundary Conditions, {\it Diff. Geom. Appl.}
{\bf 2}, 249--267.

[47] Branson T. P. (1995) Sharp Inequalities, the Functional
Determinant, and the Complementary Series, {\it Trans.
A. M. S.} {\bf 347}, 3671--3742.

[48] Branson T. P. (1996) An Anomaly Associated with
Four-Dimensional Quantum Gravity, {\it Commun. Math. Phys.}
{\bf 178}, 301--309.

[49] Branson T. P. , Gilkey P. B. and Vassilevich D. V.
(1997) Vacuum Expectation Value Asymptotics for Second-Order
Differential Operators on Manifolds with Boundary
(hep-th/9702178).

[50] Breitenlohner P. and Freedman D. Z. (1982) Stability in Gauged
Extended Supergravity, {\it Ann. Phys.} {\bf 144}, 249--281.

[51] Camporesi R. and Higuchi A. (1996) On the Eigenfunctions 
of the Dirac Operator on Spheres and Real Hyperbolic Spaces,
{\it J. Geom. Phys.} {\bf 20}, 1--18.

[52] Carow-Watamura U. and Watamura S. (1997) Chirality and
Dirac Operator on Non-Commutative Sphere, {\it Commun.
Math. Phys.} {\bf 183}, 365--382.

[53] Carslaw H. S. and Jaeger J. C. (1959) {\it Conduction
of Heat in Solids} (Oxford: Clarendon Press).

[54] Casimir H. B. G. (1948) On the Attraction Between Two
Perfectly Conducting Plates, {\it Proc. Kon. Ned. Akad.
Wetenschap.} , ser. B, {\bf 51}, 793--795.

[55] Chern S. S. (1944) A Simple Intrinsic Proof of the
Gauss-Bonnet Formula for Closed Riemannian Manifolds,
{\it Ann. of Math.} {\bf 45}, 741--752.

[56] Chern S. S. (1945) On the Curvatura Integra in a
Riemannian Manifold, {\it Ann. of Math.} 
{\bf 46}, 674--684.

[57] Chern S. S. (1979) {\it Com\-plex Ma\-ni\-folds Wi\-tho\-ut
Po\-ten\-tial The\-o\-ry} (Berlin: Springer-Verlag).

[58] Chernoff P. R. (1977) Schr\"{o}dinger and Dirac Operators
with Singular Potentials and Hyperbolic Equations,
{\it Pacific J. Math.} {\bf 72}, 361--382.

[59] Chodos A. , Jaffe R. L. , Johnson K. , Thorn C. B.
and Weisskopf V. F. (1974) New Extended Model of Hadrons,
{\it Phys. Rev.} {\bf D 9}, 3471--3495.

[60] Connes A. (1995) Non-Commutative Geometry and Physics, in
{\it Gravitation and Quantizations, Les Houches Session 
LVII}, eds. B. Julia and J. Zinn-Justin (Amsterdam: Elsevier)
805--950.

[61] D'Eath P. D. (1984) Canonical Quantization of Supergravity,
{\it Phys. Rev.} {\bf D 29}, 2199--2219.

[62] D'Eath P. D. and Halliwell J. J. (1987) Fermions in Quantum
Cosmology, {\it Phys. Rev.} {\bf D 35}, 1100--1123.

[63] D'Eath P. D. and Esposito G. (1991a) Local Boundary Conditions
for the Dirac Operator and One-Loop Quantum Cosmology,
{\it Phys. Rev.} {\bf D 43}, 3234--3248.

[64] D'Eath P. D. and Esposito G. (1991b) Spectral Boundary Conditions
in One-Loop Quantum Cosmology,
{\it Phys. Rev.} {\bf D 44}, 1713--1721.

[65] D'Eath P. D. (1996) {\it Supersymmetric Quantum Cosmology}
(Cambridge: Cambridge University Press).

[66] Deser S. and Zumino B. (1976) Consistent Supergravity,
{\it Phys. Lett.} {\bf 62 B}, 335--337.

[67] DeWitt B. S. (1965) {\it Dynamical Theory of Groups and Fields}
(New York: Gordon and Breach).

[68] DeWitt B. S. (1981) A Gauge-Invariant Effective Action, in
{\it Quantum Gravity 2, A Second Oxford Symposium}, eds. C. J.
Isham, R. Penrose and D. W. Sciama (Oxford: Clarendon Press)
449--487.

[69] Dirac P. A. M. (1928) The Quantum Theory of the Electron, 
{\it Proc. R. Soc. Lond.} {\bf A 117}, 610--624.

[70] Dirac P. A. M. (1958) {\it The Prin\-ci\-ples of Quantum Mechanics}
(Oxford: Cla\-ren\-don Press).

[71] Donaldson S. K. and Kronheimer P. B. (1990) {\it The
Ge\-o\-me\-try of Four-Ma\-ni\-folds} (Oxford: Clarendon Press).

[72] Dowker J. S. and Schofield J. P. (1990) Conformal
Transformations and the Effective Action in the Presence of
Boundaries, {\it J. Math. Phys.} {\bf 31}, 808--818.

[73] Dowker J. S. (1996) Spin on the Four-Ball, {\it Phys. Lett.}
{\bf B 366}, 89--94.

[74] Dowker J. S. , Apps J. S. , Kirsten K. and Bordag M.
(1996) Spectral Invariants for the Dirac Equation on the
d-Ball with Various Boundary Conditions, {\it Class.
Quantum Grav.} {\bf 13}, 2911--2920.

[75] Eastwood M. and Singer I. M. (1985) A Conformally
Invariant Maxwell Gauge, {\it Phys. Lett.} 
{\bf A 107}, 73--74.

[76] Eastwood M. and Rice J. (1987) Conformally Invariant
Differential Operators on Minkowski Space and their 
Curved Analogues, {\it Commun. Math. Phys.} {\bf 109},
207--228 [Erratum in {\it Commun. Math. Phys.} 
{\bf 144}, 213].

[77] Eisenhart L. P. (1926) {\it Riemannian Geometry}
(Princeton: Princeton University Press).

[78] Endo R. (1995) Heat Kernel for Spin-${3\over 2}$ 
Rarita-Schwinger Field in General Covariant Gauge,
{\it Class. Quantum Grav.} {\bf 12}, 1157--1164.

[79] Esposito G. (1994a) {\it Quantum Gravity, Quantum Cosmology and
Lorentzian Geometries}, Lecture Notes in Physics, New Series m:
Monographs, Vol. m12, second corrected and enlarged edition 
(Berlin: Springer-Verlag).

[80] Esposito G. (1994b) Gauge-Averaging Functionals for 
Euclidean Maxwell Theory in the Presence of Boundaries,
{\it Class. Quantum Grav.} {\bf 11}, 905--926.

[81] Esposito G. and Pollifrone G. (1994) Spin-Raising Operators
and Spin-${3\over 2}$ Potentials in Quantum Cosmology,
{\it Class. Quantum Grav.} {\bf 11}, 897--903.

[82] Esposito G. , Kamenshchik A. Yu. , Mishakov I. V. and Pollifrone G.
(1994a) Euclidean Maxwell Theory in the Presence of Boundaries. II,
{\it Class. Quantum Grav.} {\bf 11}, 2939--2950.

[83] Esposito G. , Kamenshchik A. Yu. , Mishakov I. V. and Pollifrone G.
(1994b) Gravitons in One-Loop Quantum Cosmology. Correspondence
Between Covariant and Non-Covariant Formalisms, 
{\it Phys. Rev.} {\bf D 50}, 6329--6337.

[84] Esposito G. (1995) {\it Complex General Relativity}, Fundamental
Theories of Physics, Vol. 69 (Dordrecht: Kluwer).

[85] Esposito G. , Gionti G. , Kamenshchik A. Yu. , Mishakov
I. V. and Pollifrone G. (1995) Spin-3/2 Potentials in
Backgrounds with Boundary, {\it Int. J. Mod. Phys.} 
{\bf D 4}, 735--747.

[86] Esposito G. , Kamenshchik A. Yu. , Mishakov I. V. and
Pollifrone G. (1995a) Relativistic Gauge Conditions in
Quantum Cosmology, {\it Phys. Rev.} {\bf D 52}, 2183--2191.

[87] Esposito G. , Kamenshchik A. Yu. , Mishakov I. V. and
Pollifrone G. (1995b) One-Loop Amplitudes in Euclidean 
Quantum Gravity, {\it Phys. Rev.} {\bf D 52}, 3457--3465.

[88] Esposito G. and Kamenshchik A. Yu. (1995) Mixed Boundary
Conditions in Euclidean Quantum Gravity, {\it Class. Quantum
Grav.} {\bf 12}, 2715--2722.

[89] Esposito G. and Kamenshchik A. Yu. (1996) One-Loop
Divergences in Simple Supergravity: Boundary Effects,
{\it Phys. Rev.} {\bf D 54}, 3869--3881.

[90] Esposito G. , Morales-T\'ecotl H. A. and Pimentel L. O.
(1996) Essential Self-Adjointness in One-Loop Quantum
Cosmology, {\it Class. Quantum Grav.} {\bf 13}, 957--963.

[91] Esposito G. and Pollifrone G. (1996) Twistors in
Conformally Flat Einstein Four-Manifolds, 
{\it Int. J. Mod. Phys.} {\bf D 5}, 481--493.

[92] Esposito G. (1996) Local Boundary Conditions in
Quantum Supergravity, {\it Phys. Lett.} 
{\bf B 389}, 510--514.

[93] Esposito G. (1996) Quantized Maxwell Theory in a 
Conformally Invariant Gauge (hep-th/9610017).

[94] Esposito G. (1997) Non-Local Boundary Conditions for
Massless Spin-${1\over 2}$ Fields, {\it Phys. Rev.}
{\bf D 55}, 3886--3888.

[95] Esposito G. , Kamenshchik A. Yu. and Pollifrone G.
(1997) {\it Euclidean Quantum Gravity on Manifolds with
Boundary}, Fundamental Theories of Physics, Vol. 85
(Dordrecht: Kluwer).

[96] Falomir H. , Gamboa Sarav\'\i\ R. E. , Muschietti 
M. A. , Santangelo E. M. and Solomin J. E. (1996) 
Determinants of Dirac Operators with Local Boundary
Conditions, {\it J. Math. Phys.} {\bf 37}, 5805--5819.

[97] Falomir H. , Gamboa Sarav\'\i\ R. E. and Santangelo E. M.
(1996) Dirac Operator on a Disk with Global Boundary 
Conditions (hep-th/9609194).

[98] Falomir H. (1996) Condiciones de Contorno Globales Para 
el Operator de Dirac (hep-th/9612155).

[99] Freed D. S. and Uhlenbeck K. K. (1995) {\it Geometry and
Quantum Field Theory} (American Mathematical Society).

[100] Fursaev D. V. (1994) Spectral Geometry and One-Loop Divergences
on Manifolds with Conical Singularities, {\it Phys. Lett.}
{\bf B 334}, 53--60.

[101] Gibbons G. W. and Hawking S. W. (1993) {\it Euclidean Quantum Gravity}
(Singapore: World Scientific).

[102] Gilkey P. B. (1975) The Spectral Geometry of a Riemannian
Manifold, {\it J. Diff. Geom.} {\bf 10}, 601--618.

[103] Gilkey P. B. (1995) {\it Invariance Theory, the Heat Equation, and
the Atiyah-Singer Index Theorem} 
(Boca-Raton: Chemical Rubber Company).

[104] Graham C. R. , Jenne R. , Mason L. J. and Sparling G. A. J.
(1992) Conformally Invariant Powers of the Laplacian. I: Existence,
{\it J. Lond. Math. Soc.} {\bf 46}, 557--565.

[105] Greiner P. (1971) An Asymptotic Expansion for the Heat Equation, 
{\it Archs. Ration. Mech. Analysis} {\bf 41}, 163--218.

[106] Greub W. , Halperin S. and Vanstone R. (1973) {\it Connections,
Curvature and Cohomology}, Vol. II (New York: Academic).

[107] Grib A. A. , Mamaev S. G. and Mostepanenko V. M. (1994)
{\it Vacuum Quantum Effects in Strong Fields} (St. Petersburg:
Friedmann Laboratory Publishing).

[108] Grubb G. (1996) {\it Functional Calculus for Boundary-Value
Problems} (Boston: Birkhauser).

[109] Hartle J. B. and Hawking S. W. (1983) Wave Function of the Universe, 
{\it Phys. Rev.} {\bf D 28}, 2960--2975.

[110] Hawking S. W. and Ellis G. F. R. (1973) {\it The Large-Scale
Structure of Space-Time} (Cambridge: Cambridge University Press).

[111] Hawking S. W. (1977) Zeta-Function Regularization of Path Integrals
in Curved Space-Time, {\it Commun. Math. Phys.} {\bf 55}, 133--148.

[112] Hawking S. W. (1979) The Path-Integral Approach to Quantum Gravity, 
in {\it General Relativity, an Einstein Centenary
Survey}, eds. S. W. Hawking and W. Israel 
(Cambridge: Cambridge University Press) 746--789.

[113] Hawking S. W. (1982) The Boundary Conditions of the 
Universe, in {\it Pontificiae Academiae Scientiarum 
Scripta Varia} {\bf 48}, 563--574.

[114] Hawking S. W. (1983) The Boundary Conditions for Gauged
Supergravity, {\it Phys. Lett.} {\bf 126 B}, 175--177.

[115] Hawking S. W. (1984) The Quantum State of the Universe, 
{\it Nucl. Phys.} {\bf B 239}, 257--276.

[116] Hawking S. W. (1996) Virtual Black Holes, {\it Phys. Rev}
{\bf D 53}, 3099--3107.

[117] H\"{o}rmander L. (1963) {\it Linear Partial Differential
Operators} (Berlin: Springer-Verlag).

[118] Hortacsu M. , Rotke K. D. and Schroer B. (1980) 
Zero-Energy Eigenstates for the Dirac Boundary Problem,
{\it Nucl. Phys.} {\bf B 171}, 530--542.

[119] Isham C. J. (1978) Spinor Fields in Four-Dimensional
Space-Time, {\it Proc. R. Soc. Lond.} {\bf A 364}, 591--599.

[120] Johnson K. (1975) The M. I. T. Bag Model, 
{\it Acta Phys. Polonica} {\bf B 6}, 865--892.

[121] Jona-Lasinio G. (1964) Relativistic Field Theories with
Symmetry-Breaking Solutions, {\it Nuovo Cimento}
{\bf 34}, 1790--1795.

[122] Kac M. (1966) Can One Hear the Shape of a Drum ? ,
{\it Amer. Math. Monthly} {\bf 73}, 1--23.

[123] Kamenshchik A. Yu. and Mishakov I. V. (1992) 
$\zeta$-Function Technique for Quantum Cosmology: The
Contributions of Matter Fields to the Hartle-Hawking Wave
Function of the Universe, {\it Int. J. Mod. Phys.} 
{\bf A 7}, 3713--3746.

[124] Kamenshchik A. Yu. and Mishakov I. V. (1993) Fermions in
One-Loop Quantum Cosmology,
{\it Phys. Rev.} {\bf D 47}, 1380--1390.

[125] Kamenshchik A. Yu. and Mishakov I. V. (1994) 
Fermions in One-Loop Quantum Cosmology II. The Problem of
Correspondence Between Covariant and Non-Covariant 
Formalisms, {\it Phys. Rev.} {\bf D 49}, 816--824.

[126] Kennedy G. (1978) Boundary Terms in the Schwinger-DeWitt
Expansion: Flat-Space Results, 
{\it J. Phys.} {\bf A 11}, L173--L178.                     

[127] Kirsten K. and Cognola G. (1996) Heat-Kernel Coefficients and
Functional Determinants for Higher-Spin Fields on the
Ball, {\it Class. Quantum Grav.} {\bf 13}, 633--644.

[128] Lamoreaux S. K. (1997) Demonstration of the Casimir Force
in the 0.6 to 6 $\mu$m Range, {\it Phys. Rev. Lett.}
{\bf 78}, 5--8.

[129] Landi G. and Rovelli C. (1996) General Relativity in
Terms of Dirac Eigenvalues (GR-QC 9612034).

[130] Lewandowski J. (1991) Twistor Equation in a Curved
Space-Time, {\it Class. Quantum Grav.} {\bf 8}, L11--L17.

[131] Luckock H. C. and Moss I. G. (1989) The Quantum Geometry of
Random Surfaces and Spinning Membranes, {\it Class. Quantum
Grav.} {\bf 6}, 1993--2027.

[132] Luckock H. C. (1991) Mixed Boundary Conditions in Quantum 
Field Theory, {\it J. Math. Phys.} {\bf 32}, 1755--1766.

[133] Magnus W. , Oberhettinger F. and Soni R. P. (1966)
{\it Formulas and Theorems for the Special Functions of
Mathematical Physics} (Berlin: Springer-Verlag).

[134] Marachevsky V. N. and Vassilevich D. V. (1996) 
A Diffeomorphism-Invariant Eigenvalue Problem for Metric
Perturbations in a Bounded Region, {\it Class. Quantum
Grav.} {\bf 13}, 645--652.

[135] Martellini M. and Reina C. (1985) Some Remarks on the
Index Theorem Approach to Anomalies, 
{\it Ann. Inst. Henri Poincar\'e} {\bf 43}, 443--458.

[136] McAvity D. M. and Osborn H. (1991a) A DeWitt Expansion of the
Heat Kernel for Manifolds with a Boundary,
{\it Class. Quantum Grav.} {\bf 8}, 603--638.

[137] McAvity D. M. and Osborn H. (1991b) Asymptotic Expansion of
the Heat Kernel for Generalized Boundary Conditions,
{\it Class. Quantum Grav.} {\bf 8}, 1445--1454.

[138] Milnor J. W. (1963) Spin-Structures on Manifolds,
{\it Enseign. Math.} {\bf 9}, 198--203.

[139] Milnor J. W. and Stasheff J. D. (1974) {\it Characteristic
Classes} (Princeton: Princeton University Press).

[140] Minakshisundaram S. and Pleijel A. (1949) Some
Properties of the Eigenfunctions of the Laplace Operator 
on Riemannian Manifolds, {\it Can. J. Math.} 
{\bf 1}, 242--256.

[141] Moniz P. (1996) Supersymmetric Quantum Cosmology:
Shaken not Stirred, {\it Int. J. Mod. Phys.}
{\bf A 11}, 4321--4382. 

[142] Moss I. G. (1989) Boundary Terms in the Heat-Kernel Expansion, 
{\it Class. Quantum Grav.} {\bf 6}, 759--765.

[143] Moss I. G. and Dowker J. S. (1989) The Correct $B_{4}$ Coefficient, 
{\it Phys. Lett.} {\bf B 229}, 261--263.

[144] Moss I. G. and Poletti S. (1994) Conformal Anomalies on 
Einstein Spaces with Boundary, {\it Phys. Lett.}
{\bf B 333}, 326--330.

[145] Moss I. G. (1996) {\it Quantum Theory, Black Holes and
Inflation} (New York: John Wiley and Sons).

[146] Moss I. G. and Silva P. J. (1997) BRST Invariant 
Boundary Conditions for Gauge Theories, {\it Phys. Rev.}
{\bf D 55}, 1072--1078.

[147] Olver F. W. J. (1954) On Bessel Functions of Large Order,
{\it Philos. Trans. Roy. Soc. London} {\bf A 247}, 328--368.

[148] Paneitz S. (1983) A Quartic Conformally Covariant 
Differential Operator for Arbitrary Pseudo-Riemannian 
Manifolds (preprint).

[149] Parker T. and Rosenberg S. (1987) Invariants of Conformal
Laplacians, {\it J. Diff. Geom.} {\bf 25}, 199--222.

[150] Penrose R. and Rindler W. (1984) {\it Spinors and Space-Time,
Vol. I: Two-Spinor Calculus and Relativistic Fields} 
(Cambridge: Cambridge University Press).

[151] Penrose R. and Rindler W. (1986) {\it Spinors and Space-Time,
Vol. II: Spinor and Twistor Methods in Space-Time Geometry}
(Cambridge: Cambridge University Press).

[152] Penrose R. (1991) Twistors as Spin-${3\over 2}$ Charges,
in {\it Gravitation and Modern Cosmology}, eds. A. Zichichi,
V. de Sabbata and N. S\'anchez (New York: Plenum Press)
129--137.

[153] Penrose R. (1994) Twistors and the Einstein Equations, in
{\it Twistor Theory}, ed. S. Huggett (New York: Marcel Dekker)
145--157.

[154] Perry M. J. (1984) The Positive Mass Theorem and Black
Holes, in {\it Asymptotic Behaviour of Mass and Space-Time
Geometry}, ed. F. J. Flaherty (Berlin: Springer-Verlag) 31--40.

[155] Piazza P. (1991) K-Theory and Index Theory on Manifolds
with Boundary, Ph.D. Thesis, M. I. T. 

[156] Piazza P. (1993) On the Index of Elliptic Operators on
Manifolds with Boundary, {\it J. Funct. Anal.}
{\bf 117}, 308--359.

[157] Polychronakos A. P. (1987) Boundary Conditions, Vacuum
Quantum Numbers and the Index Theorem, {\it Nucl. Phys.}
{\bf B 283}, 268--294.

[158] Rarita W. and Schwinger J. (1941) On a Theory of Particles
with Half-Integral Spin, {\it Phys. Rev.} {\bf 60}, 61.

[159] Reed M. and Simon B. (1975) {\it Methods of Modern Mathematical
Physics, Vol. II: Fourier Analysis and Self-Adjointness}
(New York: Academic).

[160] Rennie R. (1990) Geometry and Topology of Chiral 
Anomalies in Gauge Theories, {\it Adv. Phys.} 
{\bf 39}, 617--779.

[161] Schleich K. (1985) Semiclassical Wave Function of the Universe
at Small Three-Geometries {\it Phys. Rev.} {\bf D 32}, 1889--1898.

[162] Schwinger J. (1951) On Gauge Invariance and Vacuum Polarization, 
{\it Phys. Rev.} {\bf 82}, 664--679.

[163] Seeley R. T. (1967) Complex Powers of an Elliptic Operator, 
{\it Amer. Math. Soc. Proc. Symp. Pure Math.}
{\bf 10}, 288--307.

[164] Siklos S. T. C. (1985) Lobatchevski Plane Gravitational
Waves, in {\it Galaxies, Axisymmetric Systems and Relativity},
ed. M. A. H. MacCallum (Cambridge: Cambridge University
Press) 247--274.

[165] Spanier E. H. (1966) {\it Algebraic Topology}
(New York: McGraw Hill).

[166] Sparnay M. J. (1958) Measurements of Attractive Forces
Between Flat Plates, {\it Physica} {\bf 24}, 751--764.

[167] Stewart J. M. (1991) {\it Advanced General Relativity}
(Cambridge: Cambridge University Press).

[168] Stewartson K. and Waechter R. T. (1971) On Hearing the Shape
of a Drum: Further Results,
{\it Proc. Camb. Phil. Soc.} {\bf 69}, 353--363.

[169] Townsend P. K. (1977) Cosmological Constant in
Supergravity, {\it Phys. Rev.} {\bf D 15}, 2802--2805.

[170] Vafa C. and Witten E. (1984) Eigenvalue Inequalities
for Fermions in Gauge Theories, {\it Commun. Math. Phys.}
{\bf 95}, 257--276.

[171] van den Berg M. and Gilkey P. B. (1994) Heat Content
Asymptotics of a Riemannian Manifold with Boundary, 
{\it J. Funct. Anal.} {\bf 120}, 48--71.

[172] van Nieuwenhuizen P. and Vermaseren J. A. M. (1976) 
One-Loop Divergences in the Quantum Theory of Supergravity,
{\it Phys. Lett.} {\bf 65 B}, 263--266.

[173] van Nieuwenhuizen P. (1981) Supergravity, {\it Phys. Rep.}
{\bf 68}, 189--398.

[174] Vassilevich D. V. (1995a) Vector Fields on a Disk With
Mixed Boundary Conditions, {\it J. Math. Phys.} 
{\bf 36}, 3174--3182.

[175] Vassilevich D. V. (1995b) QED on a Curved Background and
on Manifolds with Boundary: Unitarity Versus Covariance,
{\it Phys. Rev.} {\bf D 52}, 999--1010.

[176] Waechter R. T. (1972) On Hearing the Shape of a Drum:
an Extension to Higher Dimensions, {\it Proc. Camb. Phil. Soc.}
{\bf 72}, 439.

[177] Ward R. S. and Wells R. O. (1990) {\it Twistor
Geometry and Field Theory} (Cambridge: Cambridge 
University Press).

[178] Weidmann J. (1980) {\it Linear Operators in Hilbert
Spaces} (Berlin: Springer-Verlag).

[179] Weyl H. (1946) {\it The Clas\-si\-cal Groups:
Their Invariants and Representations} (Prin\-ce\-ton:
Prin\-ce\-ton U\-ni\-ver\-si\-ty Pre\-ss).

[180] Wigner E. P. (1959) {\it Group Theory and Its Application
to the Quantum Mechanics of Atomic Spectra}
(New York: Academic).

[181] Witten E. (1982) Constraints on Supersymmetry Breaking, 
{\it Nucl. Phys.} {\bf B 202}, 253--316.

[182] Woodhouse N. M. J. (1985) Real Methods in Twistor Theory,
{\it Class. Quantum Grav.} {\bf 2}, 257--291.

[183] York J. W. (1986) Boundary Terms in the Action Principles of
General Relativity, {\it Found. Phys.} {\bf 16}, 249--257.

\bye